\documentclass[reprint,showpacs,superscriptaddress,english,twocolumn,notitlepage,longbibliography]{revtex4-2}
\usepackage[utf8]{inputenc}
\pdfoutput=1
\usepackage{times}
\usepackage{color}
\usepackage{array}
\usepackage{verbatim}
\usepackage{multirow}
\usepackage{amsmath}
\usepackage{amssymb}
\usepackage{dsfont}
\usepackage{graphicx}
\usepackage{esint}
\usepackage{makecell}
\usepackage{longtable}
\usepackage{changepage}
\usepackage[unicode=true,
 bookmarks=true,bookmarksnumbered=true,bookmarksopen=false,
 breaklinks=true,pdfborder={0 0 0},backref=false,colorlinks=true]
 {hyperref}
\usepackage{enumitem}
\usepackage[normalem]{ulem}

\usepackage{rotating}

\hypersetup{
    linkcolor=purple,          
    citecolor=blue,        
    filecolor=magenta,      
    urlcolor=magenta           
}

\usepackage{zref-clever}
\newcommand{\cref}[1]{\zcref{#1}}

\zcRefTypeSetup{equation}{
    Name-sg=\textcolor{black}{Eq.},
    name-sg=\textcolor{black}{Eq.},
    Name-pl=\textcolor{black}{Eqs.},
    name-pl=\textcolor{black}{Eqs.}
}
\zcRefTypeSetup{section}{
    Name-sg=\textcolor{black}{Sec.},
    name-sg=\textcolor{black}{Sec.},
    Name-pl=\textcolor{black}{Secs.},
    name-pl=\textcolor{black}{Secs.}
}
\zcRefTypeSetup{appendix}{
    Name-sg=\textcolor{black}{App.},
    name-sg=\textcolor{black}{App.},
    Name-pl=\textcolor{black}{Apps.},
    name-pl=\textcolor{black}{Apps.}
}
\zcRefTypeSetup{figure}{
    Name-sg=\textcolor{black}{Fig.},
    name-sg=\textcolor{black}{Fig.},
    Name-pl=\textcolor{black}{Figs.},
    name-pl=\textcolor{black}{Figs.}
}
\zcRefTypeSetup{table}{
    Name-sg=\textcolor{black}{Table},
    name-sg=\textcolor{black}{Table},
    Name-pl=\textcolor{black}{Tables},
    name-pl=\textcolor{black}{Tables}
}

\renewcommand{\paragraph}[1]{\vspace{0.2cm}{\bf \textit{#1}}}

\usepackage{simpler-wick}
\allowdisplaybreaks[1] 
\newcommand{\td}{\widetilde}
\newcommand{\ovl}{\overline}

\def\pare#1{\left( #1 \right)}
\def\brak#1{\left[#1\right]}

\def\ket#1{| #1 \rangle}

\def\Ket#1{\left| #1 \right\rangle}
\def\inn#1{\langle #1 \rangle}

\def\Im{\mathrm{Im}}
\def\Re{\mathrm{Re}}

\def\tr{\mathrm{tr}}
\def\Tr{\mathrm{Tr}}

\def\ii{\mathrm{i}}
\def\dd{\mathrm{d}}
\def\rM{\mathrm{M}}
\def\rX{\mathrm{X}}
\def\rR{\mathrm{R}}
\def\rK{\mathrm{K}}
\def\rL{\mathrm{L}}
\def\rW{\mathrm{W}}
\def\rP{\mathrm{P}}
\def\rH{\mathrm{H}}
\def\rN{\mathrm{N}}
\def\RR{\mathbf{R}}
\def\KK{\mathbf{K}}
\def\kk{\mathbf{k}}
\def\pp{\mathbf{p}}
\def\tt{\mathbf{t}}
\def\rKA{\mathrm{KA}}
\def\rPA{\mathrm{PA}}

\def\mclG{\mathcal{G}}
\def\mclT{\mathcal{T}}
\def\mclE{\mathcal{E}}
\def\mclB{\mathcal{B}}

\def\mrmL{\mathrm{L}}
\def\mrmR{\mathrm{R}}
\def\mbbA{\mathbb{A}}

\def\bG{\ovl{\Gamma}}
\def\bK{\ovl{\mathrm{K}}}
\def\bM{\ovl{\mathrm{M}}}
\def\mT{\mathcal{T}}
\def\mU{\mathcal{U}}

\def\rank{\mathrm{rank}}

\def\bpm{\begin{pmatrix}}
\def\epm{\end{pmatrix}}

\begin{document}

\title{Stable Topology in Exactly Flat Bands}

\author{Yan-Qi Li}
\thanks{These authors contributed equally to this work.}
\affiliation{International Center for Quantum Materials, School of Physics, Peking University, Beijing 100871, China}
\affiliation{Beijing Key Laboratory of Quantum Devices, Peking University, Beijing 100871, China}

\author{Yi-Jie Wang}
\thanks{These authors contributed equally to this work.}
\email{wang-yijie@pku.edu.cn}
\affiliation{International Center for Quantum Materials, School of Physics, Peking University, Beijing 100871, China}
\affiliation{Beijing Key Laboratory of Quantum Devices, Peking University, Beijing 100871, China}

\author{Pei-Han Lin}
\affiliation{International Center for Quantum Materials, School of Physics, Peking University, Beijing 100871, China}

\author{Bin Wang}
\affiliation{International Center for Quantum Materials, School of Physics, Peking University, Beijing 100871, China}
\affiliation{Beijing Key Laboratory of Quantum Devices, Peking University, Beijing 100871, China}

\author{Zhi-Da Song}
\email{songzd@pku.edu.cn}
\affiliation{International Center for Quantum Materials, School of Physics, Peking University, Beijing 100871, China}
\affiliation{Beijing Key Laboratory of Quantum Devices, Peking University, Beijing 100871, China}
\affiliation{Hefei National Laboratory, Hefei 230088, China}

\date{\today}

\begin{abstract}
Topological flat bands (FBs) offer an ideal platform for realizing exotic topological phases and exploring quantum geometric effects, yet their realization with both exact flatness and stable topology in local lattice models has been long hindered by fundamental no-go theorems. 
The obstruction to topological FBs is also manifested as the absence of exact Gaussian tensor-network state (TNS) representations for topological insulators and superconductors. 
Here, we overcome this barrier by demonstrating the existence of {\it critical topological FBs (CTFBs)} in finite-range hopping models. 
They {\it saturate} the no-go theorems via a unique structure of Bloch wavefunctions: 
While continuous over the whole Brillouin zone, the projector $P(\kk)$ onto FBs is non-analytic at isolated band touching points, thereby relaxing the inherent restrictions on the coexistence of exact flatness and stable topology. 
Filling such CTFBs yields short-range entangled topological states that exhibit power-law correlations due to the non-analyticity. 
We then establish a general symmetry-based principle to systematically construct CTFBs, as well as their parent Hamiltonians, that carry desired topological invariants in given space groups. 
It utilizes the bipartite structures and requires no further fine-tuning, and the topology is robust against {\it arbitrary} symmetry-preserving perturbations that violate the bipartite condition and gap the touching points. 
Remarkably, independent tuning of the local quantum geometry while maintaining flatness and topology is allowed. 
Examples exhibiting Chern numbers 1, 2, 3, 6, in 2D, and strong $\mathbb{Z}_2$ indices in 2D and 3D are demonstrated for concreteness. 
An automated algorithm further identifies more than 50,000 symmetry-indicated CTFBs, including crystalline and higher-order topological ones. 
In the end, we show that the bipartite structure naturally endows the filled CTFB states with {\it exact} TNS representations with finite bond dimensions. 
By bridging the topological band theory to TNS methods, CTFB provides a novel tractable starting point for exploring strongly correlated topological matter, with potential relevance to realistic material realizations. 
\end{abstract}

\maketitle

\section{Introduction}

Topology has become a cornerstone of modern condensed matter physics, with topological band theory \cite{haldane_model_1988, kane_z2_2005, bernevig_quantum_2006} representing a significant and well-established branch.  
The electronic filling of topological bands constitutes topological insulators \cite{hasan_colloquium_2010,qi_topological_2011}. 
More intriguing phenomena emerge within topological flat bands (FBs), where strong correlations arise out of interactions. 
When partially filled, topological FBs can host fractionalized states such as fractional Chern insulators (FCI) \cite{tang_high-temperature_2011,neupert_fractional_2011,sun_nearly_2011,qi_generic_2011,regnault_fractional_2011,crepel2025topologically}---a phase recently realized in moir\'e superlattices \cite{cai_signatures_2023, xu_observation_2023, lu_fractional_2024}.  
However, a fundamental theoretical challenge has persisted: the simultaneous realization of exact flatness and stable topology in local lattice models is seemingly prohibited by no-go theorems \cite{chen_impossibility_2014}.
In particular, the Dubail-Read theorem precludes the presence of gapped topological FBs in finite-range hopping models, thereby posing a fundamental obstruction to local representation of topological phases---most notably reflected by the absence of exact Gaussian tensor network states (TNS) for topological insulators and superconductors \cite{dubail_tensor_2015,read2017compactly}.

Indeed, while exact FBs exist in frustrated hopping models such as Kagome, dice, and Lieb lattices \cite{bergman_band_2008}, they generally lack stable topology, being either singular, topologically trivial, or only possessing fragile topology \cite{calugaru_general_2021, wahl_exact_2025}. 
Recently, frustration-free free-fermion theory reveals a new perspective on constructing exact FBs \cite{ono2026frustration, masaoka2026frustration}, but also with singular touching points.
For the singular case \cite{rhim_classification_2019,hwang2021flat,hwang2021general}, the Bloch wavefunctions exhibit essential discontinuities that cannot be removed by local gauge transformations, rendering the topological invariants ill-defined. 

Inspired by pioneering works such as the critical TNS for $p+\ii p$ superconductors [\onlinecite{wahl_projected_2013},  \onlinecite{dubail_tensor_2015}] and the $2\pi$-flux dice lattice model \cite{yang_fractional_2025}, we introduce the concept of {\it critical topological FBs} (CTFBs), where the fundamental constraints of the no-go theorem are {\it saturated} rather than avoided.
By allowing the projector $P(\kk)$ onto Bloch wavefunctions to be non-analytic yet continuous over the entire Brillouin zone (BZ), we show that exact flatness and stable topology can coexist (\cref{fig:Z2-main}(a)). 
Such CTFBs necessarily touch dispersive bands at isolated momenta, where jumps in the derivatives or higher-order derivatives of $P(\kk)$ release the strict constraints imposed by the Dubail-Read theorem.
This mechanism results in critical correlation functions while maintaining an area law for entanglement entropy, which in turn provides a systematic pathway to construct exact TNS representations for topological states.

Utilizing a bipartite parent Hamiltonian framework, we establish a general symmetry-based principle to construct CTFBs, which only requires to properly select the symmetry representations of the orbitals in the lattice, and does not require any fine-tuning of the model parameters. 
Notably, the symmetry data also automatically guarantee the band topology to remain robust against perturbations that break the bipartite condition, including mean-field interactions. 
The symmetry-based principle is established by combining the following two ends: (1) the topological quantum chemistry \cite{bradlyn_topological_2017} and related theories \cite{po_symmetry-based_2017,kruthoff_topological_2017}, which provides the convenient symmetry-based indicators (SIs) for the stable topology, and (2) an additional group theoretical criterion that we develop, which serves to guarantee that the projector $P(\kk)$ to the FB Bloch wave-functions is continuous near the touching points. 
Note that SIs of topology can differ from the exact topological invariant by a modulo of integers. 
Nevertheless, we find this latter condition will further allow us to recover the exact topological invariants carried by the CTFB.

Based on this principle, we demonstrate CTFB models hosting (high) Chern numbers up to $C=3$ in 2D, and the strong $\mathbb{Z}_2$ indices protected by time-reversal symmetry (TRS) in 2D and 3D. 
Leveraging the deterministic role of symmetries and the independence of microscopic details, we further implement an algorithmic search, and identify over 50,000 CTFBs across the 2D and 3D space groups (\cref{tab:statistics-main}), with the corresponding combinations of orbital (co-)representations listed in the supplementary materials (SM) \cite{SM}. 
In addition to the aforementioned strong topological invariants, these candidates also span a wide spectrum of crystalline symmetry-protected topology, such as mirror-Chern states and higher-order topological states. 
Due to the simple symmetry-guaranteed nature of CTFB, finding material realization for some of the candidates is thus plausible.

When interactions come into play in FBs, besides the global topological data, the local quantum geometric properties across the Brillouin zone can play a more intricate role in the energetics \cite{roy_2014_bandgeometryofFCI, peotta_superfluidity_2015, jackson_geometric_2015, xie_topology-bounded_2020, hu_geometric_2019, julku_superfluid_2020, ledwith_fractional_2020, wang_2021_exactlandaulevel, yu_2025_universalwilsonloop, yu_quantum_2025, hwang2021wave, hu2025anomalous, guo2025majorana}. 
Crucially, unlike the FBs with singular touching points, the quantum geometry in a CTFB is \textit{not} divergent at the touching points, but can be continuously modulated by the model parameters, akin to the gapped bands. 
Remarkably, as the band width has been pinned to zero, and the topological number is also symmetry-guaranteed, hence the tuning of quantum geometry can be carried out independently. 
This surpasses the lattice models with gapped narrow topological bands, which in general requires fine-tuning. 
As an example, for Chern CTFBs, we show a good level of holomorphic ``idealness'' in the quantum geometry and uniformity of Berry curvature can be achieved. 
CTFB thus opens a new arena to explore the quantum geometric effects. 

As for the strongly correlated topological phases, CTFB features a profound link to the TNS method. 
Each filled CTFB enjoys an exact Guassian TNS representation with finite bond dimensions, which is naturally enabled by the bipartite construction, where the larger sublattice provides the physical degrees of freedom, while the smaller sublattice turns into the tensors with virtual degrees of freedom that connect them. 
In brief, this TNS formulation simply subtracts the states coupled to the smaller sublattice from the larger one, leaving a Gaussian state that fills the FB, which we explicitly verify as topological and critical. 
Our TNS construction thus explicitly saturates the Dubail-Read theorem. 
Going beyond to the strongly correlated regime, due to the versatility of the TNS methods in studying the symmetry-protected topological (SPT) phases \cite{gu_tensor-entanlement-filtering_2009, pollmann_entanglement_2010, chen_classification_2011, schuch_classifying_2011, chen_symmetry-protected_2012, haegeman_gauging_2015, williamson_matrix_2016, wang2023exactly, ma2024variational} and topological orders (TOs) \cite{aguado_entanglement_2008, gu_tensor-entanglement_2008, buerschaper_explicit_2009, schuch_peps_2010, circ_entanglement_2011, schuch_TO_2013, csahinouglu2021characterizing, yang_chiral_2015, bultinck_anyons_2017}, CTFB thus provides a useful starting point.

This manuscript is organized as follows.
In \cref{sec:Z2-model}, we introduce the concept of CTFB through a concrete bipartite 2D lattice model exhibiting a TRS-protected $\mathbb{Z}_2$ topology. 
With this example at hand, \cref{sec:principle} establishes the general principle to systematically construct symmetry-indicated CTFBs across generic space groups. 
We further run an exhaustive enumeration of CTFBs over 2D and 3D space groups both in the absence and presence of TRS. 
\cref{sec:more-examples} showcases additional representative models. 
\cref{sec:exapmles-Chern} demonstrates 2D Chern CTFBs with Chern numbers $C=1,2,3$, with a particular emphasis on the capability of realizing high Chern numbers, and the high tunability of quantum geometry. 
\cref{sec:examples-3DZ2} presents a 3D CTFB with TRS-protected strong $\mathbb{Z}_2$ index, which is also the first construction of a 3D exactly flat band with strong topology. 
Breaking TRS for this model further leads to an axion CTFB that can exhibit higher-order edge states. 
\cref{sec:invariant} provides analytical formulations for topological invariants directly in terms of the bipartite hopping matrix $S(\kk)$, which goes beyond SIs. 
\cref{sec:criticality} relates the power-law criticality in correlation functions to the non-analyticity in projectors $P(\kk)$. The power-law criticality also indicates a power-law lower-bound to the Wannier function tails. 
\cref{sec:non-symmetry-indicated} further discusses various scenarios of CTFBs beyond SIs. 
\cref{sec:TNS} demonstrates that the many-body ground states occupying these CTFBs admit exact TNS representations with finite bond dimensions.
Finally, in \cref{sec:discussion}, we summarize our core findings, demonstrate the robustness of topology in symmetry-indicated CTFBs against arbitrary symmetry-preserving perturbations, and remark on the potential influence of the dispersive touching states in the correlated regime. 
More importantly, we discuss the plausible relevance of our CTFB theory to the actual materials, and analyze the potential applications of CTFB models in exploring the strongly correlated topological matter, including SPT and TO. 

\section{An example of CTFB with \texorpdfstring{$\mathbb{Z}_2$}{Z2} topology}
\label{sec:Z2-model}
 
\subsection{The bipartite construction}
\label{sec:Z2-model-construction}
 
\begin{figure*}[t]
\centering
\includegraphics[width=0.8\linewidth]{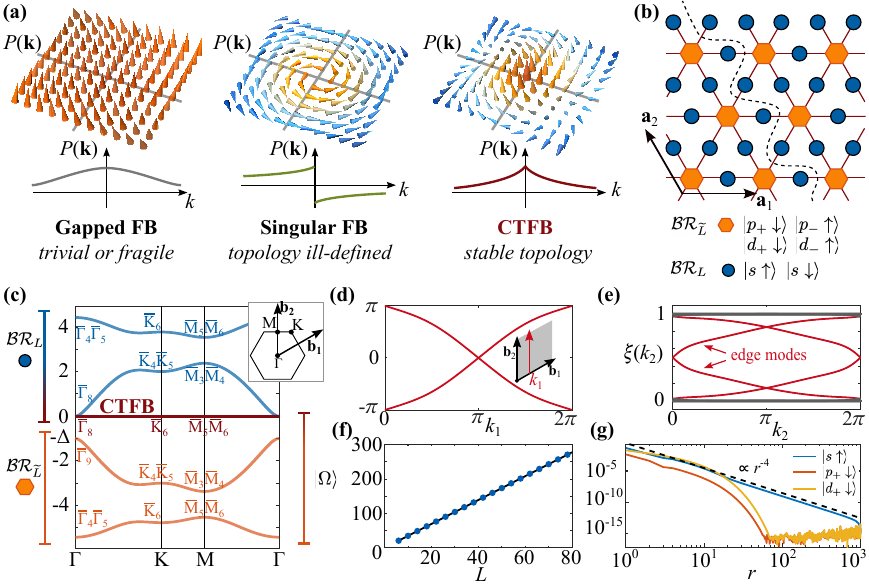}
\caption{
    Conceptual framework and realization of a $\mathbb{Z}_2$ CTFB. 
    (a) Classification of exactly flat bands in finite-range lattice models via analyticity of the projector $P(\kk)$ onto FB Bloch wavefunctions. 
    Gapped FB: $P(\kk)$ is smooth and must have trivial (or fragile) topology. 
    Singular FB: Essential discontinuity of $P(\kk)$ at the touching point renders the topology ill-defined, as seen in frustrated models like the Kagome lattice. 
    CTFB: $P(\kk)$ is continuous but non-analytic, allowing stable topological invariants well-defined and nontrivial. 
    (b) Realization of a TRS-protected $\mathbb{Z}_2$ CTFB in layer group $p\bar31m$. 
    (c) The band structure with representative parameters. 
    (d) The Wilson loop spectrum of the two-fold CTFB displaying the characteristic $\mathbb{Z}_2$ zigzag flow. 
    (e) The entanglement spectrum $\xi(k_2)$ of $\ket{\Omega}$ obtained by spatial cuts parallel to the dashed line in (b). The presence of crossing edge modes confirms the bulk-boundary correspondence. 
    (f) Entanglement entropy (EE) scaling with system size $L\times L$. The EE is calculated from $\xi(k_2)$ in (e), and scales linearly in $L$, confirming the area-law behavior. 
    (g) Real space correlation functions showing a power-law decay of $r^{-4}$. 
     }
\label{fig:Z2-main}
\end{figure*}

We first present a concrete model of 2D CTFB that hosts a TRS-protected $\mathbb{Z}_2$ topology in the layer group $p\bar{3}1m$.
We employ the bipartite lattice construction shown in \cref{fig:Z2-main}(b) to realize the FB. 
Specifically, the larger sublattice $L$ consists of $\ket{s \! \uparrow}$ and $\ket{s\!\downarrow}$ orbitals located at the $3f$ Wyckoff positions (blue) forming a Kagome lattice, while the smaller sublattice $\td L$ comprises $\ket{p_+\!\downarrow}$, $\ket{p_-\!\uparrow}$, $\ket{d_+\! \downarrow}$, and $\ket{d_-\! \uparrow}$ orbitals at the $1a$ positions (yellow) forming a triangular lattice. 
The underlying Bravais lattice is spanned by the primitive vectors $\mathbf{a}_1 = (1, 0)$ and $\mathbf{a}_2 = (-1/2, \sqrt{3}/2)$, assuming a unit lattice constant.
Here $s$, $p_{\pm} = p_x\pm \ii p_y$, and $d_{\pm} = d_{x^2-y^2} \pm \ii d_{2xy}$ denote atomic orbitals, and $\uparrow\downarrow$ represent spin. 
The bipartite Bloch Hamiltonian takes the form 
\begin{equation} \label{eq:Hk-main}
    H (\kk) = \begin{pmatrix}
        0_{6\times 6} & S(\kk) \\
        S^\dagger(\kk) & -\Delta \cdot \mathbb{I}_{4\times 4}
    \end{pmatrix}\ .
\end{equation}
where $S(\kk)$ is a $6 \times 4$ matrix describing the inter-sublattice hoppings. 
$S(\kk)$ can be constructed by defining a real-space hopping matrix from the $1a$ to $3f$ sites constrained by local mirror $M_x$ and TRS, and then generating the other terms via the three-fold rotation $C_{3z}$ and inversion $P$. 
At general momenta, the zero-energy flat bands arise from the two-dimensional kernel of the rectangular matrix $S^\dagger(\kk)$. 
(Notice that if $S^\dagger(\kk) u(\kk) = 0$, then $H(\kk) (u^T(\kk), 0_{4\times1})^T=0$.) 
For clarity, we have also assigned an intra-sublattice term to the sublattice $\td L$, which we take as a diagonal constant $-\Delta<0$. 
Nevertheless, its form can be arbitrarily modified without affecting the FB; for example, one may set $\Delta=0$, since the FB is completely decoupled from the $\td{L}$ sublattice.

Up to nearest-neighbor hoppings, the generic form of $S(\kk)$ is 
{\small
\begin{equation} \label{eq:Z2-Sk}
2\! \left(
\begin{array}{rrrr}
  -\ii t_1 \sin \! \frac{k_1}2&   \ii t_2 \sin \! \frac{k_1}2 & t_3 \cos \! \frac{k_1}2 & -t_4 \cos\!\frac{k_1}2 \\
  -\ii t_2 \sin \!\frac{k_1}2 &  -\ii t_1 \sin \!\frac{k_1}2 & t_4 \cos \!\frac{k_1}2 &  t_3 \cos \!\frac{k_1}2 \\
  -\ii t_1 \sin\! \frac{k_2}2 &   - \ii \omega t_2 \sin \!\frac{k_2}2 & -  \omega^{*} t_3 \cos \!\frac{k_2}2 & \omega^{*} t_4  \cos\!\frac{k_2}2 \\
  \ii \omega^* t_2 \sin \!\frac{k_2}2 &  -\ii t_1 \sin \!\frac{k_2}2 & -\omega t_4 \cos\! \frac{k_2}2 &  - \omega t_3 \cos \!\frac{k_2}2 \\
  -\ii t_1 \sin\! \frac{k_3}2 &   - \ii \omega^* t_2 \sin \!\frac{k_3}2 & -  \omega t_3 \cos\!\frac{k_3}2 & \omega t_4  \cos\!\frac{k_3}2 \\
  \ii \omega t_2 \sin\! \frac{k_3}2 &  -\ii t_1 \sin \!\frac{k_3}2 & -\omega^* t_4 \cos\!\frac{k_3}2 &  - \omega^* t_3 \cos \!\frac{k_3}2
\end{array}
\right) .
\end{equation}}
Here, $k_1 \!=\!  \kk \!\cdot\! \mathbf{a}_1$, $k_2 \!=\! \kk \!\cdot\! \mathbf{a}_2$, and $k_3 \!=\! -\kk \!\cdot\! (\mathbf{a}_1 \!+\! \mathbf{a}_2)$. The phase factor $\omega$ is equal to $e^{\ii \frac{\pi}3}$, and $t_{1,2,3,4}$ denote the real hopping amplitudes from the $\ket{p_+\!\downarrow}$, $\ket{p_-\!\uparrow}$, $\ket{d_+\! \downarrow}$, and $\ket{d_-\! \uparrow}$ orbitals at the origin to the $\ket{s\!\uparrow}$ orbital at $\frac12 \mathbf{a}_1$, respectively.
Importantly, as will be clear soon, different choices of parameters $t_{1,2,3,4}$, $\Delta$ will not alter the topology nor flatness of CTFB in this model.

\cref{fig:Z2-main}(c) displays the band structure with representative symmetry-allowed hopping parameters ($\Delta=t_{1,2,3,4}=1$). 
In addition to the two zero-energy FBs at generic momenta, dispersive bands also drop to zero-energy and touch the FBs at $\Gamma$ point. 

Symmetry representations of the flat bands at high-symmetry momenta can be derived following the rules established in Ref.~\cite{calugaru_general_2021}. 
Local orbitals on sublattices $L$ and $\td L$ form representations of the site-symmetry groups of their corresponding Wyckoff positions, which then induce band representations of the space group $\mclG$, denoted as $\mathcal{BR}_{L}$ and $\mathcal{BR}_{\td L}$, respectively.
In this model, $\mathcal{BR}_L = [{}^1\ovl{E}_g {}^2\ovl{E}_g]_{3f} \uparrow \mathcal{G}$ and $\mathcal{BR}_{\td L} = [\ovl{E}_{1u} \oplus {}^1\ovl{E}_g {}^2\ovl{E}_g]_{1a} \uparrow \mathcal{G}$.
The uncoupled FB representations will then be given by the formal difference $\mathcal{BR}_{L} \boxminus \mathcal{BR}_{\td L}$, where $\boxminus$ denotes the ``inverse operation'' of the direct sum of representations. 
Utilizing the band representation data from the Bilbao Crystallographic Server  \cite{bradlyn_topological_2017,elcoro_double_2017, IrRep2022}, in this model
\begin{equation} \label{eq:BR-difference}
\mathcal{BR}_{ L} \boxminus \mathcal{BR}_{\td L}
= (2\bG_8 \boxminus \bG_9;\quad \bK_6; \quad \bM_5\bM_6)\ . 
\end{equation}
Here, symbols such as $\bG_8$ and $\bM_5\bM_6$ are notations for irreducible representations (irreps) at their respective momenta, with definitions provided in SM \cite{SM}. 
The reason why $\mathcal{BR}_{ L} \boxminus \mathcal{BR}_{\td L}$ yields the uncoupled FB states is as follows: 
\begin{enumerate}[leftmargin=0.4cm]
\item If all irreps in $\mathcal{BR}_{\td L}$ are also present in $\mathcal{BR}_{L}$, these identical irreps hybridize via $S(\kk)$ and become gapped.
Consequently, only the irreps unique to $\mathcal{BR}_L$ remain at zero energy, which is  the case at the $\mathrm{K}$ and $\mathrm{M}$ points.

\item If, however, any irrep appears only in $\mathcal{BR}_{\td L}$ ({\it e.g.}, $\bG_9$), then it will not hybridize and will hence reside at energy $-\Delta$. 
In this scenario, extra irreps in $\mathcal{BR}_{L}$ ({\it e.g.}, $2\bG_8$) will also lack counterparts in $\mathcal{BR}_{\td L}$, and hence remain at zero energy. 
\end{enumerate}
In sum, all irreps with positive multiplicities in $\mathcal{BR}_{ L} \boxminus \mathcal{BR}_{\td L}$ constitute the zero-energy states at high-symmetry points, while those with negative multiplicities are located at $-\Delta$.
This logic fully explains the irrep assignment in \cref{fig:Z2-main}(c).

\subsection{Topology of the flat bands}
\label{sec:topology-of-flat-bands}

A symmetry-preserving perturbation that violates the bipartite condition can lift the $2\bG_8$ degeneracy at $\Gamma$. However, the resulting gapped band structure would always carry the irreps
\begin{equation} \label{eq:B-Z2-main}
    \mathcal{B} = (\bG_8  ;\quad \bK_6; \quad \bM_5\bM_6)\ . 
\end{equation}
Notably, $\bG_8$ has even parity, whereas $\bM_5\bM_6$ has odd parity. 
According to the Fu-Kane formula \cite{fu_topological_2007}, the 2D $\mathbb{Z}_2$ index $\delta$ is determined by the sum of the numbers of occupied odd-parity Kramers pairs, denoted as $n^-_{\mathbf{K}}$, over all time-reversal invariant momenta (TRIMs). 
In our model, the TRIMs consist of one $\Gamma$ point and three equivalent $\mathrm{M}$ points, leading to 
\begin{equation} 
    \delta =  n^-_{\Gamma} + 3n^-_{\mathrm{M}} = 1\mod 2 \ ,
\end{equation}
which indicates a nontrivial $\mathbb{Z}_2$ topology.
Since splitting the $2\bG_8$ levels in arbitrary symmetry-preserving manner results in the same nontrivial topology, a question naturally arises: is the $\mathbb{Z}_2$ index well-defined even \textit{before} the splitting? 
Answering this in the affirmative first requires demonstrating that projector $P(\kk) = \sum_{n=1,2} u_{n}(\kk) u_n^\dagger(\kk)$ onto FB wavefunctions remain continuous across the touching point at $\Gamma$. 

Rather surprisingly, we find that the continuity of $P(\kk)$ at $\Gamma$ is guaranteed by symmetry. 
To demonstrate this, we construct a low-energy k$\cdot$p theory around the $\Gamma$ point. 
Because the identical irreps present in both $\mathcal{BR}_{L}$ and $\mathcal{BR}_{\td L}$ strongly hybridize and open a large gap via the momentum-independent ($\mathcal{O}(k^0)$) terms in $S^\dagger(\kk)$, they are safely decoupled from the low-energy physics.
Therefore, we can project $S^\dagger(\kk)$ onto the active subspace spanned by the remaining representations: the two-dimensional $\bG_9$ from $\td L$ (constituting the rows) and the four-dimensional $2\bG_8$ from $L$ (constituting the columns). 
Since $\bG_8$ and $\bG_9$ have opposite parities, their coupling must be odd in $\kk$. 
Apart from parity, $\bG_9$ and $\bG_8$ share identical representations for the remaining symmetry generators: the three-fold rotation $C_{3z} = e^{-\ii \frac{\pi}3 \sigma_z}$, the mirror $M_x = -\ii\sigma_y$, and TRS $\mathcal{T}=-\ii\sigma_y K$, with $K$ denoting complex conjugation. 
Imposing these symmetries constrains the effective $2 \times 4$ off-diagonal block, up to linear order in $\kk$, to the form
\begin{equation} \label{eq:Sk-Z2-main}
    S^\dagger(\kk) = \begin{pmatrix}
        0 & \gamma_1 k_- & 0 & \gamma_2 k_- \\
        \gamma_1 k_+ & 0 & \gamma_2 k_+  & 0
    \end{pmatrix} + \mathcal{O}(k^3)\ ,
\end{equation}
where $k_{\pm} = k_x\pm \ii k_y$, and $\gamma_{1,2} \in \mathbb{R}$ are parameters determined by microscopic details (\cref{eq:Z2-Sk}). 
The two-dimensional kernel of this $2 \times 4$ matrix is spanned by the orthogonal vectors $u_1=(\gamma_2,0,-\gamma_1,0)^T + \mathcal{O}(k^2)$ and $u_2=(0,\gamma_2,0,-\gamma_1)^T + \mathcal{O}(k^2)$. 
In the limit $\kk\to 0$, these basis vectors approach constant, $\kk$-independent vectors. 
This explicitly proves that the projector $P(\kk)$ onto wavefunctions of this CTFB, in contrast to singular FBs, is continuous across the touching point at $\Gamma$, thereby allowing a well-defined $\mathbb{Z}_2$ topological index even in the absence of a gap.

To confirm the $\mathbb{Z}_2$ topology, we compute the Wilson loop spectrum of the two-fold CTFB. 
As shown in \cref{fig:Z2-main}(d), the phases of the Wilson loop eigenvalues exhibit a characteristic zigzag flow, winding nontrivially between Kramers pairs at $k_1 = 0$ and $\pi$ to provide a gauge-independent manifestation of the $\mathbb{Z}_2$ index. 
The analytic proof will presented in \cref{sec:invariant-Z2}.

Furthermore, we characterize the entanglement properties of the many-body ground state $\ket{\Omega}$ formed by filling the CTFB and all lower-energy bands. 
The entanglement spectrum is equivalent to a ``flattened'' energy spectrum, where all the bulk valence (conduction) bands are set to energy 1 (0), while edge modes form an unavoidable spectral flow.  
They can thus signify the bulk-boundary correspondence \cite{turner_entanglement_2010, hughes_inversion-symmetric_2011}. 
For this sake, we perform a real-space bipartition that divides the lattice into two subsystems. 
The entanglement spectrum $\{\xi\}$ is then obtained by diagonalizing the correlation matrix $C_{\alpha\alpha'}(\RR'-\RR)=\inn{\Omega| \psi_{\RR\alpha}^\dagger \psi_{\RR'\alpha'} |\Omega}$, where $\psi_{\RR\alpha}^\dagger$ creates a fermion in orbital $\alpha$ in the unit cell $\RR$, and $\psi_{\RR\alpha}^\dagger$, $\psi_{\RR'\alpha'}$ are restricted to the same subsystem, which is chosen as a stripe. 
Because our chosen spatial cut (\cref{fig:Z2-main}(b)) preserves translation symmetry along the lattice vector $\mathbf{a}_2$, the entanglement spectrum $\xi_n(k_2)$ can be resolved by the parallel momentum $k_2$. 
To identify the edge modes out of the bulk states in the entanglement spectrum, we compute the edge polarizations of the corresponding eigenvectors, and mark those branches with significant edge polarization as red in \cref{fig:Z2-main}(e). 
Per momentum $k_2$, there are only four entanglement eigenvalues significantly away from 0 and 1, each of which is twofold degenerate, with the corresponding modes localized near the stripe boundaries. 
They exhibit zig-zag flows between Kramer pairs, signifying the $\mathbb{Z}_2$ topology, confirming the bulk-boundary correspondence. 
In this model, due to the presence of inversion symmetry and TRS, edge modes from the two boundaries are exactly degenerate. 
All the other states extend in the bulk of the subsystem, with eigenvalues well quantized to 0 or 1, signifying empty or fully occupied states in the bulk, respectively. 
Consequently, the entanglement entropy, 
$ S = - \sum_{n, k_2} [ \xi_n(k_2) \ln \xi_n(k_2)  + (1 - \xi_n(k_2)) \ln(1-\xi_n(k_2)) ]$, 
arises entirely from the edge modes and obeys an area-law scaling (\cref{fig:Z2-main}(f)).
This identifies the ground state $\ket{\Omega}$, albeit gapless, as a short-range entangled state.

\subsection{Critical correlation functions}

Numerical calculations reveal that the spatial correlation functions $C_{\alpha \alpha}(\RR'-\RR) = \inn{\Omega| \psi_{\RR\alpha}^\dagger \psi_{\RR'\alpha} |\Omega}$ exhibit a power-law decay, scaling asymptotically as $r^{-4}$ (\cref{fig:Z2-main}(g)) for orbitals within the $L$-sublattice, {\it i.e.}, $\alpha=s\!\uparrow, \ s\!\downarrow$. 
This long-range behavior is governed by the non-analyticity of the projector $P(\kk)$ onto the CTFB. 
As proven in Ref.~\cite{SM}, under Fourier transformation, a discontinuity in the $n$-th order derivatives of $P(\kk)$ translates to an asymptotic $r^{-n-d}$ decay of the correlation function in $d$ spatial dimensions. 
In principle, one could expand $S^\dagger(\kk)$ in \cref{eq:Sk-Z2-main} to third order in $k$, solve for $u_n(\kk)$, and directly deduce the non-analyticity in $P(\kk)$. 
However, in \cref{sec:invariant}, we will provide a simpler analytical argument based on the so-called singular Wannier gauge that $P(\kk)$ in this model possesses discontinuous second-order derivatives, which leads to the $r^{-4}$ decay behavior. 

As shown in the second paragraph below \cref{eq:Z2-Sk}, the flat-band wavefunctions---and consequently $P(\kk)$---have vanishing support on the $\td L$-sublattice. 
Therefore, the correlation functions for orbitals in $\td L$, {\it i.e.}, $\alpha=p_{+}\!\!\downarrow$, $p_{-}\!\!\uparrow$, $d_{+}\!\!\downarrow$, $d_{-}\!\!\uparrow$, are only determined by the gapped dispersive bands below the CTFB, resulting in a conventional exponential decay (\cref{fig:Z2-main}(g)).

\section{Principle for CTFBs in generic space groups}
\label{sec:principle}

\subsection{Symmetry-indicated CTFB}

In bipartite constructions of the form in \cref{eq:Hk-main}, the flat bands possess well-defined topological invariants as long as $S^\dagger(\kk)$ has continuous kernels at the isolated touching points. 
If the invariants are nontrivial, we refer to such flat bands as CTFBs. 
Note the bipartite ($S(\kk)$-matrix) construction also encompasses a wide range of FB construction schemes, including the line-graphs \cite{calugaru_general_2021}. 
CTFBs generally exhibit critical correlation functions, as otherwise they would violate the Dubail-Read theorem. 
Although fine-tuning the model parameters can in some cases realize the continuity condition, in a generic lattice model it can be nontrivial or rather unnatural to do so (see \cref{sec:non-symmetry-indicated} for an example). 
Consequently, realizations of CTFBs remain extremely rare in the literature.

Therefore, in this work, we mainly focus on a specific type of CTFBs---the symmetry-indicated CTFBs---whose continuity is symmetry-guaranteed and whose topology is diagnosed by symmetry-based indicators (SIs) \cite{po_symmetry-based_2017}. 
As will be clear soon, topology of such CTFBs is robust against {\it arbitrary} symmetry-preserving perturbations, including gap-opening terms.

A group of bands is characterized by the symmetry data vector $\mclB$, {\it e.g.,} \cref{eq:B-Z2-main}, whose non-negative  components represent the multiplicities of irreps at high-symmetry momenta. 
SIs, such as the Fu-Kane formula, are linear maps from $\mclB$ to integer-valued indices, with nonzero values signifying stable topological invariants \cite{song2018quantitative,khalaf_symmetry_2018,elcoro_magnetic_2021}. 
In this framework, a topologically nontrivial $\mclB$ can only be expanded as a linear combination of band representations with fractional coefficients, while a trivial one or a fragile topological one can be expanded with integer coefficients.

For each target topological FB $\mclB$, we then search for the concrete realizations of the two sublattices, $\mathcal{BR}_{L}$ and $\mathcal{BR}_{\td L}$. 
We denote their formal difference as $\mathcal{BR}_{L} \boxminus \mathcal{BR}_{\td L} = \mclB + \Delta\mclB$, where $\Delta\mclB$ denotes the remaining Bloch states left uncoupled at the touching points, besides those already included in $\mclB$. 
Without fine-tuning, the dimension of $\mclB$ is equal to the difference of the dimensions of $\mathcal{BR}_{L}$ and $\mathcal{BR}_{\td L}$, hence $\Delta\mclB$ can be deemed as a ``zero-band'' term. 
In our previous $\mathbb{Z}_2$ example, $\Delta \mclB = (\bG_8 \boxminus \bG_9; \ 0; \ 0)$. 
In particular, the positive components and the negative components in $\Delta\mclB$ are contributed by $\mathcal{BR}_{L}$ and $\mathcal{BR}_{\td L}$, respectively, and their total dimension should be equal. 
Moreover, since $\mathcal{BR}_{L}$ and $\mathcal{BR}_{\td L}$ are both trivial atomic insulators, $\Delta\mclB$ must carry the opposite SIs to cancel that of $\mclB$. 

In practice, we reverse the searching logic. 
For a target $\mclB$ with nontrivial SIs, we first enumerate all the zero-band $\Delta\mclB$ choices that satisfy the compatibility condition \cite{bradlyn_topological_2017, po_symmetry-based_2017, kruthoff_topological_2017}, and carry the opposite SIs to $\mclB$. 
Note that one $\mclB$ may find many $\Delta\mclB$, with different touching point numbers and locations. 
Each of such $\mclB + \Delta\mclB$ can then be realized by the formal difference of many combinations of $\mathcal{BR}_{L}$ and $\mathcal{BR}_{\td L}$, and we search for the minimal ones among them. 
The detailed algorithm is presented in SM \cite{SM}.

Recall from the representation counting rule explained above (\cref{sec:Z2-model-construction}) that the zero-energy states are given by all the irreps with positive multiplicities in $\mclB + \Delta\mclB$. 
To ensure that the target SIs of $\mclB$ remain robust against arbitrary perturbations, including gap-opening terms, we further require that at each touching point, these zero-energy states must consist of the same type of irreps, such that any symmetry-preserving gap-opening always yields the same  $\mclB$, hence the same SI.
Nonetheless, to guarantee that the topology is well-defined before the gap opening, stronger restrictions are needed.

\subsection{The continuity condition}
\label{sec:continuity-unitary}

For a given $\mclB + \Delta \mclB$, we now derive group-theoretical criteria for the kernels of $S^\dagger(\kk)$ to be continuous across the touching points, which is essential to validate well-defined topological invariants of the exact FB characterized by $\mathcal{B}$. 
A specific $\mclB + \Delta\mclB$ can be realized by different bipartite constructions $\mathcal{BR}_{L} \boxminus \mathcal{BR}_{\td L}$. 
However, these different realizations differ only by the common irreps shared between $\mathcal{BR}_{L}$ and $\mathcal{BR}_{\td L}$. 
As discussed in \cref{sec:topology-of-flat-bands}, these duplicated irreps strongly hybridize and are pushed to high energies, safely decoupling from the low-energy physics. 
Consequently, the low-energy properties relevant to the band touching points are only determined by the non-common irrep components, encoded by $\mclB + \Delta \mclB$. 

Consider the expansion $\kk=\kk_0 + \pp$ around a touching point $\kk_0$.
To ensure a direction-independent limit for the kernel as $\pp\to 0$, the little group $\mclG_{\kk_0}$ of $\kk_0$ must relate momenta approaching from all directions.
This means $\mclG_{\kk_0}$ must be a super-group of point groups $3$ or $4$ for 2D models, and a super-group of $23$ or $432$ for 3D models, where $\pp$ form an irreducible real representation $v$.

Suppose that at $\kk_0$, the representation decomposition of $\mclB + \Delta \mclB$ yields $M\rho \boxminus (\bigoplus_{i=1}^N \sigma_i)$, where $\sigma_i \neq \rho$. 
Within this active subspace, the columns and rows of $S^\dagger (\kk)$ transform according to $M\rho$ and $\bigoplus_i \sigma_i$, respectively, as exemplified in \cref{eq:Sk-Z2-main}. 
If the target CTFB corresponds to $Q \rho$, then dimension matching requires 
\begin{equation} \label{eq:M-Q}
(M-Q) \dim(\rho) = \sum_i \dim(\sigma_i)\ .
\end{equation}
While the representation counting rule pins all $M$ copies of $\rho$ at zero energy precisely at $\kk_0$, we now derive the condition under which $Q$ of them connect continuously to the CTFB states in the neighborhood of $\kk_0$. 

To this end, we expand the effective $S^\dagger(\kk)$ matrix to linear order in the relative momentum $\pp$ as $\sum_\mu A_\mu p_\mu$, where $\mu=x,y$ in 2D and $\mu=x,y,z$ in 3D. 
For a specific block $S^{(i,j)}(\kk)$ connecting the irrep $\sigma_i$ to the $j$-th copy of $\rho$, the number of independent, symmetry-allowed parameters in $\{A_\mu\}$ is given by $L_i = \inn{\mathbf{1}, \rho\otimes \sigma_i^*\otimes v}$---the multiplicity of the identity representation $\mathbf{1}$ in the triple product of $\rho$ (ket), $\sigma_i^*$ (bra), and $v$ (the vector representation of $\pp$).
Consequently, $S^\dagger(\kk)$ takes the following block form: 
\begin{equation} \label{eq:S-block-form}
\begin{pmatrix}
    \sum_{a=1}^{L_1}z^{(1,1)}_{a} A^{(1)}_a(\pp) & \cdots & \sum_{a=1}^{L_1}z^{(1,M)}_{a} A^{(1)}_a(\pp) \\
    \sum_{a=1}^{L_2}z^{(2,1)}_{a} A^{(2)}_a(\pp) & \cdots & \sum_{a=1}^{L_2}z^{(2,M)}_{a} A^{(2)}_a(\pp) \\ 
    \vdots & \ddots & \vdots \\
    \sum_{a=1}^{L_N}z^{(N,1)}_{a} A^{(N)}_a(\pp) & \cdots & \sum_{a=1}^{L_N}z^{(N,M)}_{a} A^{(N)}_a(\pp)
\end{pmatrix} .
\end{equation}
Here, $z^{(i,j)}_a$ is the $a$-th independent coupling coefficient in the $S^{(i,j)}(\kk)$ block, and $A^{(i)}_{a}(\pp) = \sum_\mu A^{(i)}_{a,\mu} \ p_\mu$ is the corresponding coupling matrix.

Because all $M$ block-columns transform under the same irrep $\rho$, the coupling matrices $A^{(i)}_{a}(\pp)$ are independent of the column index $j$. 
This structure allows them to be factored out when solving for the kernel of $S^\dagger(\kk)$. 
We group these matrices into a block matrix $\mathbb{A}^{(i)}(\pp) = \big[ A^{(i)}_1(\pp) \cdots A_{L_i}^{(i)}(\pp) \big]$. 
The kernel equations provided by the $\sigma_i$ block-row can be viewed as a set of equations spanned by the basis defined by the columns of $\mathbb{A}^{(i)}(\pp)$. 
Assuming the columns of $\mathbb{A}^{(i)}(\pp)$ are all linearly independent---a condition that necessitates the bound $L_i \dim (\rho) \le \dim (\sigma_i)$---the $\sigma_i$ block-row imposes $L_i \dim(\rho)$ independent constraints on the kernel. 
Subtracting these constraints from the total degrees of freedom yields $(M-\sum_i L_i )\dim(\rho)$ {\it momentum-independent} zero-energy solutions. 
(Note that any linear dependence in $\mathbb{A}^{(i)}(\pp)$ would result in fewer constraints and correspondingly more zero-energy modes.)
These constant solutions ensure a continuous kernel of $S^\dagger(\kk)$ across $\kk_0$, provided their total number matches the target flat-band dimension $Q\dim(\rho)$, which implies $M - Q = \sum_i L_i$.

Equating this expression for $M-Q$ to that in the dimension matching condition (\cref{eq:M-Q}) leads to $\sum_i L_i \dim(\rho) = \sum_i \dim(\sigma_i)$. 
Since the inequality $L_i \dim(\rho) \le \dim(\sigma_i)$ must hold for each block-row individually, as required by the full column-rank condition of $\mathbb{A}^{(i)}(\pp)$, this sum can only be equal if all inequalities are saturated, {\it i.e.,} $L_i \dim(\rho) = \dim(\sigma_i)$ for all $i$.
Therefore, by substituting $L_i = \inn{\mathbf{1}, \rho\otimes \sigma_i^*\otimes v}$, we arrive at the generic algebraic criterion for symmetry-guaranteed continuity at a touching point:
\begin{equation} \label{eq:criterion-main}
    \forall i,\quad \inn{\mathbf{1}, \rho\otimes \sigma_i^*\otimes v}
     =  q_\rho \  \frac{\dim(\sigma_i)}{\dim(\rho)} \ ,
\end{equation}
provided that $\mathbb{A}^{(i)}(\pp)$ is always full column-rank. 
The factor $q_\rho=1$ applies for a unitary little group $\mclG_{\kk_0}$.

\subsection{The continuity condition in the presence of anti-unitary symmetries}

In the presence of TRS, the little group at $\kk_0$ may have the form of a magnetic group $\mathcal{G}_{\mathbf{k}_0} = \mathcal{G}_{\mathbf{k}_0}^U + h\mathcal{T} \mathcal{G}_{\mathbf{k}_0}^U$, where $\mathcal{G}_{\mathbf{k}_0}^U$ is the unitary subgroup and $h$ is a spatial operation or the identity. 
In this case, \cref{eq:criterion-main} remains valid with $\rho$ and $\sigma_i$ interpreted as irreducible co-representations (co-irreps).
The factor $q_\rho \in \{1, 2\}$ indicates the number of unitary irreps contained in $\rho$.

We now prove this criterion. 
The coupling matrices ($A$ tensors) must be invariant under $\mathcal{G}_{\mathbf{k}_0}^U$, transforming as the identity representation $\mathbf{1}$ within the tensor product space $\rho \otimes \sigma_i^* \otimes v$. 
Thus, any $g \in \mathcal{G}_{\mathbf{k}_0}^U$ acts trivially in this subspace. 
Since the square of the anti-unitary operation also belongs to $\mathcal{G}_{\mathbf{k}_0}^U$,  $(h\mathcal{T})^2 = 1$ for the $A$ tensors, allowing a gauge where $h\mathcal{T}$ merely complex-conjugates the coupling coefficients $z_a^{(i,j)}$.

For $q_\rho=1$, $\rho$ remains a single unitary irrep ($\rho \downarrow \mathcal{G}_{\mathbf{k}_0}^U = \rho$). 
The constraints from $h\mathcal{T}$ depend on $\sigma_i$. 
If $q_{\sigma_i}=1$, $h\mathcal{T}$ merely restricts all coupling coefficients  to be real. 
If $q_{\sigma_i}=2$ ($\sigma_i \downarrow \mathcal{G}_{\mathbf{k}_0}^U = \nu_1 \oplus \nu_2$), $h\mathcal{T}$ requires the coupling coefficients in the $\nu_2$ block-row of $S^\dagger(\kk)$ to be complex conjugates of those in the $\nu_1$ block-row, regardless of whether $\nu_1\neq\nu_2$ or not. 
Unless all coupling coefficients are fine-tuned to be real, this conjugation relation preserves the number of independent linear constraints from $\sigma_i$, which remains $\langle \mathbf{1}, \rho \otimes \sigma_i^* \otimes v \rangle$. 
Therefore, the continuity criterion naturally reduces to the unitary case (\cref{sec:continuity-unitary}).

For $q_\rho=2$ ($\rho \downarrow \mathcal{G}_{\mathbf{k}_0}^U = \mu_1 \oplus \mu_2$), the zero-energy  solutions always split into a $\mu_1$ and a $\mu_2$ sector regardless of whether $\mu_1 \neq \mu_2$ or not, and the two sectors are related by $h\mathcal{T}$.
Following \cref{sec:continuity-unitary}, the $\mu_1$ space hosts $(M - \sum_i \langle \mathbf{1}, \mu_1 \otimes \sigma_i^* \otimes v \rangle) \dim(\mu_1)$ independent zero-energy solutions.
The total number of zero-energy states is twice this. 
Using $\dim(\rho) = 2\dim(\mu_1)$ and $\langle \mathbf{1}, \rho \otimes \sigma_i^* \otimes v \rangle = 2\langle \mathbf{1}, \mu_1 \otimes \sigma_i^* \otimes v \rangle$, the total kernel dimension becomes $(M - \frac{1}{2} \sum_i \langle \mathbf{1}, \rho \otimes \sigma_i^* \otimes v \rangle) \dim(\rho)$. 
Isolating the CTFB requires this to match the target dimension $Q \dim(\rho)$, yielding $M - Q = \frac{1}{2} \sum_i \langle \mathbf{1}, \rho \otimes \sigma_i^* \otimes v \rangle$. 
This result also implies that the $\sigma_i$ block-row should impose $\frac{1}{2}  \langle \mathbf{1}, \rho \otimes \sigma_i^* \otimes v \rangle \dim(\rho)$ independent constraints. 
Equating this with the dimension matching condition (\cref{eq:M-Q}) yields:
\begin{equation}
\sum_i \frac{1}{2} \langle \mathbf{1}, \rho \otimes \sigma_i^* \otimes v \rangle \dim(\rho) = \sum_i \dim(\sigma_i)\ .
\end{equation}
Finally, as in the unitary case, requiring the grouped coupling matrices (defined below) to be full-rank ensures this equality holds term by term, which recovers \cref{eq:criterion-main} with $q_\rho = 2$.

We summarize the construction of the grouped coupling matrices $\mathbb{A}(\pp)$. 
They must be full column-rank to validate the discussions above. 
Depending on $q_\rho, q_{\sigma_i}$, they are assembled from the unitary blocks as follows. Let $L_i = \inn{\mathbf{1}, \rho \otimes \sigma_i^* \otimes v}$.
\begin{enumerate}[leftmargin=0.4cm]
\item {$q_\rho=1, q_{\sigma_i}=1$:} The grouped coupling matrix $\mathbb{A}^{(i)}(\pp)$ is constructed identically to the unitary case, with dimensions $\dim(\sigma_i) \times L_i\dim(\rho)$. 

\item {$q_\rho=1, q_{\sigma_i}=2$:} $\rho$ is a single unitary irrep, but $\sigma_i$ decomposes as $\nu_1 \oplus \nu_2$. To evaluate independent constraints, we separate the block-rows into two grouped coupling matrices, $\mathbb{A}^{(i)\prime}(\kk)$ and $\mathbb{A}^{(i)\prime\prime}(\kk)$, for the $\nu_1$ and $\nu_2$ sectors, respectively. Each has dimensions $\frac{1}{2}\dim(\sigma_i) \times \frac{1}{2} L_i \dim(\rho)$.  

\item {$q_\rho=2, q_{\sigma_i}=1$:} $\rho$ decomposes as $\mu_1 \oplus \mu_2$, while $\sigma_i$ remains a single unitary irrep. 
To ensure that the $\sigma_i$ block-row imposes $\frac{1}{2}  \langle \mathbf{1}, \rho \otimes \sigma_i^* \otimes v \rangle \dim(\rho) = 2  \langle \mathbf{1}, \mu_1 \otimes \sigma_i^* \otimes v \rangle \dim(\mu_1) $ independent constraints, as discussed above, coupling matrix columns in the $\mu_1$ and $\mu_2$ sectors must be all linearly independent, we hence group the columns from both sectors into a single matrix $\mathbb{A}^{(i)}(\kk)$. 
It has dimensions $\dim(\sigma_i) \times \frac{1}{2} L_i \dim(\rho)$. 

\item {$q_\rho=2, q_{\sigma_i}=2$:} Both co-irreps decompose. Combining cases 2 and 3, we group the columns from the $\mu_{1,2}$ sectors but separate the rows into the $\nu_{1,2}$ sectors, yielding two grouped matrices, $\mathbb{A}^{(i)\prime}(\kk)$ and $\mathbb{A}^{(i)\prime\prime}(\kk)$. Each has dimensions $\frac{1}{2}\dim(\sigma_i) \times \frac{1}{4} L_i \dim(\rho)$. 
\end{enumerate}

The $\mathbb{Z}_2$ CTFB discussed in \cref{sec:topology-of-flat-bands} exemplifies case 1. 
For the touching point at $\Gamma$, $\rho = \bG_8$ and $\sigma_1=\bG_9$ are both two-dimensional, and $q_\rho=q_{\sigma_1}=1$.
Given $\inn{\mathbf{1}, \rho \otimes \sigma_1^* \otimes v}=1$, the criterion in \cref{eq:criterion-main} is satisfied.

\subsection{Enumeration of symmetry-indicated CTFBs}

\begin{table}[h]      
\centering            
\begin{tabular}{c|l r|l r}
\hline\hline
~ & \multicolumn{2}{c|}{without TRS} &  \multicolumn{2}{c}{with TRS} 
\\
\hline\hline
 \multirow{3}{*}{~ 2D} ~&~ $|C|=1$: & 6618 ~&~ Strong $\mathbb{Z}_2$: & 1394 ~ \\  
 ~&~ $|C|=2$: & 364 ~&~ Mirror-Chern: & 1536 ~\\
 ~&~ $|C|=3$: & 120 ~&~ &  ~ \\
\hline 
  \multirow{3}{*}{~ 3D} ~&~ Axion: & 25034 ~&~  Strong $\mathbb{Z}_2$: & 5612 ~ \\ 
  ~&~ Crystalline: & 7112 ~&~ Weak $\mathbb{Z}_2$: & 696 ~ \\ 
   &  &  &~ Crystalline: & 3174 ~ \\
\hline 
\end{tabular}
\caption{Statistics of identified symmetry-indicated CTFBs. 
The table summarizes the number of CTFBs (characterized by $\mathcal{B}+\Delta \mathcal{B}$) identified across 2D and 3D space groups, categorized by their topological invariants and the presence of time-reversal symmetry (TRS). 
In 3D, the ``Crystalline'' category includes mirror-Chern states and various higher-order topological states protected by spatial symmetries. 
}
\label{tab:statistics-main}
\end{table}

With the group-theoretical continuity criteria established, we can systematically enumerate symmetry-indicated CTFBs across all space groups. 
The automated search workflow begins by filtering space groups that support non-trivial SIs and host isotropic high-symmetry momenta where $\pp$ forms an irreducible real  representation $v$. 
At these discrete momenta, we identify all pairs of $\rho$ and $\sigma$ that satisfy \cref{eq:criterion-main} and the full column-rank condition of grouped coupling matrices. 
By enumerating and connecting these local band pieces across the Brillouin zone subject to compatibility relations, we obtain valid $\mclB + \Delta \mclB$ and select those where (i) $\mclB$ carries nontrivial SIs and (ii) $\mclB + \Delta \mclB$ has the form of $M\rho \boxminus (\bigoplus_{i=1}^N \sigma_i)$ at touching points. 
We then search for the minimal bipartite construction $\mathcal{BR}_L \boxminus \mathcal{BR}_{\td L}$ that realizes $\mclB + \Delta \mclB$. 

Applying this exhaustive algorithm up to a maximum cutoff for both the total and flat-band numbers, we have tabulated a comprehensive list of symmetry-indicated CTFBs, which we classify into ten categories based on topology, spatial dimension, and the presence of  TRS, as summarized in \cref{tab:statistics-main}. 
Readers may refer to SM \cite{SM} for the comprehensive list. 
For 2D systems without TRS, we consider single-valued wallpaper groups to identify CTFBs with various Chern numbers, where the SI modulo \cite{fang_bulk_2012} naturally selects the minimal Chern number without fine-tuning. 
We will further clarify the relation between SI and Chern number in \cref{sec:exapmles-Chern}. 
For 2D systems with TRS, we consider double-valued layer groups to identify CTFBs exhibiting the strong $\mathbb{Z}_2$ index and mirror-Chern numbers. 
In 3D, we consider double-valued space groups both in the presence and absence of TRS.  
These 3D CTFBs yield a richer variety of topological invariants, including half-quantized axion angle protected by spatial symmetries, strong and weak $\mathbb{Z}_2$ indices protected by TRS,  mirror-Chern numbers,  and diverse higher order topological invariants protected by spatial symmetries.

\section{More examples}
\label{sec:more-examples}

\begin{figure*}[t]
    \centering
    \includegraphics[width=1\linewidth]{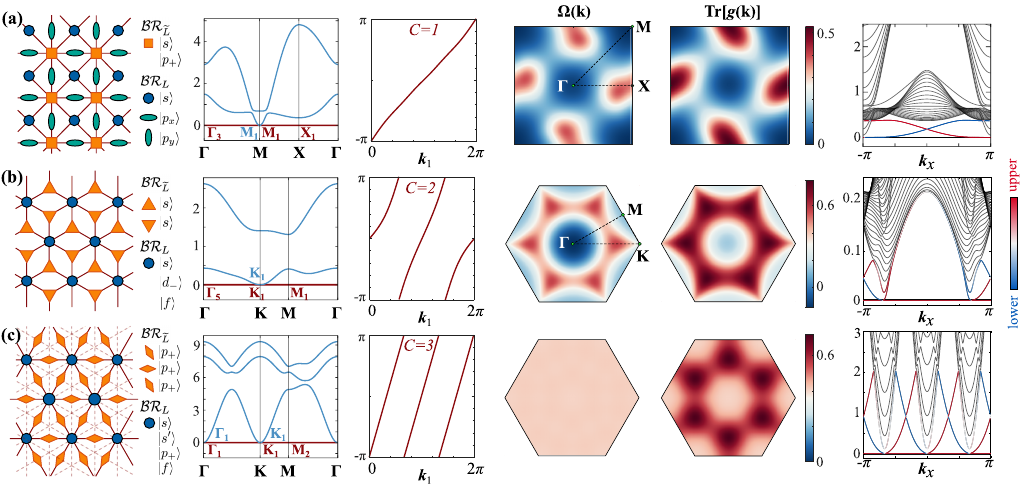}
    \caption{CTFBs with Chern numbers. 
    Three rows correspond to models with (a) $C=1$, (b) $C=2$, and (c) $C=3$. 
    Six columns from left to right display (1) the lattice structures, (2) band structures, (3) Wilson loop spectra, (4) Berry curvature $\Omega(\kk)$, (5) trace of FSM tensor $g(\kk)$, and (6) edge states in a stripe geometry.  
    The $C=1$ model respects the wallpaper group $p4$, and has CTFB characterized by $\mclB + \Delta\mclB =(\Gamma_3; \ 2\rM_1 \boxminus \rM_3; \ \rX_1)$. 
    The $C=2,3$ models respect $p6$, and have CTFBs $\mclB + \Delta\mclB =(\Gamma_5; \ 2\rK_1 \boxminus \rK_2; \ \rM_1)$ and $\mclB + \Delta\mclB =(2\Gamma_1 \boxminus \Gamma_4; \ 2\rK_1 \boxminus \rK_2; \ \rM_1)$, respectively. 
    In the (1)-st column, red bonds and dashed lines represent arbitrary symmetry-allowed hoppings. 
    Nevertheless, we remark that the dashed hoppings in (c) should not be fine-tuned to 0; otherwise, an extra $\rM_1$ irrep in the dispersive bands will touch with the flat band, which violates our CTFB target. 
    For all these models, a large $\Delta$ is chosen, so that bands from $\mathcal{BR}_{\td L}$ lie at negative energies outside the shown energy window. }
    \label{fig:Chern-main}
\end{figure*}

\subsection{2D CTFB with Chern numbers} \label{sec:exapmles-Chern}

We highlight the capacity of our framework to systematically realize CTFBs with various Chern numbers. 
With these examples, we also discuss the quantum geometric properties of CTFB, and showcase the high tunability. 

\cref{fig:Chern-main} displays symmetry-indicated Chern CTFBs that we construct with Chern numbers $C=1,2,3$, respectively, across wallpaper groups $p4$ and $p6$. 
In each case, the target Chern number is not only diagnosed by the SIs computed from the irreps at the high-symmetry points, but also definitely confirmed by the winding of the Wilson loop spectrum, the integration of Berry curvature $\Omega(\kk)$, and the existence of chiral boundary states. 
In \cref{sec:invariant}, we will further prove that the Chern numbers of these CTFBs can be analytically determined from the phase windings of $S(\mathbf{k})$-matrix without ambiguity. 

The $S(\kk)$-matrix naturally adopts the lowest order expansions in $\kk$, hence the minimal phase windings, thereby usually realizing the minimal Chern number in magnitude that is compatible with SI. 
Even so, higher Chern numbers can still be achieved in two ways. 
One is to fine-tune the $S(\kk)$-matrix to suppress the low-order windings. (See SM for an example \cite{SM}.) 
The other is to make use of the explicit expression for the Chern number (\cref{sec:invariant}) and introduce additional touching points, which allows higher Chern numbers to be realized {\it without} fine-tuning. 
We illustrate the latter construction with a concrete example realizing $C=3$ with a trivial SI in Sec.~\ref{sec:high-Chern-beyond-SI}.

For now, let us focus on the symmetry-indicated $C=3$ model in detail.
Constructions and analysis for the $C=1$ and $2$ models follow the same techniques, and can be found in SM \cite{SM}. 

In the wallpaper group $p6$, we choose the target band set as $\mclB= (\Gamma_1; \ \rK_1;\  \rM_2)$. 
The Chern number is determined by the $C_6$ indicator \cite{fang_bulk_2012}, 
\begin{equation} \label{eq:SI-Chern-C6}
e^{\ii \frac{\pi}{3} C} = \prod_{n \in \mathrm{occ}} \eta_n(\Gamma) \theta_n(\rK) \zeta_n(\rM) \ ,
\end{equation}
where $\eta$, $\theta$, and $\zeta$ denote the $C_6$, $C_3$, and $C_2$ eigenvalues of the occupied bands, respectively. 
As $\eta(\Gamma_1)=1$, $\theta(\rK_1)=1$, $\zeta(\rM_2) = -1$, the band set $\mclB$ exhibits $C=3$ mod 6. 
To realize it in the CTFB framework, we choose the augmented band set $\Delta\mclB$ so that $\mclB + \Delta\mclB= (2\Gamma_1\boxminus \Gamma_4;\  2\rK_1 \boxminus \rK_2;\  \rM_2)$. 
As $\eta(\Gamma_4) = e^{\ii\frac{\pi}3}$, $\theta(\rK_2) =e^{\ii\frac{2\pi}3} $, the augmented band set $\mclB + \Delta \mclB$ is consistent with $C=0$, and hence can be realized as a formal difference of two band representations $\mathcal{BR}_L \boxminus \mathcal{BR}_{\td L}$ (\cref{fig:Chern-main}(c)). 
The continuity criterion in \cref{eq:criterion-main} is met around all the three touching points, $\Gamma$, $\rK$ and $-\rK$. 
In particular, only the $k_x + \ii k_y$ combination in $v$ can survive in the direct product, and appear in the leading order expansion of $S(\kk)$-matrix. 

One specific lattice realization of $\mathcal{BR}_L \boxminus \mathcal{BR}_{\td L}= \mclB + \Delta\mclB$ is depicted in \cref{fig:Chern-main}(c), the $L$ sublattice is at the $1a$ Wyckoff position (the triangular lattice) with four orbitals: two $s$ orbitals, one $f$ orbital, and one $p_+=p_x + \ii p_y$ orbital. 
The $\td L$ sublattice is at the $3c$ Wyckoff positions (the Kagome lattice) with one $p_+$ orbital per site. 
They form the band representations $\mathcal{BR}_L = [2A\oplus B \oplus {}^1\!E_2]_{1a} \uparrow \mathcal{G}$, and $\mathcal{BR}_{\td L} = [B]_{3c} \uparrow \mathcal{G}$, respectively. 
For this model, we allow arbitrary non-vanishing hoppings as depicted in \cref{fig:Chern-main}(c). 
A special note is that, if the dashed (next-nearest-neighbor) hoppings in \cref{fig:Chern-main}(c) were tuned to zero, the otherwise finite gap at $\rM$ point would close \cite{SM}, and an undesired $\rM_1$ irrep from the dispersive band will touch with the flat band $\rM_2$ irrep. 
Such extra touching leads to singularity, as replacing the target $\rM_2$ with the accidental $\rM_1$ will cause the SI to flip to $C=0$ mod $6$. 
The extra touching is also incompatible with our target $\mclB + \Delta\mclB$. 
Therefore, we keep the dashed hoppings finite, albeit arbitrary, in the following discussion. 
As long as the gap at $\rM$ is finite, so that the target CTFB is realized, its Chern number will be robustly $C=3$, against any symmetry-preserving perturbations.

Constrained by the $C_6$ symmetry, the $j$-th column of $S(\kk)$ is given by 
\begin{equation} \label{eq:Sk-p6-C3}
    \pare{\begin{array}{r}
         2\ii \,t_1 \sin(\kk\cdot\boldsymbol{\tau}_j) 
         + 2\ii \,t_1' \sin(\kk\cdot\boldsymbol{\tau}_j')   \\
         2\ii \,t_2 \sin(\kk\cdot\boldsymbol{\tau}_j) 
         + 2\ii \,t_2' \sin(\kk\cdot\boldsymbol{\tau}_j') \\
        (-1)^{j-1}[2t_3 \cos(\kk\cdot\boldsymbol{\tau}_j) 
         + 2t_3' \cos(\kk\cdot\boldsymbol{\tau}_j')] \\
        e^{-\ii\frac{\pi}3(j-1)}[2t_4 \cos(\kk\cdot\boldsymbol{\tau}_j) 
         + 2t_4' \cos(\kk\cdot\boldsymbol{\tau}_j')] 
    \end{array}}\ ,
\end{equation}
where 
$\boldsymbol{\tau}_1 = \frac12 \mathbf{a}_1$, 
$\boldsymbol{\tau}_2 = \frac12 \mathbf{a}_1 + \frac12 \mathbf{a}_2$,
$\boldsymbol{\tau}_3 = \frac12 \mathbf{a}_2$ are the nearest $3c$ neighbors of the origin, and 
$\boldsymbol{\tau}_1' = \frac12 \mathbf{a}_1 + \mathbf{a}_2$, 
$\boldsymbol{\tau}_2' = - \frac12 \mathbf{a}_1 + \frac12 \mathbf{a}_2$, 
$\boldsymbol{\tau}_3' =-\mathbf{a}_1 -\frac12 \mathbf{a}_2$ are the next-nearest $3c$ neighbors of the origin. 
$\mathbf{a}_{1,2}$ are chosen as the same as in \cref{sec:Z2-model}. 
$t_{1,2,3,4}$ ($t_{1,2,3,4}'$) represent the hopping amplitudes from the $p_+$ orbital at $\boldsymbol{\tau}_1$ ($\boldsymbol{\tau}_1'$) to the four orbitals at the $1a$ position at the origin, respectively. 
The band structures and Wilson loop is depicted in \cref{fig:Chern-main}(c), the second to third panels, which shows the touching, and the well-defined Wilson loop winding, respectively.

In studying the strongly correlated states in a Chern band, a central concept is the (holomorphic) ``idealness'' of quantum geometry \cite{roy_2014_bandgeometryofFCI, ledwith_fractional_2020, wang_2021_exactlandaulevel, yu_2025_universalwilsonloop, yu_quantum_2025}, which facilitates the mapping of the interacting physics in the Chern band to the lowest Landau level (LLL). 
A flat Chern band is said to be (holomorphically) ``ideal'' if the trace inequality $|\Omega(\kk)| \le \Tr g(\kk)$ between the Berry curvature $\Omega(\kk)$ and the Fubini-Study metric (FSM) tensor $g(\kk)$ is saturated, and the Berry curvature $\Omega(\kk)$ is uniformly distributed across the Brillouin zone. 
It is thus interesting to investigate the quantum geometric properties of a Chern CTFB. 

Notably, the quantum geometric tensor (QGT) depends on the first-order derivatives of the projector, $\frak{g}_{ij}(\kk) = \mathcal{V}^\dagger(\kk) \left( \partial_{k_i} P(\kk) \right) \left( \partial_{k_j} P(\kk) \right) \mathcal{V}(\kk)$, where $\mathcal{V}(\kk)$ collects the CTFB Bloch states, and $P(\kk) = \mathcal{V}(\kk) \mathcal{V}(\kk)^\dagger$ is the corresponding projector.
The continuity of $\partial_{k_i} P(\kk)$ has not been enforced by our construction. 
Nevertheless, $\partial_{k_i} P(\kk)$ is not divergent near the touching points, which is thus able to ensure the global topology is well-defined (see \cref{sec:invariant}). 
Moreover, for Chern CTFB, we explicitly verify that both $\Omega(\kk)$ and $\Tr g(\kk)$ are still continuous around the touching points, even if the components of QGT may be discontinuous, which can be evidenced from \cref{fig:Chern-main}, with more analysis provided in SM \cite{SM}. 
Furthermore, for touching points with higher symmetries to forbid the $\mathcal{O}(k^2)$ couplings in $S(\kk)$, including the touching points discussed in \cref{sec:Z2-model}, the $\rM$ point in the $p4$ model (\cref{fig:Chern-main}(a)), and the $\Gamma$ point in the $p6$ model (\cref{fig:Chern-main}(c)), we can prove that $\partial_{k_i} P(\kk)$, hence the full QGT, will also be continuous component-wise. 
Therefore, the quantum geometric properties of CTFBs highly resemble the gapped topological bands. 
On the contrary, singular flat bands would feature divergent quantum geometric properties at the touching points, which leads to the ill-defined topology. 

Compared to the gapped topological bands, nonetheless, CTFB enjoys a unique advantage in the tunability of the quantum geometric effects, as its model parameters can be arbitrarily tuned while maintaining the exact flatness and the robust topology. 
Therefore, the tuning and the investigation of the quantum geometric effects can be neatly isolated. 
As a proof of principle, we demonstrate the optimization of QGT in the Chern CTFB with respect to the following two types of goals, which are related to the (holomorphic) ``idealness''. 
One is to minimize the deviation of the trace condition, 
\begin{align}
    \mathcal{L}_{\rm tr}[\mathbf{t}] = \frac{\int_{\mathrm{BZ}} \dd^2\kk \Big| \Tr \, g(\mathbf{k}; \mathbf{t}) - |\Omega(\mathbf{k}; \mathbf{t})|  \Big| }{\int_{\mathrm{BZ}} \dd^2\kk|\Omega(\mathbf{k}; \mathbf{t})| } 
\end{align}
where $\mathbf{t}$ denotes the collection of hopping parameters. 
We have optimized this target for the $C=1$ model (\cref{fig:Chern-main}(a)), and find $\mathcal{L}_{\rm tr} \sim 0.23$, which is good among lattice models with finite-range hoppings \cite{jackson_geometric_2015}. 
The result will be further improved if more hopping terms are turned on. 
Nevertheless, we remark that for generic models, minimizing $\mathcal{L}_{\rm tr}$ alone may lead to other pathological features, such as extreme concentration of Berry curvature or undesired decreasing of the average energy gap to remote bands. 
Therefore, in general extra constraints will be needed to avoid these issues. 
Another goal is to minimize the fluctuation of Berry curvature
\begin{equation}
    \mathcal{L}_\Omega[\mathbf{t}]
    = \sqrt{
    \int_{\mathrm{BZ}} \dd^2\kk
    \left(
    \Omega(\mathbf{k};\mathbf{t})-\frac{2\pi C }{A_{\rm BZ}}
    \right)^2 }
\end{equation}
which we apply to the $C=2$ and $C=3$ models (\cref{fig:Chern-main}(b,c)). 
Remarkably, due to the existence of more hopping parameters, the $C=3$ model is able to achieve an almost perfectly uniform distribution, while maintaining a mild trace-condition deviation of $\mathcal{L}_{\rm tr} \sim 0.29$. 

In the last column of \cref{fig:Chern-main}, we further demonstrate the energy boundary states in a stripe geometry for the $C=1,2,3$ models. 
$|C|$ chiral states appear for each boundary, which are well separated from the bulk bands, signifying that the bulk-boundary correspondence remains intact despite the touching points, which is another feature that distinguishes CTFB from the singular flat bands. 
Such bulk-boundary correspondence can also be resolved via calculating the entanglement spectrum \cite{SM}.

To close this section, we finally remark that some existing models fall in the category of Chern CTFB, including the $2\pi$-flux dice lattice model built for $\rm tMoTe_2$ \cite{yang_fractional_2025}. 
It is a $C=1$ symmetry-indicated CTFB in the space group $p3$, with $\mathcal{BR}_L=[^1E]_{1b} + [^1E]_{1c}$ and $\mathcal{BR}_{\td{L}} = [A]_{1a}$. 
Besides this, the critical Chern band constructed in Dubail and Read's work can also be cast into a fine-tuned Chern CTFB \cite{dubail_tensor_2015}, which we discuss in SM \cite{SM}.

\subsection{3D CTFB with strong \texorpdfstring{$\mathbb{Z}_2$}{Z2} index}   \label{sec:examples-3DZ2}

\begin{figure}[th]
    \centering
    \includegraphics[width=1\linewidth]{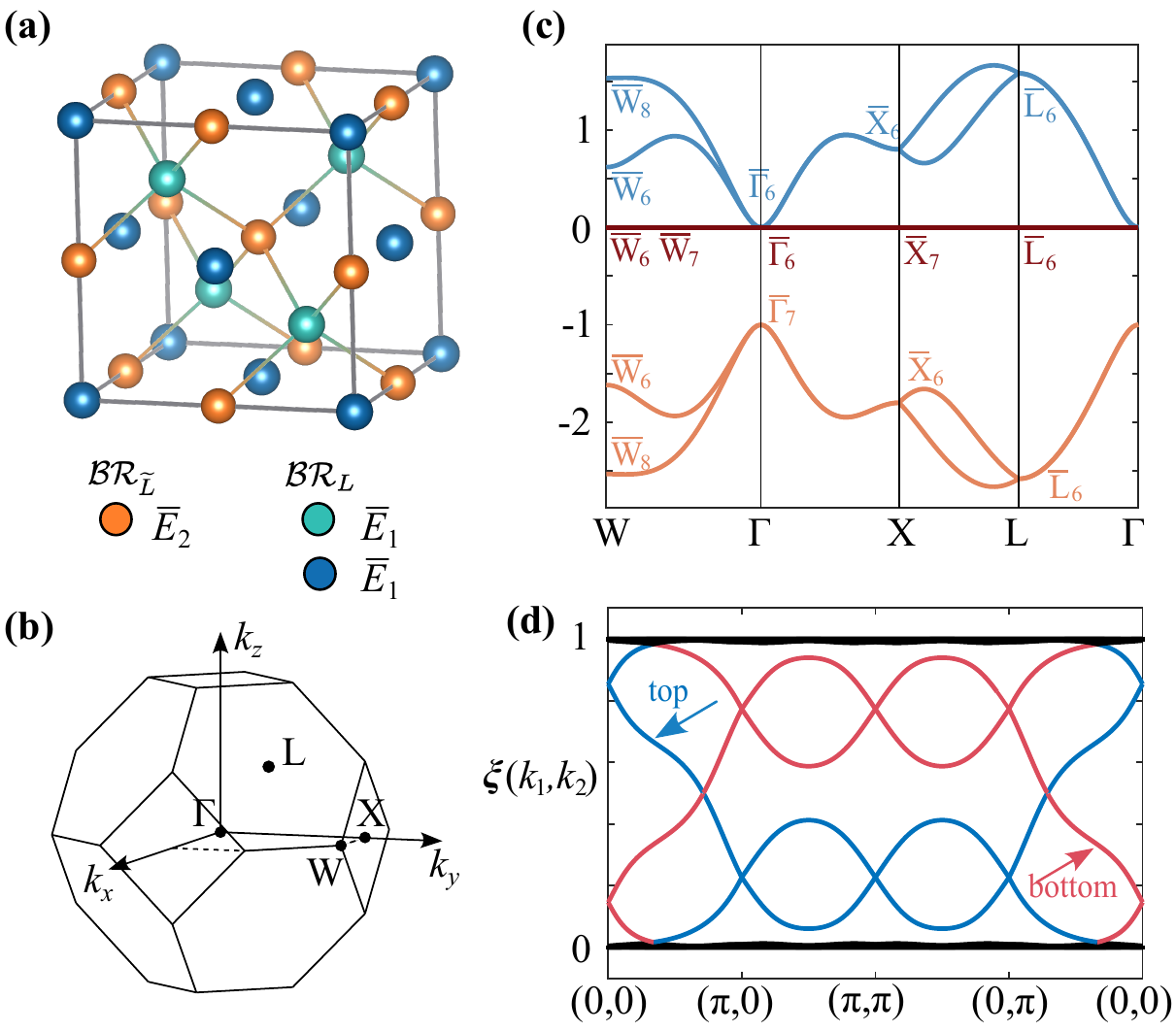}
    \caption{3D CTFB with strong $\mathbb{Z}_2$ index in space group $F\bar4 3m$. 
    (a) and (b) are the lattice structure and Brillouin zone, respectively. 
    (c) shows a band structure with representative parameters. 
    (d) is the entanglement spectrum of the Fock state $\ket{\Omega}$ occupying CTFB and lower bands with spatial cut parallel to the $\bar111$ plane. 
    $k_1$ and $k_2$ are the perpendicular momenta. 
    The blue and red spectra represent the topological boundary modes on the top and bottom surfaces, respectively.  
    }
    \label{fig:3DTI-main}
\end{figure}

A key advantage of our framework is that it can be directly applied to arbitrary dimensions, especially 3D. 
Here, we present the first construction of a 3D exactly flat band exhibiting a strong topology. 
In the presence of TRS, it possesses the 3D strong $\mathbb{Z}_2$ TI index. 
When TRS is broken, it degrades into an axion CTFB---the axion insulator with critical touching points---that can exhibit higher-order chiral hinge states. 

We consider the double-valued space group $F\bar{4}3m$, and assume the presence of TRS first. 
Its face-centered Bravis lattice is generated by $\mathbf{a}_1=(\frac12,\frac12,0)$, $\mathbf{a}_2=(\frac12,0,\frac12)$, $\mathbf{a}_3=(0,\frac12,\frac12)$, and its BZ is shown in \cref{fig:3DTI-main}(b). 
We target a flat band characterized by the symmetry data $\mathcal{B} = (\bG_6; \ \ovl{\rX}_7; \ \ovl{\rm L}_6; \ \ovl{\rm W}_6 \oplus \ovl{\rm W}_7)$. 
The topology of $\mclB$ can be diagnosed by the SI \cite{song2018quantitative,khalaf_symmetry_2018,elcoro_magnetic_2021},
\begin{equation}
z_2 = \sum_{\mathbf{K}} \Big( \frac{1}{2}n_{\mathbf{K}}^{\frac32} - \frac{1}{2}n_{\mathbf{K}}^{\frac12} \Big) \pmod 2 \ ,
\end{equation}
where $\mathbf{K}$ sums over the $S_{4z}$-invariant momenta (here, $\Gamma$ and the equivalent $\rX$ point along the $001$ direction). 
The integers $n_{\mathbf{K}}^{\frac12}$ and $n_{\mathbf{K}}^{\frac32}$ denote the number of Kramers pairs with $S_{4z}$ rotation eigenvalues $e^{\pm \ii \frac{\pi}4}$ and $e^{\pm \ii \frac{3\pi}4}$, respectively. 
Specifically, $\bG_6$ and $\ovl{\rm X}_7$ contribute to $n_{\mathbf{K}}^{\frac12}$, while $\bG_7$ and $\ovl{\rm X}_6$ contribute to $n_{\mathbf{K}}^{\frac32}$. 
Applying this formula to $\mclB$ yields $z_2 = 1$, indicating a strong $\mathbb{Z}_2$ topology protected by TRS.

We trivialize the band topology by adding the zero-band correction $\Delta\mathcal{B} = (\ovl{\Gamma}_6 \boxminus \ovl{\Gamma}_7; \ 0; \ 0; \ 0)$, where $\bG_7$ have different $S_{4z}$ eigenvalues with $\bG_6$ and hence compensate the $z_2$ SI. 
This trivialized augmented band $\mathcal{B} + \Delta\mathcal{B}$ can directly translate into the real-space bipartite configuration $\mathcal{BR}_L \boxminus \mathcal{BR}_{\td L}$ (\cref{fig:3DTI-main}(a)). 
Specifically, the $L$ sublattice comprises an $\ovl{E}_1$ doublet ($j=\frac12$ states) at the $4a$ position $(0,0,0)$, alongside an $\ovl{E}_1$ doublet at the $4c$ position $(\frac14,\frac14,\frac14)$. 
The $\td L$ sublattice comprises an $\ovl{E}_{2}$ doublet at the $4b$ position  $(\frac12,\frac12,\frac12)$, which transform identically as $\ovl{E}_1$ under proper rotations and acquires an additional minus sign under improper rotations. 
Thanks to the high symmetry, restricting $S(\kk)$ to nearest-neighbor hoppings yields a very simple form:
\begin{equation}
    S(\kk) = \begin{pmatrix}
    \ii t_1 \sum_{\boldsymbol{\tau}} e^{-\ii \kk \cdot \boldsymbol{\tau}} \boldsymbol{\tau} \cdot \boldsymbol{\sigma} \\
    \ii t_2 \sum_{\boldsymbol{\tau}'} e^{-\ii \kk \cdot \boldsymbol{\tau}'} \boldsymbol{\tau}' \cdot \boldsymbol{\sigma} 
    \end{pmatrix}\ .
\end{equation}
Here, $\boldsymbol{\tau} = (\pm\frac12,0,0)$, $(0,\pm\frac12,0)$, $(0,0,\pm\frac12)$ sum over the relative vectors from a $4b$ site to its nearest $4a$ neighbors, and $\boldsymbol{\tau}' = (-\frac14,-\frac14,-\frac14)$, $(-\frac14,\frac14,\frac14)$, $(\frac14,-\frac14,\frac14)$, $(\frac14,\frac14,-\frac14)$ sum over the relative vectors to its nearest $4c$ neighbors. 
The vector of Pauli matrices is denoted by $\boldsymbol{\sigma} = (\sigma_x, \sigma_y, \sigma_z)$, and the real parameters $t_{1,2}$ represent the respective hopping amplitudes.
The band structure with representative parameters ($t_1=1$, $t_2=1.2$, $\Delta=1$) is shown \cref{fig:3DTI-main}(c). 

Without any fine-tuning, this bipartite model robustly yields a 3D strong $\mathbb{Z}_2$ CTFB. 
The nontrivial topology is confirmed by the entanglement spectrum (\cref{fig:3DTI-main}(d)), which reveals gapless boundary modes featuring a single Dirac cone on a given surface---the hallmark of a 3D strong topological insulator.
It is remarkable that such an exact 3D strong topological flat band can be realized with merely a six-band bipartite construction.
Once the $\td L$ sublattice is integrated out, as detailed in \cref{sec:invariant}, the resulting effective Hamiltonian comprises only four bands. 

In SM~\cite{SM}, we provide another example of 3D CTFB with the strong $\mathbb{Z}_2$ topology in the space group $P\bar43m$. 

When TRS is broken, the $z_2$ index will persist, and serve to indicate the bulk axion angle $\theta$, which is still quantized into $0$ or $\pi$ by crystalline symmetries including $\ovl{4}$. 
For a generic surface that no longer preserves any crystalline symmetries, \textit{e.g.} the $(210)$ surface, it typically exhibits a gap with half-quantized Hall conductance. 
If four surfaces are cleaved to respect the $\ovl{4}$ symmetry, then $\ovl{4}$ will dictate the Hall conductance on the four surfaces to alternate in signs, hence chiral states will emerge on the hinges between them. 
This leads to a higher-order topological insulator. 
Many of the enumerated CTFBs in \cref{tab:statistics-main} also exhibit higher-order topology. 
Interested readers can obtain the complete topological invariant information by consulting the SIs of these CTFBs listed in SM \cite{SM} together with the correspondence between SIs and topological invariants established in Refs.~\cite{song2018quantitative, khalaf_symmetry_2018, elcoro_magnetic_2021}.

\section{Singular Wannier gauge and topological invariants}
\label{sec:invariant}

In this section, we formulate the topological invariants directly in terms of the bipartite hopping matrix $S(\kk)$. 
For simplicity, we now consider the effective Hamiltonian for the $L$-sublattice, 
\begin{equation} \label{eq:Heff}
H_L(\kk) = \frac{S(\kk) S^\dagger (\kk)}{\Delta} \ ,
\end{equation}
obtained by ``integrating out'' the $\td L$-sublattice. 
It hosts the same zero-energy CTFBs as the full Hamiltonian (\cref{eq:Hk-main}), since in both cases the FBs are given by the kernel of $S^\dagger(\kk)$. 
An orthonormal basis for the space spanned by dispersive bands can be constructed as $\mU(\kk) = S(\kk) \mathcal{Q}^{-\frac12}(\kk)$, where $\mathcal{Q}(\kk) = S^\dagger(\kk) S(\kk)$. 
If $\mathcal{Q}^{-\frac{1}{2}}(\kk)$ were non-singular over the entire BZ, $\mU(\kk)$ would Fourier transform to exponentially localized Wannier functions that transform as $\mathcal{BR}_{\td L}$ \cite{soluyanov_wannier_2011}. 
However, for CTFB models, the Wannierization of dispersive bands necessarily fails at the touching points where the representation of $\td L$ is not supported by $L$.
This obstruction is signified by negative components in $\mathcal{BR}_L \boxminus \mathcal{BR}_{\td L}$. 
Nevertheless, $\mU(\kk)$ provides a smooth gauge everywhere else. 
We refer to $\mU(\kk)$ as the singular Wannier gauge of the dispersive bands. 
It corresponds to Wannier functions with a power-law decay in real space.

\subsection{Chern number}

We first derive the expression for Chern number $C$ of a CTFB. 
Since $\mU(\kk)$ is smooth over the whole Brillouin zone except for the touching points, we can apply the Stokes' theorem to express the Chern number of dispersive bands in terms of Berry phases surrounding these touching points, {\it i.e.,} $-\frac1{2\pi} \sum_i \ointctrclockwise_{\partial \mathcal{D}_i} \dd\kk \cdot \mathbf{A}(\kk)$, where  $\mathcal{D}_i$ denotes an infinitesimal disk enclosing the $i$-th touching point, and $\mathbf{A}(\kk) = \ii \Tr[\mU^\dagger \partial_\kk \mU]$ is the
Berry connection of dispersive bands.
After a few steps of derivation, we find that $\mathbf{A}(\kk)$ can be directly expressed in terms of $S(\kk)$:
\begin{equation} \label{eq:A-S}
\mathbf{A}(\kk)  =  \ii \, \Tr\brak{  (S^\dagger S)^{-1} S^\dagger \partial_\kk S } 
- \frac{\ii}2 \partial_\kk \ln \det \mathcal{Q} \ .
\end{equation}
Since the topological invariants of all bands must sum to zero, the Chern number $C$ of the CTFB is opposite to that of dispersive bands.
Expressed in terms of $S(\kk)$, we obtain 
\begin{equation} \label{eq:C-Sk-main}
C = \frac{\ii}{2\pi} \sum_i \ointctrclockwise_{\partial \mathcal{D}_i} \dd\kk \cdot 
    \Tr[ (S^\dagger S)^{-1} S^\dagger \partial_\kk S ] \ .
\end{equation}
Notice that $\det \mathcal{Q}(\kk)$ in the second term of \cref{eq:A-S} is positive and  single-valued; therefore, $\partial_\kk \ln \det \mathcal{Q}(\kk) $ does not contribute to the Berry phase along $\partial \mathcal{D}_i$. 
For a CTFB, the loop integral in \cref{eq:C-Sk-main} around each touching point $i$ will individually quantize, and is referred to as the winding of $S(\kk)$-matrix, which will be shown below. 
Note that an un-quantized winding would imply that the Berry curvature at the touching points diverges, and can only occur in singular FBs. 

We now apply \cref{eq:C-Sk-main} to the $C=3$ Chern CTFB introduced in \cref{sec:exapmles-Chern} (\cref{fig:Chern-main}(c)).
The band touching at $\Gamma$ is governed by the representation $2 \Gamma_1 \boxminus \Gamma_4$, where the irreps $\Gamma_1$ and $\Gamma_4$ carry $C_6$ eigenvalues of $1$ and $e^{\ii\frac{\pi}{3}}$, respectively. 
Projected onto this basis, the relevant low-energy block of $S^\dagger(\kk)$ (\cref{eq:Sk-p6-C3}) must take the form
\begin{equation} \label{eq:Sk-winding-C=1}
S^\dagger(\pp+\Gamma) = \begin{pmatrix}
\gamma_1 (p_x+\ii p_y) & \gamma_2 (p_x+\ii p_y) 
\end{pmatrix} \ ,
\end{equation}
where the two columns transform as $2\Gamma_1$, the single row transforms as $\Gamma_4$, and $\gamma_{1,2}$ are model-dependent parameters. This form is solely determined by the $C_6$ symmetry. 
Higher-order expansions in $p$ can be omitted, as their contribution vanishes for the infinitesimal disk. 
Expressing the momentum in local polar coordinates as $p_x + \ii p_y = p e^{\ii \theta}$, we have
\begin{equation}
\Tr[ (S^\dagger S)^{-1} S^\dagger \partial_\theta S ] = -\ii \ .
\end{equation}
Substituting this into \cref{eq:C-Sk-main} yields a $+1$ contribution. 
At the other two touching points $\rK$ and $-\rK$, which are related via a $C_2$ action to each other, the low-energy $S^\dagger(\kk)$-matrices take the identical form as \cref{eq:Sk-winding-C=1}, which is enforced by the $C_3$ symmetry, hence contribute the identical windings. 
Adding them up, we get $C=3$, consistent with the Wilson loop calculation.

\subsection{TRS-protected \texorpdfstring{$\mathbb{Z}_2$}{Z2} index}
\label{sec:invariant-Z2}

We next express the TRS-protected $\mathbb{Z}_2$ invariant in 2D in terms of $S(\kk)$. 
Suppose TRS acts on $L$ and $\td L$ as $\mathcal{T} = D K$ and $B K$, respectively, where ${K}$ is complex conjugation and $DD^* = BB^* = -1$. 
Then $S(\kk)$ satisfies $S(-\kk) = D S^*(\kk) B^\dagger$. 
The orthonormal states $\mU(\kk)$ satisfy the same condition, {\it i.e.,} $\mU(-\kk) B = D \mU^*(\kk)$, where $B$ on the left-hand side can be viewed as the TRS sewing matrix. 
In a gauge where the sewing matrix is $\kk$-independent, the $\mathbb{Z}_2$ invariant manifests as an obstruction to Stokes' theorem over half of the BZ (denoted as $\mathcal{M}$) \cite{fu_time_2006}:
\begin{equation} \label{eq:Z2-S125}
\delta  = \frac{1}{2\pi} \left( \oint_{\partial \mathcal{M}} \dd\kk \cdot \mathbf{A}(\kk) - \int_{\mathcal{M}} \dd^2\kk \, \Omega(\kk) \right) \mod 2 \ ,
\end{equation}
where $\mathbf{A}$ is defined in \cref{eq:A-S} and $\Omega(\kk)$ is the corresponding Berry curvature. 
Because Stokes' theorem is valid only in regions away from the touching points, {\it i.e.,} $\mathcal{M} -  \sum_i (\mathcal{M} \cap \mathcal{D}_i)$, and the Berry curvature is everywhere finite, the second term in the equation above can be recast as the sum of Berry phases along the boundaries $\partial(\mathcal{M} - \sum_i \mathcal{M} \cap \mathcal{D}_i)$. This yields
\begin{equation} \label{eq:Z2-index-main}
    \delta = \frac{\ii}{2\pi} \sum_i \ointctrclockwise_{\partial (\mathcal{M} \cap \mathcal{D}_i)} \!\!\! \dd\kk \cdot 
    \Tr[ (S_{\boldsymbol{\eta}}^\dagger S_{\boldsymbol{\eta}})^{-1} S_{\boldsymbol{\eta}}^\dagger \partial_\kk S_{\boldsymbol{\eta}} ] \mod 2 .
\end{equation}
Here, $S_{\boldsymbol{\eta}}(\kk) = S(\kk+\ii\boldsymbol{\eta}) $ is an analytical continuation of $S(\kk)$ with $\boldsymbol{\eta} = (\eta_x, \eta_y)$ being an infinitesimal vector in a generic direction.
In practical evaluations, the limit $\boldsymbol{\eta} \!\to\! 0$ must be taken before shrinking the disks $\mathcal{D}_i$ to zero.

We have introduced $S_{\boldsymbol{\eta}}(\kk)$ to regularize touching points located at TRIMs, where $\det[S^\dagger(\kk)S(\kk)] \!=\! 0$.
By expanding $S(\kk)$ as $\sum_n \sum_{i_1 \cdots i_n} S^{(i_1 i_2 \cdots i_n)} k_{i_1} k_{i_2} \cdots k_{i_n}$, TRS enforces the condition $D [S^{(i_1 \cdots i_n)}]^* B^\dagger = (-1)^n A^{(i_1 \cdots i_n)}$. 
Using this expansion, one can directly verify that $S(-\kk + \ii\boldsymbol{\eta}) = D S^*(\kk + \ii\boldsymbol{\eta}) B^\dagger$, meaning that the regularization preserves TRS. 
Furthermore, since $S(\ii\boldsymbol{\eta})$ shares the same kernel as $\lim_{\kk \to 0} S(\kk)$, $S(\kk + \ii\boldsymbol{\eta})$ provides a well-defined construction for a TRS-symmetric critical FB. 
The vector $\boldsymbol{\eta}$ lifts the zeros of $S^\dagger S$ at the TRIMs by slightly breaking crystalline symmetries, thereby validating \cref{eq:Z2-index-main}.

We now apply \cref{eq:Z2-index-main} to the $\mathbb{Z}_2$ CTFB discussed in \cref{sec:Z2-model}.
We choose $\mathcal{M}$ to be the lower half of the hexagonal Brillouin zone (\cref{fig:Z2-main}(c)). 
For the single band touching point at $\Gamma$, we define $\mathcal{D}$ as a disk of radius $\varepsilon$. 
The boundary of their intersection decomposes as $\partial(\mathcal{M} \cap \mathcal{D}_1) = C_1 \cup C_2$, where $C_1$ is a line segment from $(\varepsilon, 0)$ to $(-\varepsilon, 0)$, and $C_2$ is a semi-circular arc of radius $\varepsilon$ in the lower half-plane. 
The low-energy block of $S^\dagger(\kk)$ relevant to band touching is given in \cref{eq:Sk-Z2-main}.
It is straightforward to see
\begin{equation}
\lim_{\varepsilon\to 0^+} \lim_{\eta\to 0} \frac{1}{2\pi} \int_{C_2} \!\! \dd\kk \cdot \ii \Tr \big[ (S^\dagger_{\boldsymbol{\eta}} S_{\boldsymbol{\eta}})^{-1} S_{\boldsymbol{\eta}}^\dagger \partial_{\kk} S_{\boldsymbol{\eta}}\big] = 0 \ ,
\end{equation}
because the two columns of $S(\kk)$ contribute opposite phase windings. 
Introducing the shift $\boldsymbol{\eta} = (\eta, 0)$, we evaluate the integral along $C_1$ as
\begin{equation}
\delta = \lim_{\varepsilon\to 0^+} \lim_{\eta\to 0} \frac{1}{2\pi} \int_{\varepsilon}^{-\varepsilon} dk \, \frac{2\ii}{k + \ii\eta}  \mod 2 = 1 \ .
\end{equation}
In this example, $\boldsymbol{\eta}$ shifts the zeros of the first and second columns of $S(\kk)$ to $\kk = (0, -\eta)$ and $\kk = (0, \eta)$, respectively.

The formula \cref{eq:Z2-index-main} can also be used to calculate the $\mathbb{Z}_2$ index for 3D CTFBs, which is defined by the difference of $\delta$'s at $k_z=\pi$ and $k_z=0$ planes.

The expressions for topological invariants in \cref{eq:C-Sk-main,eq:Z2-index-main} are particularly useful when SIs are insufficient to fully determine the topological invariants. 
For example, the SI only constrains the Chern number of the model in \cref{sec:exapmles-Chern} to $C=3 \pmod 6$, but \cref{eq:C-Sk-main} unambiguously reveals $C=3$. 

\section{Criticality of CTFB}   \label{sec:criticality}

In this section, we analyze the critical correlations of CTFB. 
The projector to the CTFB is given by $P(\kk) = \mathbb{I} - S(\kk) \mathcal{Q}^{-1}(\kk) S^\dagger(\kk)$. 
Here, $\mathbb{I}$ is the projector to the $\mathcal{BR}_L$ orbitals, and $S(\kk) \mathcal{Q}^{-1}(\kk) S^\dagger(\kk) = \mathcal{U}(\kk) \mathcal{U}(\kk)^\dagger$ is the projector to dispersive bands, following the notations of singular Wannier gauge in the previous section. 
Nonetheless, we remind that the projector $P(\kk)$ is gauge-independent. 
While $S(\kk)$ is an analytic function of $\kk$, the singularity of $\mathcal{Q}^{-1}(\kk)$ at touching points renders $P(\kk)$ non-analytic. 

The correlation functions are given by the Fourier transformation of $P(\kk)$. 
In general, a discontinuity in the $n$-th order derivatives of $P(\kk)$ Fouriers to a long-range correlation tail whose amplitude decays as $r^{-n-d}$ in the $d$ dimension \cite{SM}. 
The position of the touching point $\kk_0$, on the other hand, reflects as the phase oscillation with momentum $\kk_0$ in these long-range tails. 

To be concrete, consider the $\mathbb{Z}_2$ CTFB model in \cref{fig:Z2-main}(b). 
Based on the $S^\dagger(\kk)$ matrix in \cref{eq:Sk-Z2-main}, we have $\mathcal{Q}(\kk) = (\gamma_1^2 + \gamma_2^2) |\kk|^2  \sigma_0 + \mathcal{O}(k^4)$. 
Beyond the first-order terms in \cref{eq:Sk-Z2-main}, higher-order contributions are generically present, with their specific forms determined by symmetry. 
For example, the next leading order of $k$ contains an element of $S_{11}^\dagger(\kk) = \gamma_1'k_+^3 + \gamma_1''k_-^3$. 
Cross terms between these third-order and the first-order terms generate non-analytic contributions in the projector $P(\kk)$, such as $\frac{\gamma_1'\gamma_1}{\gamma_1^2 + \gamma_2^2} \frac{k_+^4}{|\kk|^2}$. 
These terms possess discontinuous second-order derivatives and are responsible for the $r^{-4}$ behavior of the correlation function shown in \cref{fig:Z2-main}(g).

The power-law criticality in the correlation function also implies a gauge-independent lower-bound to the long-range behavior of the orthogonal Wannier functions. 
This is manifest when we express the correlation function in the Wannier representation \cite{SM}, 
\begin{align}  \label{eq:C_via_WW}
    C_{\alpha \beta}(\mathbf{r}) = \sum_{\RR} \sum_{m \in \rm occ} W^*_{\alpha m}(\RR) W_{\beta m}(\mathbf{r}+\RR)
\end{align}
where $W_{\alpha m}(\RR)$ are the orthogonal Wannier functions constructed from the filled Bloch bands. 
There is a gauge degree of freedom in Wannierization, while the correlation function is gauge-invariant. 
\cref{eq:C_via_WW} then implies $\left|C_{\alpha \beta}(\mathbf{r})\right| \le \sum_{\RR} \sum_{m \in \rm occ} \left|W^*_{\alpha m}(\RR)\right| \left|W_{\beta m}(\mathbf{r}+\RR)\right|$, which heuristically implies that short-range Wannier functions cannot cause a long-range correlation. 

More concretely, if some gauge choice produces a set of Wannier functions that decay as $|W_{\alpha m}(\RR)| \lesssim \frac{1}{|\RR|^q}$ at large $\RR$, where $2q>d$ in order for proper normalization, then we find the correlation function cannot decay with a power law slower than 
\begin{align}   \label{eq:bound_C}
    |C_{\alpha\beta}(\mathbf{r})| \lesssim \left\{
    \begin{array}{ll }
        r^{-(2q-d)} & q<d \\
        r^{-q}\ln r & q = d \\
        r^{-q} & q>d \\
    \end{array} \right.
\end{align}
where $r = |\mathbf{r}|$ \cite{SM}. 
In 2D, for example, as we have found $|C_{\alpha\beta}(\mathbf{r})| \sim r^{-4}$ for the $\mathbb{Z}_2$ CTFB, \cref{eq:bound_C}, along with the $2q>d=2$ condition, thus implies $1<q \le 4$ for the Wannier function. 

We emphasize that this bound solely arises from the critical behavior due to the touching, and should be discerned from the topological obstruction. 
We illustrate this with the Chern CTFBs and the $\mathbb{Z}_2$ CTFBs. 
For a Chern band with Chern number $C$, choosing the singular Wannier gauge (see \cref{sec:invariant}) leads to Bloch wave-functions with the following form around the touching point, 
\begin{align}  \label{eq:uk_Chern_singular_gauge}
    \mathcal{U}(\kk) = e^{\ii C\phi_{\kk}} \mathcal{U} + \mathcal{O}(k)
\end{align}
where $\phi_\kk$ is the azimuthal angle with respect to the touching point. 
As the wave-function is smooth elsewhere, the topological obstruction manifests as a $C$-fold phase vortex centering the touching point, hence the Bloch wave-function itself is discontinuous. 
The discontinuity in the higher-order derivatives of the projector is hidden in the omitted $\mathcal{O}(k)$ dependence in \cref{eq:uk_Chern_singular_gauge}. 
Therefore, as the Fourier transformation to \cref{eq:uk_Chern_singular_gauge}, the Wannier function will decay as $|W(\mathbf{r})|\sim {r^{- n - d}}$ with $n=0$ and $d=2$ \cite{Thouless_1984_wannier,li2024constraintsrealspacerepresentations}. 
Meanwhile, the projector is discontinuous only in the derivatives, hence the correlation function decays as at least $r^{-3}$ or $r^{-4}$ \cite{SM}. 
The bound is thus loosely met, which implies that the topological obstruction enforces a longer-range behavior than the touching point for this gauge choice. 

For strong $\mathbb{Z}_2$ topological insulators, however, it is known that exponentially decaying Wannier functions can be constructed, if the locality of TRS is not enforced \cite{Brouder_exponential_2007}. 
In this case, the power-law correlation due to the touching point will enforce a stronger bound and forbid the exponentially localized Wannier function in any gauge.

\section{CTFBs beyond symmetry-based indicators}
\label{sec:non-symmetry-indicated}

The formulations for topological invariants in \cref{eq:C-Sk-main,eq:Z2-index-main} apply to generic CTFBs realized in bipartite constructions, regardless of whether they are symmetry-indicated or not. 
Therefore, they provide a guiding principle to generalize the constructions of CTFBs.
Here, we discuss three scenarios of CTFBs with stable topology that cannot be directly diagnosed by SIs, but still feature the continuous FB projector $P(\kk)$ and the well-defined topology. 
Their difference lies in the robustness to various perturbations. 
If the topology of a CTFB is robust against symmetry-preserving bipartite perturbations, as the symmetry-indicated ones we have constructed using \cref{sec:principle}, we term it as symmetry-guaranteed; otherwise we term it as fine-tuned.

\subsection{Symmetry-guaranteed high Chern numbers with trivial SI}
\label{sec:high-Chern-beyond-SI}

\begin{figure}[tb]
    \centering
    \includegraphics[width=0.96\linewidth]{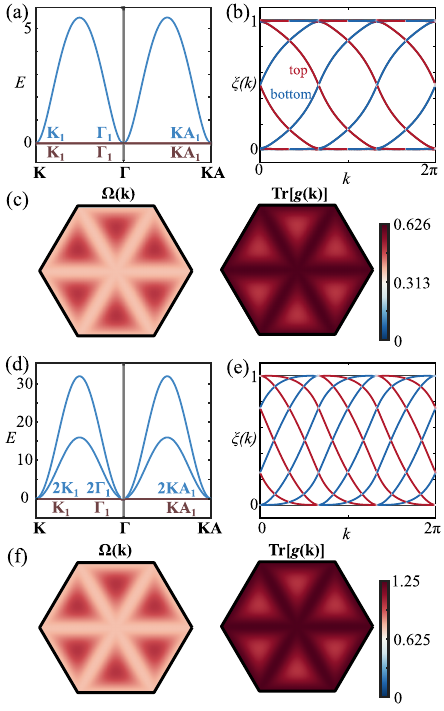}
    \caption{Symmetry-guaranteed CTFB beyond symmetry-based indicators, with Chern number $C=3N$ in wall paper group $p3$. 
    (a-c) $C=3$, with two $s$-orbitals as $L$ sublattice and one $p_+$-orbital as $\td L$ sublattice, all placed at $1a$ Wyckoff position. 
    (d-f) $C=6$, with three $s$-orbitals as $L$ sublattice and two $p_+$-orbitals as $\td L$ sublattice, all placed at $1a$ Wyckoff position. 
    (a,d) Band structures. 
    (b,e) Chiral edge modes in the entanglement spectrum, calculated in a stripe geometry, with red and blue bands polarized at the top and bottom edges, respectively. 
    (c,f) Berry curvature $\Omega(\kk)$ and FSM trace $\Tr g(\kk)$. 
    Hopping parameters have been optimized to minimize the Berry curvature fluctuation.}
    \label{fig:Chern-newC3}
\end{figure}

Recall that the SIs can differ from the exact topological invariant by a modulo of integers. 
Therefore, a non-trivial topological invariant can possess a trivial SI, hence to detect it requires using the exact formulae established in \cref{sec:invariant}. 

Such an example can be found in the $p3$ group, where the SI only determines the Chern number $C$ mod 3 (\cref{fig:Chern-newC3}(a)-(c)). 
A $C=3$ CTFB with trivial SI can be constructed by the following bipartite lattice, $\mathcal{BR}_{L} = 2 [A_1]_{1a} \uparrow \mathcal{G}$ and $\mathcal{BR}_{\td L} = [{}^2E]_{1a} \uparrow \mathcal{G}$, which puts two $s$-orbitals at the $1a$ Wyckoff position for the $L$ sublattice, and puts another $p_+$-orbital also at $1a$ for the $\td L$ sublattice. 
Their difference is
\begin{align}
\mathcal{BR}_{L} \boxminus \mathcal{BR}_{\td L} = (2\Gamma_1 \boxminus \Gamma_2 ; 2\rK_1 \boxminus \rK_2;  2\rKA_1 \boxminus \rKA_3 ) \ ,
\end{align}
hence all the three high-symmetry momenta, $\Gamma, \rK, \rKA$, are touching points.
At every touching point, the Bloch states belonging to $L$ possess angular momentum $0$ (mod 3), while the Bloch states belonging to $\td L$ possess $+1$. 
Therefore, the $S^\dagger(\kk)$ matrix at each touching point must take the identical form as \cref{eq:Sk-winding-C=1} at the leading order of $\kk$, hence contributing a winding $+1$. 
Summed up, one gets a well-defined $C=3$. 

By the same reasoning, taking $\mathcal{BR}_{L}=(N+1)[A_1]_{1a} \uparrow \mathcal{G}$ and $\mathcal{BR}_{\td L}=N[{}^2E]_{1a} \uparrow \mathcal{G}$ will yield a flat band with $C=3N$, as the winding (\cref{eq:C-Sk-main}) traces over the $N$ dispersive bands. 
We provide an example of $C=6$ in \cref{fig:Chern-newC3}(d)-(f).

Now we argue that, under a general symmetry-preserving gap-opening perturbation that violates the bipartite requirement, not only the SIs, but also the global Chern number, will remain robust, as such gap-opening does not contribute Berry curvatures. 
The same argument will also apply to the symmetry-indicated CTFBs. 
To see this, we consider the effective low-energy Hamiltonian near the touching point, written in the following diagonalized basis
\begin{align}  \label{eq:H_SS_new_p3}
    \frac{S(\kk)S^\dagger(\kk)}{\Delta} \sim \begin{pmatrix}
        \gamma^2 k^2 & 0 \\
        0 & 0 \\
    \end{pmatrix} 
\end{align}
As the two states share the same irrep, a symmetry-preserving gap-opening perturbation $H_{\rm pert}$ at the constant order will be a general $2 \times 2$ Hermitian matrix that spans all the Pauli matrices $\sigma^{x,y,z}$. 
If the $\sigma^{x}$ or $\sigma^{y}$ components are non-zero, $H_{\rm pert} + \frac{S(\kk)S^\dagger(\kk)}{\Delta}$ will possess a gap for general $\kk$. 
For this Hamiltonian, the wave-function of the lower branch does not exhibit angular dependence, hence possessing vanishing Berry curvature, and will not cause a jump of Chern number. 
Nevertheless, we remind that if the perturbations were fine-tuned so that the $\sigma^{x,y}$ components vanish at the constant order, and the $\sigma^z$ component is negative, then the band gap closing and reopening can happen at finite $\kk$, which can eventually lead to a jump of Chern number. 

\subsection{Symmetry-guaranteed CTFB with indefinite SI}

If $\mclB+\Delta\mclB$ contains multiple types of irreps at the zero-energy, and choosing different irreps into $\mclB$ leads to different SIs, then the CTFB will not possess identical topological numbers after the touching points are gapped. 
Nevertheless, symmetry still guarantees the continuity of $P(\kk)$, thereby ensuring a well-defined topology prior to gap opening.
Consider a bipartite construction in wallpaper group $p4$, where $\mathcal{BR}_L=[A\oplus {}^1E \oplus {}^2E]_{1a} \uparrow \mathcal{G}$ consists of $s$, $p_x$, and $p_y$ orbitals at the $1a$ position  $(0,0)$, and $\mathcal{BR}_{\td L}=[B\oplus {}^1E]_{1b} \uparrow \mathcal{G}$ consists of $d$ and $p_+$ orbitals at the $1b$ position $(1/2,1/2)$.  
The resulting symmetry data is $\mclB + \Delta\mclB = (\Gamma_1 \oplus \Gamma_3 \boxminus \Gamma_2;\ \rM_4;\ \rX_2)$. 
Here, $\Gamma_1$, $\Gamma_2$, $\Gamma_3$, and $\rM_4$ carry the $C_4$ rotation eigenvalues $1$, $-1$, $\ii$, and $-\ii$, respectively, while $\rX_2$ carries the $C_2$ rotation eigenvalue $-1$. 
According to the representation counting rule, the band touching at $\Gamma$ consists of two different irreps, \textit{i.e.}, $\Gamma_1 \oplus \Gamma_3$.
The ambiguity in selecting the flat band irrep at $\Gamma$ invalidates the direct application of SIs. 

Nonetheless, this construction still yields a symmetry-guaranteed CTFB whose Chern number is computable using \cref{eq:C-Sk-main}.
Constrained by the $C_4$ symmetry, the k$\cdot$p expansion of the low-energy block of $S^\dagger(\kk)$ is given by 
\begin{equation}  \label{eq:Sk_CTFB_new_p4}
    S^\dagger(\kk) = \begin{pmatrix}
        0 & \gamma k_+
    \end{pmatrix} + \mathcal{O}(k^2) \ ,
\end{equation}
where $\gamma$ is a complex parameter, the two columns transform as $\Gamma_1$ and $\Gamma_3$, and the row transforms as $\Gamma_2$. 
Crucially, as $\kk\to 0$, $S^\dagger(\kk)$ possesses a $\kk$-independent kernel $(1,0)^T$. 
This ensures that the FB has a well-defined Chern number. 
Since $S(\kk)$ has a definite phase winding around $\kk=0$, substituting it into \cref{eq:C-Sk-main} yields $C=1$.

When the symmetry-preserving gap-opening perturbations are introduced, the gap may favor either $\Gamma_1$ or $\Gamma_3$ to lie at the lower energy, while the corresponding SIs for Chern number will differ by $1$ (mod 4). 
Consequently, the gapped Chern number will depend on the sign of the gap.

\subsection{CTFB via fine-tuning}

To construct a minimal fine-tuned CTFB with Chern number $C=1$, we require an $S(\kk)$ matrix that behaves asymptotically as $(\gamma_1 k_-, \gamma_2 k_-)^T$ near $\kk=0$ and remains non-vanishing elsewhere in the Brillouin zone. 
As in a previous model (\cref{eq:Sk-winding-C=1}), this local form simultaneously guarantees kernel continuity and the requisite phase winding to realize $C=1$. 
One explicit realization is given by
{\small
\begin{equation}
    S(\kk) = \begin{pmatrix}
        t_1 \sin \frac{k_x}2 - \ii t_2 \cos \frac{k_x}2 \sin k_y \\
        t_2 \cos \frac{k_y}2 \sin k_x  - \ii t_1 \sin \frac{k_y}2
    \end{pmatrix} 
\stackrel{\kk\to 0}{\approx} \begin{pmatrix}
        \frac{t_1}2 k_x-\ii t_2 k_y \\
        t_2 k_x - \ii \frac{t_1}2 k_y
    \end{pmatrix} ,
\end{equation}}
where the second equation expands $S(\kk)$ to linear order of $\kk$ around $\kk=0$.
When $t_1=2t_2$, this expansion recovers the required asymptotic behavior. 
If $t_1\neq 2t_2$, the continuity of the kernel is lost, rendering the flat band singular. 

This model can be realized by placing $p_-$ orbitals on sublattice $L$ at $(0, 1/2)$ and $(1/2, 0)$, and an $s$ orbital on sublattice $\td L$ at $(1/2, 1/2)$. 
It respects the $C_4$ rotation symmetry: $-\ii\sigma_x S(k_x, k_y) = S(-k_y, k_x)$. 
Because $t_1$ and $t_2$ represent hoppings at different spatial distances, no crystalline symmetry can enforce the the condition $t_1 = 2t_2$. 
Thus, achieving this CTFB relies on parameter fine-tuning.

Fine-tuned CTFBs with other topological invariants can be similarly constructed. 
However, these models are in general not robust against any type of perturbations. 

\section{Tensor-network representation}
\label{sec:TNS}

Albeit critical, Fock states occupying CTFBs are generally short-range entangled (\cref{fig:Z2-main}(e), (f)), suggesting possible TNS realizations. 
Now we demonstrate that the ground state $\ket{\Omega_L}$ of the effective Hamiltonian $H_L(\kk) = S(\kk) S^\dagger(\kk) /\Delta $  (\cref{eq:Heff}) can be represented as an exact TNS. 
By annihilating the dispersive-band states from the fully filled state $\ket{\mathrm{F}}$, we can express $\ket{\Omega_L}$ as 
$\prod_{\kk} \prod_{n} (\sum_{\alpha} \psi_{\kk \alpha} \mathcal{U}_{\alpha n}^*(\kk)) \ket{\mathrm{F}}$, 
where $\psi_{\kk \alpha}$ is the annihilation operator for the $\alpha$-th orbital in sublattice $L$, and $\mathcal{U}(\kk)$ are the Bloch wavefunctions of dispersive bands in the singular Wannier gauge (\cref{sec:invariant}). 
Note that the operator product $\prod_{n} (\sum_{\alpha} \psi_{\kk \alpha} \mathcal{U}_{\alpha n}^*(\kk))$ is equal to $\prod_{n} (\sum_{\alpha} \psi_{\kk \alpha} S_{\alpha n}^*(\kk))$ up to a factor  $\sqrt{\det\mathcal{Q}(\kk)}$. 
Thus, the unnormalized ground state can be simply written as $\ket{\Omega_L} = \prod_{\kk} \prod_{n} (\sum_{\alpha} \psi_{\kk \alpha} S_{\alpha n}^*(\kk)) \ket{\mathrm{F}}$, where the {\it finite} real-space support of $S(\kk)$ naturally leads to an exact TNS representation. 
We should adopt proper boundary conditions such that the discrete momenta $\kk$ avoid the touching points where $S(\kk)$ loses rank; otherwise additional boundary operators are needed to prevent $\ket{\Omega_L}$ from vanishing. 

In real space, the ground state is equivalently expressed as $\prod_{j,b} (\sum_{i,a} \psi_{i,a} S_{ia, jb}^* ) \ket{\mathrm{F}}$, where $i\in L$ and $j\in \td{L}$ denotes sites in respective sublattices, $a,b$ label the orbitals within each site, and $S_{ia,jb}$ is the hopping matrix corresponding to $S_{\alpha n}(\kk)$. 
For each link $l \equiv\inn{i,j}$ connecting these sublattices---denoted by the bonds in \cref{fig:Z2-main,fig:Chern-main,fig:3DTI-main}---we introduce auxiliary Grassmann numbers $\eta_{l,a}$,  $\ovl\eta_{l,a}$ that carry the same orbital indices as site $i$. 
Each pair of $\eta_{l,a}$,  $\ovl\eta_{l,a}$ is equivalent to a fermionic mode with dimension 2 \cite{mortier_fermionic_2025}.
Then a standard TNS is formulated as 
{\small
\begin{equation} \label{eq:TNS-main}
\int \!\! \mathcal{D}[\ovl \eta,\eta] 
    \Bigg( \!\!\prod_{ia} e^{-\sum_{l\in \mathcal{L}_i} \!\! \ovl\eta_{la} \eta_{la} + \lambda_l \ovl\eta_{l a} \psi_{ia} }  \!\!\Bigg) 
    \Bigg( \!\! \prod_{j b} \sum_{ \substack{l\in \mathcal{L}_j \\ a} }\eta_{la} W_{la,jb} \!\! \Bigg) \ket{\mathrm{F}} 
\end{equation}}
Here $\mathcal{L}_{i,j}$ represent the links connected to $i,j$, respectively, and $\lambda_l$ and $W_{la,jb}$ are the defining data for the state. 
The second term generates a product state in the auxiliary space, while the first term projects this state back onto physical Hilbert space upon integration over the Grassmann variables. 
This ansatz reproduces $\ket{\Omega_L}$ provided $\sum_{l\in \mathcal{L}_i}\lambda_l W_{la,jb} = S_{ia,jb}$, with a canonical solution being $\lambda_l=1$, $W_{la,jb} = S_{ia,jb}$ for the link $l=\inn{i,j}$.

\cref{eq:TNS-main} provides a unified TNS realization for all CTFBs arising from bipartite constructions. 
Diagrammatically, the $W$ tensors are represented by the yellow polygons or spheres in \cref{fig:Z2-main,fig:Chern-main,fig:3DTI-main}, and the bonds are the links between them and the physical sites in $L$ sublattice. 
Consider a finite region $D$, whose boundary may cut through many $W$ tensors labeled by $j$.
Each cut tensor contributes a bond dimension $2^{\mathcal{N}_j}$, where $\mathcal{N}_j$ is the number of orbitals on site $j$.
For the constructions considered here, $\mathcal{N}_j$ is typically small.
For example, $\mathcal{N}_j=2$ for the $C=1$ CTFB in \cref{fig:Chern-main}(a), $\mathcal{N}_j=1$ the $C=3$ CTFB in \cref{fig:Chern-newC3}, and $\mathcal{N}_j=2$ the $C=6$ CTFB in \cref{fig:Chern-newC3}.

It is useful to compare our TNS ansatz \cref{eq:TNS-main} to the various existing TNS formulations for topological states, \textit{e.g.}, Refs.~\cite{wahl_projected_2013, dubail_tensor_2015}. 
In these existing approaches, auxiliary fermions live on the same sites as the physical ones in $L$, or on the bonds that connect adjacent physical sites. 
The latter leads to a projected pair-entangled states (PEPS) description \cite{mortier_fermionic_2025}. 
The key to bridge these different TNS representations is to notice that, we can equivalently express the filled FB state as
\begin{align}   \label{eq:OmegaL_move_main}
    \Ket{\Omega_L} \sim \prod_{\kk} \exp( \psi^\dagger_{\kk A} G(\kk)  \psi_{\kk\ovl{A}}) \Ket{{\rm F}_{\ovl{A}}} \, .
\end{align}
Here, we choose a group of orbitals with a total number equal to $\dim \td L$ in the $L$ sublattice as $A$, and the remaining orbitals are denoted as $\ovl A$. 
$\Ket{{\rm F}_{\ovl A}}$ fills $\ovl A$. 
While $\Ket{{\rm F}_{\ovl A}}$ is topologically trivial, $G(\kk)$ defines a $\kk$-dependent map that moves $\psi_{\kk, \ovl{\alpha} \in \ovl{A}}$ into the FB wave-function. $G(\kk)$ must possess singularities due to the non-trivial topology of the FB. 
In our bipartite approach, we show that $G(\kk) = -\Big[S_{\ovl A, :}(\kk) \big[S_{A, :}(\kk)\big]^{-1} \Big]^{T}$, where $S_{A, :}(\kk)$ and $S_{\ovl A, :}(\kk)$ denote the rows of the full $S(\kk)$ matrix that correspond to the $A$ and $\ovl A$ orbitals, respectively \cite{SM}.

One existing realization puts the auxiliary fermions $\xi, \xi^\dagger$ on the same sites as the physical ones \cite{dubail_tensor_2015}. 
Below we assume the physical sites form the simple lattices such as the triangular or square lattice, but generalization to arbitrary lattice will be direct. 
Such realization writes
\begin{align}   \label{eq:OmegaL_onsite_aux}
    |\Omega_L\rangle &\sim \int_{\mathcal{D}[\xi,\xi^\dagger]} e^{\sum_{\RR}\left(- \xi^\dagger_{\RR} K_{\mathbf{0}} \xi_{\RR} - \sum_{\mathbf{d} \in {\rm nn}} \left( \xi^\dagger_{\RR+\mathbf{d}} K_{\mathbf{d}} \xi_{\RR} \right) \right)} \\\nonumber 
    &\qquad \qquad \qquad \times e^{\sum_{\RR}\left( \psi^\dagger_{\RR  A} C \xi_{\RR} + \xi^\dagger_{\RR} B \psi_{\RR \ovl{A}} \right)} |{\rm F}_{\ovl A} \rangle
\end{align}
Here, $B$ and $C$ are on-site  couplings between physical ($\psi, \psi^\dagger$) and auxiliary ($\xi, \xi^\dagger$) modes, while
$K_{\mathbf{0}}$ and $K_{\mathbf{d}}$ for $\mathbf{d}\in{\rm nn}$ are
the on-site and nearest-neighbor couplings between auxiliary modes, respectively.
Integrating out $\xi$ and $\xi^\dagger$ in
\cref{eq:OmegaL_onsite_aux} yields \cref{eq:OmegaL_move_main}, with
$G(\mathbf{k})=C K^{-1}(\mathbf{k})B$, where
$K(\mathbf{k})=K_{\mathbf{0}}+\sum_{\mathbf{d}\in{\rm nn}}
\mathrm{e}^{-\mathrm{i}\mathbf{k}\cdot\mathbf{d}}K_{\mathbf{d}}$.
Realizing a prescribed $G(\mathbf{k})$ thus requires finding such a
nearest-neighbor $K(\mathbf{k})$ and on-site $B$, $C$.
This may require a $K(\mathbf{k})$ with larger matrix dimension than $G(\mathbf{k})$, and hence increase the bond dimension.

\cref{eq:OmegaL_onsite_aux} can be further simplified by introducing auxiliary fermions $\eta,\eta^\dagger$ living on the nearest-neighbor bonds \cite{dubail_tensor_2015}. 
We define $e = (\RR,\RR+\mathbf{d})$ as a directed edge. 
Then, with decomposition $K_{\mathbf{d}} = L_{\mathbf{d}} R_{\mathbf{d}}$, there is 
\begin{align}   \label{eq:xixi_to_xieta}
    e^{-\xi^\dagger_{\RR+\mathbf{d}} K_{\mathbf{d}} \xi_{\RR}} = \int_{\mathcal{D}[\eta,\eta^\dagger]} e^{- \eta^\dagger_{e} \eta_{e} - \xi^\dagger_{\RR+\mathbf{d}} L_{\mathbf{d}} \eta_{e} + \eta^\dagger_{e} R_{\mathbf{d}} \xi_{\RR}}
\end{align}
Inserting \cref{eq:xixi_to_xieta} into \cref{eq:OmegaL_onsite_aux}, and then integrating out $\xi,\xi^\dagger$ first \cite{SM}, we arrive at a standard PEPS representation.

We illustrate this mapping for a critical Chern TNS in SM \cite{SM}, where only one virtual orbital per unit cell is needed in the bipartite form \cref{eq:TNS-main}, while four per edge (eight per unit cell) are need in the PEPS form. 

Finally, it is worth noting that, since $\ket{\Omega_L}$ is simultaneously annihilated by the local operators $O_{j,b} =\sum_{i,\alpha} \psi_{i,a} S^*_{ia, jb}$ for all $j$ and $b$, it will be the exact ground state not only of the bilinear Hamiltonian in \cref{eq:Heff}, but any positive semi-definite many-body Hamiltonian with these annihilating local operators acting on the right, in the form of $H=g_2\sum_{ia}O^\dagger_{i,a}O_{i,a} + g_4 \sum_{ia,jb}O^\dagger_{ia}O^\dagger_{jb}O_{jb}O_{ia}+\cdots$ with $g_2,g_4 \ge 0$.

\section{Discussions}
\label{sec:discussion}

Long-standing no-go theorems expose a fundamental incompatibility between locality, exact flatness and stable topology, rooted in the analytic structure of Bloch wavefunctions in lattice systems.
Here we demonstrate that this obstruction can be saturated rather than avoided: exact flatness and stable topology coexist only when the band is necessarily critical.
By integrating topological quantum chemistry with symmetry-based indicators and the symmetry-guaranteed continuity conditions, we have then developed a unified framework to construct such CTFBs systematically. 

To clarify the terminology, it is essential to distinguish CTFBs from the other FB scenarios. 
Since any gapped FB in a local lattice model cannot exhibit stable topology, we are led to critical (gapless) FBs featuring isolated band touching points if stable topology is of concern. 
Such critical FBs with Fermi points encompass three distinct scenarios (\cref{fig:Z2-main}(a)): 
(i) FBs with a discontinuous projector $P(\kk)$ onto the Bloch wavefunctions, \textit{i.e.}, singular FBs \cite{rhim_classification_2019} exemplified by the Kagome and Lieb lattices, 
(ii) FBs with a continuous but non-analytic $P(\kk)$ and trivial topology, 
and (iii) FBs with a continuous but non-analytic $P(\kk)$ and nontrivial topology, \textit{i.e.}, CTFBs.

\begin{figure}[t]
    \centering
    \includegraphics[width=1\linewidth]{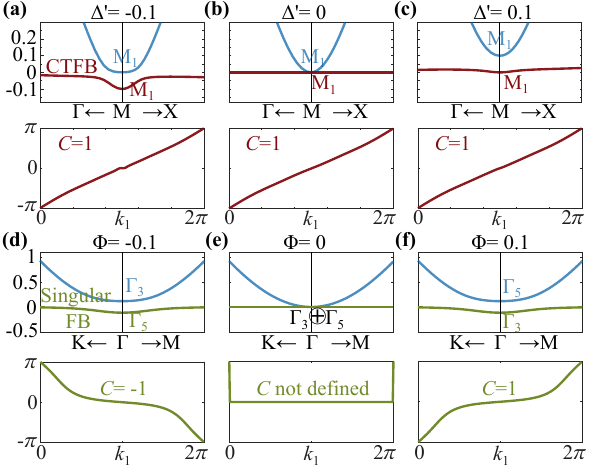}
    \caption{Robustness of symmetry-indicated CTFBs versus singular FBs. 
    (a-c) Evolution of the CTFB (defined in \cref{fig:Chern-main}(a)) under a gap-opening perturbation $\Delta'$, which is chosen as the on-site energy of the $s$ orbital in $\mathcal{BR}_{L}$. Top panels display the band structures near $\rM$. 
    Bottom panels show the corresponding Wilson loop spectra, confirming that $C$ remains robust throughout the gap-opening process. 
    (d-f) Evolution of the singular FB in Kagome lattice under a flux $\Phi$ inserted into the triangles, which respects the symmetry of $p6$. In the unperturbed case in (e), the band touching consists of distinct irreps, resulting in essential discontinuities that render $C$ ill-defined. Upon introducing $\Phi$, the resulting topology is sensitive to the sign of $\Phi$. 
    }
    \label{fig:robustness-main}
\end{figure}

A hallmark of our approach is the robustness of symmetry-indicated CTFBs against perturbations.
Since the continuity of the projector onto wavefunctions is symmetry-guaranteed, arbitrary symmetry-allowed perturbations maintaining the bipartite structure (\cref{eq:Hk-main}) preserve both the exact flatness and the topological invariants of the CTFBs. 
Moreover, as the touching points comprise multiple copies of identical irreps, generic gap-opening perturbations that break the bipartite structure (beyond \cref{eq:Hk-main}), including the mean-field interactions, must yield the same SIs, ensuring topological stability. 
As illustrated in \cref{fig:robustness-main}, this stability contrasts with singular FBs, where essential discontinuities render the topology ill-defined and sensitively dependent on the sign of the perturbation.
This robustness, combined with strict locality, positions CTFBs as a natural mechanism for nearly flat topological bands in real materials.

In stoichiometric materials, CTFB may be searched among the bipartite or split lattices \cite{regnault_catalogue_2022}, which provides a natural recipe for realizing the bipartite Hamiltonian. 
In SM \cite{SM}, we have made a comprehensive list of the symmetry data of various CTFB constructions in various space groups, which can be used to filter or locate the candidates in a high-throughput search.
More generically, if any system is \textit{a posteriori} reported to host a robust stable topology with good flatness, then CTFB is likely to appear as a natural description: among the short-range hoppings, stable topology, gap, and dispersion flatness, as the short-range hopping is more likely to be preserved in the effective models, the most likely compromise to the no-go theorem that remains would be to reduce the gap, leading to a CTFB. 

Twisted $\rm MoTe_2$ has already provided such an intriguing example. 
Although its microscopic realization has nothing to do with bipartite lattices, its low-energy effective model turns out to lie rather close to a CTFB limit, with an approximate bipartiteness in the Hamiltonian \cite{yang_fractional_2025}. 
We thus expect a unified CTFB theory will provide a general guidance for studying the similar topological flat bands, which are currently under intensive research.

In exploring the geometric or correlation effects in an interacting CTFB, while one may worry about the gapless touching of the dispersive bands, we argue its effect can be dialed down in the limit of $u \ovl\rho \to 0$, where $u$ is the interaction, and $\ovl\rho = \frac{1}{u}\int_0^{u} \dd \omega \, \rho(\omega)$ is the average density of states (DOS) of the dispersive bands that touch with FB. 
$u \ovl \rho$ measures the phase space size that can be excited by interactions outside the FB. 
As $S^\dagger(\kk) \sim \mathcal{O}(k)$, the dispersion is quadratic, which implies $\rho(\omega) \sim {\rm const}$ in 2D, and $\rho(\omega) \sim \omega^{1/2}$ in 3D. 
Therefore, one can always send $u \ovl\rho \to 0$ to suppress the excitations to the dispersive bands, while maintaining $u$ finite. 
As the flat band width is exactly zero, any non-vanishing $u$ suffices to generate the correlations in the flat bands. 
This limit guarantees that the topological FB dominates the correlation physics. 

Although the CTFB states exhibit power-law correlations inherited from the non-analytic structure, they remain short-range entangled and obey an area law for entanglement entropy, admit exact TNS representations with finite bond dimensions, hence provide a systematic starting point for exploring strongly correlated topological matter at both integer and fractional fillings. 
For example, gapped exact TNS representations for the strongly interacting TRS-protected $\mathbb{Z}_2$ SPT have only recently been formulated \cite{wang2023exactly,ma2024variational}, and connecting them back to the weak-coupling $\mathbb{Z}_2$ band insulators has remained non-trivial. 
In the CTFB approach, we are instead able to start with a non-interacting albeit gapless Gaussian TNS (\cref{sec:Z2-model}). 
It is then natural to expect that the desired gap can be opened by further concatenating tensor-network operators (TNO) that incorporate the many-body correlations, or by substituting the $W$ tensors with properly constructed many-body tensors. 
The similar logic can be generalized to obtain other non-chiral SPT phases, especially the crystalline SPT, given that the CTFB has inherently incorporated the crystalline symmetries. 
Systematic constructions of their gapped TNS representations via such a strategy are possible.

More intriguingly, our CTFB framework can also enable direct extensions to describe topological orders (TOs). 
For example, we may fractionalize the electrons using parton constructions, and consider CTFBs for the parton bands. 
Gutzwiller projection onto the physical Hilbert space can then yield candidate wavefunctions for TOs, including non-chiral phases such as fractional topological insulators and chiral phases such as FCIs and chiral spin liquids~\cite{yang_chiral_2015}.
Chiral TNS can exhibit characteristic entanglement properties and topological degeneracy despite their critical correlations~\cite{yang_chiral_2015}.
It is also shown that the gapless chiral TNS are able to efficiently approximate the gapped topological ground states on finite lattices with exponentially decreasing error \cite{wahl_projected_2013}.

For FCIs, one tractable limit that has stimulated theoretical development is the ``ideal'' holomorphic  limit of Chern bands, which connects Chern bands to the LLL~\cite{wang_2021_exactlandaulevel}, enabling analytical constructions of trial wavefunctions.
However, this limit is usually realized in continuum models such as the (chiral) twisted multilayer graphene \cite{ledwith_family_2022, wang_hierarchy_2022, dong_many-body_2023}, and is hard to realize in lattice models with a finite hopping range \cite{zhao2026ideal, li2026quantum, wang2026symmetry}. 
It also provides limited insight into the strongly correlated topological matter beyond the LLL picture. 
On the one hand, whether the (holomorphic) ``idealness'' is the best condition for stabilizing FCI remains an open question, with recent numerical works finding FCI states in rather unconventional band setups \cite{fonseca2026gradient, lu2025bosonic, liu2025topological, yang2026fractional}. 
On the other hand, a large zoo of correlated topological states exist beyond FCI. 
In this broader context, the CTFB construction explores a new limit of ``idealness'' that may provide useful starting points for investigating these phases.

A particularly interesting direction is fractionalization in flat bands with large Chern numbers, nontrivial $\mathbb{Z}_2$ topology, or crystalline-symmetry-protected topology, whose topological and geometric structures go beyond those of Landau levels.
Early theoretical studies proposed and provided numerical evidence for a rich variety of fractionalized phases in bands with Chern number $C>1$ \cite{BarkeshliQi2012TopologicalNematic}, including non-Abelian topological orders~\cite{ LuRan2012FractionalChern, Sterdyniak2013HigherChern, ZhangVishwanath2013NonAbelian}.
Recent experimental progress, including the reported fractional high-Chern insulator in twisted rhombohedral graphene~\cite{Li2026FractionalHighChern}, and signatures of fractional quantum spin Hall effect in twisted $\rm MoTe_2$~\cite{kang_evidence_2024}, further motivates this direction. 
Our framework provides a systematic construction of exactly flat bands with arbitrarily large Chern numbers, substantially expanding the family of models available for such investigations.
The resulting band Hamiltonians can serve as single-particle starting points for interacting models, while the corresponding bands can also be used as parton bands in TNS constructions of fractionalized states.
Our constructions of $\mathbb{Z}_2$ and crystalline topological flat bands extend this strategy to other symmetry classes, providing a basis for investigating the interplay between symmetry and fractionalization.

\paragraph{Note added.}
At the completion of the work, we became aware of an independent work on flat bands with strong topology \cite{liu2026theory}.
In SM \cite{SM}, we present a comparison to their real-space perspective using our $C=1$ Chern CTFB.

\begin{acknowledgements}
We are grateful to G.-D.\,Z. for useful discussions. 
Z.-D.\,S., Y.-Q.\,L., Y.-J.\,W., and B.\,W. were supported by National Natural Science Foundation of China (Grant No.\,12274005 and 12521006), National Key Research and Development Program of China (No.\,2021YFA1401900), and Quantum Science and Technology-National Science and Technology Major Project (No.\,2021ZD0302403). 
\end{acknowledgements}


%



\onecolumngrid

\tableofcontents
\clearpage

\setcounter{equation}{0}
\setcounter{figure}{0}
\setcounter{table}{0}
\setcounter{section}{0}

\renewcommand{\theequation}{S\arabic{equation}}
\renewcommand{\thefigure}{S\arabic{figure}}
\renewcommand{\thetable}{S\arabic{table}}
\renewcommand{\thesection}{S\arabic{section}}

\renewcommand{\theHequation}{S\arabic{equation}}
\renewcommand{\theHfigure}{S\arabic{figure}}
\renewcommand{\theHtable}{S\arabic{table}}
\renewcommand{\theHsection}{S\arabic{section}}

\section{Example: flat Chern band in wallpaper group \texorpdfstring{$p4$}{p4}}
\label{app:Chern}

In this section, we present a pedagogical introduction to the construction of critical topological flat bands (CTFBs), using the  single-valued wallpaper group $p4$ in the absence of time-reversal symmetry (TRS) as an example. 

\subsection{Band representation analysis}
\label{ebr_anylyse}
\begin{table}[th!]
\centering
\begin{tabular}{|c|c|c|c|c|c|c|c|c|c|c|}
\hline
  Wyckoff  & \multicolumn{4}{c|}{$1a \ (4)$} & \multicolumn{4}{c|}{$1b \ (4)$} & \multicolumn{2}{c|}{$2c \ (2)$} \\
\hline 
EBR & $ A\ (1) $ & $B \ (1)$ & ${}^1 \!E \ (1)$  & ${}^2\!E \ (1)$ & $A \ (1)$ & $B \ (1) $ & ${}^1\!E \ (1)$  & ${}^2\!E\ (1)$ & $A\ (2)$ & $B\ (2)$ \\
\hline
$\Gamma\ (0,0)$ & $\Gamma_1 $ & $\Gamma_2 $ & $\Gamma_4 $ & $\Gamma_3 $ & $\Gamma_1 $ & $\Gamma_2 $ & $\Gamma_4 $ & $\Gamma_3 $ & $\Gamma_1 \oplus \Gamma_2$ & $\Gamma_3 \oplus \Gamma_4$ \\
\hline
$\rM \ (\pi,\pi)$ &  $\rM_1$ &  $\rM_2$ &  $\rM_4$ &  $\rM_3$ &  $\rM_2$ &  $\rM_1$ &  $\rM_3$ &  $\rM_4$ &  $\rM_3\oplus \rM_4$ & $\rM_1 \oplus \rM_2$ \\
\hline 
$\rX (0, \pi)$ &  $\rX_1$ & $\rX_1$ & $\rX_2$ & $\rX_2$ & $\rX_2$ & $\rX_2$ &  $\rX_1$ & $\rX_1$ & $\rX_1\oplus \rX_2$ & $\rX_1\oplus \rX_2$ \\
\hline
\end{tabular}
\caption{The EBR data of wallpaper group $p4$.
The first row tabulates all the Wyckoff positions with the corresponding site-symmetry groups in the parentheses. 
$1a$ and $1b$ refer to the $C_4$-symmetric positions $(0,0)$ and $(\frac12,\frac12)$, respectively. $2c$ refers to the $C_2$-symmetric positions $(\frac12,0)$, $(0,\frac12)$. 
The second row tabulates the EBR names, given by the irreps of the corresponding site-symmetry groups. 
Numbers in the parentheses represent the number of bands in the corresponding EBR. 
The following three rows give the irreps at high-symmetry momenta of the EBRs.  
}
\label{tab:p4-EBR}
\end{table}

\begin{table}[h!]
\centering
\begin{tabular}{|c|r|r|r|r||c|}
\hline
\multicolumn{6}{|c|}{Irreps of the point group $4$}\\
\hline
Irreps & $E$ & $C_4$ & $C_4^2$ & $C_4^3$ & $\kk$-notation\\
\hline
 $A$ & 1 & 1 & 1 & 1  & $\Gamma_1$, $\rM_1$ \\
\hline
 $B$ & 1 & $-1$ & $1$ & $-1$ & $\Gamma_2$, $\rM_2$ \\
\hline
 ${}^2\!E$ & 1 & $\ii$ & $-1$ & $-\ii$ & $\Gamma_3$, $\rM_3$ \\
\hline
 ${}^1\!E$ & 1 & $-\ii$ & $-1$ & $\ii$ & $\Gamma_4$, $\rM_4$ \\
\hline
\end{tabular}
\hspace{1cm}
\begin{tabular}{|c|r|r||c|}
\hline
\multicolumn{4}{|c|}{Irreps of the point group $2$}\\
\hline
Irreps & $E$ & $C_2$ & $\kk$-notation \\
\hline
$A$ & 1 & $1$ & $\rX_1$ \\
\hline
$B$ & 1 & $-1$ & $\rX_2$ \\
\hline
\end{tabular}
\caption{Character tables for point groups $4$ and $2$.
Notations in the column ``irreps'' follow the convention of Ref.~\cite{bradley_mathematical_2010}, and are used for representations of site-symmetry groups of the Wyckoff positions in real space. 
At high symmetry momenta $\Gamma$, $\rM$ and $\rX$ of $p4$, the little groups are $4$ and $2$, respectively, and the energy bands can also be labeled by the point groups irreps, but following a different convention given in the ``$\kk$-notation'' columns.  
}
\label{tab:p4-irrep}
\end{table}

We tabulate the elementary band representations (EBRs) in \cref{tab:p4-EBR}, where the irreducible representations (irreps) are defined in \cref{tab:p4-irrep}. 
One may refer to \href{http://webbdcrista2.ehu.es}{Bilbao Crystallographic Server} \cite{elcoro_double_2017} or the \href{https://irrep.dipc.org/index.html}{IrRep} package \cite{IrRep2022} for the EBR data over all wallpaper and space groups. 

We target at an CTFB with irreps 
\begin{equation}
\mclB = (\Gamma_3; \quad  \rM_1; \quad \rX_1 ) \ .  
\end{equation}
Hereafter, we always use a $B$-vector to represent the irreps of a band structure. 
Its topology can be diagnosed through the generalized Fu-Kane formula \cite{fang_bulk_2012} 
\begin{equation} \label{eq:Chern-C4}
    \ii^C = \prod_{n\in \mathrm{occ}} \xi_n(\Gamma) \ \xi_n(\rM) \ \zeta_n(\rX)\ ,
\end{equation}
where $C$ is the Chern number, $\xi_n(\Gamma)$ and $\xi_n(\rM)$ are the $C_4$ eigenvalues of the $n$-th occupied band at $\Gamma$ and $M$, respectively, and $\zeta_n(\rX)$ is the $C_2$ eigenvalue of the $n$-th band at $\rX$. 
According to the character tables in \cref{tab:p4-irrep}, we find the CTFB must have a Chern number 
\begin{equation} \label{eq:p4-C-mod4}
    C = 1 \quad \mod 4\ ,
\end{equation}
provided continuity of the wavefunction over the Brillouin zone. 

To construct the CTFB from a finite-range hopping model, we introduce a $\Delta \mclB$ vector to trivialize its topology. 
First, the $\Delta \mclB$ vector must contribute $-1$ (mod 4) to the Chern number according to \cref{eq:Chern-C4}. 
Second, the band number of $\Delta \mclB$ should be zero, such that $\mclB' = \mclB + \Delta \mclB$ still represents a one-band system.
Third, $\Delta \mclB$ should be nonzero only at the momentum where the CTFB touches dispersive bands, which we choose as $\rM$. 
Fourth, $\Delta \mclB$ should satisfy the compatibility relations. 
Since there is no high-symmetry line in the Brillouin zone of the $p4$ group, no nontrivial compatibility relation beyond the band number, {\it i.e.}, all the momenta have the same number of bands, exists. 
We find the following $\Delta \mclB$ satisfies all the constraints:
\begin{equation}
    \Delta \mclB = (0; \quad \rM_1 \boxminus \rM_3; \quad 0) \ . 
\end{equation}
Here, $\boxminus$ is the ``inverse'' operation of the direct sum $\oplus$ such that $\rho \oplus \rho' \boxminus \rho' = \rho$, $\rho \oplus \rho' \boxminus \rho = \rho'$. 
One should not worry about a negative number of irreps in intermediate objects such as $\Delta \mclB$  since all the physical outputs will have nonnegative numbers of irreps. 
The trivialized $\mclB$ vector
\begin{equation} \label{eq:p4-B'}
    \mclB' = \mclB + \Delta \mclB = (\Gamma_3; \quad  2\rM_1 \boxminus \rM_3 ; \quad \rX_1 )
\end{equation}
can be written as a linear combination of EBRs with integer coefficients: 
\begin{equation} \label{eq:p4-BR-difference}
    \mclB' = [A]_{1a} \ \oplus \  [B]_{2c}  \  \boxminus \ [{}^1\! E]_{1b} \ \boxminus \ [A]_{1b}  \ ,
\end{equation}
where $[\rho]_{w}$ is a shorthand for the EBR $[\rho]_w \uparrow G$, which is a (reducible) representation of the wallpaper group $G$ induced from the irrep $\rho$ at the Wyckoff position $w$. 

According to Ref.~\cite{calugaru_general_2021}, the band structure in \cref{eq:p4-B'} admits a bipartite lattice construction: 
The bigger sublattice $L$ consists of $[B]_{2c} \oplus [A]_{1a}$, and the smaller sublattice $\td{L}$ consists of $[{}^1\!E]_{1b} \oplus [A]_{1b}$. 
All hopping and on-site terms consistent with the symmetry group are allowed, except intra-$L$-sublattice terms. 
We assume a uniform on-site energy $-\Delta$ of the $\td L$-sublattice for convenience. 
Then, there will be a single flat band at the zero energy, which is gapped from dispersive bands over the whole Brillouin zone except $\rM$. 
The band will form irreps $\Gamma_3$ and $\rX_1$ at $\Gamma$ and $\rX$, respectively, and the touching point at $\rM$ will be two-fold with both forming the irrep $\rM_1$. 

\begin{figure}[t]
    \centering
    \includegraphics[width=0.8\linewidth]{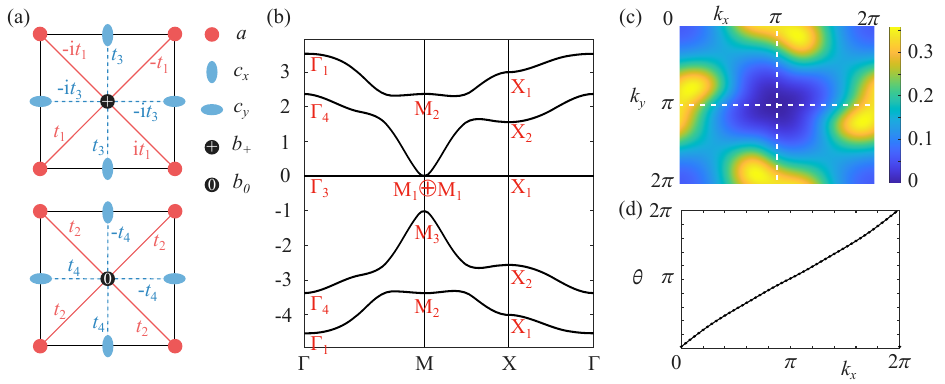}
    \caption{Critical flat band with Chern number 1 in wallpaper group $p4$.
    (a) The nearest neighbor hopping terms $\inn{\RR,\alpha| H_{\rm F} | 0, \beta}$ between $\alpha=a$, $c_x$, $c_y$ and $\beta=b_+$, $b_0$.  
    (b) The band structure, where the red characters are the irreps formed by Bloch states. 
    (c) The Berry's curvature of the flat band. 
    (d) The eigenvalue $\theta(k_x)$ of the flat band Wilson loop integrated over $k_y$, plotted as a function of $k_x$.  
    In (b)-(d),  parameters are chosen as $t_1=t_2=t_3=t_4=\Delta=1$. 
    }
    \label{fig:p4}
\end{figure}

\subsection{The hopping model and its topology}\label{subsec:p4_model}
We denote the orbital basis in real space as $\ket{\RR,\alpha}$, where $\RR\in \mathbb{Z}^2$ represents the lattice vector,  $\alpha=a$, $c_x$, $c_y$, $b_+$, $b_0$ represent $s$, $p_y$, $p_x$, $p_+=p_x+\ii p_y$, and $s$ orbitals locating at the relative positions
\begin{equation}
    \tt_{a} = (0,0),\qquad \tt_{c_x} = (\frac12,0),\qquad \tt_{c_y} = (0,\frac12),\qquad 
    \tt_{b_+} = \tt_{b_0} = (\frac12,\frac12)\ ,
\end{equation}
respectively. 
Under the $C_4$-rotation symmetry, they transform as 
\begin{equation} \label{eq:C4-action}
    C_4 \ket{\RR, \alpha} = \sum_{\beta} D_{\beta,\alpha} \ket{\RR', \beta},\qquad (\RR'+\tt_\beta = C_4\cdot (\RR + \tt_\alpha) )\ ,
\end{equation}
where the nonzero matrix elements of $D$ are given by 
\begin{equation}
D_{a,a} = 1,\qquad 
D_{c_y, c_x} = - D_{c_x, c_y} = -1,\qquad 
D_{b_+, b_+} = -\ii,\qquad 
D_{b_0, b_0} = 1\ .
\end{equation}
There are only four independent nearest-neighbor hopping terms between $L$ and $\td L$, and we choose them as 
\begin{equation}
t_1 = \inn{\RR,a| H_{\rm F} | \RR, b_+  },\qquad 
t_2 = \inn{\RR,a| H_{\rm F} | \RR, b_0  },\qquad 
t_3 = \inn{\RR,c_x| H_{\rm F} | \RR, b_+  },\qquad 
t_4 = \inn{\RR,c_x| H_{\rm F} | \RR, b_0  }\ . 
\end{equation}
They are complex numbers. 
Equivalent hopping terms can be obtained by applying the $C_4$-rotation in \cref{eq:C4-action}, as summarized in \cref{fig:p4}(a). 
The subscript ``F'' in the Hamiltonian $H_{\rm F}$ represents full model that includes both $L$ and $\td L$ orbitals. 
Define the Bloch basis 
\begin{equation}
    \ket{\phi_{\kk,\alpha}} = \frac1{\sqrt{N}} \sum_{\RR} e^{\ii \, \kk \cdot (\RR + \tt_{\alpha})} \ket{\RR,\alpha} \ ,
\end{equation}
where $N$ is the system size, we obtain the Bloch Hamiltonian 
\begin{equation}
H_{\alpha,\beta} (\kk) = \inn{\phi_{\kk,\alpha} | H_{\rm F} | \phi_{\kk,\beta}}
= \begin{pmatrix}
    0_{3\times 3} & S(\kk) \\ S^\dagger(\kk) & -\Delta \ \mathbb{I}_{2\times2}
\end{pmatrix}_{\alpha,\beta} \ . 
\end{equation}
The $S(\kk)$ matrix can be analytically derived as 
\begin{equation}\label{eq:p4_Sk}
    S(\kk) = \begin{pmatrix}
        2\ii \, t_1  \sin \frac{k_x+k_y}2 + 2t_1 \sin \frac{k_x- k_y}2 & 
        2t_2 \cos \frac{k_x+k_y}2 + 2t_2 \cos \frac{k_x-k_y}2 \\
        2t_3 \cos \frac{k_y}2 &  2\ii \, t_4  \sin \frac{k_y}2 \\
        -2\ii \, t_3 \cos \frac{k_x}2 & 2\ii \, t_4 \sin \frac{k_x}2 
    \end{pmatrix}\ . 
\end{equation}
Note that $H(\kk)$ is periodic over the Brillouin zone up to a unitary embedding matrix: 
\begin{equation} \label{eq:embedding}
    H(\kk + \mathbf{G}) = V^\dagger(\mathbf{G}) H(\kk) V(\mathbf{G}),\qquad 
    V(\mathbf{G}) = \delta_{\alpha,\beta} e^{\ii \tt_\alpha \cdot \mathbf{G}}\ ,
\end{equation}
where $\mathbf{G}$ is a reciprocal lattice. 

The band structure with $t_1=t_2=t_3=t_4=1$ is shown in \cref{fig:p4}(b). 
Since $S^\dagger(\kk)$ is a two-by-three matrix, it must have a zero mode $w(\kk)$.
Then $u(\kk) = (w^T(\kk), 0, 0)^T$ gives the flat band of $H_{\rm F}$. 
To investigate the topology of the flat band, we compute the Berry's curvature $\Omega(\kk)$ of the flat in \cref{fig:p4}(c), and find $C=\frac1{2\pi} \int \dd^2\kk \ \Omega(\kk) = 1$. 
In particular, we have exploited the following equation to compute $\Omega(\kk)$
\begin{equation}
e^{(\ii \Omega(\kk) - \mathrm{Tr}[g(\kk)]) \Delta k^2 + \mathcal{O}(\Delta k^3)} 
= \inn{u(\kk_x) | u(\kk + \vec{1}_x)}
    \inn{u(\kk + \vec{1}_x ) | u(\kk + \vec{1}_x + \vec{1}_y) }
    \inn{u(\kk + \vec{1}_x + \vec{1}_y | u(\kk+\vec{1}_y) }
    \inn{u(\kk +  \vec{1}_y | u(\kk) }\ ,
\end{equation}
where $\Delta k$ is a small quantity, $\vec{1}_x = (\Delta k, 0)$, $\vec{1}_y=(0,\Delta k)$, and $g(\kk)$ is the two-by-two Fubini-Study metric. 

We also compute the Wilson loop 
\begin{equation}
    W(k_x) = \lim_{N\to \infty} \inn{u(k_x,2\pi) | u(k_x, (N-1)\Delta k)} \cdots \inn{ u(k_x,2\Delta k) |  u(k_x,\Delta k) } \cdots \inn{ u(k_x, \Delta k) | u(k_x, 0)} \ ,
\end{equation}
where $\Delta k = \frac{2\pi}{N}$. 
Note that $\ket{u(k_x,2\pi)}$ is related to $\ket{u(k_x,0)}$ via the embedding matrix defined in \cref{eq:embedding}. 
We plot the phase $\theta(k_x) = -\ii \log W(k_x)$ in \cref{fig:p4}(d). 
Its winding gives the Chern number $C=1$. 

We now {\it analytically} calculate the Chern number. 
This calculation will also demonstrate why the touching point at $\rM$ does not affect the topology. 
Since the upper three bands together must be topologically trivial as they are equivalent to the BR of the $L$-sublattice, the Chern number $C$ of the flat band must be opposite to the total Chern number of the fourth and fifth bands. 
We hence have 
\begin{equation}
    -C = \frac1{2\pi} \int_{\mathrm{BZ}-\mathcal{D}} \dd^2\kk \ (\Omega_4(\kk) + \Omega_5(\kk) )\ ,
\end{equation}
where $\mathcal{D}$ represents an infinitesimal neighborhood of $\rM$, and $\Omega_4(\kk)$ and $\Omega_5(\kk)$ are the Berry's curvatures of the fourth and fifth bands, respectively. 
Since $S(\kk)$ is analytic and has a rank of two in $\mathrm{BZ} - \mathcal{D}$, the Bloch states $u_4(\kk)$ and $u_5(\kk)$, which are related to the orthonormalized columns of $S(\kk)$, {\it together} must have a {\it smooth gauge} in $\mathrm{BZ} - \mathcal{D}$. 
Applying Stokes' theorem, we have 
\begin{equation} \label{eq:p4-Chern-analytic}
    -C = \frac1{2\pi} \ointclockwise_{\partial \mathcal{D}} \dd\kk \cdot 
    (\ii \inn{u_4(\kk) | \partial_\kk u_4(\kk)} + \ii \inn{u_5(\kk) | \partial_\kk u_5(\kk)} ) \ . 
\end{equation}
Since $u_5(\kk)$ is gapped at $\rM$, it is smooth at $\rM$ and does not contribute to the above equation. 

To calculate the Berry's phase contributed by the fourth band, we expand the $S^\dagger(\kk)$ matrix around $\rM$: 
\begin{equation}\label{eq:unitary_on_Sk}
    S^\dagger(\pi+p_x,\pi+p_y) = 
\begin{pmatrix}
    (1+\ii) t_1 (p_x + \ii p_y) & - t_3 p_y & -\ii t_3 \ p_x \\
    0 & -2\ii t_4 & -2\ii t_4
\end{pmatrix}
\to \begin{pmatrix}
    (1+\ii) t_1 (p_x + \ii p_y) & \frac{\ii t_3}{\sqrt2}(p_x+\ii p_y) & \mathcal{O}(p) \\
    0 & 0 & -2\sqrt2 \ii t_4
\end{pmatrix}\ . 
\end{equation}
We have applied a unitary transformation to the columns such that the corresponding ket states form irreps $\rM_1$, $\rM_1$, $\rM_2$, respectively. 
The two rows correspond to bra states of the irreps $\rM_3$. $\rM_2$, respectively. 
The two $\rM_2$ states will be gapped out by the $t_4$ term and hence are irrelevant in the low energy physics. 
Then the reduced $S^\dagger(\kk)$ matrix in the remaining space reads 
\begin{equation}
    s^\dagger(\kk) = \begin{pmatrix}
        \gamma_1 (p_x + \ii p_y) & \gamma_2 (p_x + \ii p_y)
    \end{pmatrix},\qquad \gamma_1 = 1+\ii, \quad \gamma_2=\frac{\ii t_3}{\sqrt2} \ . 
\end{equation}
After further integrating out the $\rM_3$ level at $-\Delta$, the effective Hamiltonian for the third and fourth bands is $ \frac{s(\kk) s^\dagger (\kk)}{\Delta}$. 
Thus, the vector 
\begin{equation} \label{eq:p4-wavefunction-v4}
    v_4(\kk) = \frac1{|\mathbf{p}|\sqrt{|\gamma_1|^2+|\gamma_2|^2}} s(\kk)
    =\frac1{|\mathbf{p}|\sqrt{|\gamma_1|^2+|\gamma_2|^2}} 
    \begin{pmatrix}
        \gamma_1^*(p_x-\ii p_y) \\ \gamma_2^* (p_x -\ii p_y)
    \end{pmatrix}
\end{equation}
can be viewed as the wavefunction (in the rotated basis) of the fourth band that touches the flat band at $\rM$. 
Around $\rM$, it is safe to use $v_4(\kk)$ instead of $u_4(\kk)$ to compute the Berry's phase.
Substituting $v_4(\kk)$ into \cref{eq:p4-Chern-analytic} yields $C=1$. 

\begin{figure}
    \centering
    \includegraphics[width=0.8\linewidth]{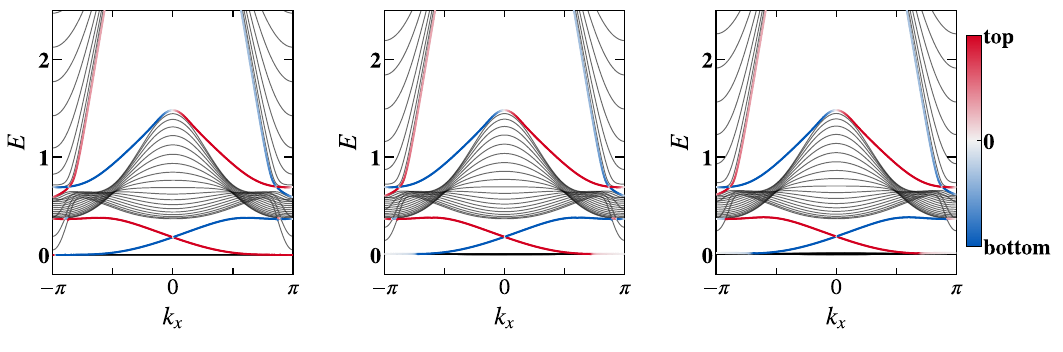}
    \caption{Evolution of the chiral edge modes in the $C=1$ model in $p4$ group, calculated in a stripe geometry. From left to right, we gradually turn on a bipartite-breaking perturbation, $\Delta'=0,0.1,0.2$. 
    The bulk touching points at $k_x = \pm \pi$ are gapped, and the bulk flat bands acquire dispersions. 
    The edge modes signifying $C=1$ remain qualitatively unchanged.}
    \label{fig:p4_edgemode}
\end{figure}

\subsection{Power-law correlation function}
\label{sec:p4-correlation}

We now determine the flat band wavefunction $u(\kk) = (w^T(\kk),0,0)^T$ around $\rM$. 
Since $s^\dagger(\kk) \cdot w(\kk) = 0$, the general solution is given by 
\begin{equation}
    w(\kk) = \frac1{\sqrt{s^\dagger(\kk) s(\kk)}} 
    \begin{pmatrix}
        -s^*_2(\kk) \\ s^*_1(\kk)
    \end{pmatrix}
    = \frac1{|\pp|\sqrt{|\gamma_1|^2+|\gamma_2|^2} } \begin{pmatrix}
        -\gamma_2 (p_x + \ii p_y)  \\ \gamma_1 (p_x + \ii p_y) 
    \end{pmatrix} + \text{higher order terms}\ ,
\end{equation}
where both $-s_2^*(\kk)$ and $s_1^*(\kk)$ are analytic because they have the form $\sum_{\RR\alpha\beta} \gamma_{\RR\alpha\beta} e^{\ii\kk\cdot(\RR+\tt_\alpha - \tt_\beta)}$ in finite-range hopping models.
It is direct to see $w(\kk)$ has the opposite Berry's phase to that of $v_4(\kk)$.
Here we also derive the higher-order dependencies on $\kk$ of $w(\kk)$. 
Notice that $w(\kk)$ satisfies the sewing matrix condition  $w(C_4\kk) = w(\kk) \cdot \ii$, which further constrains the general form of $w(\kk)$ to be 
\begin{equation}
    w(\kk) = \frac1{|\pp|\sqrt{|\gamma_1|^2+|\gamma_2|^2} + \mathcal{O}(p^3)} 
    \begin{pmatrix}
        (-\gamma_2 + \gamma_3|\pp|^2)(p_x + \ii p_y) + \gamma_4 (p_x-\ii p_y)^3 + \cdots \\ 
        (\gamma_1 + \gamma_5 |\pp|^2) (p_x + \ii p_y) + \gamma_6 (p_x-\ii p_y)^3 + \cdots
    \end{pmatrix}\ ,
\end{equation}
where $\gamma_{1,2\cdots}$ are complex coefficients. 

We can remove the discontinuity of $w(\kk)$ at $\rM$ by a local gauge transformation
\begin{equation} \label{eq:p4-bar-w}
\bar w(\kk) = e^{-\ii\theta_\pp} w(\kk) = \begin{pmatrix}
        \kappa_1 + \kappa_2 |\pp|^2 + \kappa_3 |\pp|^2 e^{-4\ii \theta_\pp} + \cdots \\
        \kappa_1' + \kappa_2' |\pp|^2 + \kappa_3' |\pp|^2 e^{-4\ii \theta_\pp} + \cdots
    \end{pmatrix}\ , \qquad 
\theta_\pp = \arccos \frac{p_x}{\sqrt{p_x^2 + p_y^2}}\ . 
\end{equation}
One may determine the coefficients $\kappa_{1,2,3}$ and $\kappa_{1,2,3}'$ from $\gamma_{1,2\cdots}$ order by order, but these explicit relations are not needed in our analysis below. 
Due to the nontrivial Chern number, the new gauge $\bar w(\kk)$ must have discontinuities in its phase at other momenta in the Brillouin zone.
But it will be convenient to use $\bar w(\kk)$ to study the local bundle around $\rM$. 
{\it Crucially, due to the singular normalization factor $\propto \frac1{|\pp|}$, the wavefunction $\bar w(\kk)$, which is already chosen continuous at $\rM$, is non-analytic at $\rM$ because  its second order derivatives are not continuous.} 
At any other momentum, one can always choose the gauge $w(\kk)$ such that the Bloch state is analytic. 
Thus, $\rM$ is the {\it only} non-analytic point in the Brillouin zone. 
Nevertheless, since $\bar w(\kk)$ is continuous, the Berry's curvature at $\rM$ is finite (\cref{fig:p4}(c)), making the Chern number of the flat band well-defined (\cref{fig:p4}(d)). 

\begin{figure}[t]
    \centering
    \includegraphics[width=1\linewidth]{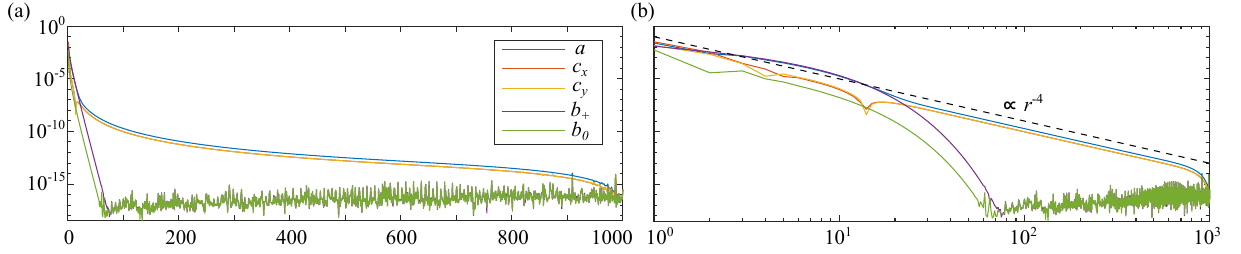}
    \caption{Correlation functions $|C_{\alpha,\alpha}(\RR)|$ of the Fock state occupying the lower three bands of the CTFB model  in the $p4$ wall paper group. $\RR$ takes values in $(n,n)$ for $n=1\cdots 1000$. The system size is $2000\times 2000$, and the parameters are given by $t_1=t_2=t_3=t_4=\Delta=1$. (a) and (b) uses linear and logarithmic coordinates for $n$, respectively. The dashed line in (b) indicates $0.1/n^4$. }
    \label{fig:p4-correlation}
\end{figure}

The finite differentiability leads to power-law decaying correlation functions. 
We consider the Fock state $\ket{\Omega_{\rm F}}$ occupying the lower three bands. 
We denote the annihilation operator of the orbital $\ket{\RR,\alpha}$ as $\psi_{\RR,\alpha}$. 
Then the correlation function can be computed as 
\begin{equation} \label{eq:correlation-def}
C_{\alpha\beta}(\RR-\RR') = \inn{\Omega | \psi_{\RR'\alpha}^\dagger \psi_{\RR \beta} |\Omega} = 
\frac1{N} \sum_{\kk} e^{\ii \kk \cdot(\RR + \tt_\beta - \RR' - \tt_\alpha)} P_{\alpha\beta}(\kk),\qquad 
    P_{\alpha\beta}(\kk) = \sum_{n\in \mathrm{occ}}
    u_{\alpha n}^*(\kk) u_{\beta n}(\kk) 
\end{equation}
with $u_{\alpha n}(\kk)$ being the Bloch wavefunction of the $n$-th band, $P(\kk)$ the projector to occupied bands. 
In this example, $\sum_{n\in \mathrm{occ}} = \sum_{n=1}^3$. 
We present the diagonal correlation functions, where $\alpha=\beta$, in \cref{fig:p4-correlation}. 
It can be seen that, for $\alpha=b_+, b_0 \in \td{L}$, the correlation functions decay exponentially; whereas for $\alpha=a, c_x, c_y \in L$, the correlation functions decay as $1/r^4$. 
We provide an analytical proof in \cref{sec:correlation-analytical}. 

\subsection{Entanglement spectrum}

To study the entanglement features of the filled CTFB state, we regard the correlation function $C_{\alpha, \beta}(\RR-\RR')$ (\cref{eq:correlation-def}) as a matrix $C_{\alpha\RR', \beta\RR}$, where $(\RR,\beta)$ and $(\RR',\alpha)$ are the right and left indices, respectively. 
For an interested subsystem $A$, we introduce the sub-matrix $C^{(A)}$ of $C$ by restricting $(\RR,\beta)$, $(\RR',\alpha) \in A$. 
The entanglement between the subsystem $A$ and its complement $\bar A$ is characterized by the eigenvalues $\{\xi_i\}$ of $C^{(A)}$, which are also referred to as entanglement spectrum. 
In particular, the entanglement entropy is given by
\begin{equation}
    S_{A,\bar A} = - \sum_i [ \xi_i \ln \xi_i + (1-\xi_i) \ln (1-\xi_i)]\ .
\end{equation}
Since $C_A$ is a density matrix, all its eigenvalues lie between 0 and 1, and $S_{A,\bar A}$ is nonnegative.

\begin{figure}[h]
\centering
 \includegraphics[width=0.7\linewidth]{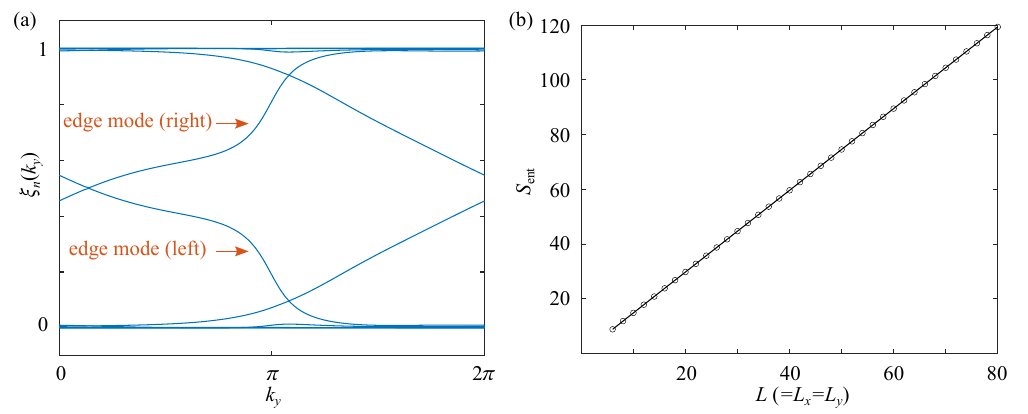}
 \caption{Entanglement property of the critical flat Chern number in wallpaper group $p4$. 
  A system of the size $L\times L$ ($L\in 2\mathbb{N}$) is cut into two subsystems, $R_x=1\cdots L/2$ and $R_x=L/2+1 \cdots L$. 
  Since the subsystems respect the translation symmetry along $y$, the entanglement spectrum $\{\xi_i (k_y)\}$ can be labeled by the momentum $k_y$. 
  We plot the entanglement spectrum in (a), where the parameters are chosen as $t_1=t_2=t_3=t_4=\Delta=1$. 
  We also calculate the entanglement entropy $S_{\rm ent}$ at different system sizes, as shown in (b). 
  The linear dependence of $S_{\rm ent}$ on $L$ reveal the area-law nature of the entanglement. 
  Note that in (a), the reason why the spectrum at ($+k_y$, right edge) is not symmetric to ($-k_y$ , left edge) is because our choice of edge has not obeyed the $C_2$ symmetry.
 }
 \label{fig:entanglement_p4}
\end{figure}

We consider an $N = L_x \times L_y$ system with $L_x$ being an even integer, and choose the subsystem as the left half, {\it i.e.}, $A = \{ (\RR,\beta) \ | \ 1\le R_x \le L_x/2\}$. 
Since $A$ respects the translation symmetry along the $y$ direction, we can apply Fourier transformation to the coordinates in $y$ direction: 
\begin{equation}
C^{(A)}_{\alpha R_x', \beta R_x}(k_y) = \frac1{L_y} \sum_{R_y R_y'}
e^{-\ii k_y(R_y - R_y')} C_{\alpha\RR', \beta \RR}^{(A)} 
= \frac1{L_x} \sum_{k_x } e^{\ii k_x (R_x - R_x') + \ii \kk\cdot (\tt_\beta - \tt_\alpha)} P_{\alpha \beta} (\kk)
\end{equation}
where $\kk=(k_x,k_y)$, $\RR=(R_x,R_y)$. 
We denote the eigenvalues of $C^{(A)}_{\alpha R_x', \beta R_x}(k_y)$ as $\{ \xi_i (k_y) \}$. 
As demonstrated in Ref.~\cite{hughes_inversion-symmetric_2011}, $\{\xi_i(k_y)\}$ can be viewed as energy bands of the subsystem $A$ with a flattened bulk Hamiltonian, where valence and conduction bands have the energies 1 and 0, respectively. 
The topology is manifested as spectral flows in $\{\xi_i(k_y)\}$, in correspondence to the topological edge modes. 
We plot the entanglement spectrum and entropy in \cref{fig:entanglement_p4}. The linear dependence of $S_{\rm ent}$ on $L$ ($=L_x = L_y$) reveal the area-law nature of the entanglement entropy.

\subsection{Real-space perspective: compact localized states and loop states}

In this section, we relate the CTFB wave-functions in the momentum space to the compact localized states (CLS) and the loop states in the real space \cite{liu2026theory}. 

First, we obtain the kernel states of $S^\dagger(\kk)$. 
As $S^\dagger(\kk)$ is a $2 \times 3$ matrix, its kernel state must be orthogonal to both of its rows. 
Since both rows are three dimensional vectors, the orthogonal vector can be easily expressed as their cross product. 
Therefore, the un-normalized kernel state at arbitrary $\kk$ is given by
\begin{align}  \label{eq:p4_kernel_forallk}
    \Ket{\kk} = \begin{pmatrix}
        \Ket{\kk, a} & \Ket{\kk, c_x} & \Ket{\kk, c_y} \\
    \end{pmatrix} \begin{pmatrix}
        t_3 t_4 \Big( -\ii  \sin\frac{k_x}{2}\cos\frac{k_y}{2}  - \cos\frac{k_x}{2} \sin\frac{k_y}{2} \Big) \\
        \ii t_2 t_3  \Big( \cos\frac{k_x}{2} \cos\frac{k_x+k_y}{2} + \cos\frac{k_x}{2} \cos\frac{k_x-k_y}{2} \Big) + t_1 t_4 \Big( \sin\frac{k_x}{2} \sin\frac{k_x+k_y}{2} + \ii \sin\frac{k_x}{2} \sin\frac{k_x-k_y}{2} \Big) \\
        -t_1t_4 \Big( \sin\frac{k_y}{2} \sin\frac{k_x+k_y}{2} + \ii \sin\frac{k_y}{2} \sin\frac{k_x-k_y}{2} \Big) - t_2t_3 \Big( \cos\frac{k_y}{2}\cos\frac{k_x+k_y}{2} + \cos\frac{k_y}{2} \cos\frac{k_x-k_y}{2} \Big) \\
    \end{pmatrix}
\end{align}
where we recall that the Fourier transformation is defined as (up to a constant normalization factor)
\begin{align}
    |\kk,a\rangle = \sum_{\RR} |\RR, a\rangle e^{\ii \kk \cdot \RR} \qquad 
    |\kk,c_x\rangle = \sum_{\RR} |\RR,c_x\rangle e^{\ii \kk \cdot \RR} e^{\ii \frac{k_x}{2}} \qquad 
    |\kk,c_y\rangle = \sum_{\RR} |\RR , c_y\rangle e^{\ii \kk \cdot \RR} e^{\ii \frac{k_y}{2}} 
\end{align}
To obtain a CLS, we Fourier transform $\Ket{\kk}$. 
In principle, a $\kk$-dependent prefactor can be assigned to each $\Ket{\kk}$. 
For the CLS to have a finite support, we want the number of Fourier components contained in $|\kk\rangle$ to be finite, and as few as possible. Therefore, we choose to define
\begin{align}  \label{eq:p4_CLS}
    \Ket{{\rm CLS}; \RR} = \sum_{\kk} e^{-\ii \kk \cdot \RR} e^{-\ii \frac{k_x+k_y}{2}} |\kk\rangle
\end{align}
where $|\kk\rangle$ is given by \cref{eq:p4_kernel_forallk} with no additional $\kk$-dependent factors, and the ``center'' for this CLS appears to be located at $(\frac{1}{2}, \frac{1}{2})$. Substituting \cref{eq:p4_kernel_forallk} into \cref{eq:p4_CLS} leads to 
\begin{align}   \label{eq:p4_CLS_final}
    \Ket{{\rm CLS}; \RR}
    &= t_3t_4\Big[
        (-1+\ii)\Ket{\RR, a}
        +(1+\ii)\Ket{\RR+(1,0), a}
        -(1+\ii)\Ket{\RR+(0,1), a}
        +(1-\ii)\Ket{\RR+(1,1), a}
    \Big] \nonumber\\
    &+ \Big[
        (-t_1t_4+\ii t_2t_3)\big(\Ket{\RR+(1,1), c_x} + \Ket{\RR+(-1,0), c_x}\big)
        \nonumber\\
        &\qquad
        +\ii(-t_1t_4+t_2t_3)\big(\Ket{\RR+(1,0), c_x}+\Ket{\RR+(-1,1), c_x}\big)
        \nonumber\\
        &\qquad
        +\big((1+\ii)t_1t_4+2\ii t_2t_3\big)
        \big(\Ket{\RR+(0,0), c_x}+\Ket{\RR+(0,1), c_x}\big)
    \Big] \nonumber\\
    &+ \Big[
        (t_1t_4-t_2t_3)\big(\Ket{\RR+(1,1), c_y}+\Ket{\RR+(0,-1), c_y}\big)
        \nonumber\\
        &\qquad
        +\big((-1+\ii)t_1t_4-2t_2t_3\big)
        \big(\Ket{\RR+(1,0), c_y}+\Ket{\RR+(0,0), c_y}\big)
        \nonumber\\
        &\qquad
        +(-\ii t_1t_4-t_2t_3)
        \big(\Ket{\RR+(1,-1), c_y}+\Ket{\RR+(0,1), c_y}\big)
    \Big].
\end{align}
It spans a region $2 \times 2$ unit cells (with the boundary included). 
Its $C_4$ eigenvalue (with $C_4$ center located at $\RR+(\frac{1}{2}, \frac{1}{2})$) is $+\ii$. 

\cref{eq:p4_CLS_final} obeys a ``Stokes theorem'', 
\begin{align}  \label{eq:p4_Stokes}
    \sum_{\RR \in V} e^{\ii (\pi,\pi) \cdot \RR} \Ket{{\rm CLS}; \RR} = \sum_{\RR^* \in \partial V} \Ket{{\rm  edge}; \RR^*}
\end{align}
where $\Ket{{\rm edge}; \RR^*}$ denotes some states that are supported only near the boundary of $V$, $\partial V$, and $\RR^*$ labels the spatial degrees of freedom at the boundary. 
\cref{eq:p4_Stokes} can be explicitly verified by noting that for $\Ket{{\rm CLS}; \RR} = \sum_{\Delta\RR,\alpha} \psi_{\Delta\RR,\alpha}(\RR) \Ket{\RR+\Delta\RR, \alpha}$ [\cref{eq:p4_CLS_final}], the expansion coefficients obey $\sum_{\Delta\RR} e^{\ii (\pi,\pi) \cdot \Delta\RR} \psi_{\Delta\RR, \alpha}(\RR)$. 
When $V$ is the whole two-dimensional periodic space, the right-hand-side vanishes. 
This is consistent with the fact that $\Ket{\kk} = \sum_{\RR} e^{\ii \kk \cdot \RR} \Ket{{\rm CLS}; \RR} = 0$ at and only at $\kk = \mathbf{M}= (\pi, \pi)$ [\cref{eq:p4_kernel_forallk}], where the touching point is located. 

Consider $V$ is a stripe that runs along the $x$ direction. Suppose its width along $y$ direction is large enough ($\ge 3$ unit cells). Then by Stokes theorem, 
\begin{align}
    \sum_{\RR \in V} e^{\ii (\pi,\pi) \cdot \RR} \Ket{{\rm CLS}; \RR}  = \theta_{y}(y_1) - \theta_y(y_2)
\end{align}
where $y_1, y_2$ label the uppermost and lowest row of unit cells contained in the stripe $V$, respectively. Calculation shows that 
\begin{align}
    \theta_y(y_1) = \sum_{\RR|\RR_y=y_1} e^{\ii (\pi, \pi) \cdot \RR} \Big[ 2 t_3t_4 \Ket{\RR, a} - 2(1+\ii) t_1 t_4 \Ket{\RR, c_x}
    + (1+\ii) t_1 t_4 \Ket{\RR, c_y} - (1+\ii) t_1 t_4 \Ket{\RR-(0,1), c_y} \Big] 
\end{align}
where $\sum_{\RR|\RR_y=y_1}$ sums over $\RR$ with fixed $\RR_y=y_1$. 
In fact, an individual edge, \textit{e.g.} $\theta_y(y_1)$, must also be a zero mode, as the Hamiltonian is strictly local, hence unable to mix $\theta_y(y_1)$ and $\theta_y(y_2)$ when the stripe is wide enough. An individual edge cannot be expressed as a summation of CLS's. For example, we can examine the following (summation over) loop state, which can also be expressed as a summation over CLS's and one certain $\theta_y(y_1)$, 
\begin{align}
    \Theta_y = \sum_{y} \theta_y(y_1) = 2t_3t_4 \Ket{\mathbf{M}, a} - 2(1-\ii) t_1 t_4 \Ket{\mathbf{M}, c_x} + 2(1-\ii) t_1 t_4 \Ket{\mathbf{M}, c_y}
\end{align}
Such $\Theta_y$ has momentum $\mathbf{M}$, but $\Ket{\kk}$ [\cref{eq:p4_kernel_forallk}] that spans CLS's vanishes at $\kk = \mathbf{M}$, hence $\Theta_y$ does not belong to CLS. 
$\Theta_y$ forms a $C_4$ irrep $\rm M_1$ with eigenvalue $1$, in agreement with the previous momentum space analysis of the touching point irreps. 

Similarly, one can obtain the stripe edge states running along the $y$ direction. 
Summing \cref{eq:p4_CLS_final} along the stripe gives
\begin{align}
    \sum_{\RR \in V} e^{\ii(\pi,\pi)\cdot\RR}\Ket{{\rm CLS};\RR}
    =\theta_x(x_1)-\theta_x(x_2).
\end{align}
where $x_1$ and $x_2$ are the leftmost and rightmost edges, with
\begin{align}
    \theta_x(x_1)
    =\ii \sum_{\RR|\RR_x=x_1} e^{\ii(\pi,\pi)\cdot\RR}
    \Big[
        2 t_3t_4\Ket{\RR,a} 
        -(1+\ii)t_1t_4\Ket{\RR,c_x}
        +(1+\ii)t_1t_4\Ket{\RR-(1,0),c_x}
        +2(1+\ii)t_1t_4\Ket{\RR,c_y}
    \Big]
\end{align}
Therefore, $\Theta_x = \sum_{x_1} \theta_x(x_1) = \ii \Theta_y$ spans the same Hilbert space. 
Crucially, they are also equal to the touching point wave-function, as obtained previously. 
Such a property is not enjoyed by singular flat bands, \textit{e.g.} the Lieb lattice, where $\Theta_x$ and $\Theta_y$ are linearly independent. 
Note that in the above real-space definitions, the relative gauge between $\Theta_x$ and $\Theta_y$ has not been fixed with $\partial_{k_x}|\kk\rangle$ and $\partial_{k_y}|\kk\rangle$, hence the sign of $+\ii$ in the relation $\Theta_x = \ii \Theta_y$ cannot determine the phase winding around the touching point yet.

On the other hand, we can examine the derivative $\partial_{\kk}\Ket{\kk}$ near $\mathbf{M}$. 
We expand $|\kk\rangle$ to the linear order of $\pp=\kk-(\pi, \pi)$, 
\begin{align}  \label{eq:p4_kernel_nearM}
    \Ket{\pp+(\pi,\pi)} = \begin{pmatrix}
        \Ket{\pp+(\pi,\pi), a} & \Ket{\pp+(\pi,\pi), c_x} & \Ket{\pp+(\pi,\pi), c_y} \\
    \end{pmatrix} \begin{pmatrix}
        t_3 t_4 \Big( \frac{\pp_x+\ii \pp_y}{2} \Big) \\
        t_1 t_4 (-1+\ii) \Big( \frac{\pp_x + \ii \pp_y}{2} \Big) \\
        t_1 t_4 (1 - \ii) \Big( \frac{\pp_x + \ii \pp_y}{2} \Big) \\
    \end{pmatrix}
\end{align}
As the Bloch ``spinor'' (the $3 \times 1$ vector in \cref{eq:p4_kernel_nearM}) vanishes at $\mathbf{M}$, the total derivative is non-vanishing only when the derivative acts on this spinor. Therefore, $\partial_{p_x}|\pp+(\pi,\pi)\rangle \big|_{\pp=0}$ and $\partial_{p_y}|\pp+(\pi,\pi)\rangle \big|_{\pp=0}$ are given by
\begin{align}
    \partial_{p_x}|\pp+(\pi,\pi)\rangle \big|_{\pp=0} = 
    (-\ii) \times \partial_{p_y}|\pp+(\pi,\pi)\rangle \big|_{\pp=0} = 
    t_3 \Ket{\mathbf{M}, a} - t_1 (1-\ii) \Ket{\mathbf{M}, c_x} + t_1 (1-\ii) \Ket{\mathbf{M}, c_y}
\end{align}
which are also proportional to $\Theta_y$ (and $\Theta_x$). 
Here the phase $-\ii$ in relation $\partial_{p_x}|\pp+(\pi,\pi)\rangle \big|_{\pp=0} = (-\ii) \times \partial_{p_y}|\pp+(\pi,\pi)\rangle \big|_{\pp=0}$ is directly associated with the phase winding hence the Chern number. 
This wave-function furnishes the CTFB touching point.

\clearpage

\section{General algorithm for constructing CTFBs}
\label{app:algr}

We develop a systematic framework to construct symmetry-indicated CTFBs, using the bipartite crystalline lattice \cite{calugaru_general_2021}. 
We first introduce the notations on space groups and topological quantum chemistry in \cref{app:algr-def}, then we investigate the condition of symmetry-guaranteed continuity of flat band wave-function near a touching point in \cref{app:algr-touching}, finally we present the general work flow to search for symmetry-indicated CTFBs in \cref{app:algr-workflow}. 

\subsection{Definitions and notations}
\label{app:algr-def}

\paragraph{Space group and (co-)irreps.}
3D materials with negligible SOC or significant SOC are described by single-valued and double-valued space groups, respectively. 
We denote their unitary part as $\mclG^U = \Big\{g = \{ R_g | \tt_g  \} \Big\}$ and $\mclG^U = \Big\{g = \{ R_g | \tt_g | s_g \} \Big\}$, respectively, where $R_g$ is the $\mathrm{O}(3)$ spatial action, $\tt_g$ is the spatial translation, and $s_g$ implements the same rotation as $R_g$ in the spin-$1/2$ Hilbert space. 
In the momentum space, $g$ brings $\kk$ to $R_g \kk$. 
If time-reversal symmetry $\mclT$ is also present, the full space group reads $\mclG = \mclG^U \bigcup \mclG^U \mclT$. 
For 2D materials, we consider the layer groups $\mclG_{\rm LG}$. 
Each layer group $\mclG_{\rm LG}$ can be deemed as the normal subgroup of some 3D space group $\mclG$ modulo the translations along $z$-direction, which we summarize in \cref{tab:isoTP_2D}. 
We will treat all these scenarios on equal footing in below. 

We label different high-symmetry points in the Brillouin zone by $\KK$ ($\KK= \rm \Gamma,A,K,\cdots$). 
Symmetry actions that leave $\KK$ invariant up to reciprocal lattice vectors form the little group $\mclG_{\KK}$, which either takes the form of $\mclG_{\KK} = \mclG_{\KK}^U \subset \mclG^U$ or $\mclG_{\KK} = \mclG_{\KK}^U \bigcup \mclG^U_{\KK} \cdot h\mclT$. 
$h\mclT$ with $h \in \mclG^U$ generates the anti-unitary coset. 

Bloch states at $\KK$ are classified into different (co-)irreps of $\mclG_{\KK}$, which we label by $\gamma$. 
In below, the term ``irrep'' will specifically refer to the representations of the unitary part $\mclG^U_{\KK}$, while ``(co-)irrep'' will refer to the representations of the full little group $\mclG_{\KK}$, with or without the anti-unitary part. 
A (co-)irrep $\gamma$ takes the form of $\gamma = \bigoplus_{i=1}^{q_\gamma} \rho_i$, where $q_\gamma = 1$ or $2$, and $\rho_i$ is an irrep. 
If $q_{\gamma}=1$, $h\mclT$ (if present) acts within the irrep $\rho_1 \subset \gamma$; and if $q_{\gamma}=2$, $h\mclT$ maps $\rho_1 \subset \gamma$ to $\rho_2 \subset \gamma$. 
Notice that, the $q_\gamma=2$ case can be further divided into two subcases -- either $\rho_1 = \rho_2$ or $\rho_1 \not=\rho_2$, but we do not need to distinguish them in this work. 

Bloch states at $\KK$ will be labeled by $|\KK,\rho_i\subset \gamma, m, \alpha\rangle$, where $m=1,2,\cdots$ labels the different copies (multiplicities) of the same (co-)irrep $\gamma$, and $\alpha = 1,\cdots,\dim(\rho_i)$ labels the degenerate Bloch states within the same irrep $\rho_i \subset \gamma$. 
A unitary symmetry action $g \in \mclG^U_{\KK}$ acts as 
\begin{align}  \label{eq:Dg}
    g |\KK, \rho_i \subset \gamma, m, \alpha\rangle = \sum_{\alpha'} |\KK, \rho_i \subset \gamma, m, \alpha'\rangle \Big[ D^{\rho_i}(g) \Big]_{\alpha',\alpha}  \ .
\end{align}
We fix the representation matrix $D^{\rho_i}$ as the same across all the $m=1,2,\cdots$ multiplicities. 
$h\mclT$ (if present) acts as
\begin{align}    \label{eq:DhT_rho}
    &h\mclT |\KK, \rho \subset \gamma, m, \alpha\rangle = \sum_{\alpha'} |\KK, h\mclT(\rho) \subset \gamma, m, \alpha'\rangle \Big[ D^{h\mclT(\rho) \leftarrow \rho}(h\mclT) \Big]_{\alpha',\alpha} 
\end{align}
Notice that $h\mclT$ furnishes a one-to-one map between the irrep components contained in $\gamma$ -- for $q_\gamma = 1$, we denote $h\mclT(\rho_1) = \rho_1$, and for $q_\gamma = 2$, we denote $h\mclT(\rho_i) = \rho_{3-i}$. 

The momentum deviation from $\KK$ will be denoted by $\kk$. 
We denote the vector representation formed by $\kk$ under the \textit{unitary} symmetry actions $\mclG^U_{\KK}$ as $v$. 
If $v$ is real irreducible (note this property is \textit{not} related to the presence or absence of $h\mclT$), we refer to such $\KK$ as an isotropic high-symmetry point. 
In 3D, $\KK$ is isotropic if $v$ is also an irrep over complex numbers, or equivalently, if $\mclG^U_{\KK}$ is isomorphic to $O, O_h, T, T_h$ or $T_d$. 
In 2D, $\KK$ is isotropic if any $g \in \mclG^U_{\KK}$ has $[R_g]_{x,y} \not=0$. 
We list the isotropic high-symmetry points in \cref{tab:isoTP_2D, tab:isoTP_3D} for reference. 
Only isotropic high-symmetry points serve as touching points in our construction of the symmetry-indicated CTFBs.

\paragraph{Elementary band representation and symmetry-based indicators of topological bands.}
Symmetry-based indicators (SI) of band topology in space group $\mclG$ is encoded by a data matrix $\mclE$ of size $N_{R} \times N_{EBR}$. 
Each column (labeled by $a=1,\cdots,N_{EBR}$) is termed as an elementary band representation (EBR), which represents one of the ``smallest'' realization of atomic insulators, and each row (labeled by a composite index $(\KK,\gamma)$) of this column indicates how many times of the (co-)irrep $\gamma$ at $\KK$ is occupied when this EBR is Fouriered to the momentum space. 
The entries $\mclE_{(\KK,\gamma),a} \in \mathbb{Z}_{\ge 0}$ are hence all non-negative integers. 
In Smith normal form, $\mclE = \mrmL \cdot \Lambda \cdot \mrmR$, where $\mrmL$ and $\mrmR$ are unimodular integer matrices, and $\Lambda$ is a diagonal matrix of non-negative integers $\lambda_p \in \mathbb{Z}_{\ge 0}$. 
Here, $p$ indexes the diagonal positions. 
The total number of $\lambda_p > 0$ is the rank of $\mclE$, which we denote as $r$. 
We sort $\lambda_p$ such that $p=1,\cdots,r$ are positive. 
In particular, any $\lambda_p > 1$ corresponds to an SI \cite{po_symmetry-based_2017}. 

Any electron band structure can be similarly expressed as an integer column data vector $\mclB$, indexed by $\mclB_{(\KK,\gamma)}$, where each row element indicates how many times of $\gamma$ is occupied at $\KK$. 
For $\mclB$ to represent a physical band structure, there should be $\mclB_{(\KK,\gamma)} \ge 0$. 
If $d_{\mclB \big|_{\KK}} = \sum_{\gamma \in \KK} \mclB_{(\KK,\gamma)} \times \dim(\gamma)$ is equal at all high-symmetry points $\KK$, then $\mclB$ has a well-defined band dimension which we dub as $d_{\mclB}$. 

The compatibility relations. and SI of $\mclB$ are examined by $\mrmL^{-1} \cdot \mclB$. 
(i) If $(\mrmL^{-1} \cdot \mclB)_p \not= 0$ for some $p > r$, then $\mclB$ must violate the compatibility relations. 
Such $\mclB$ cannot be expressed by any linear combination of EBR (including fractional combinations). According to Ref.~\cite{po_symmetry-based_2017}, such $\mclB$ must exhibit symmetry-protected nodal points (lines) in high-symmetry lines (planes). 
If $\mathcal{B}$ preserves the compatibility relations, it belongs to one of the following two cases: 
(ii) If $(\mrmL^{-1} \cdot \mclB)_p = 0$ mod $\lambda_p$ for all $p \le r$, then $\mclB$ is either topologically trivial, or carries a fragile topology. 
(iii) If $(\mrmL^{-1} \cdot \mclB)_p \not=0$ mod $\lambda_p$ for some $p \le r$, then $\mclB$ carries a non-trivial SI indicating a stable topology. 
$\mclB$ belonging to case (iii) is the focus of this work.

\paragraph{Bipartite crystalline lattice.}
\label{sec: Bipartite lattice}
The larger and smaller sublattices in the bipartite structure are denoted as $L$ and $\td{L}$. 
The band representation they form are denoted as $\mathcal{BR}_L$ and $\mathcal{BR}_{\td{L}}$, with band dimensions $d_{\mathcal{BR}_L}$ and $d_{\mathcal{BR}_{\td{L}}}$, respectively. 
The CTFB is characterized by $\mathcal{BR}_L \boxminus \mathcal{BR}_{\td{L}}$ which we rewrite as $\mathcal{BR}_L \boxminus \mathcal{BR}_{\td{L}} = \mclB + \Delta\mclB$. Here, $\mclB$ is the target topological flat band with definite dimension $d_{\mclB}$, and $\Delta\mclB$ describes the rest of the (co)-irreps entering the touching points with $d_{\Delta\mclB} = 0$. 
We also define $D_{\Delta\mclB \big|_\KK} = \sum_{\gamma \in \KK} \big| \Delta\mclB_{\KK,\gamma} \big| \times \dim(\gamma)$ and $D_{\Delta\mclB} = \max_{\KK} D_{\Delta\mclB \big|_\KK}$. 
The total band dimension of the bipartite lattice bounded from below by $d = d_{\mathcal{BR}_L} + d_{\mathcal{BR}_{\td{L}}} \ge d_{\mclB} + D_{\Delta\mclB}$.

\subsection{Theory at touching point}
\label{app:algr-touching}

The kernel of the off-diagonal operator $S^\dagger(\kk)$ in the Bloch Hamiltonian forms the flat bands. 
This section investigates how to guarantee the continuity of flat band wave-functions near a touching point $\KK$, using only the symmetry data there. 
For this sake, we study the $k \cdot p$ expansion of $S^\dagger(\kk)$ near $\KK$. 
In below, $\kk$ denotes the momentum deviation from $\KK$, and we will omit the subscripts regarding $\KK$ if not causing confusion. 
We will establish the continuity condition at the $O(\kk)$ order of $k\cdot p$ expansion, and dictate that $O(\kk^2)$ corrections do not violate it. 

\subsubsection{Form of \texorpdfstring{$S^\dagger(\kk)$}{Sk}}   \label{app:algr-touching-Sk}

We classify the Bloch states belonging to $L$ and $\td{L}$ sublattices into (co-)irreps, denoted by $\gamma$ and $\td{\gamma}$, respectively, and organize $S^\dagger(\kk)$ into blocks denoted by $S^\dagger(\kk; \td{\gamma} \leftarrow \gamma)$. 
If and only if $\td{\gamma} = {\gamma}$, $S^\dagger(\kk; \td{\gamma} \leftarrow \gamma) \sim \mathrm{const}$, which will push them into the remote bands, hence do not enter the touching point construction. 
We thus analyze the form of $S^\dagger(\kk; \td{\gamma} \leftarrow \gamma)$ for $\td{\gamma} \not= \gamma$ at the leading order $O(\kk)$. 

\paragraph{Unitary symmetries.}
We first analyze the irrep subblocks $S^\dagger(\kk; \td{\rho} \leftarrow \rho)$ where $\td{\rho}\subset\td{\gamma}$ and ${\rho}\subset{\gamma}$, which are constrained by the unitary symmetries.  
By \cref{eq:Dg}, unitary symmetry $g \in \mclG^U_{\KK}$ requires that
\begin{align}    \label{eq:g_Sk}
    \Big[S^\dagger(R_g\kk; \td{\rho} \leftarrow \rho)\Big]_{\td{\alpha}',\alpha'} = \Big[D^{\td{\rho}}(g)\Big]_{\td{\alpha}',\td{\alpha}} \Big[S^\dagger(\kk;\td{\rho} \leftarrow \rho) \Big]_{\td{\alpha};\alpha} \Big[D^{\rho\dagger}(g) \Big]_{\alpha,\alpha'} 
\end{align}
Notice that we do not need the multiplicity label $m$ here. 
Repeated indices are contracted implicitly. 
At $O(\kk)$ order, $S^\dagger(\kk) = \kk_x A^x + \kk_y A^y + \kk_z A^z = \kk_\mu A^\mu$ (with $\mu=x,y,z$), where $[A^{\mu}]_{\td{\alpha}, \alpha}$ are matrices independent of $\kk$. 
Substituting this expansion into \cref{eq:g_Sk}, we get
\begin{align} \label{eq:g_kA}
    (R_g\kk)_\nu \Big[A^\nu(\td{\rho} \leftarrow \rho)\Big]_{\td{\alpha}',\alpha'} = \kk_\mu \Big[D^{\td{\rho}}(g)\Big]_{\td{\alpha}',\td{\alpha}} \Big[A^\mu(\td{\rho} \leftarrow \rho) \Big]_{\td{\alpha};\alpha} \Big[D^{\rho\dagger}(g) \Big]_{\alpha,\alpha'} 
\end{align}
Recall that $R_g \in O(3)$, and $(R_g\kk)_\nu = [R_g]_{\nu \mu} \kk_\mu $. Take derivative with respect to $\kk_\mu$, invert \cref{eq:g_kA} with $R_g^T$, then we obtain 
\begin{align} \label{eq:g_A}
    \Big[A^\nu(\td{\rho} \leftarrow \rho)\Big]_{\td{\alpha}',\alpha'} = \Big[D^{\td{\rho}}(g)\Big]_{\td{\alpha}',\td{\alpha}} \Big[A^\mu(\td{\rho} \leftarrow \rho) \Big]_{\td{\alpha};\alpha} \Big[D^{\rho\dagger}(g) \Big]_{\alpha,\alpha'} \Big[ R_g^T \Big]_{\mu; \nu}
\end{align}

$A$ can be deemed as a tensor (or reshaped into a vector) with three indices, $\mu \otimes \td{\alpha} \otimes \alpha$, and $g \in \mclG^U_{\KK}$ transforms it with $\big[R_g\big]_{\nu;\mu} \otimes \big[ D^{\td{\rho}}(g) \big]_{\td{\alpha}',\td{\alpha}} \otimes \big[ D^{\rho}(g) \big]^*_{\alpha',\alpha}$, which is a direct product representation $v \otimes \td{\rho} \otimes \rho^*$, where $\rho^*$ is the complex conjugation of $\rho$. 
\cref{eq:g_A} then dictates that $A$ must form the identity irrep of $\mclG^U_{\KK}$. 
We denote the multiple of times that the identity irrep $\mathbf{1}$ appears in $v \otimes \td{\rho} \otimes \rho^*$ as $L(\td{\rho} \leftarrow \rho) = \langle \mathbf{1}, v \otimes \td{\rho} \otimes \rho  \rangle\in \mathbb{Z}_{\ge 0}$, which indicates the number of free parameters in this Hamiltonian block at order $O(\kk)$. 
We label the $L(\td{\rho} \leftarrow \rho)$ independent solutions to \cref{eq:g_A} as $A(n; \td{\rho} \leftarrow \rho)$ with $n = 1,\cdots,L(\td{\rho} \leftarrow \rho)$. 
We orthonormalize them as $\sum_{\mu} \mathrm{Tr} \big[ A^{\mu \dagger}(n;\td{\rho} \leftarrow \rho) A^{\mu}(n';\td{\rho} \leftarrow \rho) \big] = \delta_{n,n'}$, so that entries of $A$ are dimensionless. 
The hopping Hamiltonian in this subblock then takes the form 
\begin{align}   \label{eq:Sk_tdrho_rho_general}
    S^\dagger(\kk;\td{\rho} \leftarrow \rho) = \sum_{n} Z(n;\td{\rho} \leftarrow \rho) \times \kk_\mu A^\mu(n;\td{\rho} \leftarrow \rho)
\end{align}
where $Z(n;\td{\rho} \leftarrow \rho)$ for $n=1,\cdots,L(\td{\rho} \leftarrow \rho)$ are complex numbers that represent the hopping energy scales associated with each set of coupling matrices $A^\mu(n; \td{\rho}\leftarrow\rho)$.

\paragraph{Anti-unitary symmetries.}
We now show $h\mclT$ (if present) simply imposes constraints that relate $Z(n;\td{\rho} \leftarrow \rho)$ to $Z^*(n';h\mclT(\td{\rho}) \leftarrow h\mclT(\rho))$ [\cref{eq:hT_Z}]. 
Given \cref{eq:DhT_rho}, the symmetry of $h\mclT$ requires that
\begin{align}  \label{eq:hT_Sk}
    \Big[S(R_{h\mclT}\kk; h\mclT(\td{\rho}) \leftarrow h\mclT(\rho)) \Big]_{\td{\alpha}';\alpha'} = \Big[D^{h\mclT(\td{\rho}) \leftarrow \td{\rho}}(h\mclT) \Big]_{\td{\alpha}',\td{\alpha}} \Big[S(\kk;\td{\rho} \leftarrow \rho) \Big]^{*}_{\td{\alpha};\alpha} \Big[D^{h\mclT(\rho) \leftarrow \rho\dagger}(h\mclT) \Big]_{\alpha,\alpha'} \ ,
\end{align}
where $R_{h\mclT} = - R_h \in O(3)$ as $\mclT$ reverses momenta. Substitute \cref{eq:Sk_tdrho_rho_general} into \cref{eq:hT_Sk}. 

First, notice that as any $g \in \mclG^U_{\KK}$ is represented as an identity action in the linear space formed by $A(n; \td{\rho}\leftarrow\rho)$, $h\mclT$ commutes with all $g$ \textit{in this representation space}. 
Also, as $(h\mclT)^2 \in \mclG^U_{\KK}$ is represented as an identity action as well, $(h\mclT)^2 = 1$ \textit{in this representation space}. 
These facts allow us to always \textit{choose} the basis of $A$ tensors according to
\begin{align}  \label{eq:hT_A}
    \Big[A^{\nu}(n;h\mclT(\td{\rho}) \leftarrow h\mclT(\rho)) \Big]_{\td{\alpha}';\td{\alpha}} = \Big[D^{h\mclT(\td{\rho}) \leftarrow \td{\rho}}(h\mclT) \Big]_{\td{\alpha}', \alpha'} \Big[A^{\mu}(n; \td{\rho} \leftarrow \rho) \Big]^{*}_{\td{\alpha};\alpha} \Big[D^{h\mclT(\rho) \leftarrow \rho\dagger}(h\mclT) \Big]_{\alpha,\alpha'} \Big[ R_{h\mclT}^T \Big]_{\mu\nu}   \ \ . 
\end{align}
Then, to meet \cref{eq:hT_Sk} (where $S^\dagger(\kk)$ is given by \cref{eq:Sk_tdrho_rho_general}), there must be
\begin{align}   \label{eq:hT_Z}
    Z^*(n; h\mclT(\td{\rho}) \leftarrow h\mclT(\rho)) = Z(n;\td{\rho}\leftarrow\rho)  \ .
\end{align}
Note \cref{eq:hT_A} also implies that $L(h\mclT(\td{\rho}) \leftarrow h\mclT(\rho)) = L(\td{\rho}\leftarrow\rho)$. 
We also define $L(\td{\gamma} \leftarrow \gamma) = \sum_{\td{\rho} \subset \td{\gamma}, \rho \subset \gamma} L(\td{\rho}\leftarrow\rho)$.

\subsubsection{Symmetry-indicated continuity at touching point}   \label{app:algr-touching-CKC}

To achieve a symmetry-indicated CTFB, we require that at the touching point, the $L$ sublattice only contains one type of (co-)irrep $\gamma$; otherwise there will be an ambiguity in dividing $\mathcal{BR}_L \boxminus \mathcal{BR}_{\td{L}}$ into $\mclB$ and $\Delta\mclB$, unless special care is taken. 
Therefore, we deal with the following form of CTFB in this work, 
\begin{align}  \label{eq:form_BDB}
    \mclB \big|_\KK = Q \gamma \qquad \textrm{and} \qquad (\mclB + \Delta\mclB) \big|_\KK = (M \gamma) \boxminus (\bigoplus_{j=1}^{J} \td{\gamma}_j) \quad \textrm{with}\quad \td{\gamma}_j \not= \gamma . 
\end{align}
$Q,M,J$ here are positive integers. By notation $\bigoplus^J_{j=1}$ or $\sum_{j=1}^{J}$, we allow $\td{\gamma}_j = \td{\gamma}_{j'}$ even if $j\not=j'$. 
The (co-)irrep dimensions must obey $\sum_{j=1}^{J} \dim (\td{\gamma}_j) = (M-Q) \times \dim (\gamma)$ so that $d_{\Delta\mclB} = 0$. 
For clarity, we first examine the two scenarios, (i) $q_\gamma=1$ and (ii) $q_\gamma = 2$, separately, and then write the continuity condition in a unified manner. 

\paragraph{Case (i), without $\mclT$.} 
It suffices to look at irreps $(M \rho) \boxminus (\bigoplus_{j=1}^J \td{\rho}_j)$ (where $\td{\rho}_j \not= \rho$). 
For the kernel wave-functions to be continuous near $\KK$, they must form $Q$ copies of the $\rho$ irrep when approaching $\KK$, namely, up to $O(\kk)$ corrections, the kernel space must be spanned by the following basis
\begin{align}   \label{eq:continuous_kernel}
    |ker; \kk, \rho, q', \alpha\rangle = \sum_{m=1}^M |\kk, \rho, m, \alpha\rangle \times Y_{m;q'}  \ ,
\end{align}
where $q'=1,\cdots,Q$, and $Y_{:,q'}$ are orthonormal column vectors for different $q'$. 
In particular, if no extra kernel states exist other than \cref{eq:continuous_kernel}, then the $O(\kk^2)$ contributions in $S^\dagger(\kk; \bigoplus_{j=1}^{J} \td{\rho}_j \leftarrow M\rho)$ and the hybridization with remote bands can only perturb \cref{eq:continuous_kernel} with some $O(\kk)$ corrections, which will not violate the continuity. 
We thus require that no extra kernel states exist at arbitrary $\kk \not= 0$ other than \cref{eq:continuous_kernel}, in order to guarantee the continuity. 

In terms of the irrep blocks [\cref{eq:Sk_tdrho_rho_general}], $S^\dagger(\kk)$ at order $O(\kk)$ takes the following form, 
\begin{align}   \label{eq:general_Sdag}
    S^\dagger(\kk) = \sum_{j=1}^{J} \sum_{m=1}^{M} | 
    \kk, \td{\rho}_j,\td{\alpha}_j \rangle \Bigg( \sum_{n_j=1}^{L_j}  Z_{(j,n_j), m} \times \Big[  k_\mu A_{(j,n_j)}^\mu \Big]_{\td{\alpha}_j;\alpha} \Bigg) \langle \kk,\rho,m,\alpha| \ .
\end{align}
Since we are dealing with a fixed $\rho$, above we have abbreviated some notations. 
First, we have abbreviated $L_j = L(\td{\rho}_j\leftarrow\rho)$. 
Second, $(n; \td{\rho}_j \leftarrow \rho)$ is abbreviated as $(j, n_j)$, where $j$ labels the irrep $\td{\rho}_j$, while $n_j = 1,\cdots,L_j$ labels the different free parameters. 
In \cref{eq:general_Sdag}, the hopping energies $Z$ also acquire another index $m=1,\cdots,M$, as the coupling strengths of the same $\td{\rho}_j$ to the $M$ copies of $\rho$ will be independent without fine-tuning. 
However, the coupling matrix $k_\mu A^\mu$ must take the same form across all the $M$ copies. 
In \cref{eq:general_Sdag}, within each block at the $j$-th row ($j=1,\cdots,J$) and the $m$-th ($m = 1,\cdots,M$) column, the $n_j = 1,\cdots,L_j$ index is contracted. 

It will be convenient to collect $Z_{(j,n_j);m}$ into a matrix, where rows are labeled by the composite index $(j,n_j)$, while columns are labeled by $m$. 
Heuristically, the desired flat-band wave-functions [\cref{eq:continuous_kernel}] only depend on $m$ but not on $\alpha$, which suggests that it suffices to examine the kernel of the $Z$, as it is the only quantity in $S^\dagger(\kk)$ that depends on $m$. 
The $Z$ matrix is of size $(\sum_{j=1}^J L_j) \times M$. 
Without fine-tuning, the entries $Z_{(j,n_j);m}$ can be regarded as independently generated, hence $\rank Z = \min\{\sum_{j=1}^J L_j, M\}$ is ``maximal''. 
Its kernel dimension is expected to be $Q$, hence we should henceforth \textit{require} $M-Q = \sum_{j=1}^J L_j$. 
We now solve $\ker Z$ explicitly. 
By QR decomposition, $Z = \Xi Y^{-1}$, where $Y_{m;q}$ is a unitary matrix of size $M \times M$, and $\Xi_{(j,n_j),q}$ is a lower-triangle matrix of size $(M-Q) \times M$ that takes the form
\begin{align}   \label{eq:Xi}
    \Xi =\left( \begin{array}{cccccc||cc}
        \Xi_{(1,1),1} &  &  &  & & & 0 & \cdots \\
        \Xi_{(1,2),1} & \Xi_{(1,2),2} & & & & & 0 & \cdots \\
        \cdots & \cdots & \cdots & & & & 0 & \cdots \\
        \Xi_{(1,L_1),1} & \Xi_{(1,L_1),2} & \cdots & \Xi_{(1,L_1),L_1} & & & 0 & \cdots \\
        \cdots & \cdots & \cdots & \cdots & \cdots & & 0 & \cdots \\
        \Xi_{(J,L_J),1} & \Xi_{(J,L_J),2} & \cdots & \cdots & \cdots & \Xi_{(J,L_J),M-Q} & 0 & \cdots \\
    \end{array} \right)
\end{align}
In above, we have ordered $(j,n_j)$ in the following way --- ascending $n_j$ for fixed $j$ first, then ascending $j$. 
The last $Q$ columns are all zero, implying that the last $Q$ columns of $Y$, denoted by $Y_{:,q}$ for $q=M-Q+1, \cdots, M$, will form $\ker Z$. 
Inserting these $Y_{:,q}$ into \cref{eq:continuous_kernel}, we get the desired continuous flat band wave-functions, which are kernel to $S^\dagger(\kk)$ [\cref{eq:general_Sdag}]. 

Recall that we should also require no extra kernel states other than \cref{eq:continuous_kernel} exist. 
To do this, we insert the first $M-Q$ non-vanishing columns of $\Xi$ back into $S^\dagger(\kk)$ [\cref{eq:general_Sdag}] (which is also equivalent to directly carrying out a $Y Y^{-1}$ transformation to the contracted $m$ index in \cref{eq:general_Sdag}), 
\begin{align}   \label{eq:general_Sdag_Xi}
    S^\dagger(\kk) = \sum_{j=1}^{J} \sum_{q=1}^{M-Q} | 
    \kk, \td{\rho}_j,\td{\alpha}_j \rangle \Bigg( \sum_{n_j=1}^{L_j}  \Xi_{(j,n_j), q} \times  \Big[  k_\mu A_{(j,n_j)}^\mu \Big]_{\td{\alpha}_j;\alpha} \Bigg) \langle \kk,\rho,q,\alpha| \ .
\end{align}
Here, $|\kk,\rho,q,\alpha\rangle = \sum_{m=1}^M | \kk,\rho,m,\alpha \rangle Y_{m;q}$ for $q=1,\cdots,M-Q$ are orthogonal to the continuous kernel states found above. 
Now, \cref{eq:general_Sdag_Xi} forms a square matrix, and should be required to be full-rank. 
To impose this, we carry out a further column transformation $X$ to the $q$ index, which can eventually bring $\Xi$ into a diagonal form, $\Xi = {\Xi}^{\rm diag} \cdot X^{-1}$. 
\begin{align}
    {\Xi}^{\rm diag} = \left( \begin{array}{cccccc||cc}
        \Xi^{\rm diag}_{(1,1)} &  &  & & & & 0 & \cdots \\
         & \cdots & & & & & 0 & \cdots \\
         &  & \Xi^{\rm diag}_{(1,L_1)} & & & & 0 & \cdots \\
         &  &  & \Xi^{\rm diag}_{(2,1)} & & & 0 & \cdots \\
         &  &  &  & \cdots & &  0 & \cdots \\
         &  &  &  &  & \Xi^{\rm diag}_{(J,L_J)} & 0 & \cdots \\
    \end{array} \right) \ ,
\end{align}
where these diagonal elements can be equally indexed by $(j,n_j)$ or $q = 1,\cdots,M-Q$. 
$X$ will not be unitary, but it preserves the matrix rank. 
Insert $\Xi^{\rm diag}$ back into $S^\dagger(\kk)$ [\cref{eq:general_Sdag_Xi}] (which is equivalently to further inserting $X X^{-1}$ to the contracted $q$ index in \cref{eq:general_Sdag_Xi}), and we get
\begin{align}   \label{eq:general_Sk_diag}
    S^\dagger(\kk) &= \sum_{j=1}^J |\kk,\td{\rho}_j,\td{\alpha}_j\rangle \Bigg( \sum_{n_j=1}^{L_{j}} \Xi^{\rm diag}_{(j,n_j)} \times \left[ \kk_\mu A^\mu_{(j,n_j)} \right]_{\td{\alpha}_j,\alpha} \Bigg) \langle \kk,\rho,(j,n_j),\alpha| 
\end{align}
\cref{eq:general_Sk_diag} takes the following block structure,  
{\footnotesize
\begin{align}  \label{eq:general_Sk_diag_block}
    S^\dagger(\kk) 
    &= \left( \begin{array}{c|c|c|c|c|c|c|c|c}
        \Xi^{\rm diag}_{(1,1)} \times k_\mu A^\mu_{(1,1)} & \cdots & \Xi^{\rm diag}_{(1,L_1)} \times k_\mu A^\mu_{(1,L_1)} &  &  & & & &  \\
    \hline
        & & & \Xi^{\rm diag}_{(2,1)} \times k_\mu A^\mu_{(2,1)} & \cdots & \Xi^{\rm diag}_{(2,L_2)} \times k_\mu A^\mu_{(2,L_2)} & & &  \\
    \hline
        & & & & & & \cdots & &  \\
    \hline
        & & & & & & & \cdots & \Xi^{\rm diag}_{(J,L_J)} \times k_\mu A^\mu_{(J,L_J)}  \\
    \end{array} \right)  \ .
\end{align} }
At this stage, all $\Xi^{\rm diag}_{(j,n_j)}$ can be factored out without affecting the rank, hence the the full-rank condition to be derived does not depend on the model hopping parameters, but only to the total number and specific form of the $A(n; \td{\rho}_j \leftarrow \rho)$ tensors. 
Such information are determined solely by symmetry. 
Recall that \cref{eq:general_Sk_diag_block} is a square matrix. Each row represents a $\td{\rho}_j$, and each colum represents a copy of $\rho$. 
For \cref{eq:general_Sk_diag_block} to be full-rank, each superblock $\td{\rho}_j \leftarrow L_j \times \rho$ must be individually full-rank. 
We denote this superblock in a more compact way, 
\begin{align}    \label{eq:mrmA_i}
    \mbbA \big(\kk; \td{\rho}_j,\rho \big) = \Bigg(
        k_\mu A^\mu \Big(1; \td{\rho}_j \leftarrow \rho \Big) ~~ \Bigg| ~~ \cdots ~~ \Bigg| ~~
        k_\mu A^\mu\Big(L(\td{\rho}_j \leftarrow \rho); \td{\rho}_j \leftarrow \rho\Big) 
    \Bigg) \ .
\end{align}
One of the necessary conditions for $\mbbA(\kk; \td{\rho}_j , \rho)$ to be full-rank is that it is square, $\dim (\td{\rho}_j) = L(\td{\rho}_j \leftarrow \rho) \times  \dim (\rho)$. 
Therefore, we examine if $\det \mbbA(\kk; \td{\rho}_j , \rho) \not= 0$ for all $\kk \not= 0$. 
Since these quantities are inborn to the irreps themselves, we conclude that only irrep pairs with a square and full-rank $\mbbA(\kk;\td{\rho} , \rho)$ can be used in the touching point construction, with a form of $\Big( L(\td{\rho} \leftarrow \rho) \times \rho \Big) \boxminus \td{\rho}$. 
We will discuss several instances of $\mbbA(\kk;\td{\rho},\rho)$ in \cref{app:algr-touching-eg}.

\paragraph{Case (i), with $\mclT$.}
If $h \mclT$ is present and $q_\gamma=1$ ($\gamma=\rho_1$), then in order for $|ker;\kk,\rho_1 \subset \gamma,q',\alpha\rangle$ (for each $q'$) to transform in a closed way as $|\kk,\rho_1 \subset \gamma,m,\alpha\rangle$ (for each $m$) under $h\mclT$, $Y_{:,q'}$ must be real-valued. 
This property can also be double checked from the $Z$ matrix. 
By \cref{eq:hT_Z}, rows of the $Z$ matrix are always ``paired'' (including ``self-paired'') as $Z_{(j,n_j),:} = Z_{(j',n_{j'}),:}^*$, where $\td{\rho}_j = h\mclT(\td{\rho}_{j'})$ and $n_j = n_{j'}$. 
The kernel of such $Z$ can always be chosen as real-valued. 
Also notice that, such complex conjugation relations do not reduce the rank of $Z$ matrix in general. 
Therefore, there should still be $M-Q = \sum_{j=1}^J L(\td{\gamma}_j \leftarrow \gamma) = \sum_{j=1}^J \sum_{\td{\rho}_i \subset \td{\gamma}_j} L(\td{\rho}_i \leftarrow \rho)$ to match the desired kernel dimension. 
Summing over all (co-)irrep $\td{\gamma}_j$ is equivalent to summing over all the irrep components $\td{\rho}_i \subset \td{\gamma}_j$ contained in all $\td{\gamma}_j$. 

In examining the rank of the remaining square part in $S^\dagger(\kk)$, we still reduce $Z = \Xi^{\rm diag} (YX)^{-1}$ into a diagonal form, using a \textit{complex-valued} column transformation $XY$, as all extra kernel states should be forbidden at $\kk \not= 0$, regardless of whether they can approach a full (co-)irrep $\gamma$ at $\KK$ or not. 
Therefore, all the remaining analysis will follow the previous case. 
For $\td{\gamma}_j$ and $\gamma$ (where $\gamma = \rho$ but $\td{\gamma}_j = \bigoplus_{i=1}^{q_{\td{\gamma}_j}} \td{\rho}_i$ can take the general form), we separately examine for each $\td{\rho}_i \subset \td{\gamma}_j$ if $\mbbA(\kk; \td{\rho}_i, \rho)$ [defined in \cref{eq:mrmA_i}] is square and $\det \mbbA(\kk; \td{\rho}_i, \rho) \not= 0$ for all $\kk \not= 0$. 
In particular, the square condition $\dim(\td{\rho}_i) = L(\td{\rho}_i \leftarrow \rho) \times \dim(\rho)$ can be equivalently written as $\dim(\td{\gamma}_j) = L(\td{\gamma}_j \leftarrow \gamma) \times \dim(\gamma)$, as $\gamma = \rho$, and $h\mclT$ enforces both $\dim \td{\gamma}_j = q_{\td{\gamma}_j} \times \dim(\td{\rho}_i)$ and $L(\td{\gamma}_j \leftarrow \gamma) = q_{\td{\gamma}_j} \times L(\td{\rho}_i \leftarrow \gamma)$ for each $\td{\rho}_i \subset \td{\gamma}_j$. 
Only (co-)irrep pairs that obey these conditions can be used in the touching point construction, with a form of $\Big( L(\td{\gamma}_j \leftarrow \gamma) \times \gamma \Big) \boxminus \td{\gamma}_j$ (where $\gamma = \rho$). 

\paragraph{Case (ii).} 
If $h\mclT$ is present and $q_\gamma=2$ ($\gamma = \rho_1 \oplus \rho_2$), then for the flat-band wave-functions to be continuous, the necessary and sufficient condition is that the kernel space is spanned by the following basis
\begin{align}   \label{eq:continuous_kernel_ii}
    |ker; \kk, \rho_1\subset\gamma, q', \alpha\rangle &= \sum_{m=1}^M |\kk, \rho_1\subset\gamma, m, \alpha\rangle \times Y_{m;q'}  \\\nonumber
    |ker; \kk, \rho_2\subset\gamma, q', \alpha\rangle &= \sum_{m=1}^M |\kk, \rho_2\subset\gamma, m, \alpha\rangle \times Y_{m;q'}^* 
\end{align}
up to $O(\kk)$ corrections. 
Notice that, each half of \cref{eq:continuous_kernel_ii} takes the same form as Case (i) [\cref{eq:continuous_kernel}], while the two halves are simply related by $h\mclT$. 
This property can also be double checked from the $Z$ matrices. 
The first half $\bigoplus_{j=1}^J \td{\gamma}_j \leftarrow M \rho_1$ generates one $Z$ matrix, while $\bigoplus_{j=1}^J \td{\gamma}_j \leftarrow M \rho_2$ simply generates $Z^*$ [\cref{eq:hT_Z}, up to a proper ordering of the rows that do not affect kernel]. 
$\ker Z$ and $\ker Z^*$ can always be chosen as the complex conjugate of one another. 
To find $Q$ copies of (co-)irreps as $\kk$ approaching $\KK$, there thus must be $M-Q = \sum_{j=1}^J L(\td{\gamma}_j \leftarrow \rho_1) = \sum_{j=1}^J \frac{L(\td{\gamma}_j \leftarrow \gamma)}{q_{\gamma}}$, where $q_\gamma=2$ in this case, and $q_{\gamma} \times L(\td{\gamma} \leftarrow \rho_1) = L(\td{\gamma} \leftarrow \gamma)$ for $\rho_1 \subset \gamma$.

To examine the rank of the remaining square matrix in $S^\dagger(\kk)$, we still carry out the further column transformations for the $\rho_1$ half such that $Z = \Xi^{\rm diag} (XY)^{-1}$. For the $\rho_2$ half, there is automatically $Z^* = \Xi^{\rm diag *} (X^*Y^*)^{-1}$. 
In terms of $S^\dagger(\kk)$, it will be superblock diagonalized similarly to \cref{eq:general_Sk_diag_block}. 
Still, $\Xi^{\rm diag}$ can be factored out, allowing us to focus on the $A$ tensors. 
Each diagonal superblock at $\td{\gamma}_j \leftarrow \frac{L(\td{\gamma}_j \leftarrow \gamma)}{q_\gamma} \times \gamma$ (where $\gamma = \rho_1 \oplus \rho_2$) takes the form of
\begin{align}  \label{eq:mrmA_ii}
    \mbbA(\kk;\td{\gamma}_j, \gamma) = \Bigg(
        \mbbA(\kk;\td{\gamma}_j, \rho_1) \Bigg| \mbbA(\kk;\td{\gamma}_j, \rho_2)
    \Bigg) \ .
\end{align}
where $\mbbA(\kk;\td{\gamma}_j, \rho)$ juxtaposes $\mbbA(\kk;\td{\rho}_i, \rho)$ for all $\td{\rho}_i \subset \td{\gamma}_j$ together, while $\mbbA(\kk;\td{\gamma}_j,\gamma)$ further juxtaposes $\mbbA(\kk;\td{\gamma}_j, \rho_1)$ and $\mbbA(\kk;\td{\gamma}_j, \rho_2)$. 
Now, each superblock \cref{eq:mrmA_ii} must be square, namely $\dim(\td{\gamma}_j) = \frac{L(\td{\gamma}_j \leftarrow \gamma)}{q_\gamma} \times \dim(\gamma)$, and $\det \mbbA(\kk; \td{\gamma}_j, \gamma) \not= 0$ for all $\kk \not= 0$. 
Only pairs of $\td{\gamma}_j$ and $\gamma$ that meet these requirements can contribute to the touching point construction. 

\paragraph{Summary of touching point construction.}
\phantomsection
\label{par:touch}
We collect all the identical $\td{\gamma}_j$ together and introduce the notation $\bigoplus^{\not=}_{\td{\gamma}}$, so that $\td{\gamma} \not= \gamma$ are not repeated in summation. The touching point construction with symmetry-indicated continuity proceeds as follows.  
\begin{enumerate}   
    \item For all the (co-)irreps of the little group $\mclG_{\KK}$, compute $L(\td{\gamma} \leftarrow \gamma) = \langle \mathbf{1} ,  v \otimes \td{\gamma} \otimes \gamma^* \rangle = \sum_{\td{\rho} \subset \td{\gamma}, \rho \subset \gamma} \langle \mathbf{1} ,  v \otimes \td{\rho} \otimes \rho^* \rangle$. 
    \item Examine if $\dim (\td{\gamma}) = \frac{L(\td{\gamma} \leftarrow \gamma)}{q_\gamma} \times  \dim (\gamma)$. 
    \item If true, solve the superblock $\mbbA\left( \kk; \td{\gamma}_j , \gamma \right)$ explicitly [\cref{eq:mrmA_i,eq:mrmA_ii}]. 
    Examine if $\det \mbbA(\kk) \not= 0$ for all $\kk \not= 0$. 
    \item If true, $\frac{L(\td{\gamma}\leftarrow\gamma)}{q_{\gamma}} \gamma  \boxminus \td{\gamma}$ will serve as a basic unit for $\Delta\mclB \big|_{\KK}$. 
    Enumerate positive integer $Q$ and non-negative integers $b_{\td{\gamma}}$, then all the following constructions can guarantee the continuity
    \begin{align}   \label{eq:BDB_construction}
        \mclB \big|_\KK = Q \gamma \qquad \textrm{and} \qquad 
        \Delta \mclB \big|_\KK =\bigoplus_{\td{\gamma}}^{\not=} b_{\td{\gamma}} \Bigg( \frac{L(\td{\gamma}\leftarrow\gamma)}{q_{\gamma}} \gamma  \boxminus \td{\gamma} \Bigg)  \ .
    \end{align}
    $b_{\td{\gamma}}$ represents the multiplicity of $\td{\gamma}$. 
\end{enumerate}

We finally remark on why only isotropic high symmetry point can serve as the touching point. 
If $v = v_{x,y} \otimes v_z$ is real reducible, for example, then the $A$ tensor solved from $v_z \otimes \td\gamma \otimes \gamma^*$ for any $\td\gamma,\gamma$ will be vanishing in $A^x$ and $A^y$, hence $k_\mu A^\mu = 0$ for any $\kk \not=0$ but $k_z = 0$. 
Such $\mbbA(\kk)$ is not full-rank, and it is not guaranteed in general that $O(\kk^2)$ contributions in $S^\dagger(\kk)$ and hybridization to remote bands always lift these extra kernel states along these directions. 

\subsubsection{Examples of touching point Hamiltonian} 
\label{app:algr-touching-eg}

We examine several examples of the $\mbbA(\kk;\td{\gamma},\gamma)$ matrix. 

(1) SG215, $\ovl{\Gamma}_6 \leftarrow \ovl{\Gamma}_7$. 
\begin{align}
\mbbA(\kk; \ovl{\Gamma}_6 \leftarrow \ovl{\Gamma}_7) &= \begin{pmatrix}
k_z & k_- \\
k_+ & -k_z
\end{pmatrix}   \ , \qquad \det \mbbA(\kk; \ovl{\Gamma}_6 \leftarrow \ovl{\Gamma}_7) = - \kk^2 \ .
\end{align}
where $k_\pm = k_x \pm \ii k_y$. 
As $\mbbA(\kk; \ovl{\Gamma}_6 \leftarrow \ovl{\Gamma}_7)$ is full-rank as long as $\kk \not= 0$, $\ovl{\Gamma}_7 \boxminus \ovl{\Gamma}_6$ can enter the touching point construction. 

(2) SG200, $\ovl{\Gamma}_5 \leftarrow \ovl{\Gamma}_9$. 
\begin{align}
\mbbA(\kk; \ovl{\Gamma}_5 \leftarrow \ovl{\Gamma}_9) = \begin{pmatrix}
k_z & \omega^* k_x - \ii \omega k_y \\
\omega^* k_x + \ii \omega k_y & -k_z
\end{pmatrix} \ , \qquad \det \mbbA(\kk; \ovl{\Gamma}_5 \leftarrow \ovl{\Gamma}_9) = - \Big( \omega k_x^2 + \omega^* k_y^2 + k_z^2 \Big) \ .
\end{align}
where $\omega = e^{\ii \frac{2\pi}{3}}$. $\det \mathbb{A}(\mathbf{k}; \ovl{\Gamma}_5 \leftarrow \ovl{\Gamma}_9) = 0$ along $|k_x| = |k_y| = |k_z|$ directions, hence we avoid using $\ovl{\Gamma}_9 \boxminus \ovl{\Gamma}_5$ in the touching point construction. 

(3) SG220, $\ovl{P}_7 \leftarrow \ovl{P}_8$. 
\begin{align}
\mbbA(\kk; \ovl{P}_7 \leftarrow \ovl{P}_8) = \begin{pmatrix}
0 & -k_z & -k_y \\
k_z & 0 & -k_x \\
k_y & k_x & 0
\end{pmatrix} \ , \qquad \det \mbbA(\kk; \ovl{P}_7 \leftarrow \ovl{P}_8) = 0 \ .
\end{align}
Notice that
\begin{align}
-i\mbbA(\kk) \overset{U^\dagger\mbbA(\kk)U}\sim \td{\mbbA}(\kk) = \begin{pmatrix} k_z & k_-/{\sqrt{2}} & 0 \\ k_+/\sqrt{2} & 0 & k_-/\sqrt{2} \\ 0 & k_+/\sqrt{2} & -k_z\end{pmatrix} = \kk \cdot \mathbf{S}
\end{align}
with the unitary matrix $U$ $\begin{pmatrix} 1/\sqrt{2} & 0 & -1/\sqrt{2} \\ -i/\sqrt{2} & 0 & -i/\sqrt{2} \\ 0 & -1 & 0\end{pmatrix}$. $\td{\mbbA}(\kk)$ is $\kk \cdot \mathbf{S}$, with $\mathbf{S}$ is the $L=1$ angular momentum operator. 

(4) SG223, $\ovl{R}_5 \leftarrow \ovl{R}_6$. 
\begin{align}
\mbbA(\kk; \ovl{R}_5 \leftarrow \ovl{R}_6) &= \begin{pmatrix}
0 & 0 & - (\omega k_x - \omega^* \ii k_y) & \ii k_z \\
0 & 0 & k_z &\ii (\omega k_x + \ii \omega^* k_y) \\
-\ii (\omega^* k_x + \ii \omega k_y) & \ii k_z & 0 & 0 \\
k_z & (\omega^* k_x - \ii \omega k_y) & 0 & 0
\end{pmatrix} \ , \\\nonumber
\det \mbbA(\kk; \ovl{R}_5 \leftarrow \ovl{R}_6) &= (-\ii)^2 \Big| k_z^2 + \omega k_x^2 + \omega^* k_y^2 \Big|^2 = - (k_x^4 + k_y^4 + k_z^4) + (k_x^2k_y^2 + k_x^2k_z^2 + k_y^2k_z^2) \ .
\end{align}
$\det \mathbb{A}(\mathbf{k};\ovl{R}_5 \leftarrow \ovl{R}_6)) = 0$ along $|k_x| = |k_y| = |k_z|$ directions. 

In the above four examples, case (1) satisfies the full-rank condition, but cases (2)-(4) do not, which are discarded in the calculation.  

For 3D cases not mentioned here, we found it is sufficient to examine $\mbbA(\kk)$ is full rank or not by checking special $\kk$ directions $\mbbA(1,0,0)$ and $\mbbA(1,1,1)$. And we use this criterion in our numerical calculation. 
We have also examined the 2D cases listed in \cref{{tab:DoubleTR_2D_summary}}, and all $\mbbA(\kk;\td{\gamma},\gamma)$ super-blocks are full-rank. 
For 2D Chern band cases listed in \cref{{tab:2D_Chern_band}}, $\mbbA(\kk)$ is inborn full rank for $\dim[\mbbA(\kk)]$ are all $1 \times 1$.

\subsection{Work flow}
\label{app:algr-workflow}

With the symmetry-indicated continuity condition established in \cref{app:algr-touching}, we now summarize the overall work flow to search for symmetry-indicated CTFB in all space groups. 

\begin{enumerate}
    \item For space group $\mclG$, examine if its EBR data contains non-trivial SI. 
    \item If true, examine if it contains isotropic high symmetry point. 
    \item If true, solve $L(\td{\gamma} \leftarrow \gamma)$ for all $\td{\gamma} \not= \gamma$ at the isotropic high symmetry point. 
    If $\dim(\td{\gamma}) = \frac{L(\td{\gamma} \leftarrow \gamma)}{q_\gamma} \times \dim(\gamma)$, solve $\mbbA(\kk;\td{\gamma},\gamma)$ and evaluate if $\det \mbbA(\kk;\td{\gamma},\gamma) \not= 0$ for all $\kk \not= 0$. 
    If all true, store the combination $\frac{L(\td{\gamma} \leftarrow \gamma)}{q_\gamma} \gamma \boxminus \td{\gamma}$. 
    \item Determine the target flat band dimension $d_{\mclB}$ and a touching point dimension cutoff $D_{\Delta \mclB}$. 
    \item For all high-symmetry points $\KK$, enumerate flat band ``pieces'' $\mclB \big|_{\KK} = \bigoplus_{\gamma \in \KK} c_{\gamma} \gamma$ that satisfy $d_{\mclB \big|_{\KK}} = d_{\mclB}$, with $c_\gamma$ being arbitrary non-negative integers. 
    Each $\mclB \big|_{\KK}$ piece will then be associated with a set of $\Delta\mclB \big|_{\KK}$ pieces. 
    If $\KK$ is isotropic, and the current flat band piece reads $\mclB \big|_{\KK} = Q \gamma$, then enumerate $\Delta\mclB \big|_{\KK} = \bigoplus_{\td{\gamma}}^{\not=} b_{\td{\gamma}} \Big( \frac{L(\td{\gamma} \leftarrow \gamma)}{q_\gamma} \gamma \boxminus \td{\gamma} \Big)$ for all non-negative integers $b_{\td{\gamma}}$ that satisfy $D_{\Delta\mclB \big|_{\KK}} \le D_{\Delta\mclB}$. 
    If $\KK$ is not isotropic or $\mclB \big|_{\KK} \not= Q\gamma$, $\Delta\mclB \big|_{\KK} = 0$. 
    \item Connect $\mclB \big|_{\KK}$ pieces at different $\KK$. Examine if $\mclB$ respects compatibility relations and carries non-trivial SI. 
    \item If true, further find all possible ways to connect the $\Delta\mclB \big|_{\KK}$ pieces asscociated with the chosen $\mclB \big|_{\KK}$ pieces, such that $\Delta\mclB$ respect  compatibility relations and carries the opposite SI than $\mclB$. 
    \item Such $\mclB + \Delta\mclB = \mathcal{BR}_L \boxminus \mathcal{BR}_{\td{L}} =  \mclE \cdot y$ contributes one symmetry-indicated CTFB, where $y$ is an integer vector to be solved that represents the bipartite construction. 
    It must take the form of $y = \sum_{p=1}^{N_{EBR}} t_p \big[ \mrmR^{-1} \big]_{:,p}$, where $t_p = \frac{\big(\mrmL^{-1} \cdot (\mclB + \Delta\mclB)\big)_p }{\lambda_p}$ are fixed for $p \le r$, and $t_p$ for $p > r$ are arbitrary integers that can be used to minimize the total band dimension $d$. 
    Search in this parameter space to find the minimal real-space construction. 
\end{enumerate}

\clearpage
\begin{table}[h]
    \centering
    \begin{tabular}{l|l|l}
        LG ID & SG ID & IHSP \\ \hline
        49 ($p4$) & 75 & [GM, M] \\
        50 ($p\bar{4}$) & 81 & [GM, M] \\
        51 ($p4/m$) & 83 & [GM, M] \\
        52 ($p4/n$) & 85 & [GM, M] \\
        53 ($p422$) & 89 & [GM, M] \\
        54 ($p42_12$) & 90 & [GM, M] \\
        55 ($p4mm$) & 99 & [GM, M] \\
        56 ($p4bm$) & 100 & [GM, M] \\
    \end{tabular} \quad 
    \begin{tabular}{l|l|l}
        LG ID & SG ID & IHSP \\ \hline
        57 ($p\bar{4}2m$) & 111 & [GM, M] \\
        58 ($p\bar{4}2_1m$) & 113 & [GM, M] \\
        59 ($p\bar{4}m2$) & 115 & [GM, M] \\
        60 ($p\bar{4}b2$) & 117 & [GM, M] \\
        61 ($p4/mmm$) & 123 & [GM, M] \\
        62 ($p4/nbm$) & 125 & [GM, M] \\
        63 ($p4/mbm$) & 127 & [GM, M] \\
        64 ($p4/nmm$) & 129 & [GM, M] \\
    \end{tabular} \\ 
    \vspace{0.5cm}
    \begin{tabular}{l|l|l}
    LG ID & SG ID & IHSP \\ \hline
        65 ($p3$) & 143 & [GM, K, KA] \\
        66 ($p\bar{3}$) & 147 & [GM, K] \\
        67 ($p312$) & 149 & [GM, K] \\
        68 ($p321$) & 150 & [GM, K, KA] \\
        69 ($p3m1$) & 156 & [GM, K] \\
        70 ($p31m$) & 157 & [GM, K, KA] \\
        71 ($p\bar{3}1m$) & 162 & [GM, K] \\
        72 ($p\bar{3}m1$) & 164 & [GM, K] \\
    \end{tabular}\quad 
    \begin{tabular}{l|l|l}
    LG ID & SG ID & IHSP \\ \hline
        73 ($p6$) & 168 & [GM, K] \\
        74 ($p\bar{6}$) & 174 & [GM, K, KA] \\
        75 ($p6/m$) & 175 & [GM, K] \\
        76 ($p622$) & 177 & [GM, K] \\
        77 ($p6mm$) & 183 & [GM, K] \\
        78 ($p\bar{6}m2$) & 187 & [GM, K] \\
        79 ($p\bar{6}2m$) & 189 & [GM, K, KA] \\
        80 ($p6/mmm$) & 191 & [GM, K] \\
    \end{tabular}
    \caption{Isotropic high symmetry points (ISHPs) in 2D layer groups. 
    Each 2D layer group (LG ID) can be understood as the quotient of a 3D space group (SG ID) over translations along $z$ direction. } 
    \label{tab:isoTP_2D}
\end{table}

\vspace{1em}

\begin{table}[h!]
    \centering
    \begin{tabular}{l|l}
        SG & IHSP \\ \hline
        195 ($P23$) & [GM, R] \\
        196 ($F23$) & [GM] \\
        197 ($I23$) & [GM, H, P, PA] \\
        198 ($P2_13$) & [GM, R] \\
        199 ($I2_13$) & [GM, H, P, PA] \\
        200 ($Pm\bar{3}$) & [GM, R] \\
        201 ($Pn\bar{3}$) & [GM, R] \\
        202 ($Fm\bar{3}$) & [GM] \\
        203 ($Fd\bar{3}$) & [GM] \\
    \end{tabular} \quad 
    \begin{tabular}{l|l}
        SG & IHSP \\ \hline
        204 ($Im\bar{3}$) & [GM, H, P] \\
        205 ($Pa\bar{3}$) & [GM, R] \\
        206 ($Ia\bar{3}$) & [GM, H, P] \\
        207 ($P432$) & [GM, R] \\
        208 ($P4_232$) & [GM, R] \\
        209 ($F432$) & [GM] \\
        210 ($F4_132$) & [GM] \\
        211 ($I432$) & [GM, H, P] \\
        212 ($P4_332$) & [GM, R] \\
    \end{tabular} \quad 
    \begin{tabular}{l|l}
        SG & IHSP \\ \hline
        213 ($P4_132$) & [GM, R] \\
        214 ($I4_132$) & [GM, H, P] \\
        215 ($P\bar{4}3m$) & [GM, R] \\
        216 ($F\bar{4}3m$) & [GM] \\
        217 ($I\bar{4}3m$) & [GM, H, P, PA ] \\
        218 ($P\bar{4}3n$) & [GM, R] \\
        219 ($F\bar{4}3c$) & [GM] \\
        220 ($I\bar{4}3d$) & [GM, H, P, PA ] \\
        221 ($Pm\bar{3}m$) & [GM, R] \\
    \end{tabular} \quad 
    \begin{tabular}{l|l}
        SG & IHSP \\ \hline
        222 ($Pn\bar{3}n$) & [GM, R] \\
        223 ($Pm\bar{3}n$) & [GM, R] \\
        224 ($Pn\bar{3}m$) & [GM, R] \\
        225 ($Fm\bar{3}m$) & [GM] \\
        226 ($Fm\bar{3}c$) & [GM] \\
        227 ($Fd\bar{3}m$) & [GM] \\
        228 ($Fd\bar{3}c$) & [GM] \\
        229 ($Im\bar{3}m$) & [GM, H, P] \\
        230 ($Ia\bar{3}d$) & [GM, H, P] \\
    \end{tabular}
    \caption{Isotropic high symmetry points (IHSPs) in 3D space groups. Modulo translations, the unitary part of little groups $\mclG^U_{\KK}$ at these touching points $\KK$ are isomorphic to one of the following point groups $O, O_h, T, T_h, T_d$. }  
    \label{tab:isoTP_3D}
\end{table}

\clearpage

\section{Finite differentiability and power-law correlation function}
\label{sec:correlation-analytical}

We now analytically demonstrate that a discontinuity in the $n$-th order derivatives of the projector [\cref{eq:correlation-def}]
\begin{equation}
  P_{\alpha\beta}(\kk) = \sum_{n\in \mathrm{occ}} u^*_{\alpha n}(\kk) u_{\beta n}(\kk)  
\end{equation}
will lead to a $r^{-n-d}$ tail in the correlation function 
\begin{equation}   \label{eq:Corr_Proj}
  C_{\alpha\beta}(\RR) = \frac1{N} \sum_{\kk} e^{\ii \kk \cdot (\RR + \tt_\beta - \tt_\alpha)} P_{\alpha\beta}(\kk) \ ,
\end{equation}
where $d$ is the spatial dimension. 

Heuristically, the long-range power-law tail will solely be determined by the properties near the touching points, as without the touching points, the correlation would be exponentially decaying, as in all the gapped bands. 
Therefore, it suffices to focus on the contributions to the Fourier integral from the vicinity of the touching points. 
For each touching point $\kk_0$, we write the momenta near it as $\kk=\kk_0+\delta\kk$, then
\begin{equation}  
  C_{\alpha\beta}(\RR) \sim \sum_{\kk_0} \frac{e^{\ii \kk_0 \cdot (\RR + \tt_\beta - \tt_\alpha)} }{N}  \sum_{|\delta\kk| \lesssim x^{-1}_c} e^{\ii \delta\kk \cdot (\RR + \tt_\beta - \tt_\alpha)} P_{\alpha\beta}(\kk_0+\delta\kk) \ ,
\end{equation}
where $x_c^{-1}$ is a small momentum cutoff near the touching points. 
Note that the pre-factor $e^{\ii \kk_0 \cdot (\RR + \tt_\beta - \tt_\alpha)}$ simply dictates a ``short-wave'' phase oscillation pattern with large momentum $\kk_0$. 
The long-range decay of $|C_{\alpha\beta}(\RR)|$ is determined by the integrals over $\delta\kk$, which we analyze below in more detail. 
If there are more than one touching points that share the same power-law tail, then the phase oscillations with different momenta will interfere. 
The number and locations of the touching points are thus encoded in these phase oscillation patterns.

\subsection{The 2D case}  \label{sec:correlation-analytical-2D}

Matrix elements of the projector $P(\kk)$ generally have the form ${|\pp|^{-2s}}p_+^{l_+} p_-^{l_-} = p_+^{l_+ - s} p_-^{l_- - s}$ when $\kk = \mathbf{K} + \pp$ approaches a touching point $\mathbf{K}$.
Here $p_{\pm} = p_x \pm \ii p_y$, and $s\ge 1$, $l_{\pm} \ge 0$ are integers. 
There must be $l_+ + l_- -2s >0$ for the projector to be continuous. 
When $l_\pm <s$ or $l_- <s$, the matrix element has discontinuities in its $(l_+  + l_- -2s)$-th order derivatives. 
We now define $n=l_++ l_- - 2s$, $m = l_+ - l_-$, and consider a general integral of the form 
\begin{equation}
    I_{n,m}(r) = \int_0^\infty \dd p \ p \int \dd \theta \ p^n 
    e^{\ii p r \cos \theta - p x_c}
    e^{\ii m\theta}\ , \qquad (m-n=0 \mod 2)\ ,
\end{equation}
that will appear in the correlation function. 
The $n$-th order derivatives of $p^{n} e^{\ii m\theta}$ is discontinuous if $n < |m|$. 
Here $x_c \gg 1$ is a soft cutoff introduced to screen short-distance behaviors, $p = |\pp|$, $r=|\RR|$, and $\theta$ is the azimuthal angle between $\pp$ and $\RR$. 
We will take the $x_c/|\RR| \to 0^+$ limit in the end.

Since $I_{n,-m}(r) = I_{n,m}^*(-r)$, we only need to calculate the integral for  the $m\ge 0$ case. 
Carrying out the integral over $p$, we obtain
\begin{equation}
    I_{n,m}(r) = (n+1)! \int \dd \theta \frac{ e^{\ii m \theta}  }{(x_c - \ii r \cos\theta)^{n+2}} 
\end{equation}
To evaluate the integral over $\theta$, we introduce the complex variable $z = e^{\ii\theta}$ and write $\dd \theta$ as $\frac{dz}{\ii z}$. 
Then we have 
\begin{equation} \label{eq:Inm-general-prev}
I_{n,m}(r) = \frac{(n+1)! \, (2\ii)^{n+2} }{r^{n+2}} 
    \ointctrclockwise \frac{\dd z}{\ii z} 
    \frac{ z^m }{(z + z^{-1} + 2\ii \varepsilon )^{n+2}} 
= \frac{(n+1)! \, (2\ii)^{n+2} }{r^{n+2}} 
    \ointctrclockwise \frac{\dd z}{\ii} 
    \frac{ z^{m+n+1} }{(z -\ii + \ii \varepsilon )^{n+2} (z+\ii + \ii\varepsilon)^{n+2}} 
\end{equation}
where $\varepsilon = x_c/r$ is a small quantity, and $\ointctrclockwise$ integrates $z$ along the unit circle centered at $z=0$. 
Higher order terms in $\varepsilon$ have been omitted.
Since $\varepsilon \to 0^+$, we can replace the integrand by 
    $f(z) = \frac{ z^{m+n+1} }{(z -\ii )^{n+2} (z+\ii)^{n+2}} $ 
and simultaneously deform the integral contour to a circle enclosing the pole at $z=\ii$. 
Applying the residue theorem, we have
\begin{equation} \label{eq:Inm-general}
    I_{n,m}(r) = \frac{(n+1)! \, (2\ii)^{n+2} }{r^{n+2}} 2\pi\  \mathrm{Res}[f,\ii]\ . 
\end{equation}
Since $n=m$ mod 2, $f(z)$ is an odd function, and there must be $\mathrm{Res}[f,\ii] = \mathrm{Res}[f,-\ii]$. 
As a meromorphic function, all residues of $f(z)$ (including the one at infinity) must sum to zero.
Thus, 
\begin{equation}
    \mathrm{Res}[f,\ii] =  \mathrm{Res}[f,-\ii] = -\frac12 \mathrm{Res}[f,\infty] 
    = \frac12 \mathrm{Res} \brak{ \frac1{z^2} f\pare{\frac1{z}},0}
    = \frac12 \mathrm{Res} \brak{ \frac{z^{n-m+1}}{(1-\ii z)^{n+2} (1+\ii z)^{n+2}} ,0}\ .
\end{equation}
Given $n=m$ mod 2, it is nonzero only if $n<m$.  
{\it 
Therefore, for 2D systems, if $P(\kk)$ has a discontinuity in its $n$-th order derivatives, the correlation function will have a component decaying as ${r^{-n-2}}$. 
}

\subsection{The 3D case}

Matrix elements in the projector $P(\kk)$ generally have the form $ |\pp|^{-2s} g(\pp)$ when $\kk = \mathbf{K} + \pp$ approaches a touching point $\mathbf{K}$.
Here $s\ge 1$ is an integer, and $g(\pp)$ is an analytic function. 
Its contribution to  the correlation function reads $\int \dd^3\pp \ e^{\ii \pp\cdot \mathbf{r} - x_c |\mathbf{p}|} |\pp|^{-2s} g(\pp)$, where $x_c\gg 1$ is a soft cutoff introduced to screen short-distance behaviors.
We now fix the direction of $\mathbf{r}$ and consider the correlation function's dependence on $r=|\mathbf{r}|$. 
Without loss of generality, we expand the angular dependence of $g(\pp)$ in spherical harmonics, choosing the principal axis along $\mathbf{r}$. 
A general component in $g(\pp)$ then takes the form $|\pp|^{l} Y_{l}^m(\theta, \varphi)$, with $\theta = \inn{\pp,\mathbf{r}}$ and $\varphi$ being the azimuthal angle of $\mathbf{p}$ in the chosen coordinate system. 
Define $n = l-2s$, $P(\kk)$ has discontinuities in $n$-th order derivatives if $s>0$. 
We need to evaluate the integral
\begin{equation}
    J_{n,l}(r) = \int_0^{\infty} \dd p \ p^2 \int_0^\pi \dd\theta  \ \sin\theta \ 
    e^{\ii p r \cos\theta - x_c p} p^{n}  P_l(\cos\theta) ,\qquad (n-l=0 \mod 2)\ ,
\end{equation}
to obtain the $r$-dependence of the correlation function.
Here $P_l(\theta)$ is the Legendre polynomial in $Y_l^0(\theta,\varphi)$, and we have assumed $m=0$ since otherwise the integral vanishes.  

Carrying out the integral over $p$, we obtain 
\begin{equation}
    J_{n,l}(r) = \frac{(n+1)! (\ii)^{n+3}}{r^{n+3}} \int_{-1}^1 \dd t \ \frac{ P_l(t)  }{( t + \ii\varepsilon )^{n+3}} \ ,
\end{equation}
where $t=\cos\theta$, $\varepsilon = x_c/r$. 
We will take the $\varepsilon\to 0^+$ limit in the end. 
To simplify the calculation, we interpret $J_{n,l}(r)$ as an integral 
\begin{equation}
    J_{n,l}(r) = - \frac{(n+1)! (\ii)^{n+3}}{r^{n+3}} \ointctrclockwise_{\mathcal{C}} 
    \frac{\dd z}{2\pi\ii} \ 
    \frac{ P_l(z)  }{( z + \ii\varepsilon )^{n+3}} \ln \pare{\frac{z-1}{z+1} }
\end{equation}
along the contour $\mathcal{C}:$ $(-1-\ii \eta) \to (1-\ii \eta) \to (1+\ii \eta) \to (-1+\ii\eta) \to (-1 - \ii \eta)$ with $\eta$ being an infinitesimal positive quantity.
There are $\ln \pare{\frac{z-1}{z+1} } = \pm \pi \ii$ for $z=x \pm \ii \eta$ with $x\in(-1,1)$. 
All residues of the integrand $f(z) = \frac{ P_l(z)  }{( z + \ii\varepsilon )^{n+3}} \ln \pare{\frac{z-1}{z+1} }$, including those at infinity and the branch-cut, must sum to zero, hence 
\begin{equation} 
J_{n,l}(r) 
=  \frac{(n+1)! (\ii)^{n+3}}{r^{n+3}} \pare{ \mathrm{Res}[f, -\ii\varepsilon] + \mathrm{Res}[f, \infty] }\ . 
\end{equation}
The residue at $-\ii \varepsilon$ is given by 
\begin{equation}
     \mathrm{Res}[f, -\ii\varepsilon]
= \frac{1}{(n+2)!} \frac{\dd^{n+2}}{\dd z^{n+2}} \pare{ P_l(z) \ln \pare{\frac{z-1}{z+1}   } } \Bigg|_{z=-\ii\varepsilon} \ .
\end{equation}
Since $\ln \pare{\frac{z-1}{z+1}} = -\pi \ii - 2z + \frac23 z^3  + \cdots$ around $z=-\ii\varepsilon$, we have 
\begin{equation}
    \Im[\mathrm{Res}[f,-\ii\varepsilon]] = - \pi \frac{1}{(n+2)!} \frac{\dd^{n+2}}{\dd z^{n+2}} P_l(z) \Bigg|_{z=0} ,\quad 
    \Re[\mathrm{Res}[f,-\ii\varepsilon]] = \frac{1}{(n+2)!} \frac{\dd^{n+2}}{\dd z^{n+2}} \pare{ P_l(z) \pare{2z + \frac23 z^3 + \cdots}  }\Bigg|_{z=0} \ .
\end{equation}
As $P_l(z) = c_1 z^{l} + c_2 z^{l-2} + \cdots$ and $n=l$ mod 2, $\Im[\mathrm{Res}[f,-\ii\varepsilon] $ is generally nonzero if $l\ge n+2$. 
On the other hand, $\Re[\mathrm{Res}[f,-\ii\varepsilon]]$ is always zero because $P_l(z) \pare{2z + \frac23 z^3 + \cdots}$ has the parity of $(n+1)$ mod 2 whereas $ \frac{\dd^{n+2}}{\dd z^{n+2}}$ has the parity of $n$ mod 2. 

To analyze the residue at infinity, we notice that
\begin{equation}
    f(z) \sim \frac{1+ \sum_{n=1}^{\infty} c_n z^{-n} }{ z^{n-l+4} } , 
\end{equation}
where $c_n$'s are coefficients. 
It has expansion in $z^{-1}$ if $n-l+4 \le 1$. 
Provided $n=l$ mod 2, $ \mathrm{Res}[f, \infty]$ is generally nonzero if $l\ge n+4$. 

In summary, when $l>n$, $J_{n,l}$ is nonzero and contributes a $r^{-n-3}$ tail to the correlation function. 
{\it Therefore, for 3D systems, if $P(\kk)$ has a discontinuity in its $n$-th order derivatives, the correlation function will have a component decaying as $r^{-n-3}$.} 

\subsection{Wannier function}

We investigate the (orthonormal) Wannier functions of CTFB. 
We show that, while Wannier functions are in general gauge dependent, the power-law tail in the gauge-invariant correlation function enforces a lower bound to the long-range tail of the Wannier functions in any gauge. 
Therefore, a CTFB with power-law correlation cannot find localized Wannier functions that exponentially decay. 

The Wannier function is defined as 
\begin{align}    \label{eq:defi_Wannier}
    W_{\alpha m}(\RR-\RR_0) &= \frac{1}{N} \sum_{\kk, n} e^{\ii \kk \cdot (\RR-\RR_0+\boldsymbol{\tau}_{\alpha})} u_{\alpha n}(\kk) [U(\kk)]_{n m} \\\nonumber
    u_{\alpha n}(\kk) &=  \sum_{\Delta \RR, m} e^{-\ii \kk \cdot (\Delta\RR + \boldsymbol{\tau}_{\alpha})} W_{\alpha m}(\Delta\RR) [U^\dagger(\kk)]_{m n}
\end{align}
Here, $\RR_0$ labels the center of the Wannier function, while $\RR-\RR_0$ labels the spread of the Wannier function over unit cells. $U(\kk)$ is a unitary transformation within the interested bands. Wannier functions defined in such a way are orthogonal, but are not localized for topological bands. 
The gauge degree of freedom originates from the choice of $U(\kk)$. 

Inserting \cref{eq:defi_Wannier} into \cref{eq:Corr_Proj}, we find the correlation function can be alternatively expressed via the Wannier functions
\begin{align}
	   C_{\alpha \beta}(\RR) = \sum_{\RR', m} W^*_{\alpha m}(\RR') W_{\beta m}(\RR+\RR')
\end{align}
We thus find that the asymptotic behavior of Wannier functions (of arbitrary gauge) will be lower-bounded by the behavior of the gauge-independent correlation functions, 
\begin{align}  \label{eq:C_W_inequality}
	\left|C_{\alpha \beta}(\RR) \right| \le \sum_{\RR',m} \left| W^*_{\alpha m}(\RR') W_{\beta m}(\RR+\RR') \right|
\end{align}
In particular, localized Wannier functions with finite support or with exponential decay cannot lead to a power-law correlation. 

Next, for power-law decaying Wannier functions, we relate the powers to those of the correlation functions via the inequality \cref{eq:C_W_inequality}. 
Assume that 
\begin{align}
    \left| W_{\alpha m}(\RR) \right|  \le  M \frac{1}{r(\RR)^q}
\end{align}
where $r(\RR) = \sqrt{1 + \RR^2}$ regularizes the behavior at short-range, while maintains the power-law behavior at long-range. $M$ can be any finite positive number. For $W$ to be normalizable, we further assume $2q>d$, where $d$ is the spatial dimension. 
Then, \cref{eq:C_W_inequality} implies 
\begin{align}
    \left|C_{\alpha \beta}(\RR) \right|  \le M^2 d_{\rm FB}   \sum_{\RR'} \frac{1}{r(\RR')^q r(\RR+\RR')^q}
\end{align}
To investigate the summation at sufficiently large $|\RR| = R$, we divide it into three regions. (1) $|\RR'| \lesssim \frac{R}{2}$, (2) $|\RR'+\RR| \lesssim \frac{R}{2}$, and (3) remaining. 

The analysis to regions (1) and (2) are similar, so we focus on region (1). 
\begin{align}
    \sum_{\RR'\in (1)} \frac{1}{r(\RR')^q r(\RR+\RR')^q} \le \frac{M_1}{R^q} \sum_{\RR'\in (1)}  \frac{1}{r(\RR')^q} \sim \frac{1}{R^q} \times \left\{
    \begin{array}{ll }
        R^{d-q} & q<d \\
        \ln R & q = d \\
        \mathcal{O}(1) & q>d \\
    \end{array} \right. \sim 
    \left\{
    \begin{array}{ll }
        R^{-(2q-d)} & q<d \\
        R^{-q}\ln R & q = d \\
        R^{-q} & q>d \\
    \end{array} \right.
\end{align}
The first step identifies that $r(\RR+\RR')$ is smooth in region (1), hence approximates it as a constant of order $\frac{1}{R^q}$. The second step estimates the summation $\sum_{\RR'\in (1)}$ with an integral $\sim \int_{0}^{R/2} |\RR'|^{d-1} d|\RR'|$. 
On the other hand, for region (3), 
\begin{align}
    \sum_{\RR'\in (3)} \frac{1}{r(\RR')^q r(\RR+\RR')^q} \le M_3 \sum_{\RR'\in (3)}  \frac{1}{r(\RR')^{2q}} \sim \int_{R}^{\infty} \dd r \, r^{d-1} \, r^{-2q} \sim \frac{1}{R^{2q-d}}
\end{align}

Therefore, the asymptotic behavior of the correlation function $C_{\alpha \beta}(\RR)$ is controlled by the dominant contribution of all three regions, 
hence
\begin{align}   \label{eq:Corr_bound_Wannier}
    |C_{\alpha\beta}(\RR)| \lesssim \left\{
    \begin{array}{ll }
        R^{-(2q-d)} & q<d \\
        R^{-q}\ln R & q = d \\
        R^{-q} & q>d \\
    \end{array} \right.
\end{align}
where symbols $\sim$ and $\lesssim$ ignore $\mathcal{O}(1)$ coefficients. 
Reading \cref{eq:Corr_bound_Wannier} in the opposite way, the power-law tail of the correlation functions thus indicates a lower bound for the power-law tail of Wannier functions $|W(\RR)| \sim \frac{1}{|\RR|^q}$. 

We illustrate the bound with a Chern CTFB. For simplicity, suppose it has one touching point, and well-defined Chern number $C$. One can always choose the following gauge (the singular gauge), that the normalized flat band wave-function $|u(\kk)\rangle$ is smooth everywhere except for the touching point $\kk=0$, where the topology enforces a $C$-fold phase vortex
\begin{align}
    |u(\kk)\rangle = e^{\ii C\phi_{\kk}} |u\rangle + \mathcal{O}(k)
\end{align}
The discontinuity in the projector will only manifest in the omitted higher orders contributions, and gets hidden by the stronger singularity of the phase vortex. 
The Wannier function is the Fourier transformation of $|u(\kk)\rangle$. Following the same reasoning in \cref{sec:correlation-analytical-2D}, its asymptotic behavior at large distance will be described by $I_{0,C}(r)$ [\cref{eq:Inm-general-prev,eq:Inm-general}]. The residue to be computed is $\frac{C \ii^C}{(2\ii)^2}$, which is a finite constant. 
Therefore, the Wannier function decays as $\frac{1}{r^2}$ in the radial direction. 
Notice that the analysis in \cref{sec:correlation-analytical-2D} has implicitly fixed the azimuthal angle of $\mathbf{r}$ as $\phi_{\mathbf{r}} = 0$. 
The full angular dependence can be easily recovered as $\frac{e^{\ii C \phi_{\mathbf{r}}}}{r^2}$, which is the same as a gapped Chern band \cite{Thouless_1984_wannier,li2024constraintsrealspacerepresentations}. 
In 2D, $d=2$, the bound we derived dictates $|C_{\alpha \beta}(r)| < \frac{1}{r^2} \ln r$, which is indeed satisfied by $|C_{\alpha \beta}(r)| \sim \frac{1}{r^4}$ we found. 
Therefore, for Chern CTFB, (assuming certain gauge choice), topology leads to a longer-range Wannier function than the requirement of the touching point and the critical behavior of the correlation function. 

For $\mathbb{Z}_2$ TI, however, it has been shown that exponentially decaying Wannier functions can be constructed, if the locality of time-reversal symmetry is not enforced \cite{Brouder_exponential_2007}. 
In this case, the non-existence of exponentially decaying Wannier functions in the corresponding CTFB is thus solely dictated by the touching point and the critical behavior of the correlation function. 

\clearpage

\section{Tensor-network state: bipartite approach versus Dubail-Read's approach}

In this section, we convert the Chern band TNS constructed by Dubail and Read (DR) \cite{dubail_tensor_2015} into our bipartite form, and show the two approaches are equivalent. 
Remarkably, the DR form of TNS \cite{dubail_tensor_2015} can be directly converted into a PEPS, where auxiliary fermions live on the edges that connect the adjacent physical degrees of freedom. 
The calculation thus establishes a link between different TNS forms. 

\subsection{A Chern band example}

The Chern band model considered by DR \cite{dubail_tensor_2015} is defined on a square lattice, spanned by $(m,n) \in \mathbb{Z}^2$. 
There are 2 orbitals on each site, denoted by $c^\dagger_{\RR,\alpha}$ ($\alpha=1,2$). 
Their Fourier transformation is given by convention $c^\dagger_{\kk, \alpha} \sim \sum_{\RR} e^{\ii \kk \cdot \RR} c^\dagger_{\RR,\alpha}$, where $\sim$ omits the normalization factors. 
In the DR approach, the auxiliary fermions live on the same sites as the physical ones. 
While the original construction utilized auxiliary Majoranas, here in order to explicitly conserve the fermion charge, we use complex fermions. 
Let there be two auxiliary complex fermions per site, $\xi^\dagger_{\RR,\beta}$ ($\beta=1,2$), whose Fourier transformations are alike $\xi^\dagger_{\kk,\beta} \sim \sum_{\RR} e^{\ii \kk \cdot \RR} \xi^\dagger_{\RR,\beta}$. 
Below, auxiliary fermions are also treated as Grassmann numbers, as in our bipartite construction in the main text. Choose coupling matrices
\begin{align}
    B = \begin{pmatrix}
        1 \\
        0 \\
    \end{pmatrix} \qquad C = \begin{pmatrix}
        1 &
        0 \\
    \end{pmatrix} \qquad K(\kk) = \begin{pmatrix}
        \sin k_x + \ii \sin k_y & -\lambda (2 - \cos k_x - \cos k_y) \\
        \lambda (2 - \cos k_x - \cos k_y)  & \sin k_x - \ii \sin k_y \\
    \end{pmatrix}
\end{align}
Here, $\lambda$ can be deemed as variational parameters. 
$B, C$ will be the coupling between physical ones and auxiliary ones, and are chosen as the on-site form, while $K(\kk)$ is the coupling between auxiliary ones, and includes both the on-site and nearest-neighbor coupings, 
\begin{align}
    K(\kk) &= \sum_{\mathbf{d} = (0,0),\pm(1,0),\pm(0,1)} e^{-\ii \kk \cdot \mathbf{d}} K_{\mathbf{d}} \qquad \qquad \qquad \textrm{with} \qquad 
    K_{(0,0)} =
    \begin{pmatrix}
        0&-2\lambda\\
        2\lambda&0
    \end{pmatrix} \\\nonumber
    K_{(1,0)} &=\frac{1}{2}
    \begin{pmatrix}
        \ii&\lambda\\
        -\lambda&\ii
    \end{pmatrix} \qquad 
    K_{-(1,0)} =\frac{1}{2}
    \begin{pmatrix}
        -\ii&\lambda\\
        -\lambda&-\ii
    \end{pmatrix} \qquad 
    K_{(0,1)} =\frac{1}{2}
    \begin{pmatrix}
        -1&\lambda\\
        -\lambda&1
    \end{pmatrix} \qquad
    K_{-(0,1)} =\frac{1}{2}
    \begin{pmatrix}
        1&\lambda\\
        -\lambda&-1
    \end{pmatrix}.
\end{align}
Then, the Gaussian TNS is written as
\begin{align}   \label{eq:TNS_DR}
    |\Omega_L\rangle &\sim \int_{\mathcal{D}[\xi,\xi^\dagger]} \exp\left(-\sum_{\kk} \xi^\dagger_\kk K(\kk) \xi_{\kk} + \sum_{\kk} c^\dagger_{\kk,2} C \xi_{\kk} + \sum_{\kk} \xi^\dagger_{\kk} B c_{\kk,1} \right)  |1,0\rangle \\\nonumber
    &\sim \int_{\mathcal{D}[\xi,\xi^\dagger]} \exp\left(-\sum_{\RR} \left( \sum_{\mathbf{d}} \xi^\dagger_{\RR+\mathbf{d}} K_{\mathbf{d}} \xi_{\RR} \right) +\sum_{\RR} c^\dagger_{\RR,2} C \xi_{\RR} +\sum_{\RR} \xi^\dagger_{\RR} B c_{\RR,1} \right) |1,0\rangle
\end{align}
where $|1,0\rangle \sim \prod_{\kk} c^\dagger_{\kk,1} |\mathrm{vac}\rangle$, and $\xi^\dagger_{\kk} = (\xi^\dagger_{\kk,1} , \xi^\dagger_{\kk,2})$. 
Note $\det K(\kk) \sim k_x^2+k_y^2+\mathcal{O}(\kk^4)$ vanishes around $\kk = (0,0)$.

To convert \cref{eq:TNS_DR} into a PEPS, for each directed edge $e=(\RR,\mathbf{d})$ with $\mathbf{d} \not= (0,0)$, pointing from $\RR$ to $\RR+\mathbf{d}$, choose a factorization $K_{\mathbf{d}}=L_{\mathbf{d}}R_{\mathbf{d}}$. 
One choice is $L_{\mathbf{d}}=K_{\mathbf{d}}$ and $R_{\mathbf{d}}=\mathbb{I}$. 
Introduce independent complex Grassmann variables $\eta_e$ and $\bar\eta_e$ on every such directed edge. 
The nearest-neighbor couplings between $\xi, \xi^\dagger$ can be decoupled via couplings between $\xi$ to $\eta^\dagger$. 
For each directed edge $e = (\RR,\mathbf{d})$ pointing from $\RR$ to $\RR+\mathbf{d}$, 
\begin{align}
    \textrm{for}~ \mathbf{d}\not=(0,0) \qquad 
    \exp\left( 
    -\xi^\dagger_{\RR+\mathbf{d}} K_{\mathbf{d}} \xi_{\RR}\right) \sim 
    \int_{\mathcal{D}[\eta,\eta^\dagger]}  \exp\left(
        - \eta_{(\RR,\mathbf{d})}^\dagger \eta_{(\RR,\mathbf{d})} -\xi^\dagger_{\RR+\mathbf{d}} L_{\mathbf{d}} \eta_{(\RR,\mathbf{d})}
        + \eta_{(\RR,\mathbf{d})}^\dagger R_{\mathbf{d}} \xi_{\RR}
    \right)  \label{eq:decouple_to_edge}
\end{align}
The oppositely directed term $\xi^\dagger_{\RR} K_{-\mathbf{d}} \xi_{\RR+\mathbf{d}}$ on the same link may be equivalently grouped together. 
After applying \cref{eq:decouple_to_edge} to every directed edge (every link with two opposite directions), only on-site terms as $\xi^\dagger_{\RR} K_{(0,0)} \xi^\dagger_{\RR}$ remain, 
\begin{align}  
    |\Omega_L\rangle &\sim \int_{\mathcal{D}[\xi,\xi^\dagger,\eta,\eta^\dagger]} \exp\Bigg(\sum_{\RR}- \xi^\dagger_{\RR} K_{(0,0)} \xi_{\RR} - \sum_{e} \eta^\dagger_{e} \eta_{e} \\\nonumber
    &\qquad \qquad \qquad - \sum_{\RR} \sum_{e}^{e \to \RR} \xi^\dagger_{\RR} L_{\mathbf{d}} \eta_{e} + \sum_{\RR} \sum_{e}^{\RR \to e} \eta^\dagger_{e} R_\mathbf{d} \xi_{\RR} + \sum_{\RR} c^\dagger_{\RR,2} C \xi_{\RR} + \sum_{\RR} \xi^\dagger_{\RR} B c_{\RR,1} \Bigg) |1,0\rangle
\end{align}
Here, $\sum_{e}^{e \to \RR}$ sums over directed edges $e$ that point to $\RR$, while $\sum_{e}^{\RR \to e}$ sums over directed edges $e$ that leave $\RR$. Define for shortness
\begin{align}
    P^\dagger_{\RR} = c^\dagger_{\RR,2} C
    +\sum_{e}^{\RR \to e} \eta_{e}^\dagger R_{\mathbf{d}} \qquad \qquad 
    Q_{\RR} = B c_{\RR,1}
    -\sum_{e}^{e\to \RR}  L_{\mathbf{d}} \eta_e 
\end{align}
where $\mathbf{d}$ is uniquely determined by $e$. Then, 
\begin{align}  
    |\Omega_L\rangle &\sim \int_{\mathcal{D}[\xi,\xi^\dagger,\eta,\eta^\dagger]} \exp\Bigg(\sum_{\RR}- \xi^\dagger_{\RR} K_{(0,0)} \xi_{\RR} - \sum_{e} \eta^\dagger_{e} \eta_{e} + \sum_{\RR} \xi^\dagger_{\RR} Q_{\RR} + \sum_{\RR} P^\dagger_{\RR} \xi_{\RR} \Bigg) |1,0\rangle
\end{align}
Now, integrating over $\xi,\xi^\dagger$ (gives asssuming $\lambda\neq0$ so that $K_{(0,0)}$ is invertible), 
\begin{align}
    |\Omega_L\rangle
    &\sim \int_{\mathcal{D}[\eta,\eta']}
    \exp\left(-\sum_e \eta_e^\dagger \eta_e
     + \sum_{\RR} P^\dagger_{\RR}K_{(0,0)}^{-1}Q_{\RR} \right) \Ket{1,0} \\\nonumber
     &\sim \int_{\mathcal{D}[\eta,\eta']}
    \exp\left(-\sum_e \eta_e^\dagger \eta_e
     + \sum_{\RR} \left(c^\dagger_{\RR,2} C
    +\sum_{e}^{\RR \to e} \eta_{e}^\dagger R_{\mathbf{d}}\right) K_{(0,0)}^{-1} \left( B c_{\RR,1}
    -\sum_{e}^{e\to \RR}  L_{\mathbf{d}} \eta_e  \right) \right) \Ket{1,0}
\end{align}
Crucially, note that $\eta_e$ has coupling to $c_{\RR}$ only if $\RR \to e$ or $e \to \RR$, and to another $\eta_{e'}$ only if $e \to \RR \to e'$ or vice versa. $c^\dagger_{\RR}$ only has on-site couplings.

Now we relate the DR TNS \cref{eq:TNS_DR} to our bipartite TNS. 
The key bridge is that, after integration over the auxiliary fermion fields, the Gaussian state reads
\begin{align}  \label{eq:psi_filled_explicit}
    |\psi\rangle \sim \exp\left( \sum_{\kk} g_{\kk} c^\dagger_{\kk,2} c_{\kk,1} \right) \Ket{1,0} \sim \prod_{\kk} \left( c^\dagger_{\kk,1} + g_\kk c^\dagger_{\kk,2} \right) |\mathrm{vac}\rangle
\end{align}
with 
\begin{align}
    g_{\kk} = C K^{-1}(\kk) B = \frac{\sin k_x - \ii \sin k_y}{(\sin k_x)^2 + (\sin k_y)^2 + \lambda^2(2 - \cos k_x - \cos k_y)^2}
\end{align}
$g_{\kk}$ has one pole at $\kk=(0,0)$, near which $g_{\kk} \sim \frac{1}{k_x+\ii k_y} \left(1 + \mathcal{O}(\kk^2)\right)$. 
Here, $c^\dagger_{\kk,1} + g_\kk c^\dagger_{\kk,2}$ is the FB wave-function, and it can be directly verified that the normalized wave-function tends to the same limit $c^\dagger_{\kk,2}$ (up to a global phase factor) as $\kk$ approaches $(0,0)$ in various directions. 
Therefore, by definition, $|\psi\rangle$ corresponds to a filled CTFB. 

Next, it is obvious that the same FB is the zero mode of the following $S^\dagger(\kk)$ matrix, 
\begin{align}   \label{eq:DR_Sk}
    S^\dagger(\kk) = \begin{pmatrix}
        -\Big(\sin k_x - \ii \sin k_y\Big) \\
        (\sin k_x)^2 + (\sin k_y)^2 + \lambda^2(2 - \cos k_x - \cos k_y)^2\\
    \end{pmatrix}^T
\end{align}
We can Fourier this $S(\kk)$ matrix back into the real space. 
For clarity, we take the location of the $\td{L}$ orbital (to be denoted as $d^\dagger_{\RR}$) in each unit cell also coincides with $c^\dagger_{\RR, \alpha}$. 
Then, any factor with the form of $e^{\ii m k_x} e^{\ii n k_y}$ in the expansion of \cref{eq:DR_Sk} represents a hopping process $d^\dagger_{\RR+(m,n)} c_{\RR,\alpha}$. 
One can then deem $d^\dagger_{\RR}$ and $c^\dagger_{2}$ as $s$-wave orbitals, and $c^\dagger_{1}$ as $p_-$-wave. 
However, one immediately sees that multiple hopping processes including $d^\dagger_{\RR+(1,0)} c_{\RR,1}$, $d^\dagger_{\RR} c_{\RR,2}$, $d^\dagger_{\RR+(1,0)} c_{\RR,2}$, $d^\dagger_{\RR+(2,0)} c_{\RR,2}$, and $d^\dagger_{\RR+(1,1)} c_{\RR,2}$ (and their $C_4$ counterparts that are not listed here) are involved, while there are only two individual parameters ($\lambda$ and an overall hopping strength). 
Therefore, this form is highly fine-tuned. 
For example, as long as the $d^\dagger_{\RR} c_{\RR,2}$ amplitude is turned up, (currently its strength is $1 + 5\lambda^2$), then the touching point at $\kk=(0,0)$ will be immediately gapped, and with this amplitude further increasing, the FB will eventually evolve into an atomic insulator formed by $c^\dagger_{\RR,1}$. 

With the $S(\kk)$ matrix given, the bipartite TNS form can be immediately obtained following our discussion in the main text. 
For every auxiliary tensor located at $\RR$ (centered around $d^\dagger_{\RR}$), it has bonds connected to the physical sites ($c^\dagger_{\RR',\alpha}$) at $\RR$, $\RR+(1,0)$, $\RR+(2,0)$, $\RR+(1,1)$, and other $C_{4}$ partners. 

\subsection{General relation}

In principle, the above process can be reversed. 
Consider the filled CTFB state in the main text, 
\begin{align}   \label{eq:OmegaL_exclude_holes}
    \Ket{\Omega_L} \sim \prod_{\kk} \prod_{n = 1}^{\dim \td{L}} \left( \sum_{\alpha} \psi_{\kk \alpha} S^*_{\alpha n}(\kk) \right) \Ket{{\rm F}}
\end{align}
For $\kk$ not at the touching point, $S(\kk)$ will have the maximal rank. 
\cref{eq:OmegaL_exclude_holes} excludes all the dispersive electrons from the fully filled Gaussian state of $L$ sublattice, then the remaining ones will only fill the FB. 
Without loss of generality, we can choose $\dim \td{L}$ rows in $S(\kk)$, and denote the set of these real-space orbitals as $A$. The complementary set is $\ovl{A}$, with $\dim L - \dim \td{L} = d_{\rm FB}$ rows. 
Then, $S(\kk)$ takes the block form
\begin{align}
    S(\kk) = \begin{pmatrix}
        S_{A}(\kk) \\ 
        S_{\ovl{A}}(\kk) \\
    \end{pmatrix}
\end{align}
Without loss of generality, we can assume that $S_A(\kk)$, which is a square matrix, is full rank hence invertible if $\kk$ is not the touching point. Then the matrix below shares the same kernel space (as well as dispersive space) with $S(\kk)$, 
\begin{align}
    \td S(\kk) = S(\kk) S_{A}(\kk)^{-1} = \begin{pmatrix}
        1_{\dim \td{L} \times \dim \td{L}} \\ 
        G(\kk) \\
    \end{pmatrix} \qquad \qquad \mathrm{where} ~~ 
    G(\kk)  =  S_{\ovl{A}}(\kk) S^{-1}_{A}(\kk)
\end{align}
Then we can equivalently write 
\begin{align} 
    \Ket{\Omega_L} \sim \prod_{\kk} \prod_{\alpha \in A} \left( \psi_{\kk \alpha} + \sum_{\ovl\alpha\in \ovl{A}} \psi_{\kk \ovl\alpha } [G(\kk)]_{\ovl\alpha \alpha} \right) \Ket{{\rm F}}
\end{align}
Next, we collect all $\psi^\dagger_{\kk \alpha}$ with $\alpha \in A$ into a row vector $\psi^\dagger_{\kk A}$, and similarly define $\psi^\dagger_{\kk \ovl{A}}$. There is further
\begin{align}   \label{eq:OmegaL_move}
    \Ket{\Omega_L} \sim \prod_{\kk} \exp(- \psi^\dagger_{\kk A} [G(\kk)]^T  \psi_{\kk\ovl{A}}) \Ket{{\rm F}_{\ovl{A}}}
\end{align}
where $\Ket{{\rm F}_{\ovl{A}}} =  \prod_{\kk} \prod_{\alpha \in A} \psi_{\kk \alpha} \Ket{{\rm F}}$ fills all orbitals in $\ovl A$. 

Next, one can introduce various auxiliary fermions $\eta$, and bilinear couplings between $\eta$ and the physical ones $\psi$. The aim is to design these couplings so as to reproduce \cref{eq:OmegaL_move}, namely, searching for $B, C, K$ such that
\begin{align}
    \exp\left( -\psi^\dagger_{\kk A} [G(\kk)]^T \psi_{\kk \ovl{A}} \right) \sim \int_{\mathcal{D}[\eta,\ovl\eta]} \exp\left( - \eta^\dagger_{\kk} K(\kk) \eta_{\kk} + \psi^\dagger_{\kk A} B(\kk) \eta_{\kk} + \eta^\dagger_{\kk} C(\kk) \psi_{\kk \ovl{A}} \right)
\end{align}
with $G(\kk) = B(\kk) K^{-1}(\kk) C(\kk)$. 
Note that, for particular realizations such as PEPS, $\eta$ is required to live on the nearest-neighbor bonds between physical sites, hence the general form of $B, C, K$ will be largely constrained. 
Therefore, one may need to introduce more fermion modes, and a larger auxiliary dimension, in order to reproduce the desired $G$. 
The PEPS bond dimension, though, should still remain finite, as the state $\Ket{\Omega_L}$ has an area-law entanglement entropy.

\clearpage

\section{More examples of CTFBs}

The $2\pi$-flux dice lattice model provides an example of CTFB with $C=1$ in the wallpaper group $p3$ \cite{yang_fractional_2025}. 
It also appears in our automated construction, as tabulated in \cref{sec:2D-wallpaper}. 
In the rest of this section, we focus on CTFBs in other groups. 

\subsection{Fine-tuned higher Chern numbers}
\label{subsec:higher-Chern}

Our approach can be used to construct CTFBs with higher Chern numbers. 
In \cref{{tab:2D_Chern_band}}, we tabulate the symmetry-guaranteed critical flat Chern bands that are diagnosed by the symmetry-based indicators over all wallpaper groups. 
In groups $p4$ and $p6$, we find multiple CTFB constructions with large Chern numbers, {\it e.g.}, $C=2$ (mod 4) and $C=3$ (mod 6). 

\begin{figure}[th]
    \centering
    \includegraphics[width=0.8\textwidth]{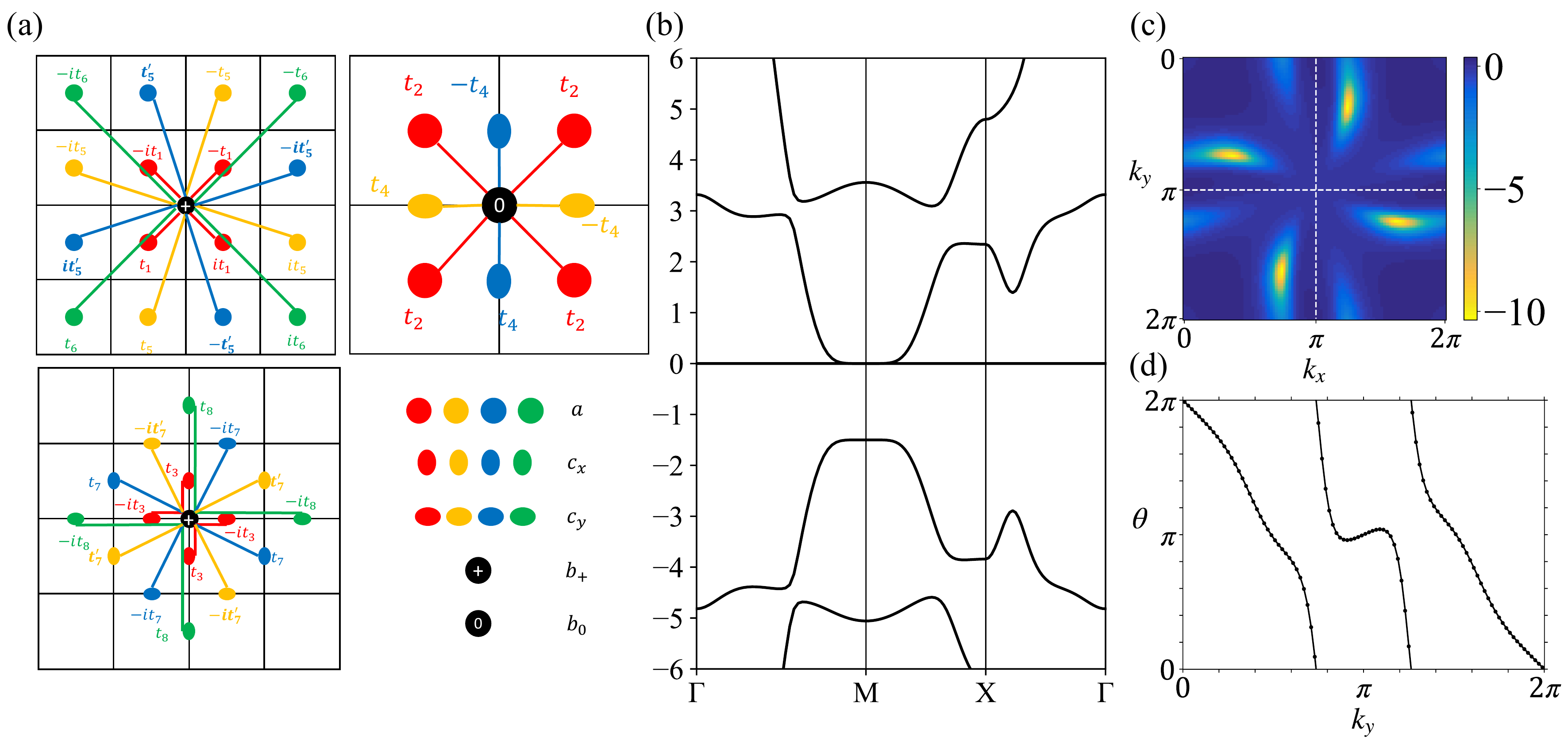}
    \caption{Critical flat band with Chern number $-3$ in wallpaper group $p4$. (a) The hopping terms. (b) Band structure. The hopping parameters are set as $t_1=t_2=1$, $t_3=t_4=1.5$, $\Delta=1.5$, and the additional parameters are given by \cref{eq:chern3-additional_vals}. (c) Berry curvature of the flat band. (d) Wilson loop of the flat band, showing a winding number of $-3$.
    }
    \label{fig:chern3}
\end{figure}

In this section we present an example of CTFB with $C=-3$ using the orbitals introduced in \cref{app:Chern}. 
Given the representation analysis in \cref{eq:p4-B',eq:p4-BR-difference}, the generalized Fu-Kane formula only constrains $C=1$ mod 4 (\cref{eq:p4-C-mod4}).
The model in \cref{app:Chern} has realized the $C=1$ CTFB.
We now construct a $C=-3$ CTFB by introducing longer-range hopping terms. 
For simplicity, we only consider additional hoppings from the $a$, $c_x$ and $c_y$ orbitals to the $b_+$ orbital: 
\begin{equation}
\begin{gathered}
    t_5=\langle \RR,a|H_{\rm F}|\RR+\hat{x},b_+\rangle,\quad t_5'=\langle \RR,a|H_{\rm F}|\RR-\hat{x},b_+\rangle,\quad t_6=\langle \RR,a|H_{\rm F}|\RR+\hat{x}+\hat{y},b_+\rangle, \\
    t_7=\langle \RR,c_x|H_{ \rm F}|\RR+\hat{x},b_+\rangle,\quad t_7'=\langle \RR,c_x|H_{\rm F}|\RR-\hat{x},b_+\rangle,\quad t_8=\langle \RR,c_x|H_{\rm F}|\RR+\hat{y},b_+\rangle. 
\end{gathered}
\end{equation}
These hopping terms and their symmetry partners are shown in \cref{fig:chern3}(a). 

The additional hopping terms lead to a correction to the first column of the $S(\kk)$ matrix in \cref{eq:p4_Sk} 
{\small
\begin{equation}
    [\Delta S(\kk)]_{\cdot,1}=\begin{pmatrix}
         2t_5\left(\ii\sin\tfrac{k_x+3k_y}{2}+\sin\tfrac{3k_x-k_y}{2}\right) +2t_5'\left(\ii\sin\tfrac{k_x-3k_y}{2}+\sin\tfrac{-3k_x-k_y}{2}\right)
         +2t_6\left(\ii\sin\tfrac{3k_x+3k_y}{2}+\sin\tfrac{3k_x-3k_y}{2}\right) \\
        2t_7\cos\frac{2k_x-k_y}{2}+2t_7'\cos\frac{2k_x+k_y}{2}+2t_8\cos\frac{3k_y}{2}\\
        -2\ii t_7\cos\frac{k_x+2k_y}{2}-2\ii t_7'\cos\frac{k_x-2k_y}{2}-2\ii t_8\cos\frac{3k_x}{2}
    \end{pmatrix}\ . 
\end{equation}}
Repeating the analysis around \cref{eq:p4-wavefunction-v4} in \cref{subsec:p4_model}, we derive the un-normalized Bloch state $s(\kk)$  for the fourth band around $\rM$ under a rotated basis: 
\begin{align}
    s_1(\kk) =& (1-\ii)(t_1-(2+\ii)t_5-(1+2\ii)t_5'+3t_6)( p_x-\ii p_y) \nonumber \\
    &-\frac{1}{24}(1-\ii)(t_1-(14+13\ii)t_5-(13+14\ii)t_5'+27t_6)(p_x^3-\ii p_y^3) \nonumber \\
    &-\frac{1}{8}(1-\ii)(t_1-(6-3\ii)t_5+(3-6\ii)t_5'+27t_6)(p_y^2p_x-\ii p_x^2p_y)+O(p^4) \ ,
\end{align}
\begin{align}
    s_2(\kk) =& -\frac{1}{\sqrt2}\ii(t_3-(1+2\ii)t_7-(1-2\ii)t_7'-3t_8)(p_x-\ii p_y) \nonumber \\
    &+\frac{1}{24\sqrt2}\ii(t_3-(1+2\ii)t_7-(1-2\ii)t_7'-3t_8)(p_x^3-\ii p_y^3) \nonumber \\
    &+\frac{1}{4\sqrt2}\ii((2+\ii)t_7+(2-\ii)t_7')(p_y^2p_x-\ii p_x^2p_y)+O(p^4) \ ,
\end{align}
where $p_x = k_x - \pi$, $p_y = k_y - \pi$. 
The Chern number is determined by the phase winding of $v_4(\kk) = \frac1{\sqrt{s^\dagger(\kk) s(\kk)}} s(\kk)$ around $\rM$. 
To realize $C=-3$, we should choose parameters such that the linear-in-$p$ terms in $s(\kk)$ vanish. 
Then the leading terms in $s(\kk)$ would be $(p_x + \ii p_y)^3$ and lead to a phase winding of $-3\times 2\pi$. 
This condition requires
\begin{equation}
    t_1=2t_5+t_5'+3t_6,\quad t_5+2t_5'=0,\quad t_5=4t_6,\quad 
    t_3=t_7+t_7'+3t_8,\quad t_7-t_7'=0,\quad t_7=-3t_8 \ .
\end{equation}
Solving these equations determine the additional hopping parameters
\begin{equation}\label{eq:chern3-additional_vals}
    t_5=\frac{4t_1}{3},\quad t_5'=-\frac{2t_1}{3},\quad t_6=\frac{t_1}{3},\quad t_7=t_7'=t_3,\quad t_8=-\frac{t_3}{3} \ .
\end{equation}

With these hopping parameters, we obtain the CTFB with $C=-3$ in \cref{fig:chern3}.
We remark that $C=-3$ is a result of the fine-tuning condition in \cref{eq:chern3-additional_vals}, without which the general symmetry-guaranteed CTFB with the chosen orbitals will have $C=1$. 
In \cref{{sec:Robust-higher-Chern-numbers}}, we provide CTFBs with large Chern numbers, {\it e.g.}, $C=2$ mod 6 and $C=3$ mod 6, that do not rely on such fine-tuning conditions.

\subsection{Robust higher Chern numbers}
\label{sec:Robust-higher-Chern-numbers}
Now we construct high-Chern-number CTFBs in wallpaper group $p6$ with $C=2$ and $C=3$.
These examples are symmetry-indicated, and the Chern numbers can be diagnosed via symmetry-indicators in the $p6$ group, which is $C$ mod 6. 

\begin{table}[th!]
\centering
\begin{tabular}{|c|c|c|c|c|c|c|c|c|c|c|}
\hline
  Wyckoff  & \multicolumn{4}{c|}{$1a \ (6)$} & \multicolumn{1}{c|}{$2b \ (3)$} & \multicolumn{1}{c|}{$3c \ (2)$} \\
\hline 
EBR & $ A\ (1) $ & $B \ (1)$ & ${}^1 \!E_{1} \ (1)$  & ${}^1\!E_{2} \ (1)$ & $A_{1} \ (2)$ & $B \ (3) $  \\
\hline
$\Gamma\ (0,0)$ & $\Gamma_1 $ & $\Gamma_2 $ & $\Gamma_5 $ & $\Gamma_6 $ &  $\Gamma_1 \oplus \Gamma_2$ & $\Gamma_2 \oplus \Gamma_4 \oplus \Gamma_6 $  \\
\hline 
$\rK (\frac{2\pi}{3},\frac{2\sqrt{3}}{3}\pi)$ &  $\rK_1$ & $\rK_1$ & $\rK_3$ & $\rK_3$ & $\rK_2\oplus \rK_3$ & $\rK_1 \oplus \rK_2 \oplus \rK_3 $  \\
\hline
$\rM \ (\pi,\frac{\sqrt{3}}{3}\pi)$ &  $\rM_1$ &  $\rM_2$ &  $\rM_1$ &  $\rM_2$ &  $\rM_1\oplus \rM_2$ &  $2\rM_1 \oplus \rM_2 $  \\
\hline
\end{tabular}
\caption{The EBR data of wallpaper group $p6$.
The first row tabulates all the Wyckoff positions with the corresponding site-symmetry groups in the parentheses. 
$1a$ refer to the $C_6$-symmetric positions $(0,0)$, respectively. $2b$ refers to the $C_3$-symmetric positions $(\frac23,\frac13)$,$(\frac13,\frac23)$. $3c$ refers to the $C_2$-symmetric positions $(\frac12,0)$,$(\frac12,\frac12)$,$(0,\frac12)$. 
The second row tabulates the EBR names, given by the irreps of the corresponding site-symmetry groups. 
Numbers in the parentheses represent the number of bands in the corresponding EBR. 
The following three rows give the irreps at high-symmetry momenta of the EBRs.  
}
\label{tab:p6-EBR}
\end{table}
\vspace{0.3cm}
\begin{table}[h!]
\centering  
\begin{tabular}{|c|c|c|c|c|c|c||c|}
\hline
\multicolumn{8}{|c|}{Irreps of the point group $6$}\\
\hline
Irreps & $E$ & $C_6$ & $C_6^2$ & $C_6^3$ & $C_6^4$ & $C_6^5$ & $\mathbf{k}$-notation \\
\hline
$A$ & 1 & 1 & 1 & 1 & 1 & 1 & $\Gamma_1$ \\
\hline
$B$ & 1 & -1 & 1 & -1 & 1 & -1 & $\Gamma_2$\\
\hline
${}^2\!E_1$ & 1 & $e^{-\ii\frac{2\pi}{3}}$ & $e^{\ii\frac{2\pi}{3}}$ & 1 & $e^{-\ii\frac{2\pi}{3}}$ & $e^{\ii\frac{2\pi}{3}}$ & $\Gamma_3$ \\
\hline
${}^2\!E_2$ & 1 & $e^{\ii\frac{\pi}{3}}$ & $e^{\ii\frac{2\pi}{3}}$ & -1 & $e^{-\ii\frac{2\pi}{3}}$ & $e^{-\ii\frac{\pi}{3}}$ & $\Gamma_4$ \\
\hline
${}^1\!E_1$ & 1 & $e^{\ii\frac{2\pi}{3}}$ & $e^{-\ii\frac{2\pi}{3}}$ & 1 & $e^{\ii\frac{2\pi}{3}}$ & $e^{-\ii\frac{2\pi}{3}}$ & $\Gamma_5$\\
\hline
${}^1\!E_2$ & 1 & $e^{-\ii\frac{\pi}{3}}$ & $e^{-\ii\frac{2\pi}{3}}$ & -1 & $e^{\ii\frac{2\pi}{3}}$ & $e^{\ii\frac{\pi}{3}}$ & $\Gamma_6$ \\
\hline
\end{tabular}
\hspace{0.5cm}  
\begin{tabular}{|c|c|c|c||c|}
\hline
\multicolumn{5}{|c|}{Irreps of the point group $3$}\\
\hline
Irreps & $E$ & $C_3$  & $C_3^2$& $\mathbf{k}$-notation \\
\hline
$A_1$ & 1 & $1$ & $1$ & $\rK_1$ \\
\hline
${}^2\!E$ & 1 & $e^{\ii\frac{2\pi}{3}}$ & $e^{-\ii\frac{2\pi}{3}}$& $\rK_2$ \\
\hline
${}^1\!E$ & 1 & $e^{-\ii\frac{2\pi}{3}}$&$e^{\ii\frac{2\pi}{3}}$ & $\rK_3$ \\
\hline
\end{tabular}
\hspace{0.5cm}
\begin{tabular}{|c|c|c||c|}
\hline
\multicolumn{4}{|c|}{Irreps of the point group $2$}\\
\hline
Irreps & $E$ & $C_2$ & $\mathbf{k}$-notation \\
\hline
$A$ & 1 & $1$ & $\rM_1$ \\
\hline
$B$ & 1 & $-1$ & $\rM_2$ \\
\hline
\end{tabular}
\caption{Character tables for point groups $6$, $3$ and $2$.  
Notations in the column ``irreps'' follow the convention of Ref.~\cite{bradley_mathematical_2010}, and are used for representations of site-symmetry groups of the Wyckoff positions in real space. 
At high symmetry momenta $\Gamma$, $\rK$ and $\rM$ of $p6$, the little groups are $6$, $3$ and $2$, respectively, and the energy bands can also be labeled by the point groups irreps, but following a different convention given in the ``$k$-notation'' columns. }
\label{tab:p6-irrep}  
\end{table}

\subsubsection{CTFB with \texorpdfstring{$C=2$}{C=2}}
\label{sec:C=2}
We construct the CTFB with  with the band irreps 
\begin{equation}
    \mclB = (\Gamma_5;\quad  \rK_1; \quad \rM_1) \ ,
\end{equation}
whose Chern number $C$ can be diagnosed by the generalized Fu-Kane formula \cite{fang_bulk_2012}
\begin{equation}
    e^{\ii\frac{\pi}{3}C}= \prod_{n\in occ} \eta_n(\Gamma)\theta_n(\rK)\zeta_n(\rM) \ .
    \label{eq:FK_p6}
\end{equation}
$\eta_n(\Gamma)$, $\theta_n(\rK)$ and $\zeta_n(\rM)$ are the $C_6$, $C_3$  and $C_2$ eigenvalues of the $n$-th occupied band at $\Gamma$, $\rK$ and $\rM$ respectively. For this model, it can be verified $C=2 \mod 6$ according to \cref{tab:p6-EBR} and \cref{eq:FK_p6}. 

\begin{equation}
    \Delta \mclB=(0; \quad \rK_1\boxminus \rK_2; \quad 0) \ ,
\end{equation}
can be chosen to trivialize the B vector as follows,
\begin{equation}
    \mclB'=\mclB+\Delta \mclB=  (\Gamma_5; \quad 2\rK_1\boxminus \rK_2; \quad \rM_1) \ .
\end{equation}
which satisfies the compatibility relation and continuity condition for the wavefunction at the touching point according to \cref{{par:touch}}. 

The corresponding real space construction is
\begin{equation}
    \mclB'= [{}^1E_1\oplus A\oplus B]_{1a} \boxminus [A_1]_{2b}.
\end{equation}

In wallpaper group $p_6$, the two lattice vectors are $\mathbf{a}_1= (1,0)$ and $\mathbf{a}_2=(-\frac{1}2,\frac{\sqrt{3}}2)$ and the sublattice vectors are 
\begin{equation}
    \tt_{a} = (0,0),\qquad 
    \mathbf{t}_{b_1} = (\frac12,\frac{\sqrt{3}}6), \qquad \mathbf{t}_{b_2} = (0, \frac{\sqrt{3}}3) \ . 
\end{equation}
We denote the five orbital basis in each unit cell as $\ket{\alpha}=\ket{a, d_-}$, $\ket{a,s}$ and $\ket{a,f}$ for irreps $[{}^1E_1]$, $[A]$ and $[B]$ at the $1a$ Wyckoff position, and $\ket{\alpha}=\ket{b_1, s}$, $\ket{b_2,s}$ for irrep $[A_1]$ at the $2b$ Wyckoff position. 

There are only three independent nearest-neighbor hopping terms between $\ket{d_-}$, $\ket{s}$ and $\ket{f}$ orbitals on $1a$ Wyckoff position and $\ket{s}$ orbitals on one of $2b$ Wyckoff position:
\begin{equation}
t_1 = \inn{\RR,a, d_-| H_{\rm F} | \RR, b_1, s } ,\qquad 
t_2 = \inn{\RR,a, s | H_{\rm F} | \RR, b_1, s } ,\qquad 
t_3 = \inn{\RR,a, f | H_{\rm F} | \RR, b_1, s }   \ , 
\end{equation}
with bra basis from $L$ sublattice, and ket basis from $\widetilde{L}$ sublattice. 
The $S(\mathbf{k})$ matrix is 
\begin{equation}
    S(\mathbf{k}) = \begin{pmatrix}
        t_1\alpha_1(\mathbf{k})& 
       t_1 e^{\ii\frac{2\pi}{3}} \alpha_2(\mathbf{k}) \\
        t_2 \beta_1(\mathbf{k})  &  t_2 \beta_2(\mathbf{k}) \\
        t_3 \beta_1(\mathbf{k})  &  -t_3 \beta_2(\mathbf{k}) 
    \end{pmatrix}\ . 
\end{equation}
where the matrix elements are
\begin{align}
\begin{split}
 \alpha_1(\mathbf{k}) &= e^{\ii\left(\frac{k_x}{2}+\frac{\sqrt{3}}{6}k_y\right)}+
     e^{-\ii\frac{2\pi}{3}}e^{\ii(\frac{\sqrt{3}}{6}k_y - \frac{k_x}2)}  +
     e^{\ii\frac{2\pi}{3}}e^{-\ii\frac{\sqrt{3}}{3}k_y},\\
 \alpha_2(\mathbf{k}) &= e^{\ii\frac{\sqrt{3}}{3}k_y}\left[ 1 + 2e^{-\ii\frac{\sqrt{3}}{2}k_y}\cos\left(\frac{k_x}{2}+\frac{2\pi}{3}\right)\right],\\
 \beta_1(\mathbf{k}) &= e^{\ii\left(\frac{k_x}{2}+\frac{\sqrt{3}}{6}k_y\right)}+
     e^{\ii(\frac{\sqrt{3}}{6}k_y - \frac{k_x}2)}  +
     e^{-\ii\frac{\sqrt{3}}{3}k_y},\\
 \beta_2(\mathbf{k}) &= e^{\ii\frac{\sqrt{3}}{3}k_y}\left[ 1 + 2e^{-\ii\frac{\sqrt{3}}{2}k_y}\cos\left(\frac{k_x}{2}\right)\right] \ .
 \end{split}
\end{align}

\subsubsection{CTFB with \texorpdfstring{$C=3$}{C=3} }
\label{sec:C=3}
Another CTFB example has the band irreps as
\begin{equation}
    \mclB = (\Gamma_1;\quad \rK_1;\quad \rM_2) \ ,
\end{equation}
with Chern number $C=3$ diagnosed by the generalized Fu-Kane formula \cref{eq:FK_p6}. 

\begin{equation}
    \Delta \mclB=(\Gamma_1 \boxminus \Gamma_4; \quad \rK_1\boxminus \rK_2; \quad 0) \ ,
\end{equation}
is added to gap the flat band $B$ with other bands.
\begin{equation}
    \mclB'=\mclB+\Delta \mclB=  (2\Gamma_1 \boxminus \Gamma_4; \quad 2\rK_1\boxminus \rK_2; \quad \rM_2) \ ,
\end{equation}
which gives us the real space construction as
\begin{equation}
    \mclB'=\mclB+\Delta \mclB=  [2A\oplus B \oplus {}^1E_2]_{1a} \boxminus [B]_{3c}.
\end{equation}

There are seven orbitals in one unit cell: $\ket{\alpha}=\ket{a,s}$, $\ket{a,s'}$, $\ket{a,f}$, $\ket{a,p_+}$ for the irreps at the $1a$ Wyckoff position, and $\ket{\alpha}=\ket{c_1, p_+}$, $\ket{c_2, p_+}$, $\ket{c_3, p_+}$ for the irrep at the $3c$ Wyckoff position. 
The corresponding sublattice vectors are
\begin{equation}
    \mathbf{t}_{a} = (0,0), \qquad 
    \mathbf{t}_{c_1} = (\frac12,0), \qquad \mathbf{t}_{c_2} = (-\frac14,-\frac{\sqrt{3}}4),\qquad \mathbf{t}_{c_3} = (-\frac14,\frac{\sqrt{3}}4)\ . 
\end{equation}
There are only four independent nearest neighbor hopping terms: 
\begin{equation}
\begin{aligned}
& t_1 = \inn{\RR,a,s | H_{\rm F} | \RR, c_1, p_+},\qquad 
t_2 = \inn{\RR,a,s' | H_{\rm F} | \RR, c_1, p_+},\nonumber\\
& t_3 = \inn{\RR,a,f | H_{\rm F} | \RR, c_1, p_+},\qquad
t_4 = \inn{\RR,a,p_+ | H_{\rm F} | \RR, c_1, p_+}  \ , 
\end{aligned}
\end{equation}
and four independent next-nearest-neighbor hopping terms:
\begin{equation}
\begin{aligned}
t_1'&=\inn{\RR, a, s | H_{\rm F} | \RR+\mathbf{a}_2, c_1, p_+},\qquad
t_2' =\inn{\RR,a, s' | H_{\rm F} | \RR+\mathbf{a}_2, c_1, p_+},\qquad\\
t_3' &=\inn{\RR,a, f | H_{\rm F} | \RR+\mathbf{a}_2, c_1, p_+},\qquad 
t_4' =\inn{\RR,a, p_+ | H_{\rm F} | \RR+\mathbf{a}_2, c_1, p_+}
\ . 
\end{aligned}
\end{equation}

The $S(\mathbf{k})$ matrix is 
\begin{equation}
\label{eq:Sk_C3}
    S(\mathbf{k}) = \begin{pmatrix}
        t_1f_1(\mathbf{k})+t_1'f_1'(\mathbf{k})& 
       t_1 g_1(\mathbf{k})+t_1'g_1'(\mathbf{k}) &
       t_1h_1(\mathbf{k}) +t_1'h_1'(\mathbf{k}) \\
        t_2 f_2(\mathbf{k}) +t_2'f_2'(\mathbf{k}) &  
        t_2 g_2(\mathbf{k}) +t_2'g_2'(\mathbf{k}) & 
        t_2 h_2(\mathbf{k})+t_2'h_2'(\mathbf{k}) \\
        t_3 f_3(\mathbf{k}) +t_3'f_3'(\mathbf{k}) &  
        -t_3 g_3(\mathbf{k})-t_3'g_3'(\mathbf{k}) &
         t_3 h_3(\mathbf{k}) +t_3'h_3'(\mathbf{k}) \\
         t_4 f_4(\mathbf{k})+t_4'f_4'(\mathbf{k}) &
         e^{-\ii\frac{\pi}{3}}[t_4 g_4(\mathbf{k})+t_4'g_4'(\mathbf{k})] &
         e^{-\ii\frac{2\pi}{3}}[t_4 h_4(\mathbf{k})+t_4'h_4'(\mathbf{k})]
    \end{pmatrix}\ . 
\end{equation}
where 

\begin{equation}
\begin{aligned}
 f_i(\mathbf{k}) &= e^{\ii\frac{k_x}{2}}(1-\eta_i e^{-\ii k_x}),\qquad 
 f_i'(\mathbf{k})=e^{\ii\frac{k_x}{2}}(e^{\ii(-\frac{k_x}{2}+\frac{\sqrt{3}}{2}k_y)}-\eta_ie^{-\ii(\frac{k_x}{2}+\frac{\sqrt{3}}{2}k_y)}),\\
 g_i(\mathbf{k}) &= e^{\ii(\frac{k_x}{4}+\frac{\sqrt{3}}{4}k_y)}(1-\eta_i e^{-\ii(\frac{k_x}{2}+\frac{\sqrt{3}}{2}k_y)}),\qquad 
 g_i'(\mathbf{k})=e^{\ii(\frac{k_x}{4}+\frac{\sqrt{3}}{4}k_y)}(e^{-\ii k_x}-\eta_i e^{\ii(\frac{k_x}{2}-\frac{\sqrt{3}}{2}k_y)}),\\
 h_i(\mathbf{k}) &= e^{\ii(-\frac{k_x}{4}+\frac{\sqrt{3}}{4}k_y)}(1-\eta_i e^{\ii(\frac{k_x}{2}-\frac{\sqrt{3}}{2}k_y)}),\qquad 
 h_i'(\mathbf{k})=e^{\ii(-\frac{k_x}{4}+\frac{\sqrt{3}}{4}k_y)}(e^{-\ii(\frac{k_x}{2}+\frac{\sqrt{3}}{2}k_y)}-\eta_i e^{\ii k_x}) \ ,
\end{aligned}
\end{equation}
with $\eta_{i}=[1,1,-1,-1]^T$.

If $t_1'=t_2'=0$, then the first and second rows of $S(\kk)$ [\cref{eq:Sk_C3}] will become linearly dependent at arbitrary $\kk$, which implies that $(-t_2^*, t_1^*, 0, 0)^T$ (up to a normalization factor) will serve as the kernel state of $S^\dagger(\kk)$ at arbitrary $\kk$. 
At $\rM$, it brings about an extra zero mode, with transforms as an $\rM_1$ irrep, besides the existing $\rM_2$ irrep (which originates from the third and fourth rows) that is required by our target CTFB construction. 
This is an accidental degeneracy that is not protected by any symmetry, and it will lead to singularity --- if one chooses the $\rM_1$ irrep to enter the flat band, then one obtains a trivial atomic insulator ($C=0$) with constant Bloch wave-function across the full Brillouin zone; and if one chooses $\rM_2$, then one gets a discontinuity (divergent quantum geometry) near $\rM$, as for any infinitesimal $\delta\kk$ away from $\rM$, the Bloch state approaches the $\rM_1$ irrep. 
Importantly, for any $t_1'$ and $t_2'$ away from this accidental limit, the $\rM_1$ irrep will immediately get gapped, and the discontinuity will evolve into a high concentration of Berry curvature and quantum geometry, akin to singular flat bands. 
Therefore, we introduce general $t_1', t_2'$ to keep the gap open at the $\rM$ point, in order to furnish the target CTFB.

\subsection{Symmetry-indicated CTFB beyond symmetry-indicators: CTFB with \texorpdfstring{$C=3$}{C=3} in \texorpdfstring{$p3$}{p3} group}   \label{app:C3inp3}

We present another example of symmetry-indicated CTFB, with Chern number $C=3$ in the $p3$ group. 
Interestingly, the Chern number cannot be diagnosed by SIs, which is $C$ mod 3 in this group, and hence is not included in our automated search. 
Nonetheless, it is still a symmetry-indicated CTFB, in the sense that the Chern number is fully determined by the symmetry data, and can be computed from the winding analysis presented in the main text Sec.V.A. 
Remarkably, it only contains $2+1$ bands, which is simpler than the $C=3$ CTFB in $p6$ group that we discussed previously. 

\begin{table}[th!]
\centering
\begin{tabular}{|c|c|c|c|c|c|c|c|c|c|}
\hline
  Wyckoff  & \multicolumn{3}{c|}{$1a \ (3)$} & \multicolumn{3}{c|}{$1b \ (3)$} & \multicolumn{3}{c|}{$1c \ (3)$} \\
\hline 
EBR & $ A_1\ (1) $ & ${}^1E \ (1)$ & ${}^2 \!E \ (1)$  & $A_1 (1)$ & $^1 E \ (1)$ & $^2E (1) $ & $A_1 (1)$ & $^1E (1)$ & $^2E (1)$  \\
\hline
$\Gamma\ (0,0)$ & $\Gamma_1 $ & $\Gamma_3 $ & $\Gamma_2 $ & $\Gamma_1 $ &  $\Gamma_3 $ & $\Gamma_2 $ & $\Gamma_1$ & $\Gamma_3$ & $\Gamma_2$  \\
\hline 
$\rK (\frac{2\pi}{3},\frac{2\sqrt{3}}{3}\pi)$ &  $\rK_1$ & $\rK_3$ & $\rK_2$ & $\rK_2$ & $\rK_1$ & $\rK_3$ & $\rK_3$ & $\rK_2$ & $\rK_1$ \\
\hline
$\rKA (-\frac{2\pi}{3},-\frac{2\sqrt{3}}{3}\pi)$ &  $\rKA_1$ & $\rK A_2$ & $\rK A_3$ & $\rK A_2$ & $\rK A_3$ & $\rK A_1$ & $\rK A_3$ & $\rK A_1$ & $\rK A_2$ \\
\hline
$\rM \ (\pi,\frac{\sqrt{3}}{3}\pi)$ &  $\rM_1$ &  $\rM_1$ &  $\rM_1$ &  $\rM_1$ &  $\rM_1$ &  $\rM_1 $ & $\rM_1$ & $\rM_1$ & $\rM_1$ \\
\hline
\end{tabular}
\caption{The EBR data of wallpaper group $p3$.
The first row tabulates all the Wyckoff positions with the corresponding site-symmetry groups in the parentheses. 
$1a$, $1b$ and $1c$ refer to the $C_3$-symmetric positions $(0,0)$, $(\frac13,\frac23)$ and $(\frac23,\frac13)$, respectively. 
The second row tabulates the EBR names, given by the irreps of the corresponding site-symmetry groups. 
Numbers in the parentheses represent the number of bands in the corresponding EBR. 
The following four rows give the irreps at high-symmetry momenta of the EBRs.  
}
\label{tab:p3-EBR}
\end{table}

\begin{table}[h!]
\centering
\begin{tabular}{|c|r|r|r||c|}
\hline
\multicolumn{5}{|c|}{Irreps of the point group $3$}\\
\hline
Irreps & $E$ & $C_3$ & $C_3^2$ & $\kk$-notation\\
\hline
 $A_1$ & 1 & 1 & 1 & $\Gamma_1$, $\rK_1$, $\rKA_1$  \\
\hline
 $^2E$ & 1 & $e^{\ii \frac{2\pi}3}$ & $e^{-\ii \frac{2\pi}3}$ & $\Gamma_2$, $\rK_2$, $\rKA_3$ \\
\hline
 ${}^1\!E$ & 1 & $e^{-\ii \frac{2\pi}3}$ & $e^{\ii \frac{2\pi}3}$ & $\Gamma_3$, $\rK_3$, $\rKA_2$ \\
\hline
\end{tabular}
\caption{Character tables for point groups $3$. Notations in the column ``irreps'' follow the convention of Ref.~\cite{bradley_mathematical_2010}, and are used for representations of site-symmetry groups of the Wyckoff positions in real space. 
At high symmetry momenta $\Gamma$, $\rK$ and $\rKA$ of $p3$, the little groups are $3$, and the energy bands can also be labeled by the point groups irreps, but following a different convention given in the ``$\kk$-notation'' columns.  
}
\label{tab:p3-irrep}
\end{table}

In real-space, define the primitive lattice vectors, $\mathbf{a}_1=(1,0)$ and $\mathbf{a}_2=(-\frac12,\frac{\sqrt{3}}2)$. 
Also, define $\mathbf{a}_3 = -(\mathbf{a}_1+\mathbf{a}_2) = (-\frac12, -\frac{\sqrt{3}}2)$. $\mathbf{a}_1$ to $\mathbf{a}_3$ are separated by $120^\circ$. 
In momentum space, define $\mathbf{b}_1=2\pi(1,\frac1{\sqrt{3}})$ and $\mathbf{b}_2=2\pi(0,\frac{2}{\sqrt{3}})$. 

Consider the model with 
\begin{align}
    \mathcal{BR}_{L} = 2 [A_1]_{1a} \, , \qquad \mathcal{BR}_{\td L} = [{}^2E]_{1a}
\end{align}
namely, the $L$ sublattice contains two $s$-orbitals, and the $\td L$ sublattice contains one $p_+$-orbital. The CTFB is described by
\begin{align}
\mathcal{BR}_{L} \boxminus \mathcal{BR}_{\td L} = \mclB + \Delta\mclB = (2\Gamma_1 \boxminus \Gamma_2 ; \quad 2\rK_1 \boxminus \rK_2; \quad 2\rKA_1 \boxminus \rKA_3 ; \quad \rM_1)
\end{align}
with a robust choice of $\mclB = (\Gamma_1;  \rK_1; \rKA_1; \rM_1)$. 
All the three high-symmetry points are touching points. 
The SI for Chern number (the generalized Fu-Kane formula) in $p3$ group reads
\begin{align}
e^{\ii \frac{2\pi C}3} = \prod_{i \in {\rm  occ}}\theta_i(\Gamma)\theta_i(\rK)\theta_i({\rm KA})
\end{align}
where $\theta$ denotes $C_3$ eigenvalue at high symmetry points. 
Notably, for such $\mclB$, one finds $C=0$ mod 3, hence the SI is insufficient to determine whether $\mclB$ is topological. 
Nevertheless, the general form, hence the winding, of the $S(\kk)$ matrix near each touching point is fixed by symmetry, and the topology is still symmetry-indicated. 
We derive it below. 

Define independent inter-sublattice hoppings, 
\begin{align}
    &t_{1,1} =\langle \RR, s,1 | H |\RR+\mathbf{a}_1,p_+\rangle \qquad t_{1,2} =\langle \RR, s,1 | H |\RR+\mathbf{a}_1+\mathbf{a}_2,p_+\rangle \\\nonumber
    &t_{2,1} =\langle \RR, s,2 | H |\RR+\mathbf{a}_1,p_+\rangle \qquad t_{2,2} =\langle \RR, s,2 | H |\RR+\mathbf{a}_1+\mathbf{a}_2,p_+\rangle 
\end{align}
where $\Ket{\RR,s,\beta}$ with $\beta=1,2$ labels the two $s$-orbitals. Notice that, intra-unit-cell ``hoppings'' like $\langle \RR, s,1 | H |\RR,p_+\rangle$ are forbidden by symmetry.

The hopping matrix is 
\begin{align}
S(\kk) = \begin{pmatrix}
t_{1,1} \, f(\kk) + t_{1,2} \, g(\kk) \\
t_{2,1} \, f(\kk) + t_{2,2} \, g(\kk)
\end{pmatrix}, \quad \textrm{with} \quad f(\mathbf k) = \omega \sum_{j=1}^{3} \omega^{-j} e^{\ii \kk \cdot \mathbf{a}_j} \, , \quad 
g(\mathbf k) = \sum_{j=1}^{3} \omega^{-j} e^{-\ii \kk \cdot \mathbf{a}_j} \, .
\end{align}
with $\omega=e^{\ii \frac{2\pi}{3}}$. Near the three touching points, up to the leading order
\begin{align}
S(\pp+\Gamma) &= \frac{3\ii}{2} p_- \begin{pmatrix}
t_{1,1} - \omega^* t_{1,2} \\
t_{2,1} - \omega^* t_{2,2} 
\end{pmatrix} \\\nonumber
S(\pp+\rK) &= \frac{3\ii}{2} p_- \begin{pmatrix} \omega^* t_{1,1} - t_{1,2} \\ \omega^* t_{2,1} - t_{2,2} \end{pmatrix} \\\nonumber
S(\pp+\rKA) &= \frac{3\ii}{2} p_- \begin{pmatrix} 
 \omega t_{1,1} - \omega t_{1,2}  \\ 
\omega t_{2,1} - \omega t_{2,2} \\
\end{pmatrix}
\end{align}
Each touching point thus contributes a definite winding $+1$, leading to total Chern number $+3$. 

We remark that, although the SI is trivial, the Chern number in this model is still robust against perturbations that break the bipartite conditions, akin to the former models. 
To see this, we consider the effective Hamiltonian near the touching point in the presence of some perturbation. 
As one can always carry out a unitary transformation to bring $S^\dagger(\kk) \sim  (k_-, 0)$, 
\begin{align}
    H_{\rm gapped}(\kk) = H_{\rm pert}(\kk) + \frac{1}{\Delta} S(\kk) S^\dagger(\kk) 
    \sim h_0 \sigma^0 + h_{x} \sigma^x + h_{y} \sigma^y + h_{z} \sigma^z +  \begin{pmatrix}
        k^2 & 0 \\
        0 & 0 \\
    \end{pmatrix} \ .
\end{align}
Here, we only keep the leading constant terms in the perturbation $H_{\rm pert}(\kk)$. 
Higher order of $H_{\rm pert}(\kk)$ starts at $\mathcal{O}(k^2)$, and must be proportional to $k^2 = \kk_x^2 +\kk_y^2$, as both states in this low-energy space enjoy the same irrep. 
For such a Hamiltonian, the Bloch states have no angular dependence, and will contribute no Berry curvature. 
Therefore, gapping the touching points in a CTFB does not contribute additional Berry curvatures.

\subsection{Strong \texorpdfstring{$\mathbb{Z}_2$}{Z2} CTFB in 2D}
\label{subsec:Z2-CTFB}

\begin{table}[th!]
\centering
\begin{tabular}{|c| c|c|c|c| c|c|c|c|}
\hline
  Wyckoff  & \multicolumn{4}{c|}{$1a \ (\bar3m)$} & \multicolumn{2}{c|}{$2c \ (32)$} & \multicolumn{2}{c|}{$3f \ (2/m)$} \\
\hline 
EBR & ${}^1\!\ovl{E}_g {}^2\!\ovl{E}_g$ (2) & ${}^1\!\ovl{E}_u {}^2\!\ovl{E}_u$ (2) & $\ovl{E}_{1g}$ (2) & $\ovl{E}_{1u}$ (2) & ${}^1\!\ovl{E} {}^2\!\ovl{E}$ (4) & $\ovl E_{1}$ (4) &  ${}^1\!\ovl{E}_g {}^2\!\ovl{E}_g$ (6) & ${}^1\!\ovl{E}_u {}^2\!\ovl{E}_u$ (6) \\
\hline
$\Gamma \ (0,0)$ & $\bG_4\bG_5$ & $\bG_6\bG_7$ & $\bG_8$ & $\bG_9$  & 
    $\bG_4\bG_5 \oplus \bG_6\bG_7$ & $\bG_8\oplus\bG_9$ &
    $\bG_4\bG_5\oplus 2\bG_8$ & $\bG_6\bG_7\oplus 2\bG_9$ \\
\hline
$\mathrm{K} \ (\frac{2\pi}3,\frac{2\pi}3)$  & 
    $\bK_4\bK_5$ & $\bK_4\bK_5$ & $\bK_6$ & $\bK_6$ & 
    $2\bK_6$ & $\bK_4\bK_5\oplus \bK_6$ & 
    $\bK_4\bK_5\oplus 2\bK_6$ & $\bK_4\bK_5\oplus 2\bK_6$ \\
\hline 
$\mathrm{M} \ (\pi,0,0)$ & 
    $\bM_3\bM_4$ & $\bM_5\bM_6$ &  $\bM_3\bM_4$ & $\bM_5\bM_6$ & 
    $\bM_3\bM_4 \oplus \bM_5\bM_6$ & $\bM_3\bM_4 \oplus \bM_5\bM_6$ &
    $\bM_3\bM_4 \oplus 2 \bM_5\bM_6$ & $2\bM_3\bM_4 \oplus \bM_5\bM_6$ \\
\hline
\end{tabular}
\caption{EBRs of the double-valued layer group $p\bar31m$ with time-reversal symmetry, which is taken from the $z=0$ and $k_z=0$ data of the space group $P\bar31m$.
The first row tabulates all the Wyckoff positions with the corresponding site-symmetry groups in the parentheses. 
$1a$, $2c$, and $3f$ are positions at the triangular, honeycomb, and Kagome lattices, respectively. 
The second row tabulates the EBR names, given by the irreps of the corresponding site-symmetry groups. 
Numbers in the parentheses represent the number of bands in the corresponding EBR. 
The following three rows give the co-irreps at high-symmetry momenta of the EBRs.  
}
\label{tab:p-31m-EBR}
\end{table}

\begin{table}[h]
\centering
\begin{tabular}{|c|c|r|r|r|r|r|r||c|}
\hline
\multicolumn{9}{|c|}{Co-irreps of the point group $\bar3m$} \\
\hline
Co-irreps & A.H. & \ $1$ & $2\{3_{001}^+\}$ & $3\{2_{1\bar10}\}$ & $\bar1$ & 2$\{-3_{001}^+\}$ & $3\{m_{1\bar10}\}$ & $\kk$-notation \\
\hline 
${}^1\!\ovl{E}_g {}^2\! \ovl{E}_g$ & $ E_{\frac32 g}$ & $2$ & $-2$ & $0$ & $2$ & $-2$ & $0$ & $\bG_4\bG_5$ \\
\hline 
${}^1\!\ovl{E}_u {}^2\! \ovl{E}_u$ & $E_{\frac32 u}$ & $2$ & $-2$ & $0$ & $-2$ & $2$ & $0$ &  $\bG_6\bG_7$ \\
\hline 
$\ovl{E}_{1g}$ & $E_{\frac12 g}$ & $2$ & $1$ & $0$ & $2$ & $1$ & $0$ & $\bG_8$ \\
\hline 
$\ovl{E}_{1u}$ & $E_{\frac12 u}$ & $2$ & $1$ & $0$ & $-2$ & $-1$ & $0$ & $\bG_9$ \\
\hline
\end{tabular}
\vspace{0.1cm}
\\ 
\begin{tabular}{|c|c|r|r|r||c|}
\hline
\multicolumn{6}{|c|}{Co-irreps of the point group $3m$} \\
\hline
Co-irreps & A.H. & \ $1$ & $2\{3_{001}^+\}$ & $3\{m_{1\bar10}\}$ & $\kk$-notation \\
\hline 
${}^1\!\ovl{E} {}^2\! \ovl{E}$ & $ E_{\frac32}$ & $2$ & $-2$ & $0$  & $\bK_4\bK_5$ \\
\hline 
$\ovl{E}_{1}$ & $E_{\frac12}$ & $2$ & $1$ & $0$ & $\bK_6$ \\
\hline 
\end{tabular}
\hspace{0.5cm}
\begin{tabular}{|c|c|r|r|r|r||c|}
\hline
\multicolumn{7}{|c|}{Co-irreps of the point group $2/m$} \\
\hline
Co-irreps & A.H. & \ $1$ & $2_{1\bar10}$ & \ $\bar1$ & $m_{1\bar10}$ & $\kk$-notation \\
\hline 
${}^1\!\ovl{E}_g {}^2\! \ovl{E}_g$ & $ E_{\frac12g}$ & $2$ & $0$ & $2$  & $0$ & $\bM_3\bM_4$ \\
\hline 
${}^1\!\ovl{E}_u {}^2\! \ovl{E}_u$ & $ E_{\frac12u}$ & $2$ & $0$ & $-2$  & $0$ & $\bM_5\bM_6$ \\
\hline 
\end{tabular}
\caption{Character tables for co-irreps of double-valued point groups $\bar3m$, $32$, and $2/m$ with time-reversal symmetry.
Notations in the ``co-irreps'' columns follow the convention of Ref.~\cite{bradley_mathematical_2010}, and are used for representations of site-symmetry groups of the Wyckoff positions in real space. 
We also tabulate the corresponding notations in the convention of Altmann and Herzig \cite{altmann_point-group_1994} in the second columns since they directly reflect the effective angular momenta. 
The high-symmetry momenta $\Gamma$, $\mathrm{K}$, and $\rM$ of $p\bar31m$ have little groups $\bar3m$, $32$, and $2/m$, respectively.
They further respect an anti-unitary symmetry formed by time-reversal followed by inversion, which promotes the irreps to co-irreps. 
Energy bands at these momenta can also be labeled by the point groups co-irreps, but following a different convention given in the ``$\kk$-notation'' columns.  
}
\label{tab:p-31m-irrep}
\end{table}

In this subsection we present a CTFB exhibiting the $\mathbb{Z}_2$ topological invariant protected by time-reversal symmetry $\mT$. 
We consider the double-valued layer group $p\bar31m$ as an example. 
As tabulated in table \cref{{tab:p-31m-EBR}}, the $\mathbb{Z}_2$ CTFB can be constructed as 
\begin{equation} \label{eq:Bp-p-31m}
\mclB' = [{}^1\!\ovl{E}_{g} {}^2\ovl{E}_{g}]_{3f} \; \boxminus \; 
    [ \ovl{E}_{1u} \oplus {}^1\!\ovl{E}_{g} {}^2\ovl{E}_{g} ]_{1a}
= ( 2\bG_8 \boxminus \bG_9 ; \quad \bK_6; \quad \bM_5\bM_6 )\ . 
\end{equation}
The EBRs and co-irreps are defined in \cref{tab:p-31m-EBR,tab:p-31m-irrep}, respectively. 
According to the counting rules established in previous sections, the flat band will form the co-irreps 
\begin{equation}
    \mclB = (\bG_8; \quad \bK_6;\quad \bM_5\bM_6) \ ,
\end{equation}
where the $\bG_8$ level belongs to a touching point of two degenerate $\bG_8$ levels. 
The $\mathbb{Z}_2$ topology is diagnosed by the Fu-Kane formula
\begin{equation}
z_2 = \sum_{\KK \in \mathrm{TRIM}} n_{\KK}^- \mod 2
    \quad = n_{\Gamma}^- + 3n_{\rM}^- \mod 2\ .
\end{equation}
Here $n_{\KK}^-$ represents the number of occupied Kramer pairs with odd-parity at the time-reversal invariant momentum (TRIM) $\KK$. 
Since $\bG_8$ has even parity and $\bM_5\bM_6$ has odd parity, the Fu-Kane formula  predicts a nontrivial $\mathbb{Z}_2$ topology, provided the Bloch wavefunction is continuous at $\Gamma$.

The continuity at $\Gamma$ is guaranteed by symmetry. 
According to \cref{{par:touch}}, the $S(\kk)$ matrix will have a continuous kernel at $\Gamma$ if $\bG_8^* \otimes \bG_9 \otimes E_u$ only contains a single identity representation, where $E_u$ is the (single-valued) representation formed by the $\kk$-vector. 
This is indeed the case, and one can refer to Table 42.8 of Ref.~\cite{altmann_point-group_1994} for the direct products of representations.

\begin{figure}[th]
    \centering
    \includegraphics[width=1.0\linewidth]{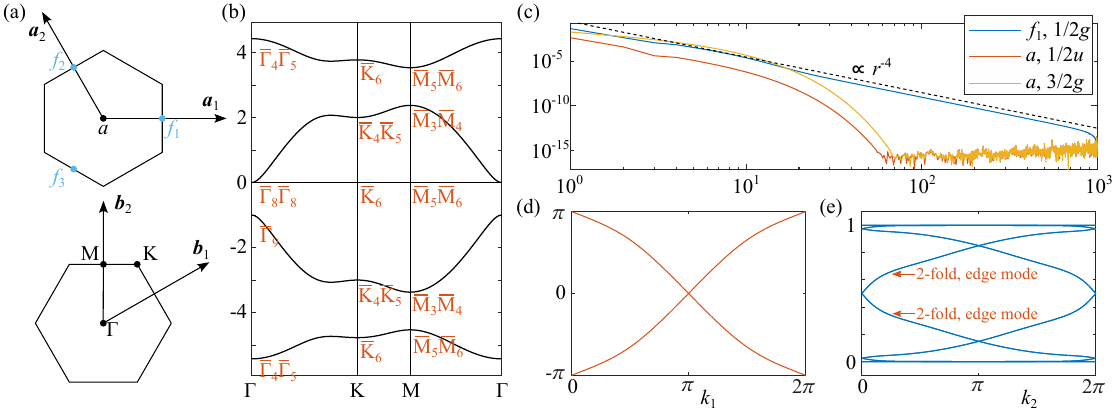}
    \caption{The $\mathbb{Z}_2$ CTFB in layer group $p\bar31m$ with time-reversal symmetry. (a) The lattice structure and Brillouin zone.
    (b) The band structure with parameters $t_1=t_2=t_3=t_4=\Delta=1$, where each band is two-fold degenerate due to the $P\mT$ symmetry. 
    (c) Correlation functions in the Fock state $\ket{\Omega}$ that occupies the lowest six bands.
    We only present diagonal components for orbitals $\alpha=(f_1,\frac12g)$, $(a,\frac12u)$, $(a,\frac32g)$ along $\RR=n\mathbf{a}_1$ with $n=1,2\cdots 1000$. The system size is $2000\times 2000$, defined by the basis vectors $\mathbf{a}_{1,2}$. The dashed line indicates $0.3/n^4$. 
    (d) The Wilson loop spectrum of the flat band, where the Wilson loop operator is integrated along $\mathbf{b}_2$, and the spectrum is plotted along $\mathbf{b}_1$.
    (e) The entanglement spectrum for the subsystem $A=\{ (\RR,\alpha) \;|\; 1\le \frac1{2\pi} \RR \cdot \mathbf{b}_1 \le 50\}$ of the $\ket{\Omega}$  state on a $100\times 100$ lattice with periodic boundary conditions. 
    Since the subsystem respects the translational symmetry along $\mathbf{a}_2$, the momentum $\kk=k_2 \mathbf{b}_2$ is preserved, and the spectra are plotted as functions of $k_2$. 
    }
    \label{fig:p-31m}
\end{figure}

We now construct the hopping model. 
The lattice vectors are given by $\mathbf{R}= R_1 \mathbf{a}_1 + R_2 \mathbf{a}_2$, where $R_{1,2}\in \mathbb{Z}$ and 
\begin{equation}
    \mathbf{a}_1 = (1,0),\qquad \mathbf{a}_2 = (-\frac12,\frac{\sqrt3}2) 
\end{equation}
are the basis (\cref{fig:p-31m}(a)). 
In each unit cell, $1a$ locates at the origin and $3f$ locate at 
\begin{equation}
    \tt_{1} = (\frac12,0), \qquad 
    \tt_{2} = (-\frac14,\frac{\sqrt3}4),\qquad 
    \tt_{3} = (-\frac14,-\frac{\sqrt3}4)\ . 
\end{equation}
We denote the orbitals in $[\ovl E_{1u}]_{1a}$ and $[ {}^1\ovl E_{1g} {}^2\ovl E_{1g} ]_{1a}$ as  $\alpha=(p_+,\downarrow)$, $(p_-,\uparrow)$, and $\alpha=(d_+,\downarrow)$, $(d_-,\uparrow)$, respectively. 
We denote the orbitals at the $3f$ positions as $\alpha=(s,\uparrow)$, $(s, \downarrow)$. 
Symmetry generators act on the orbitals at the $1a$ and $3f$ positions as 
\begin{equation} \label{eq:p-31m-symmetry}
    C_{3z}^{(a)} = e^{-\ii \frac{\pi}3 \sigma_z } \oplus (-\sigma_0),\qquad 
    M_x^{(a)} = (-\ii\sigma_y) \oplus (-\ii\sigma_y),\qquad 
    P^{(a)} = (-\sigma_0)\oplus (\sigma_0),\qquad 
    \mT^{(a)} = (-\ii\sigma_y) \oplus (-\ii\sigma_y) K\ ,
\end{equation}
and
\begin{equation}
    C_{3z}^{(f)} = e^{-\ii\frac{\pi}3\sigma_z},\qquad 
    M_x^{(f)} = -\ii\sigma_y,\qquad 
    P^{(f)} = \sigma_0,\qquad 
    \mT^{(f)} = -\ii\sigma_y K \ , 
\end{equation}
respectively, where $K$ is the complex conjugation. 
These operations also change the positions of the orbitals accordingly. 

We first consider the two-by-four hopping matrix $h_{i,j}(\tt_{1}) = \inn{\RR,f_1,i |H_{\rm F}|\RR,a,j}$ with $i=(s,\uparrow)$, $(s,\downarrow)$, $j=(p_+,\downarrow)$, $(p_-,\uparrow)$, $(d_+,\downarrow)$, $(d_-,\uparrow)$. 
Here $\tt_{1}$ is the vector from the position of the ket state to the position of the bra state. 
Constrained by the $M_x$ and $\mT$ symmetries, it must have the form 
\begin{equation}
    h(\tt_{1}) = \begin{bmatrix}
        t_1\sigma_0 + \ii t_2 \sigma_y,\qquad t_3 \sigma_0 + \ii t_4\sigma_y
    \end{bmatrix} \ ,\qquad t_{1,2,3,4}\in \mathbb{R}\ . 
\end{equation}
$h(\tt_{2})$ and $h(\tt_{3})$ can be obtained by applying the $C_{3z}$ symmetry:
\begin{equation}
    h(\tt_{2}) = C_{3z}^{(f)} \cdot h(\tt_{1}) \cdot C_{3z}^{(a)\dagger} ,\qquad 
    h(\tt_{3}) = C_{3z}^{(f)} \cdot h(\tt_{2}) \cdot C_{3z}^{(a)\dagger}\ . 
\end{equation}
$h(-\tt_{1,2,3})$ can be obtained by applying the inversion symmetry: 
\begin{equation}
    h(-\tt_{1,2,3}) = P^{(f)} \cdot h(\tt_{1,2,3}) \cdot P^{(a)\dagger} \ \ . 
\end{equation}
We hence have the $S(\kk)$ matrix 
\begin{equation}
    S(\kk) = \begin{pmatrix}
        h(\tt_{1}) e^{-\ii\kk\cdot\tt_{1}} + h(-\tt_{1}) e^{\ii\kk\cdot\tt_{1} } \\
        h(\tt_{2}) e^{-\ii\kk\cdot\tt_{2}} + h(-\tt_{2}) e^{\ii\kk\cdot\tt_{2} } \\
        h(\tt_{3}) e^{-\ii\kk\cdot\tt_{3}} + h(-\tt_{3}) e^{\ii\kk\cdot\tt_{3} }
    \end{pmatrix}\ . 
\end{equation}
The full Hamiltonian in momentum space is 
\begin{equation}
    H(\kk) = \begin{pmatrix}
        0_{6\times 6} & S(\kk) \\
        S^\dagger(\kk) & -\Delta \cdot \mathbb{I}_{4\times 4}
    \end{pmatrix}\ . 
\end{equation}
Here $-\Delta$ is an on-site energy at the $1a$ position. 

The band structure with $t_1=t_2=t_3=t_4=\Delta=1$ is shown in \cref{fig:p-31m}(b). 
Following the calculations in \cref{app:Chern}, we obtain the correlation functions,  Wilson loop spectrum, and entanglement spectrum in \cref{fig:p-31m}(c), (d), and (e), respectively. 
The correlation functions for orbitals at the $3f$ positions decay as $r^{-4}$, and the Wilson loop and entanglement spectrum reveal the  $\mathbb{Z}_2$ topology.

\subsection{Strong \texorpdfstring{$\mathbb{Z}_2$}{Z2} CTFB in 3D}
\label{subsec:3D Z2-CTFB}

\begin{table}[!ht]
\centering
\begin{tabular}{|c|c|c|c|c|c|c|c|c|c|c|c|c|}
\hline
Wyckoff  
& \multicolumn{3}{c|}{$1a\;(\bar{4}3m)$} 
& \multicolumn{3}{c|}{$1b\;(\bar{4}3m)$} 
& \multicolumn{2}{c|}{$3c\;(\bar{4}2m)$} 
& \multicolumn{2}{c|}{$3d\;(\bar{4}2m)$} \\
\hline 
EBR & $ \ovl{E}_1\ (2) $ & $\ovl{E}_2 \ (2)$ & $ \ovl{F} \ (4)$  & $\ovl{E}_1 \ (2)$ & $\ovl{E}_2 \ (2)$ & $\ovl{F} \ (4)$ & $\ovl{E}_1 \ (6) $ & $\ovl{E_2} \ (6)$  & $\ovl{E}_1\ (6)$ & $\ovl{E}_2\ (6)$ \\
\hline
$\Gamma\ (0,0,0)$ & $\ovl{\Gamma}_6 $ & $\ovl{\Gamma}_7 $ & $\ovl{\Gamma}_8 $ & $\ovl{\Gamma}_6 $ & $\ovl{\Gamma}_7 $ & $\ovl{\Gamma}_8 $ & $\ovl{\Gamma}_6 \oplus \ovl{\Gamma}_8 $ & $\ovl{\Gamma}_7 \oplus \ovl{\Gamma}_8 $ & $\ovl{\Gamma}_6 \oplus \ovl{\Gamma}_8$ & $\ovl{\Gamma}_7 \oplus \ovl{\Gamma}_8$ \\
\hline
$\rR \ (\pi,\pi,\pi)$ &  $\ovl{\rR}_6$ &  $\ovl{\rR}_7$ &  $\ovl{\rR}_8$ & $\ovl{\rR}_6$ &  $\ovl{\rR}_7$ &  $\ovl{\rR}_8$ &  $\ovl{\rR}_7 \oplus \ovl{\rR}_8$ &  $\ovl{\rR}_6 \oplus \ovl{\rR}_8$ & $\ovl{\rR}_7 \oplus \ovl{\rR}_8$ & $\ovl{\rR}_6 \oplus \ovl{\rR}_8$ \\
\hline
$\rM \ (\pi,\pi,0)$ &  $\ovl{\rM}_7$ &  $\ovl{\rM}_6$ &  $\ovl{\rM}_6 \oplus \ovl{\rM}_7$ &  $\ovl{\rM}_6$ &  $\ovl{\rM}_7$ &  $\ovl{\rM}_6 \oplus \ovl{\rM}_7$ &  $2\ovl{\rM}_6 \oplus \ovl{\rM}_7$ &  $\ovl{\rM}_6 \oplus 2 \ovl{\rM}_7$ &  $\ovl{\rM}_6\oplus 2\ovl{\rM}_7$ & $2\ovl{\rM}_6 \oplus \ovl{\rM}_7$ \\
\hline 
$\rX (0, \pi, 0)$ &  $\ovl{\rX}_7$ & $\ovl{\rX}_6$ & $\ovl{\rX}_6 \oplus \ovl{\rX}_7$ & $\ovl{\rX}_6$ & $\ovl{\rX}_7$ & $\ovl{\rX}_6 \oplus \ovl{\rX}_7$ &  $\ovl{\rX}_6 \oplus 2 \ovl{\rX}_7$ & $2\ovl{\rX}_6 \oplus \ovl{X}_7$ & $2\ovl{\rX}_6 \oplus \ovl{X}_7$ & $\ovl{\rX}_6\oplus 2\ovl{\rX}_7$  \\
\hline
\end{tabular}\\
\vspace{0.2cm}
\begin{tabular}{|c|c|r|r|r|r|r||c|}
\hline
\multicolumn{8}{|c|}{Co-irreps of the double point group $\bar{4}3m$}\\
\hline
Co-irreps & A.H. & $E$ & $6\{2_{001}\}$ & $8\{3^+_{111}\}$ & $12\{m_{1-10}\}$ & $6\{\ovl{4}^+_{001}\}$ & $\kk$-notation\\
\hline
 $\ovl{E}_1$ & $E_{1/2}$ & 2 & 0 & 1 & 0  & $\sqrt{2}$ & $\ovl{\Gamma}_6$, $\ovl{\rR}_6$ \\
\hline
 $\ovl{E}_2$ & $E_{5/2}$ & 2 & $0$ & $1$ & $0$ & $-\sqrt{2}$ & $\ovl{\Gamma}_7$, $\ovl{\rR}_7$ \\
\hline
 $\ovl{F}$ & $F_{3/2}$ & 4 & 0 & $-1$ & $0$ & $0$ & $\ovl{\Gamma}_8$, $\ovl{\rR}_8$ \\
\hline
\end{tabular} \\
\vspace{0.2cm}
\begin{tabular}{|c|c|r|r|r|r|r||c|}
\hline
\multicolumn{8}{|c|}{Co-irreps of the double point group $\ovl{4}2m$}\\
\hline
Co-irreps & A.H. & $E$ & $2\{2_{001}\}$ & $2\{\ovl{4}^+_{001}\}$ & $4\{2_{010}\}$ & $4\{m_{110}\}$ & $\kk$-notation \\
\hline
$\ovl{E}_1$ & $E_{3/2}$ & 2 & $0$ & $-\sqrt{2}$ & $0$ & $0$ & $\ovl{\rM}_6$, $\ovl{\rX}_6$ \\
\hline
$\ovl{E}_2$ & $E_{1/2}$ & 2 & $0$ & $\sqrt{2}$ & $0$ & $0$ & $\ovl{\rM}_7$, $\ovl{\rX}_7$ \\
\hline
\end{tabular}
\caption{The EBR data and character tables for double-valued space group $P\bar{4}3m$ with time-reversal symmetry.
The first table is the EBR data, where  $1a$ refers to the Wyckoff position $(0,0,0)$, $1b$ refers to  $(\frac12,\frac12,\frac12)$,  $3c$ refers to $(\frac12,\frac12,0)$, $(0,\frac12,\frac12)$,$(\frac12,0,\frac12)$, and $3d$ refers to $(\frac12,0,0)$, $(0,\frac12,0)$, $(0,0,\frac12)$. 
The second and third tables are character tables for point groups $\ovl 43m$ and $\ovl 42m$ with time-reversal symmetry, respectively.
Notations in the  ``co-irreps'' columns follow the convention of Ref.~\cite{bradley_mathematical_2010}, and are used for representations of site-symmetry groups of the Wyckoff positions in real space. 
We also tabulate the corresponding notations in the convention of Altmann and Herzig \cite{altmann_point-group_1994} in the ``A.H.'' columns since they directly reflect the effective angular momenta. 
The high symmetry momenta $\Gamma$, $\rR$ and $\rM$, $\rX$ respect the little groups $\ovl{4}3m$ and $\ovl{4}2m$, respectively.
And they further respect the time-reversal symmetry. 
Energy bands at these momenta can also be labeled by the point groups co-irreps, but following a different convention given in the ``$\kk$-notation'' columns. 
}
\label{tab:p-43m-irrep}
\end{table}

Now we construct a 3D $\mathbb{Z}_2$ CTFB in the double-valued space group $P\bar43m$ with time-reversal symmetry. 
We consider a band structure characterized by 
\begin{equation}
\mclB = (\ovl{\Gamma}_6; \quad  \ovl{\rR}_6; \quad \ovl{\rM}_6; \quad \ovl{\rX}_7 ) \ .  
\end{equation}
Its topology is diagnosed by the symmetry-based indicator \cite{song2018quantitative} 
\begin{align}
z_2 = \sum_\KK \frac{1}{2} n_\KK^{3/2} - \frac12n_\KK^{1/2} \; \mod 2 \ ,
\end{align}
where $\KK$ summed over four $S_4$ invariant momenta, $n_\KK^{1/2}$ the number of Kramer pairs at $\KK$ with $\tr[D(S_4)]=\sqrt{2}$, $n_\KK^{3/2}$ the number of Kramer pairs at $\KK$ with $\tr[D(S_4)]=-\sqrt{2}$. 

In this example, we choose the zero-band correction 
\begin{align}
\Delta \mclB &= \left(  \ovl{\Gamma}_6 \boxminus \ovl{\Gamma}_7; \quad 0; \quad 0; \quad 0 \right) 
\end{align}
to trivialize the topology of the flat band $\mclB$.  
The augmented band structure $\mclB + \Delta \mclB$ can be realized as
\begin{align}
\mclB + \Delta \mclB = [\ovl{E}_1]_{1a} \oplus [\ovl{E}_1 \oplus \ovl{F}]_{1b} \boxminus [\ovl{E}_2]_{3c} \ .
\end{align}

In order to construct the lattice model explicitly, we first define the sublattice vectors 
\begin{equation}
\tt_{a}=(0,0,0),\qquad 
\tt_{b}=(\frac12,\frac12,\frac12),\qquad 
\tt_{c_1}=(\frac12, \frac12, 0), \qquad 
\tt_{c_2}=(0,\frac12, \frac12),\qquad 
\tt_{c_3}=(\frac12, 0, \frac12)
\end{equation}
for the Wyckoff positions $1a$, $1b$, and $3c$, respectively. 
We then fix the gauge of representation matrices for the involved co-irreps.
Following the convention of the Bilbao Crystallographic Server, the $\ovl43m$ generators in both $[\ovl{E}_1]_{1a}$ and $[\ovl{E}_1]_{1b}$ are
\begin{align}
\label{eq:3D E1}
D^{(\ovl{E}_1)}(2_z) = 
\bpm -\ii & 0 \\ 0 & \ii \epm,\quad 
D^{(\ovl{E}_1)}(2_y) = \bpm 0 & -1 \\ 1 & 0 \epm, \quad 
D^{(\ovl{E}_1)}(3_{111}^+) = \frac 1 {\sqrt{2}}
\bpm e^{-\ii\frac \pi 4} & e^{-\ii \frac {3\pi}{4}} \\ e^{-\ii\frac{\pi}4} & e^{\ii \frac{\pi}4} \epm, \quad
D^{(\ovl{E}_1)} (m_{1\bar10}) = 
\bpm 0 & e^{\ii\frac{3 \pi}{4}} \\ e^{\ii\frac{\pi}{4}} & 0 \epm \ .
\end{align}
The $\ovl43m$ generators in $[\ovl{F}]_{1b}$ are
\begin{align}
\label{eq:3D F}
& D^{(\ovl{F})}(2_z) = -\ii\sigma_z \oplus \ii \sigma_z \ , \quad
D^{(\ovl{F})}(2_y) = -\ii \sigma_y \oplus \ii \sigma_x \ , \quad
D^{(\ovl{F})}(3_{111}^+) = \frac{e^{\ii\frac{5\pi}{12}}}{\sqrt{2}} \bpm 1 & -\ii \\ 1 & \ii \epm  \oplus \frac{e^{-\ii\frac{5\pi}{12}}}{\sqrt{2}} \bpm 1 & 1 \\ \ii & -\ii \epm \ , \nonumber \\
& D^{(\ovl{F})}(m_{1\bar10}) = -\ii \sigma_y \otimes \mathbb{I}_{2\times2} \ .
\end{align}
The $\bar42m$ generators in the $[\ovl{E}_2]_{c_1}$ representation at $\tt_{c_1}$  position are
\begin{align}
\label{eq:3D E2}
D^{(\ovl{E}_2)}(2_z) = -\ii\sigma_z, \qquad
D^{(\ovl{E}_2)}(2_y) = \bpm 0 & e^{-\ii\frac{\pi}{4}} \\ e^{-\ii \frac{3\pi}{4}} & 0 \epm, \qquad
D^{(\ovl{E}_2)}(m_{1\ovl{1}0}) = \ii\sigma_x \ .
\end{align}
One should not confuse the notation of the two-dimensional co-irrep $[\ovl E_2]_{c_1}$ at $c_1$ with the six-dimensional induced representation $[\ovl E_2]_{3c} \equiv [\ovl E_2]_{c_1} \uparrow \bar43m$.
Orbitals on the three $3c$ positions are related by the $3^+_{111}$ operation. 
We choose the gauge where 
\begin{align}
\label{eq:3D C3_L1}
D^{(c)}(3^+_{111}) = 
\bpm  {0}_{2\times2} & {0}_{2 \times 2} & \mathbb{I}_{2\times2} \\ \mathbb{I}_{2\times2} & {0}_{2 \times 2} & {0}_{2 \times 2} \\ {0}_{2 \times 2} & -\mathbb{I}_{2\times2} & {0}_{2 \times 2} \epm \ . 
\end{align}
The representation matrices for TRS within $[\ovl{E}_1]_{1a,1b}$, $[\ovl{F}]_{1b}$, and $[\ovl{E}_2]_{c1}$ are  
\begin{align}
\label{eq:3D T}
D^{(\ovl{E}_1)}(\mclT) = \ii\sigma_y , \qquad
D^{(\ovl{F})}(\mclT) = \bpm 0 & 0 & e^{-\ii3 \pi/4} & 0 \\ 0 & 0 & 0 & e^{-\ii\pi/4} \\ e^{\ii\pi/4} & 0 & 0 & 0 \\ 0 & e^{\ii3 \pi/4} & 0 & 0\epm \ ,\qquad 
D^{(\ovl{E}_2)}(\mclT) = \ii\sigma_y  \ , 
\end{align}
respectively. 
Note that the TRS operator commutes with all crystalline symmetries. 

We first look at the hopping matrix between $(c_1,i)$ and $(w,j)$ orbitals within the same unit cell:  
\begin{equation}
h_{ij}^{(w)}
=
\langle \mathbf{R}, c_1, i \,|\, H_{\rm F} \,|\, \mathbf{R}, w, j \rangle \ ,
\end{equation}
where $w=a,b$ labels the ket site, and $i,j$ label orbitals at the given site.
The vector from bra site to ket site is given by $\Delta\tt_w = \tt_{w} - \tt_{c_1}$. 
All the other nearest-neighbor hopping terms can be generated by applying symmetries to $h_{ij}^{(w)}$.  
The hopping term $h^{(a)}$ is only constrained by the point group  $H_a = \{1, m_{1\bar{1}0}\}$ and $\mclT$ symmetries, and its generic form is 
\begin{align}
h^{(a)} = e^{-\ii\frac{\pi}8}
\begin{pmatrix}
t_1 & e^{\ii\frac{3\pi}4}t_2 \\
\ii t_2 & e^{\ii\frac{\pi}4} t_1
\end{pmatrix} \ , \quad t_1, t_2 \in \mathbb{R} \ .
\end{align}
The hopping term $h^{(b)}$ is constrained by the point group $H_b = \{ 1, C_{2z}, m_{1\bar10} , m_{110}\}$  and $\mclT$ symmetries. 
Its generic form is 
\begin{equation}
h^{(b)} = \pare{\begin{array}{cc|cccc}
   e^{-\ii\frac{\pi}8} t_3  & 0 & e^{\ii\frac{\pi}8} t_4 & 0 & 0 &  e^{-\ii \frac{\pi}8} t_5 \\
    0 &  e^{\ii\frac{\pi}8} t_3 & 0 & e^{-\ii \frac{5\pi}8} t_5 & e^{\ii \frac{5\pi}8} t_4 & 0
\end{array}},\qquad 
    t_{3,4,5} \in \mathbb{R} \ ,
\end{equation}
where the two column blocks correspond to the co-irreps $[\ovl E_1]_{1b}$ and $[\ovl F]_{1b}$, respectively. 
Then we can construct the $S^\dagger(\kk)$ matrix as 
{\footnotesize
\begin{equation}
S^\dagger(\kk) = \begin{pmatrix} 
    \sum_{g\in G_{c_1}/H_a } D^{(c_1)}(g) h^{(a)} D^{(a)\dagger}(g) e^{\ii (g \Delta \tt_{a})\cdot \kk}  & 
    \sum_{g\in G_{c_1}/H_b } D^{(c_1)}(g) h^{(b)} D^{(b)\dagger}(g) e^{\ii (g \Delta \tt_{b})\cdot \kk} \\
    \sum_{g\in G_{c_1}/H_a } D^{(c_1)}(g) h^{(a)} D^{(a)\dagger}(g) D^{(a)\dagger}(3_{111}^+) e^{\ii (C_{3} g \Delta \tt_{a})\cdot \kk} & 
    \sum_{g\in G_{c_1}/H_b } D^{(c_1)}(g) h^{(b)} D^{(b)\dagger}(g) D^{(b)\dagger}(3^+_{111}) e^{\ii (C_{3} g \Delta \tt_{b})\cdot \kk} \\
    -\sum_{g\in G_{c_1}/H_a } D^{(c_1)}(g) h^{(a)} D^{(a)\dagger}(g) D^{(a)\dagger2}(3_{111}^+) e^{\ii (C_{3}^2 g \Delta \tt_{a})\cdot \kk} & 
    -\sum_{g\in G_{c_1}/H_b } D^{(c_1)}(g) h^{(b)} D^{(b)\dagger}(g) D^{(b)\dagger2}(3^+_{111}) e^{\ii (C_{3}^2 g \Delta \tt_{b})\cdot \kk}
\end{pmatrix} .
\end{equation}}
Here $G_{c1}$ is the little group $\bar42m$ of the $c_1$ position, and the cosets $G_{c1}/H_a$ and $G_{c1}/H_b$ can be represented by $\{ 1, 2_{x}, 2_{y}, 2_z\}$ and $\{1, 2_y\}$, respectively.
$D^{(w)}$ is the representation matrix at the Wyckoff position $w$, {\it e.g.,} $D^{(b)}(g) = D^{(\ovl E_1)}(g) \oplus D^{(\ovl F)}(g)$. 
$C_{3}$ is the $3^+_{111}$ rotation matrix.

\begin{figure}[h!]
    \centering
    \includegraphics[width=1.0\linewidth]{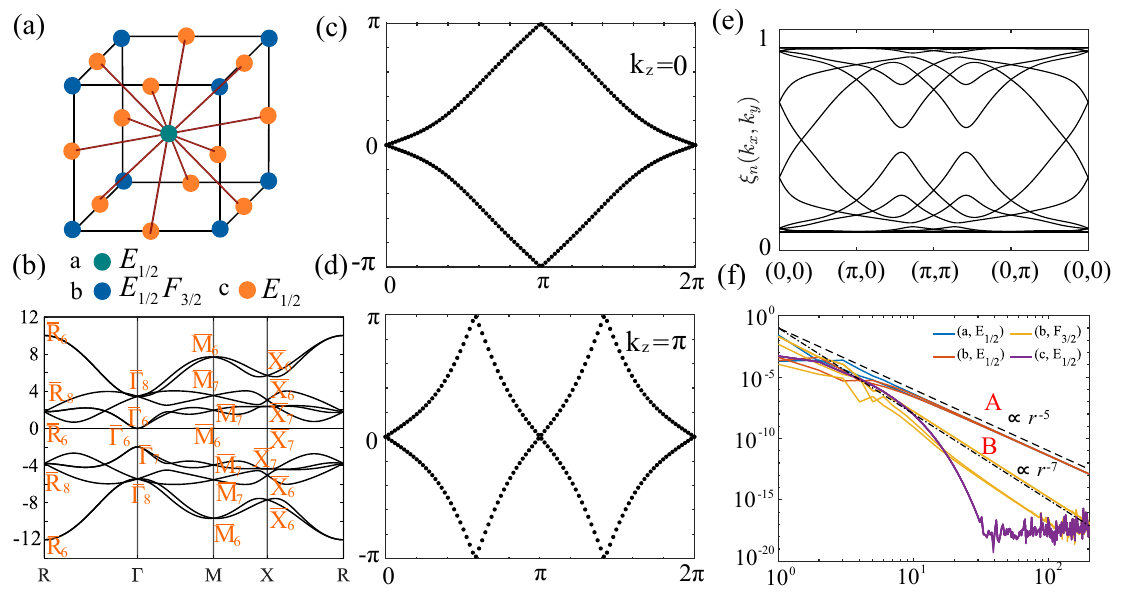}
    \caption{The $\mathbb{Z}_2$ CTFB in double space group $P\bar{4}3m$ with time-reversal symmetry. 
    (a) The lattice structure of space group $P\bar{4}3m$ and the Wyckoff positions $1a$, $1b$ and $3c$ in one unit cell are presented. (b) The band structure is calculated with parameters $t_1 = -0.5$, $t_2 = -0.5$, $t_3=3$, $t_4 = 0.5$, $t_5 = 2$ and $\Delta = 2$. The Wilson loop spectrum at $k_z = 0$ and $k_z = \pi$ planes are shown in $(c)$ and (d), respectively. (e) The entanglement spectrum for subsystem $A=\{ (\RR,\alpha) | 1 \leq \frac{1}{2\pi}\RR \cdot \mathbf{b}_3 \leq 100 \}$ of the $\ket{\Omega}$ state on a $200 \times 200 \times 200$ lattice with periodic boundary condition. The translation symmetry exists along $\mathbf{a}_1$ and $\mathbf{a}_2$ direction, and thus $k_x$, $k_y$ are good quantum numbers. We plot the entanglement spectrum along $(0,0) \rightarrow (\pi,0) \rightarrow (\pi,\pi) \rightarrow (0,\pi)\rightarrow (0,0)$ high symmetry line on $(k_x,k_y)$ plane. 
    (f) Correlation functions in the Fock state $\ket{\Omega}$ that occupies the lowest eight bands. We present the diagonal elements for orbitals $\alpha = (a,E_{1/2})$, $(b,E_{1/2})$, $(b,F_{3/2})$ and $(c,E_{1/2})$ along $\RR = n(\mathbf{a}_1+\mathbf{a}_2+\mathbf{a}_3)$, with $n = 1, \cdots 200$. The system size is $800 \times 800 \times 800$ in basis $\mathbf{a}_1$, $\mathbf{a}_2$ and $\mathbf{a}_3$. The fitting $A$ line scales as $0.1/n^5$ and $B$ scales as $0.1/n^7$. }
    \label{fig:3D Z2 TI sup}
\end{figure}

The full Hamiltonian is 
\begin{align}
H(\kk) &= 
\bpm
0_{8\times8} & S(\kk) \\
S^\dagger(\kk) & -\Delta \cdot \mathbb{I}_{6 \times 6}
\epm \ .
\end{align}
In the main text, the parameters are chosen as $t_1 = -0.5$, $t_2 = -0.5$, $t_3 = 3$, $t_4=1$, $t_5=6$ and $\Delta = 2$ for the band structure and entanglement spectrum shown in Fig. 3 in the main text.  In the following, we adopt $t_1 = -0.5$, $t_2=-0.5$, $t_3=3$, $t_4=0.5$, $t_5 =2$ and $\Delta = 2$ to compute the band structure, Wilson loop spectrum, correlation function, 
and entanglement spectrum presented in \cref{fig:3D Z2 TI sup}. The Wilson loop spectra at $k_z = 0$ and $k_z = \pi$ planes indicate the non-trivial topology. The entanglement shows zig-zag-type topological boundary modes on both top and bottom surfaces. The correlation function exhibits power-law decay $\propto r^{-5}$ for orbitals $[{E}_{1/2}]_{1a}$ and $[{E}_{1/2}]_{1b}$, and $\propto r^{-7}$ for orbitals $[F]_{3/2}$ (the orbital labels follow the ``A.H.'' convention).
 
\subsection{Mirror Topological Crystalline CTFB in 3D}

\begin{table}[th!]
\centering
\begin{tabular}{|c|c|c|c|c|c|c|c|c|}
\hline
Wyckoff & \multicolumn{6}{c|}{$2a \ (m \bar{3})$} & \multicolumn{2}{c|}{$6b \ (mmm)$} \\ \hline
EBR & $\ovl{E}_g (2)$ & $\ovl{E}_u (2)$ & ${}^1\ovl{F}_g (2)$ & ${}^1\ovl{F}_u (2)$ & ${}^2\ovl{F}_g(2)$ & ${}^2\ovl{F}_u (2)$ & $\ovl{E}_g(6)$ & $\ovl{E}_u (6)$ \\ \hline
$\Gamma \ (0,0,0)$ & $\ovl{\Gamma}_5$ & $\ovl{\Gamma}_8$ & $\ovl{\Gamma}_7$ & $\ovl{\Gamma}_{10}$ & $\ovl{\Gamma}_6$ & $\ovl{\Gamma}_9$ & $\ovl{\Gamma}_5 \oplus \ovl{\Gamma}_6 \oplus \ovl{\Gamma}_7$ & $\ovl{\Gamma}_8 \oplus \ovl{\Gamma}_9 \oplus{\Gamma}_{10}$ \\ \hline
$\rH \ (2\pi,2\pi,2\pi)$ & $\ovl{\rH}_5$ & $\ovl{\rH}_8$ 
& $\ovl{\rH}_7$ & $\ovl{\rH}_{10}$ & $\ovl{\rH}_6$ & $\ovl{\rH}_9$ & $\ovl{\rH}_5 \oplus \ovl{\rH}_6 \oplus \ovl{\rH}_7$ & $\ovl{\rH}_8 \oplus \ovl{\rH}_9 \oplus \ovl{\rH}_{10}$ \\ \hline
$\mathrm{P} \ (\pi,\pi,\pi)$ & 
$\ovl{\mathrm{P}}_5$ & $\ovl{\mathrm{P}}_5$ & $\ovl{\mathrm{P}}_7$ & $\ovl{\mathrm{P}}_7$ & $\ovl{\mathrm{P}}_6$ & $\ovl{\mathrm{P}}_6$ & 
$\ovl{\mathrm{P}}_5 \oplus \ovl{\mathrm{P}}_6 \oplus \ovl{\mathrm{P}}_7$ & 
$\ovl{\mathrm{P}}_5 \oplus \ovl{\mathrm{P}}_6 \oplus \ovl{\mathrm{P}}_7$  \\ \hline
$\rN \ (\pi,\pi,0)$ & 
$\ovl{\rN}_3 \oplus \ovl{\rN}_4$ & $\ovl{\rN}_5 \oplus \ovl{\rN}_6$ & 
$\ovl{\rN}_3 \oplus \ovl{\rN}_4$ & 
$\ovl{\rN}_5 \oplus \ovl{\rN}_6$ & 
$\ovl{\rN}_3 \oplus \ovl{\rN}_4$ & 
$\ovl{\rN}_5 \oplus \ovl{\rN}_6$ & 
$\ovl{\rN}_3 \oplus \ovl{\rN}_4 \oplus 2 \ovl{\rN}_5 \oplus 2 \ovl{\rN}_6$ & 
$2\ovl{\rN}_3 \oplus 2\ovl{\rN}_4 \oplus \ovl{\rN}_5 \oplus \ovl{\rN}_6$
\\ \hline
\end{tabular}
\caption{EBRs of the double-valued layer group $Im\bar3$ without time-reversal symmetry.
The first row tabulates $2a$ and $6b$ Wyckoff positions with the corresponding site-symmetry groups in the parentheses. 
The second row tabulates the EBR names, given by the irreps of the corresponding site-symmetry groups. 
Numbers in the parentheses represent the number of bands in the corresponding EBR. 
The following three rows give the co-irreps at high-symmetry momenta of the EBRs.  
}
\label{tab:Im-3-EBR}
\end{table}
\begin{table}[h!]
\centering  
\begin{tabular}{|c|c|c|c|c|c|c|c|c||c|}
\hline
\multicolumn{10}{|c|}{Co-irreps of the double point group $m\bar{3}$}\\
\hline
Co-irreps & $E$ & $6\{2_{001}\}$ & $4\{ 3^+_{111} \}$ & $4\{ 3^-_{111}\}$ & $i$ & $6\{m_{001}\}$ & $4\{-3^+_{111} \}$ & 
$4\{ -3^-_{111} \}$ & 
$\mathbf{k}$-notation \\
\hline
${}^2\ovl{F}_g$ & $2$ & $0$ & $e^{\ii \frac{2\pi}3}$ & 
$-e^{\ii \frac{\pi}3}$ & $2$ & $0$ & $e^{\ii \frac{2 \pi}3}$ & $-e^{\ii \frac{\pi}3}$ & $\ovl{\Gamma}_6$, $\ovl{\rH}_6$\\ 
\hline
${}^1\ovl{F}_g$ & $2$ & $0$ & $-e^{\ii \frac{\pi}3}$ & 
$e^{\ii \frac{2\pi}3}$ & $2$ & $0$ & $-e^{\ii \frac{\pi}3}$
& $e^{\ii \frac{2\pi}3}$ & $\ovl{\Gamma}_7$, $\ovl{\rH}_7$\\
\hline
$\ovl{E}_u$ & $2$ & $0$ & $1$ & $1$ & $-2$ & $0$ & $-1$ & $-1$ & $\ovl{\Gamma}_8, \ovl{\rH}_8$
\\ 
\hline
\end{tabular} \\
\vspace{0.2cm}
\begin{tabular}{|c|c|r|r|r|r|r|r|r|}
\hline
\multicolumn{9}{|c|}{Co-irreps of the double point group $mmm$}\\
\hline
Co-irreps & $E$ & $2\{2_{001} \}$ & $2 \{2_{010} \}$
& $2\{ 2_{100}\}$ & $i$ & $2\{m_{001}\}$ & $2\{m_{010}\}$
& $2\{m_{100} \}$ 
\\ \hline
$\ovl{E}_g$ & $2$ & $0$ & $0$ & $0$ & $2$ & $0$ & $0$ & $0$ 
\\
\hline
$\ovl{E}_u$ & $2$ & $0$ & $0$ & $0$ & $-2$ & $0$ & $0$ & $0$ \\
\hline
\end{tabular} 
\caption{Character tables for co-irreps of double-valued point group $m\bar{3}$ and $mmm$. Notations in the "co-irreps" columns follow the convention of Ref.~\cite{bradley_mathematical_2010}, and are used for representations of site-symmetry groups of the Wyckoff positions $2a$ and $6b$ in real space, respectively. }
\end{table}

We give a 3D example for Mirror Topological Crystalline CTFB in double-valued space group $Im\bar{3}$ without time reversal symmetry. The flat band representation is chosen as,
\begin{align}
B = (\Gamma_8; \quad \rH_8; \quad \rP_5; \quad \rN_3 \oplus \rN_4 )
\end{align}
with a zero-band vector  
\begin{align}
\Delta B = (\Gamma_8 \boxminus \Gamma_5; \quad \rH_8 \boxminus \rH_5; \quad 0; \quad 0) \ .
\end{align}
This band structure $B+\Delta B$ can be realized by the real space construction 
\begin{align}
[{}^2\ovl{F}_g + {}^1\ovl{F}_g + 2\ovl{E}_u]_{2a} \boxminus [E_g]_{6b}  \ .
\end{align}
The primitive vectors for the body-centered lattice are 
\begin{align}
\mathbf{a}_1 = (-\frac12, \frac12,\frac12), \quad \mathbf{a}_2 = (\frac12,-\frac12,\frac12), \quad \mathbf{a}_3 = (\frac12,\frac12,-\frac12) \ .
\end{align}

For clarity, we define the sublattice vectors
\begin{align}
\mathbf{t}_{a} = (0,0,0), \quad
\mathbf{t}_{b_1} = (0,0,\frac{1}2), \quad
\mathbf{t}_{{b}_2} = (\frac12,0,0), \quad 
\mathbf{t}_{{b}_3} = (0,\frac12,0)
\end{align}
for $2a$ and $6b$ Wyckoff positions, respectively. 

The gauge of co-irrep matrices follows the convention of Bilbao Crystallographic Server. The generator matrices involved for site symmetry group $m\bar{3}$ on $2a$ Wyckoff position are 
\begin{align}
D^{(^2\ovl{F}_g)}(2_z) = 
\begin{pmatrix}
-\ii & 0 \\ 0 & \ii
\end{pmatrix}, \quad
D^{(^2\ovl{F}_g)}(2_y) = 
\begin{pmatrix}
0 & -1 \\ 1 & 0
\end{pmatrix}, \quad
D^{(^2\ovl{F}_g)}(i) = 
\begin{pmatrix}
1 & 0 \\ 0 & 1
\end{pmatrix}, \quad
D^{(^2\ovl{F}_g)}(3^+_{111}) = \frac{e^{\ii\frac{5\pi}{12}}}{\sqrt{2}}
\begin{pmatrix}
1 & -\ii \\ 1 & \ii
\end{pmatrix}
\end{align}
\begin{align}
D^{(^1\ovl{F}_g)}(2_z) = 
\begin{pmatrix}
-\ii & 0 \\ 0 & \ii
\end{pmatrix}, \quad
D^{(^1\ovl{F}_g)}(2_y) = 
\begin{pmatrix}
0 & -1 \\ 1 & 0
\end{pmatrix}, \quad
D^{(^1\ovl{F}_g)}(i) = 
\begin{pmatrix}
1 & 0 \\ 0 & 1
\end{pmatrix}, \quad
D^{(^1\ovl{F}_g)}(3^+_{111}) = \frac{e^{-\ii\frac{5\pi}{12}}}{\sqrt{2}}
\begin{pmatrix}
-\ii & -1 \\ -\ii & 1
\end{pmatrix}
\end{align}
\begin{align}
D^{(\ovl{E}_u)}(2_z) = 
\begin{pmatrix}
-\ii & 0 \\ 0 & \ii
\end{pmatrix}, \quad
D^{(\ovl{E}_u)}(2_y) = 
\begin{pmatrix}
0 & -1 \\ 1 & 0
\end{pmatrix}, \quad
D^{(\ovl{E}_u)}(i) = 
\begin{pmatrix}
-1 & 0 \\ 0 & -1
\end{pmatrix}, \quad
D^{(\ovl{E}_u)}(3^+_{111}) = \frac{e^{-\ii\frac{\pi}{4}}}{\sqrt{2}}
\begin{pmatrix}
1 & -\ii \\ 1 & \ii
\end{pmatrix} \ .
\end{align}
The generators of $mmm$ point group in $[\ovl{E}_g]$ representation at $6b$ Wyckoff position are
\begin{align}
D^{(\ovl{E}_g)}(2_z) = 
\begin{pmatrix}
0 & -1 \\ 1 & 0
\end{pmatrix}, \quad
D^{(\ovl{E}_g)}(2_y) = 
\begin{pmatrix}
0 & -\ii \\ -\ii & 0
\end{pmatrix}, \quad
D^{(\ovl{E}_g)}(i) = 
\begin{pmatrix}
1 & 0 \\ 0 & 1
\end{pmatrix} \ .
\end{align}
The $3^+_{111}$ irrep matrix for $[\ovl{E}_g]_{6b}$ induced from $[\ovl{E}_g]_{b1}$ can be chosen as 
\begin{align}
\label{eq:3D C3_L1_}
D^{(b)}(3^+_{111}) = 
\bpm  {0}_{2\times2} & {0}_{2 \times 2} & \mathbb{I}_{2\times2} \\ \mathbb{I}_{2\times2} & {0}_{2 \times 2} & {0}_{2 \times 2} \\ {0}_{2 \times 2} & -\mathbb{I}_{2\times2} & {0}_{2 \times 2} \epm \ . 
\end{align}

The hopping matrix between $(b_1,i)$ and $(a,j)$ orbitals within the same unit cell is:
\begin{align}
h_{ij}^{(a)} = \langle \RR,b_1,i| H_F|\RR,a,j \rangle
\end{align}
with $i,j$ denotes orbitals at $b_1$ and $a$ site, respectively.

The hopping term between $b_1$ and $a$ site is constrained by the point group $H_a = \{ 1,2_z,m_{010},m_{100} \}$. It has the following form
\begin{equation}
h^{(a)} = \pare{\begin{array}{cc|cc|cc|cc}
t_1 & t_1 & t_2 & t_2 & t_3 & -t_3 & t_4 & -t_4 \\
\ii t_1 & -\ii t_1 & \ii t_2 & -\ii t_2 & \ii t_3 & \ii t_3 & \ii t_4 & \ii t_4
\end{array}},\qquad t_{1,2,3,4} \in \mathbb{C}
\end{equation}
where the four column blocks corresponds to the co-irreps $[^2\ovl{F}_g]$, $[{}^1\ovl{F}_g]$, $[\ovl{E}_u]$ and $[\ovl{E}_u]$, respectively. The $S_N^\dagger(\kk)$ matrix is as 
\begin{align}
S_N^\dagger(\kk) &= 
\begin{pmatrix}
\sum_{g \in G_{b_1}/H_a} D^{(b_1)}(g)h^{(a)}D^{(a) \dagger}(g) e^{\ii(g \Delta \mathbf{t}_a) \cdot \kk}  \\
\sum_{g \in G_{b_1}/H_a} D^{(b_1)}(g)h^{(a)}D^{(a) \dagger}(g) D^{(a)^\dagger} (3^+_{111}) e^{\ii(C_3 g \Delta \mathbf{t}_a) \cdot \kk} \\
-\sum_{g \in G_{b_1}/H_a} D^{(b_1)}(g) h^{(a)} D^{(a) \dagger}(g) D^{(a) \dagger 2}(3^+_{111}) e^{\ii(C_3^2 g \Delta \mathbf{t}_a) \cdot \kk} 
\end{pmatrix}
\end{align}
with $\Delta \tt_a = \tt_a - \tt_{b_1}$, the little group $G_{b_1}=mmm$. The cosets is $G_{b_1} / H_a = \{ 1, 2_y\}$. 

\begin{figure}[h!]
    \centering
    \includegraphics[width=1.05\linewidth]{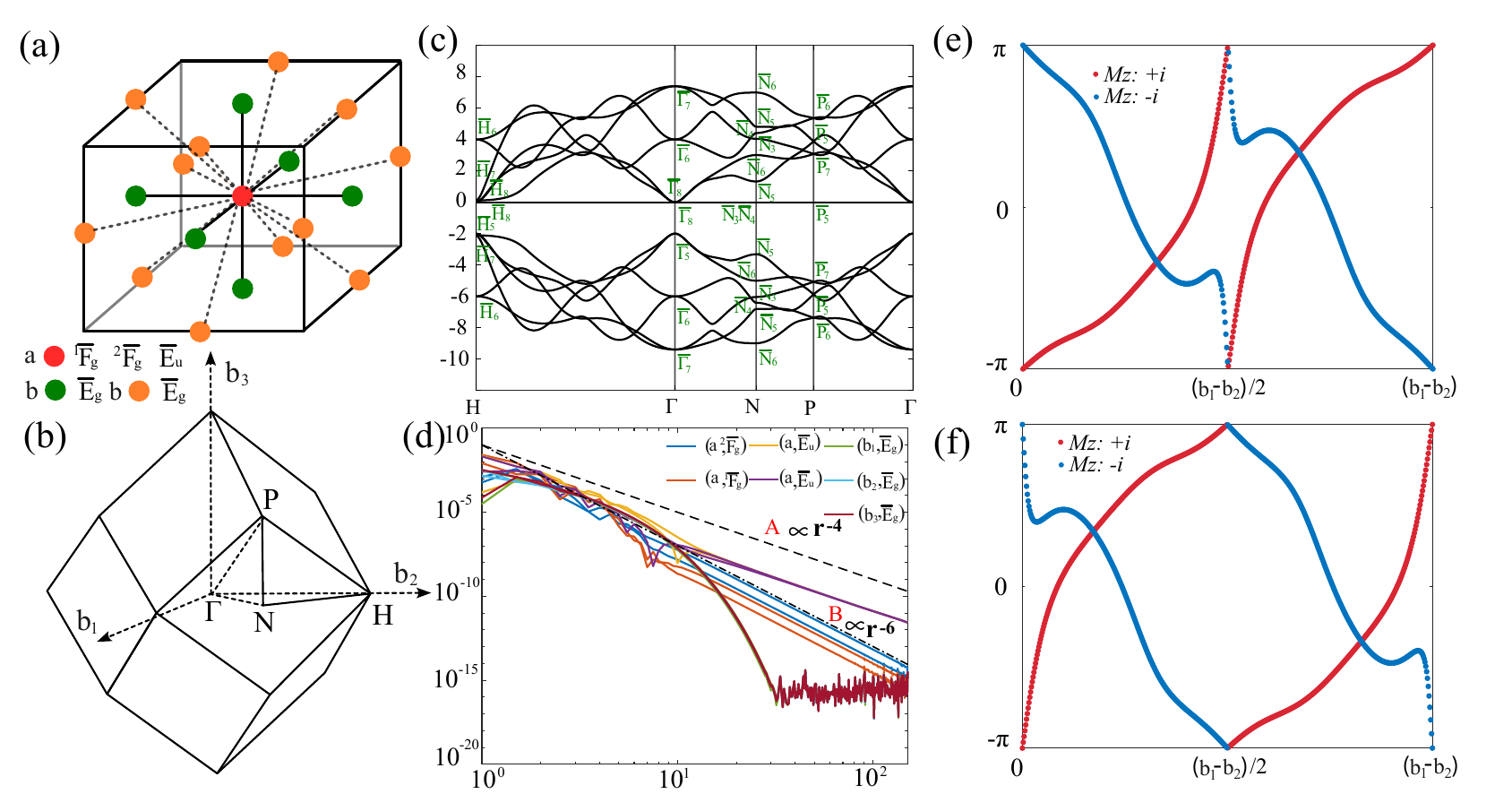}
    \caption{The Mirror Crystalline CTFB in double group $Im\bar{3}$ without time-reversal symmetry. (a) The lattice structure of space group $Im\bar{3}$ and the Wyckoff positions $2a$ and $6b$ in one unit cell, where green circle denotes the nearest and next-nearest $6b$ Wyckoff positions, respectively. (b) The Brillouin zone of $Im\bar{3}$. (c) The band structure and irreps at high symmetry points. (d) The correlation functions in the Fock state $\ket{\Omega}$ that occupies the lowest eight bands. We present the diagonal elements for orbitals $\alpha = (a,{}^2\bar{F}_g), (a,{}^1\bar{F}_g), (a,\bar{E}_u), (b,\bar{E}_g)$ along $\RR=n(\mathbf{a}_1+\mathbf{a}_2+\mathbf{a}_3)$ direction, with $n=1,2\cdots,150$. The system size is $800 \times 800 \times 800$ in basis $\mathbf{a}_1$, $\mathbf{a}_2$ and $\mathbf{a}_3$. The fitting line A scales as $0.1/n^4$ and B scales as $0.1/n^6$. The 2D Wilson loop spectra calculated along $\mathbf{b}_1-\mathbf{b}_2$ and integrated along $\mathbf{b}_3$, evaluated at (e) $k_z=0$ and (f) $\frac{1}{2}(\mathbf{b}_1+\mathbf{b}_2 - \mathbf{b}_3)$ plane, respectively. } 
    \label{fig:3D MTCI sup}
\end{figure} 

To remove the accidental touching between the flat band and dispersive bands, we also add the next nearest hopping terms into the model. One of the next-nearest hopping term between $(b_1,i)$ and $(a,j)$ orbitals is
\begin{align}
\td{h}_{ij}^{(a)} = \langle \RR, b_1, i| H_F | \RR+\mathbf{a}_1, a, j \rangle \ ,
\end{align}
with the hopping vector from bra to ket is $\td{\Delta} \tt_a = \tt_a - \tt_{b_1} + \mathbf{a}_1$. 

The hopping term between $b_1$ site at $\RR$ unit cell and $a$ site at $\RR + \mathbf{a}_1$ unit cell is constrained by the point group $\td{H}_{a} = \{ 1, m_{001}\}$. The general matrix is as
\begin{equation}
\td{h}^{(a)} = \pare{\begin{array}{cc|cc|cc|cc}
t_1' & t_2' & t_3' & t_4' & t_5' & t_6' & t_7' & t_8'\\
\ii t_1' & -\ii t_2' & \ii t_3' & -\ii t_4' & -\ii t_5' & \ii t_6' & -\ii t_7' & \ii t_8'
\end{array}},\qquad t'_{1,2,3,4,5,6, 7, 8} \in \mathbb{C} \ .
\end{equation}
The ${S}^\dagger_{N.N}(\kk)$ matrix is 
\begin{align}
{S}_{N.N}^\dagger(\kk) &= 
\begin{pmatrix}
\sum_{g \in G_{b_1}/\td{H}_a} D^{(b_1)}(g)\td{h}^{(a)}D^{(a) \dagger}(g) e^{\ii(g \td{\Delta} \mathbf{t}_a) \cdot \kk}  \\
\sum_{g \in G_{b_1}/\td{H}_a} D^{(b_1)}(g)\td{h}^{(a)}D^{(a) \dagger}(g) D^{(a)^\dagger} (3^+_{111}) e^{\ii(C_3 g \td{\Delta} \mathbf{t}_a) \cdot \kk} \\
-\sum_{g \in G_{b_1}/\td{H}_a} D^{(b_1)}(g) \td{h}^{(a)} D^{(a) \dagger}(g) D^{(a) \dagger 2}(3^+_{111}) e^{\ii(C_3^2 g \td{\Delta} \mathbf{t}_a) \cdot \kk} 
\end{pmatrix} \ ,
\end{align}
where the coset is $G_{b_1}/\td{H}_a = \{1,2_x, 2_y, 2_z \}$.

The hopping matrix is $S^\dagger(\kk) = S^\dagger_N(\kk) + S^\dagger_{N.N}(\kk)$, and the complete Hamiltonian is
\begin{align}
H(\kk) &= \begin{pmatrix}
0_{8 \times 8} & S(\kk) \\
S^\dagger(\kk) & -\Delta \mathbb{I}_{6 \times 6}
\end{pmatrix} \ .
\end{align}

The band structure, correlation function and Wilson loop spectrum are plotted in \cref{fig:3D MTCI sup} with parameters $t_1=0.2+0.4 \ii$, $t_2=-0.8+0.4 \ii$, $t_3=0.2+0.3\ii$, $t_4=-0.1-0.9\ii$, $t_1'=-0.9-0.2\ii$, $t_2'=0.1+0.6\ii$, $t_3'=-0,4+0.6\ii$, $t_4'=-0.3-0.2\ii$, $t_5'=0.1+0.4\ii$, $t_6'=-0.8-0.8\ii$, $t_7'=0.3+0.9\ii$, $t_8'=0.2-0.5 \ii$ and $\Delta = 2$. 
The correlation function exhibits power-law decay $r^{-4}$ for orbitals $[^2\ovl{F}_g]$, $[^1\ovl{F}_g]$ and $[\ovl{E}_u]$, and $\propto r^{-6}$ for orbitals $[\ovl{E}_g]_{6b}$.
To diagnose the topology, we compute the 2D Wilson loop spectra on two mirror-invariant planes: $k_z=0$ and $\frac{1}{2}(\mathbf{b}_1+\mathbf{b}_2-\mathbf{b}_3)=2\pi(0,0,1)$ [\cref{fig:3D MTCI sup}(e) and (f)]. In each plane, the spectrum is resolved by the mirror eigenvalues $\pm i$. The opposite chiralities of the Wilson loop flow between the $+i$ and $-i$ sectors directly manifest a mirror Chern number $C_M = \frac{1}{2}(C_+ - C_-) = 1$ on both mirror-invariant planes.

\clearpage
\section{Quantum geometry and optimization}

The (complex) quantum geometric tensor (QGT) for a multi-band system of $d$ bands, with Bloch function $U(\kk)$, is defined as
\begin{align}
    \frak{g}_{ij}(\kk) = \Big(\partial_{i} U^\dagger(\kk) \Big)  \Big( \mathbb{I} - P(\kk) \Big) \Big( \partial_{j} U(\kk) \Big) = U^\dagger(\kk) \Big( \partial_i P(\kk) \Big) \Big( \partial_j P(\kk) \Big) U(\kk)
\end{align}
where $i,j$ are spatial directions in momentum space, $P(\kk) = U(\kk) U^\dagger(\kk)$ and $U^\dagger(\kk) U(\kk) = \mathbb{I}_{d\times d}$. 
Tracing over the band indices, one obtains the Fubini-Study metric (FSM) $g_{ij}(\kk)$ and the Berry curvature $\Omega(\kk)$
\begin{align}  \label{eq:def_chiij}
    &\chi_{ij}(\kk) = \Tr\left[ P(\kk) \Big( \partial_i P(\kk) \Big) \Big( \partial_j P(\kk) \Big) \right] 
    & g_{ij}(\kk) = \Re \chi_{ij}(\kk) \qquad \Omega_{ij}(\kk) = -2\Im \chi_{ij}(\kk)
\end{align}
Note that the FSM can be regarded as a metric, in the sense that it measures the distance of projectors
\begin{align} \label{eq:gij_as_metric}
    \frac{1}{2} || P(\kk+\dd\kk) - P(\kk) ||^2 = \sum_{ij} g_{ij} \dd\kk_i \dd \kk_j
\end{align}

The QGT depends on the $1$-st order derivatives of the projector, $\partial_{i} P(\kk)$, which is not dictated to be continuous by our CTFB construction. 
In this section, we will demonstrate the following conclusions. 
(1) $\partial_{i} P(\kk)$ (hence QGT) is not divergent near the touching point of a CTFB. 
(2) If $\mathcal{O}(k^2)$ terms are forbidden in $S(\kk)$ matrix by symmetry, then $\partial_{i} P(\kk)$ (hence QGT) will also be continuous. 
(3) For Chern CTFBs that we focused on, the Berry curvature $\Omega(\kk)$, and the trace and determinant of FSM $g(\kk)$, will also be continuous, even if $\partial_i P(\kk)$ is not continuous.

\subsection{Continuity of quantum geometric tensor}

To begin with, we remark that, the total projector to all the low-energy states near the touching points (including FB and the dispersive bands that touch with FB) should not contain any singularity, as the remote bands are all gapped. 
Therefore, it suffices to drop the gapped remote bands in analysis.  

We focus on the Chern CTFBs in 2D first, and discuss the generalization to other cases later. 

For a Chern CTFB, suppose the touching point has the lowest symmetry that can still guarantee the ``isotropy'', the $C_{3}$ rotation symmetry. 
Then to the $\mathcal{O}(k^2)$ order, there is in general
\begin{align}
    S^\dagger(\kk) = \begin{pmatrix}
        \alpha k_+ + \beta_1 k_-^2 & \beta_2 k_-^2 \\
    \end{pmatrix} + \mathcal{O}(k^3)
\end{align}
Note we can always carry out a $2 \times 2$ $\kk$-independent unitary transformation to the columns of $S^\dagger(\kk)$ to bring the $\mathcal{O}(k)$ dependence to the above form. 
Also, note that if the touching point has higher symmetries, $C_4$ or $C_6$, then the $\mathcal{O}(k^2)$ order will be forbidden, $\beta_1= \beta_2=0$. 
For such $S^\dagger(\kk)$, 
\begin{align}
    \mathcal{Q}(\kk) &= |\alpha|^2 k^2 + \Big(\alpha \beta_1^* k_+^3  + c.c. \Big) + \mathcal{O}(k^4) \\\nonumber
    \mathcal{Q}^{-1}(\kk) &= \frac{1}{|\alpha|^2 k^2} \left(  1 - \frac{\Big(\alpha \beta_1^* k_+^3  + c.c. \Big) }{|\alpha|^2 k^2} + \mathcal{O}(k^2) \right) 
\end{align}
where $c.c.$ denotes complex conjugate. The projector to the dispersive band reads
\begin{align}   
    \mathbb{I}_{2 \times 2} - P(\kk) &= S(\kk) \mathcal{Q}^{-1}(\kk) S^\dagger(\kk) = \mathcal{Q}^{-1}(\kk) \left[\begin{pmatrix}
        |\alpha|^2 k^2 + \Big(\alpha \beta_1^* k_+^3  + c.c. \Big)  & c.c. \\
       \alpha \beta_2^* k_+^3 & 0 \\
    \end{pmatrix} + \mathcal{O}(k^4) \right] \\\nonumber
    &= \begin{pmatrix}
         1 & c.c. \\
         \frac{\beta_2^*}{\alpha^*} |\kk| e^{\ii 3 \phi_{\kk}} & 0 \\
    \end{pmatrix} + \mathcal{O}(k^2)
\end{align}
hence the projector to FB is
\begin{align}  \label{eq:Pk_1stderivative}
    P(\kk) =  \begin{pmatrix}
         0 & c.c. \\
         -\frac{\beta_2^*}{\alpha^*} |\kk| e^{\ii 3 \phi_{\kk}} & 1 \\
    \end{pmatrix} + \mathcal{O}(k^2)
\end{align}
The 1st-order derivative to the $\mathcal{O}(k^2)$ terms in the projector will vanish at the touching point as $\kk \to 0$. 
It it then obvious that $[P(\kk)]_{2,1}$ contains a discontinuity at the $1$-st order derivative if and only if $\beta_2 \not= 0$ (without fine-tuning, this is equivalent to), 
\begin{align}
    \partial_{k_x} P(\kk) 
    = -\frac{\beta_2^*}{\alpha^*} \begin{pmatrix}
        0 & c.c. \\
        f_x(\kk) & 0 \\
    \end{pmatrix} + \mathcal{O}(k) \qquad  f_x(\kk) = e^{\ii 3 \phi_{\kk}} (\cos\phi_{\kk} - 3\ii \sin \phi_{\kk}) \\\nonumber
    \partial_{k_y} P(\kk) 
    = -\frac{\beta_2^*}{\alpha^*}\begin{pmatrix}
        0 & c.c. \\
        f_y(\kk) & 0 \\
    \end{pmatrix} + \mathcal{O}(k)  \qquad f_y(\kk)= e^{\ii 3 \phi_{\kk}} (\sin\phi_{\kk} + 3\ii \cos \phi_{\kk}) 
\end{align}
Note that $\partial_i P(\kk)$ is indeed finite. 
Via \cref{eq:def_chiij}, we further compute
\begin{align}   \label{eq:geometry_p3_touch}
    \chi_{ij}(\kk) = f_i(\kk) f^*_j(\kk) + \mathcal{O}(k) \quad \overset{\rm matrix~form}{\longrightarrow}  \quad  \chi = \frac{|\beta_2|^2}{|\alpha|^2}
    \begin{pmatrix}
        5 - 4\cos2\phi_{\kk} & c.c.  \\
        -4 \sin2\phi_{\kk} + 3\ii  & 5 + 4\cos 2\phi_\kk  \\
    \end{pmatrix}  + \mathcal{O}(k)
\end{align}
Therefore, $\Tr g_{ij}(\kk) \sim 10  \frac{|\beta_2|^2}{|\alpha|^2}$, $\det g_{ij}(\kk) \sim 0$, $\Omega(\kk) \sim 6  \frac{|\beta_2|^2}{|\alpha|^2}$, which are all continuous around the touching point. 
This explicitly justifies the previous proposals for Chern CTFB. 

\begin{figure}[bt]
    \centering
    \includegraphics[width=0.9\linewidth]{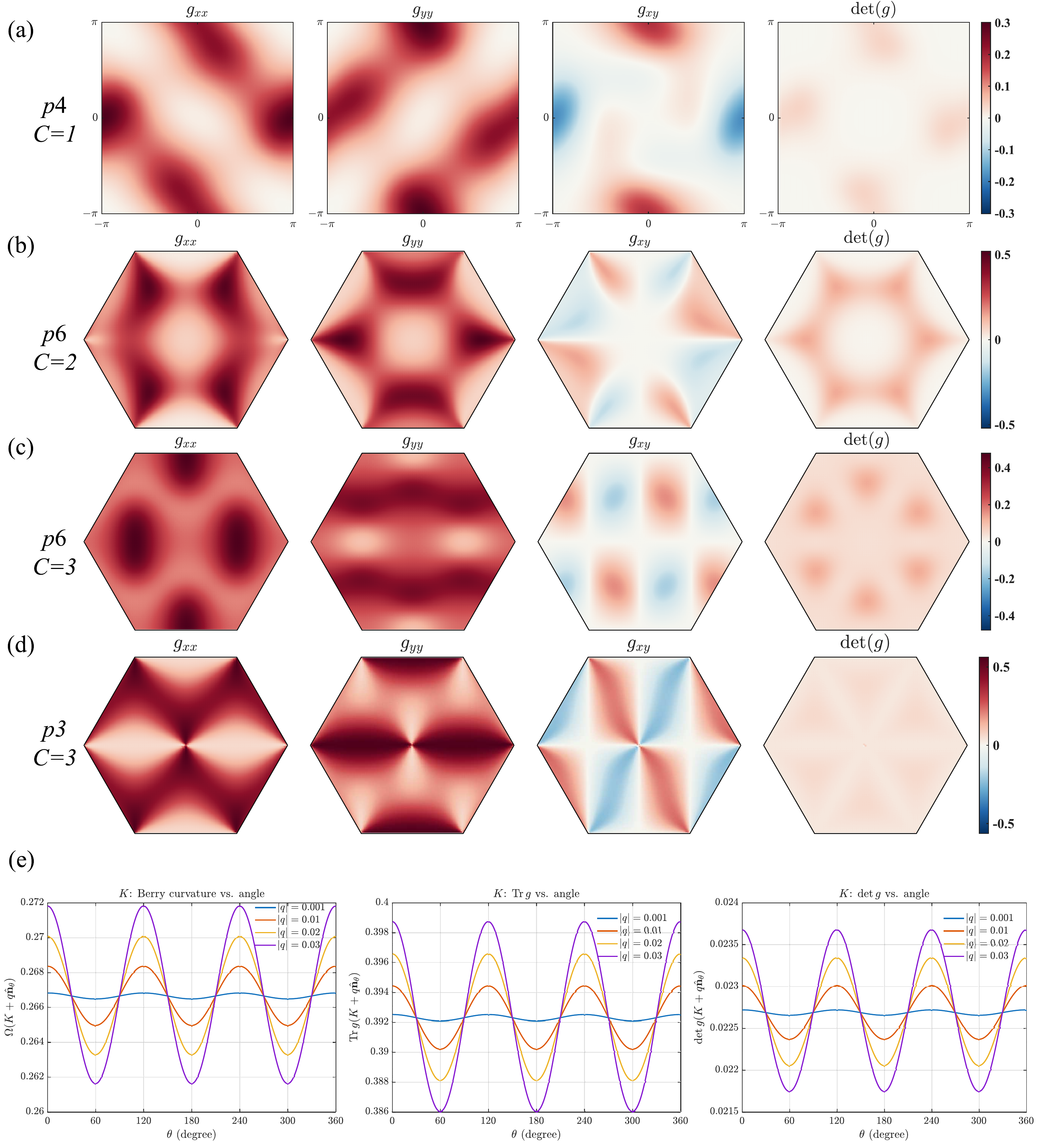}
    \caption{Components and determinant of the Fubini-Study metric in the Chern CTFB models we have discussed. 
    (a) $C=1$ in $p4$ group (\cref{app:Chern}). 
    (b) $C=2$ in $p6$ group (\cref{sec:C=2}). 
    (c) $C=3$ in $p6$ group (\cref{sec:C=3}).
    (d) $C=3$ in $p3$ group (\cref{app:C3inp3}).
    (e) Zoom-in of $\Omega(\kk), \Tr g(\kk), \det g(\kk)$ around the $C_3$ touching point $\rK$ in the $p6$, $C=2$ model, where individual FSM components can be discontinuous. 
    Each colored line is calculated along a small circle around $\rK$, illustrating the angular dependence of these quantities. From purple to blue, the circle radius diminishes, and the displayed quantities approach the same limit along different directions, which justifies the continuity. }
    \label{fig:geometry_components}
\end{figure}

Numerically, we illustrate the symmetry-allowed discontinuity of FSM components at $C_3$ touching points in \cref{fig:geometry_components}, including $\pm \rK$ in $p6$ models, and $\Gamma, \rK, \rKA$ in $p3$ models. 
(Note that the actual discontinuity for the $p6,C=3$ model is very weak for the chosen set of parameters. ) 
Nevertheless, if we zoom in to the vicinity of such $C_3$ touching points (\cref{fig:geometry_components}(e), we can see that $\Omega$, $\Tr g$ and $\det g$ are indeed continuous. 
On the other hand, at the $C_4$ and $C_6$ invariant touching points, including $\rM$ in the $p4$ model, and $\Gamma$ in the $p6,C=3$ model, symmetry guarantees that no discontinuity exists even for the individual FSM components.
We also remark that, besides the analytical calculations that focus on the singularities \cref{eq:geometry_p3_touch}, the numerical results in \cref{fig:geometry_components} also contain a smooth modification to the quantum geometric properties from the remote bands.

For general CTFBs, the projector still takes a similar form to \cref{eq:Pk_1stderivative}, 
\begin{align}
    P(\kk) = P_{\rm const} + P_{\mathcal{O}(k)}(\kk) + \mathcal{O}(k^2)
\end{align}
In particular, the continuity at the constant order is symmetry-guaranteed by construction. 
All the $\kk$-dependence originates from $S(\kk)$, and takes the form as powers of $|\kk| \mathcal{Y}(\theta_{\kk}, \phi_{\kk})$, where $\mathcal{Y}(\theta_{\kk}, \phi_{\kk})$ are finite functions with regular angular dependence. 
At the $\mathcal{O}(k)$ order in $P(\kk)$, $1$-st order derivative to $|\kk| \mathcal{Y}(\theta_{\kk}, \phi_{\kk})$ can thus be discontinuous in general, but must still be finite. 
As these $\mathcal{O}(k)$ terms in $P(\kk)$ are also contributed by the $\mathcal{O}(k^2)$ couplings in $S(\kk)$, they will not exist as long as the symmetries at the touching points are high enough to forbid such couplings. 
For the $\mathcal{O}(k^2)$ orders in $P(\kk)$, the $1$-st order derivative vanishes. 
Therefore, the QGT for generic CTFB will be continuous if symmetries forbid $\mathcal{O}(k^2)$ terms in $S(\kk)$, and finite albeit discontinuous otherwise. 

\subsection{Optimization}

The multiple hopping parameters in our CTFB models can be varied to tune the local quantum geometry without affecting the global topology. 
We consider two optimization schemes:
minimization of the local trace-condition deviation and minimization of the Berry-curvature fluctuation.
\\
\begin{enumerate}
    \item 
\textbf{Local trace condition.}
The functional to be minimized is chosen as
\begin{align}
 \mathcal{L}_{\mathrm{tr}}(\mathbf{t})
    =
    \frac{
    \displaystyle
    \int_{\mathrm{BZ}}
    \left|
    \operatorname{Tr}[g(\mathbf{k};\mathbf{t})]
    -
    |\Omega(\mathbf{k};\mathbf{t})|
    \right| d^2k
    }{
    \displaystyle
    \int_{\mathrm{BZ}}
    |\Omega(\mathbf{k};\mathbf{t})|\,d^2k
    }.
\end{align}
where $\mathbf{t}$ represents all hopping parameters. 

\item
\textbf{Berry curvature uniformity.}
For a Chern CTFB with fixed Chern number $C$, the functional to be minimized is 
\begin{equation}
    \delta_\Omega(\mathbf{t})
    = 
    \sqrt{\int_{\mathrm{BZ}}
    \left(
    \Omega(\mathbf{k};\mathbf{t})-\frac{2\pi C }{A_{\rm BZ}}
    \right)^2 d^2k }
\end{equation}
where $\frac{2\pi C}{A_{\rm BZ}}$ is the average Berry curvature. 
\end{enumerate}
After optimization, the hopping parameters for the Chern CTFBs displayed in the main text Fig. 2 are as following. 
(a) $p4,C=1$ model: $t_1 = 2.3$, $t_2 = 1.2$, $t_3 = 1.1$, $t_4 = 0.7$, while controlling $\Delta = 5$;
(b) $p6,C=2$ model: $t_1 = 1$, $t_2 = 0.362$, $t_3 = 1.055$, while controlling $\Delta = 5$;
(c) $p6,C=3$ model: $t_1  = 1$, $t_2  = 3.738$, $t_3  = 3.869$, $t_4  = 0.001$, $t_1' = 2.918  \ii$, $t_2' = -0.781 \ii$, $t_3' = -0.004 \ii$, $t_4' = 3.442 \ii$, while controlling $\Delta  = 10$. 
For the $p3,C=3$ model, $t_{1,1}=1$, $t_{2,1}=1.27-2.83\ii$, $t_{1,2}=3+0.77\ii$, $t_{2,2}=-0.62+0.78\ii$, while controlling $\Delta  = 10$. For the $p3,C=6$ model, $t^+=[1,1;-\ii,\ii;-\ii,\ii], t^-=[-1,1;-\ii,-\ii;\ii,\ii]$.

\clearpage

\section{Summary tables of CTFBs}
\label{sec:table-CTFB}
\label{lab:table guide}

We summarize the constructions of symmetry-indicated CTFBs in \cref{tab:2D_Chern_band, tab:DoubleTR_2D_summary, tab:DoubleTR_summary, tab:DoubleNoTR_summary}. 
We consider single-valued wallpaper groups for Chern bands (\cref{tab:2D_Chern_band}), double-valued layer groups with TRS for CTFBs with strong $\mathbb{Z}_2$ index and mirror-Chern numbers (\cref{tab:DoubleTR_2D_summary}), double-valued 3D space groups with (\cref{ tab:DoubleTR_summary}) and without TRS (\cref{tab:DoubleNoTR_summary}) for various 3D topological invariants including strong $\mathbb{Z}_2$ index, mirror-Chern numbers, and higher-order topological indices, {\it etc}.  

For every space group and each nontrivial SI, we tabulate the following information:
\begin{enumerate}
\item The entry ``Number of $\mclB$'' counts distinct CTFBs ($\mclB$) with dimension up to the cutoff $d_{\mclB}$, regardless of whether they are accompanied by a $\Delta\mclB$.
\item The entry ``Number of $\mclB + \Delta\mclB$'' counts distinct  $\mclB+\Delta\mclB$ with $\mclB$ subject to the dimension cutoff $d_{\mclB}$ and $\Delta \mclB$ subject to the dimension cutoff  $D_{\Delta \mclB}$, as defined in the ``Bipartite Crystalline Lattice'' part of \cref{sec: Bipartite lattice}. 
The combination $\mclB + \Delta\mclB$ satisfies the compatibility relations, carries trivial SI, and obeys the symmetry-indicated continuity conditions near all touching points derived in \cref{app:algr-touching}.
Since a single $\mclB$ may correspond to multiple $\Delta\mclB$, number of $\mclB+\Delta\mclB$ is usually much larger than the number of $\mclB$. 
\item The entry ``Minimal band dimension $d$'' gives the dimension of the minimal models realizing the CTFBs of the desired SI, minimized over all realizations $\mathcal{BR}_L \boxminus \mathcal{BR}_{\td L}$ of a given $\mclB + \Delta \mclB$ and over all $\mclB + \Delta \mclB$ of the desired SI.
\end{enumerate}
Only space groups with nontrivial SI and isotropic high symmetry momenta are considered. 
The convention of SI follows Refs.~\cite{song2018quantitative,elcoro_magnetic_2021}.

In Table 1 of the main text we have classified all the CTFBs into ten categories: 
\begin{enumerate}
\item The ``$|C|=1$'' category in 2D systems without TRS comprises CTFBs with SIs $z_{3R}=1,2$ mod 3 of wallpaper group $p3$,  $z_{4R}=1,3$ mod 4 of wallpaper group $p4$, and $z_{6R}=1,5$ mod 6 of wallpaper group $p6$. 
These SIs are Chern numbers module 3, 4, 6, respectively. 
Due to the modulo operation, a given SI can in principle correspond to multiple distinct Chern numbers. However, as discussed in \cref{subsec:higher-Chern}, in the absence of additional fine-tuning, a symmetry-based CTFB realizes the Chern number with the smallest absolute value because only the minimal winding is guaranteed. 
Thus, the above values of SIs are interpreted as $C=1,-1$.
\item  For the same reason, the ``$|C|=2$'' category in 2D systems without TRS comprises CTFBs with SIs  $z_{4R}=2$ mod 4 of wallpaper group $p4$, and $z_{6R}=2,4$ mod 6 of wallpaper group $p6$.
\item  For the same reason, the ``$|C|=3$'' category in 2D systems without TRS  comprises CTFBs with the SI $z_{6R}=3$ mod 6 of wallpaper group $p6$.
\item The ``Strong $\mathbb{Z}_2$'' category in 2D systems with TRS comprises CTFBs where the Fu-Kane index $\delta_{\rm t}=1$ or the mirror-Chern number is odd, {\it i.e.}, $C_{\rm m}=1,2$ mod 3, $C_{\rm m}=1,3$ mod 4, $C_{\rm m}=1,3,5$ mod 6.  
In particular, for the case of $C_{\rm m} \bmod 3$, according to discussions in \cref{subsec:higher-Chern}, without fine tuning, our construction typically realizes the smallest absolute value of mirror-Chern number. 
Thus we interpret $C_{\rm m}=1,2$ mod 3 as $C_{\rm m}=1,-1$, respectively. 
\item The ``mirror-Chern'' category in 2D systems with TRS comprises CTFBs with even mirror-Chern numbers, {\it i.e.}, $C_{\rm m}=2$ mod 4, $C_{\rm m}=2,4$ mod 6. 
\item The ``Axion'' category in 3D systems without TRS comprises CTFBs where $\eta_{2I}'=1$ or one of $z_2$, $z_4$, $z_8$ is odd (if present), regardless of values of other SIs. In all the tabulated space groups, the 3D Chern numbers are forbidden, thus these SIs always correspond to well-defined half-quantized magnetoelectric response. 
\item The ``Crystalline'' category in 3D systems without TRS comprises all CTFBs with nontrivial SIs that do not fall into the ``Axion'' category. It includes 3D mirror-Chern states ({\it e.g.,} $z_{2w}=1$, $z_{\rm 4m,\pi}=1,2,3$) and higher order topological states ({\it e.g.,} $z_8=4$). 
\item The ``Strong $\mathbb{Z}_2$'' category in 3D systems with TRS comprises CTFBs where one of $z_2$, $z_4$, $z_8$ is odd (if present), regardless of values of the other SIs.
\item The ``Weak $\mathbb{Z}_2$'' category in 3D systems with TRS comprises CTFBs with $z_{2w}=1$ or $z_{\rm 4m,\pi}=1,3$ that do not fall into the ``Strong $\mathbb{Z}_2$'' category. In all space groups under consideration, the three weak indices are identical, {\it i.e.}, $z_{2w,1}=z_{2w,2}=z_{2w,3}$, and we hence denote them as $z_{2w}$. 
\item The ``Crystalline'' category in 3D systems with TRS comprises CTFBs with nontrivial SIs that do not fall into the ``Strong $\mathbb{Z}_2$'' or ``Weak $\mathbb{Z}_2$'' categories. It includes 3D mirror-Chern states ({\it e.g.,} $z_{\rm 4m,\pi}=2$) and higher order topological states ({\it e.g.,} $z_8=4$). 
\end{enumerate}
For 3D systems without TRS, odd values of SIs $\eta_{4I}$ (defined by inversion eigenvalues) and $\delta_{2S}$ (defined by $S_4$ eigenvalues) correspond to Weyl semimetals \cite{elcoro_magnetic_2021}. 
If they present, the corresponding flat bands would exhibit singular touching points with dispersive bands and hence are by definition singular FBs. 
However, in all the tabulated space groups these two SIs are enforced to be even. 
First, only $\eta_{2I}'$ rather than $\eta_{4I}$ appears in tabulated centrosymmetric space groups, meaning $\eta_{4I}$ must be even because the integer-valued $\eta_{2I}'$ is defined as $\eta_{4I}/2$. 
Second, in space groups \#215--\#220, $\delta_{2S}$ is well-defined but always trivial. 
If $\delta_{2S}$ was 1, there would be $4n+2$ ($n\in\mathbb{Z}_{\ge 0}$) Weyl points carrying the same chirality between the $k_z=0$ and $k_z=\pi$ planes. 
Contradictorily, vertical mirror or glide symmetries, which always present in \#215-\#220, enforce a vanishing net chirality between the $k_z=0$ and $k_z=\pi$ planes.

\begin{table}[htp] \footnotesize
    \centering
    \begin{tabular}{l|l|r|r|r}
    \hline
        SG ID & SI & Number of $\mclB$ within $d_{\mclB} \le 1$ & Number of $\mclB + \Delta\mclB$ within $D_{\Delta \mclB} \le 6$ & Minimal band dimension $d$ \\
    \hline
        $p3$ & ${z_{3R}}$ & $\begin{array}{rrr} & 9 & 9 \end{array}$ & $\begin{array}{rr} 2997 & 2997
        \end{array}$ & 
        $\begin{array}{rrr}
        & 3 & 3
        \end{array}$\\
    \hline
        $p4$ & ${z_{4R} }$ & $\begin{array}{rrrr} & 8 & 8 & 8 \end{array}$ & $\begin{array}{rrrr} 
        192 & 208 & 192 \end{array}$ & 
        $\begin{array}{rrrr}
        & 5 & 5 & 5
        \end{array}$\\
    \hline
        $p6$ & ${z_{6R}}$ & $\begin{array}{rrrrrr} & 6 & 6 & 6 & 6 & 6 \end{array}$ & 
        $\begin{array}{rrrrrr} 120 & 78 & 120 & 78 & 120
        \end{array}$ &
        $\begin{array}{rrrrrr}
        & 7 & 5 & 7 & 5 & 7
        \end{array}$\\
    \hline
    \end{tabular}
    \caption{Summary of CTFB constructions in wallpaper groups ${p3}$, ${p4}$ and ${p6}$. 
    The SI $z_{nR}$ equals $C$ mod $n$ for $n=3,4,6$. 
    In the last three columns, we only list results corresponding to non-trivial SIs, and SIs are sorted in ascending order. 
    In searching for $\mathcal{B} + \Delta \mathcal{B}$ combinations, we restrict to $d_{FB} \le 1$ and $D_{\Delta \mclB}\le 6$. 
    Explicit constructions for the minimal models are provided in \cref{{sec:2D-wallpaper}}. }
    \label{tab:2D_Chern_band}
\end{table}

\begin{table}[htp] \footnotesize
    \centering
    \begin{tabular}{l|l|l|r|r|r}
    \hline
        LG ID (SG ID) & LG & SI & Number of $\mclB$ within $d_{\mclB} \le 4$ & Number of $\mclB + \Delta\mclB$ within $D_{\Delta \mclB} \le 8$ & Minimal band dimension $d$ \\
    \hline
        51 (83) & $p 4/m$ & $\rm C_m$ mod $4$ & 
        $\begin{array}{rrrr}
             & 80 & 86 & 80 \\
        \end{array}$ & $\begin{array}{rrrr}
             & 112 & 368 & 112 \\
        \end{array}$ & $\begin{array}{rrrr}
             & 10 & 10 & 10 \\
        \end{array}$ \\
        52 (85) & $p4/n$ & $\rm \delta_t$ & $\begin{array}{rr}
             & 6 \\
        \end{array}$ & $\begin{array}{rr}
             & 8 \\
        \end{array}$ & $\begin{array}{rr}
             & 20 \\
        \end{array}$ \\
        61 (123) & $p4/mmm$ & $\rm C_m$ mod 4 & $\begin{array}{rrrr}
             & 80 & 86 & 80 \\
        \end{array}$ & $\begin{array}{rrrr}
             & 112 & 368 & 112 \\
        \end{array}$ & $\begin{array}{rrrr}
             & 10 & 10 & 10 \\
        \end{array}$ \\
        62 (125) & $p4/n bm$ & $\rm \delta_t$ & $\begin{array}{rr}
             & 6  \\
        \end{array}$ & $\begin{array}{rr}
             & 8 \\
        \end{array}$ & $\begin{array}{rr}
             & 20 \\
        \end{array}$ \\
        63 (127) & $p4/m b m$ & $\rm C_m$ mod 4 & $\begin{array}{rrrr}
             & 4 & 6 & 4 \\
        \end{array}$ & $\begin{array}{rrrr}
             & 0 & 8 & 0 \\
        \end{array}$ & $\begin{array}{rrrr}
             & - & 20 & - \\
        \end{array}$ \\
        64 (129) & $p4/n mm$ & $\rm \delta_t$ & $\begin{array}{rr}
             & 6  \\
        \end{array}$ & $\begin{array}{rr}
             & 8 \\
        \end{array}$ & $\begin{array}{rr}
             & 20 \\
        \end{array}$ \\
        66 (147) & $p\ovl{3}$ & $\rm \delta_t$ & $\begin{array}{rr}
             & 50  \\
        \end{array}$ & $\begin{array}{rr}
             & 10 \\
        \end{array}$ & $\begin{array}{rr}
             & 10 \\
        \end{array}$ \\
        71 (162) & $p\ovl{3}1m$ & $\rm \delta_t$ & $\begin{array}{rr}
             & 50  \\
        \end{array}$ & $\begin{array}{rr}
             & 36 \\
        \end{array}$ & $\begin{array}{rr}
             & 10 \\
        \end{array}$ \\
        72 (164) & $p\ovl{3} m1$ & $\rm\delta_t$ & $\begin{array}{rr}
             & 50  \\
        \end{array}$ & $\begin{array}{rr}
             & 36 \\
        \end{array}$ & $\begin{array}{rr}
             & 10 \\
        \end{array}$ \\
        74 (174) & $p\ovl{6}$ & $\rm C_m$ mod 3 & $\begin{array}{rrr}
             & 81 & 81 \\
        \end{array}$ & $\begin{array}{rrr}
             & 90 & 90 \\
        \end{array}$ & $\begin{array}{rrr}
             & 6 & 6 \\
        \end{array}$ \\
        75 (175) & $p6/m$ & $\rm C_m$ mod 6 & $\begin{array}{rrrrrr}
             & 66 & 72 & 66 & 72 & 66 \\
        \end{array}$ & $\begin{array}{rrrrrr}
             & 54 & 198 & 48 & 198 & 54 \\
        \end{array}$ & $\begin{array}{rrrrrr}
             & 14 & 10 & 14 & 10 & 14 \\
        \end{array}$ \\
        78 (187) & $p\ovl{6}m2$ & $\rm C_m$ mod 3 & $\begin{array}{rrr}
             & 81 & 81  \\
        \end{array}$ & $\begin{array}{rrr}
             & 90 & 90 \\
        \end{array}$ & $\begin{array}{rrr}
             & 6 & 6 \\
        \end{array}$ \\
        79 (189) & $p\ovl{6}2m$ & $\rm C_m$ mod 3 & $\begin{array}{rrr}
             & 15 & 15  \\
        \end{array}$ & $\begin{array}{rrr}
             & 84 & 84 \\
        \end{array}$ & $\begin{array}{rrr}
             & 6 & 6 \\
        \end{array}$ \\
        80 (191) & $p6/mmm$ & $\rm C_m$ mod 6 & $\begin{array}{rrrrrr}
             & 66 & 72 & 66 & 72 & 66 \\
        \end{array}$ & $\begin{array}{rrrrrr}
             & 54 & 198 & 48 & 198 & 54 \\
        \end{array}$ & $\begin{array}{rrrrrr} 
             & 14 & 10 & 14 & 10 & 14 \\
        \end{array}$ \\
    \hline
    \end{tabular}
    \caption{Summary of CTFB constructions in 2D double-valued layer groups with TRS. 
    $\rm \delta_t$ is the 2D $\mathbb{Z}_2$ TI index, and $\rm C_m$ is the mirror-Chern number. 
    In the last three columns, we only list results corresponding to non-trivial SIs, and SIs are sorted in ascending order. 
    In searching for $\mclB + \Delta\mclB$ combinations, we restrict $d_{\mclB} \le 4$ and $D_{\Delta \mclB} \le 8$. 
    Explicit constructions for the minimal models are provided in \cref{{sec:2D-layer}}. 
    }
    \label{tab:DoubleTR_2D_summary}
\end{table}

\clearpage

\begin{table}[htp] \footnotesize
    \centering
    \begin{tabular}{l|l|l|r|r|r}
    \hline
        ID & SG & SI & Number of $\mclB$ within $d_{\mclB} \le 8$ & Number of $\mclB + \Delta\mclB$ within $D_{\Delta \mclB} \le 16$ & Minimal band dimension $d$ \\
    \hline
        200 & $Pm\ovl{3}$ & $(z_{\rm 2w}, \textcolor{red}{z_4})$ & 
        $\begin{array}{rrrr}
             & 297 & 336 & 297 \\
            264 & 297 & 264 & 297 \\
        \end{array}$ & {$\begin{array}{rrrr}
             & 101 & 192 & 101 \\
            40 & 101 & 40 & 101 \\
        \end{array}$} & $\begin{array}{rrrr}
             & 16 & 20 & 16 \\
            14 & 16 & {14} & 16 \\
        \end{array}$ \\
    \hline
        201 & $Pn\ovl{3}$ & $(z_{\rm 2w}, \textcolor{red}{z_4})$ & 
        $\begin{array}{rrrr}
             & 8 & 16 & 8 \\
            8 & 8 & 8 & 8 \\
        \end{array}$ & $\begin{array}{rrrr}
             & 1 & 4 & 1 \\
             4 & 1 & 4 & 1 \\
        \end{array}$ & $\begin{array}{rrrr}
             & 52 & 56 & 52 \\
            28 & 52 & 28 & 52 \\
        \end{array}$ \\
    \hline
        202 & $Fm\ovl{3}$ & $\textcolor{red}{z_4}$ & 
        $\begin{array}{rrrr}
             & 165 & 184 & 165 \\
        \end{array}$ & $\begin{array}{rrrr}
             & 23 & 24 & 23 \\
        \end{array}$ & $\begin{array}{rrrr}
             & 26 & 28 & 26 \\
        \end{array}$ \\
    \hline
        203 & $Fd \ovl{3}$ & $\textcolor{red}{z_4}$ & 
        $\begin{array}{rrrr}
            20 & 40 & 20 \\
        \end{array}$ & $\begin{array}{rrrr}
            3 & 6 & 3 \\
        \end{array}$ & $\begin{array}{rrrr}
             & 28 & 56 & 28 \\
        \end{array}$ \\
    \hline
        204 & $Im \ovl{3}$ & $(z_{\rm 2w}, \textcolor{red}{z_4})$ & 
        $\begin{array}{rrrr}
            & 55 & 80 & 55 \\
            {64} & 55 & {72} & 55 \\
        \end{array}$ & $\begin{array}{rrrr}
            & 16 & 36 & 16 \\
            {4} & 16 & {32} & 16 \\
        \end{array}$ & 
        $\begin{array}{rrrr}
            & 26 & 28 & 26 \\
            {28} & 26 & {14} & 26 \\
        \end{array}$ \\
    \hline
        205 & $Pa\ovl{3}$ & $\textcolor{red}{z_4}$ & $\begin{array}{rrrr}
            & 2 & 6 & 2 \\
        \end{array}$ & $\begin{array}{rrrr}
            & 0 & 0 & 0 \\
        \end{array}$ & - \\
    \hline
        206 & $Ia\ovl{3}$ & $(z_{\rm 2w}, \textcolor{red}{z_4})$ & $\begin{array}{rrrr}
            & 4 & 12 & 4 \\
            2 & 4 & 2 & 4 \\
        \end{array}$ & $\begin{array}{rrrr}
            & 0 & 0 & 0 \\
            0 & 0 & 0 & 0 \\
        \end{array}$ & - \\
    \hline
        215 & $P\ovl{4}3m$ & $\textcolor{red}{z_2}$ & $\begin{array}{rr}
            & 648 \\
        \end{array}$ & $\begin{array}{rr}
            & 1720 \\
        \end{array}$ & $\begin{array}{rr}
            & 14 \\
        \end{array}$ \\
    \hline
        216 & $F\ovl{4}3m$ & $\textcolor{red}{z_2}$ & $\begin{array}{rr}
            & 816 \\
        \end{array}$ & $\begin{array}{rr}
            & 444 \\
        \end{array}$ & $\begin{array}{rr}
            & 6 \\
        \end{array}$ \\
    \hline
        217 & $I\ovl{4}3m$ & $\textcolor{red}{z_2}$ & $\begin{array}{rr}
            & 124 \\
        \end{array}$ & $\begin{array}{rr}
            & 920 \\
        \end{array}$ & $\begin{array}{rr}
            & 22 \\
        \end{array}$ \\
    \hline
        218 & $P\ovl{4}3n$ & $\textcolor{red}{z_2}$ & $\begin{array}{rr}
            & 18 \\
        \end{array}$ & $\begin{array}{rr}
            & 12 \\
        \end{array}$ & $\begin{array}{rr}
            & 28 \\
        \end{array}$ \\
    \hline
        219 & $F\ovl{4}3c$ & $\textcolor{red}{z_2}$ & $\begin{array}{rr}
            & 18 \\
        \end{array}$ & $\begin{array}{rr}
            & 12 \\
        \end{array}$ & $\begin{array}{rr}
            & 28 \\
        \end{array}$ \\
    \hline
        220 & $I\ovl{4}3d$ & $\textcolor{red}{z_2}$ & $\begin{array}{rr}
            & 0 \\
        \end{array}$ & $\begin{array}{rr}
            & 0 \\
        \end{array}$ & - \\
    \hline
        221 & $Pm\ovl{3}m$ & $(z_{\rm 4m,\pi}, \textcolor{red}{z_8})$ &  
            {$\begin{array}{rrrrrrrr}
                & 3452 & 3488 & 3452 & 3396 & 3452 & 3488 & 3452 \\
            3394 & 3452 & 3418 & 3449 & 3418 & 3452 & 3394 & 3449 \\
            3510 & 3449 & 3488 & 3449 & 3510 & 3449 & 3488 & 3449 \\
            3394 & 3449 & 3394 & 3452 & 3418 & 3449 & 3418 & 3452 \\
        \end{array}$}  & {$\begin{array}{rrrrrrrr}
               & 112 & 142 & 112 & 464 & 112 & 142 & 112 
            \\
            16 & 112 & 20 & 16 & 20 & 112 & 16 & 16 
            \\
            28 & 16 & 142 & 16 & 28 & 16 & 142 & 16 
            \\
            16 & 16 & 16 & 112 & 20 & 16 & 20 & 112 \\
        \end{array}$}  &  
        {$\begin{array}{rrrrrrrr}
            & 26 & 28 & 26 & 56 & 26 & 28 & 26 \\
        26 & 26 & 34 & 44 & 34 & 26 & 26 & 44 \\
        36 & 44 & 28 & 44 & 36 & 44 & 28 & 44 \\
        26 & 44 & 26 & 26 & 34 & 44 & 34 & 26 \\
        \end{array}$}  \\
    \hline
        222 & $Pn\ovl{3}n$ & $\textcolor{red}{z_4}$ &  $\begin{array}{rrrr}
            & 6 & 8 & 6 \\
        \end{array}$  &  $\begin{array}{rrrr}
            & 2 & 4 & 2 \\
        \end{array}$  &  $\begin{array}{rrrrrrrr}
            & 100 & 104 & 100 \\
        \end{array}$  \\
    \hline
        223 & $Pm\ovl{3}n$ & $\textcolor{red}{z_4}$ &  $\begin{array}{rrrr}
            & 30 & 30 & 30 \\
        \end{array}$  &  $\begin{array}{rrrr}
            & 0 & 0 & 0 \\
        \end{array}$  & - \\
    \hline
        224 & $Pn\ovl{3}m$ & $(z_{\rm 2w}, \textcolor{red}{z_4})$ &  $\begin{array}{rrrr}
            & 20 & 24 & 20 \\
            20 & 20 & 20 & 20 \\
        \end{array}$  &  $\begin{array}{rrrr}
            & 0 & 0 & 0 \\
            0 & 0 & 0 & 0 \\
        \end{array}$  & - \\
    \hline
        225 & $Fm\ovl{3}m$ & $\textcolor{red}{z_8}$ &  $\begin{array}{rrrrrrrr}
            & 2179 & 2195 & 2179 & 2276 & 2179 & 2195 & 2179  \\ 
        \end{array}$  &  $\begin{array}{rrrrrrrr}
            & 46 & 48 & 46 & 156 & 46 & 48 & 46 \\
        \end{array}$  &  $\begin{array}{rrrrrrrr}
            & 34 & 36 & 34 & 42 & 34 & 36 & 34 \\
        \end{array}$  \\
    \hline
        226 & $Fm\ovl{3}c$ & $\textcolor{red}{z_8}$ &  $\begin{array}{rrrrrrrr}
            & 39 & 39 & 39 & 48 & 39 & 39 & 39  \\ 
        \end{array}$  &  $\begin{array}{rrrrrrrr}
            & 2 & 2 & 2 & 8 & 2 & 2 & 2 \\
        \end{array}$  &  $\begin{array}{rrrrrrrr}
            & 56 & 52 & 56 & 84 & 56 & 52 & 56 \\
        \end{array}$  \\
    \hline
        227 & $Fd\ovl{3}m$ & $\textcolor{red}{z_4}$ &  $\begin{array}{rrrr}
            & 92 & 134 & 92 \\
        \end{array}$  &  $\begin{array}{rrrr}
            & 0 & 0 & 0 \\
        \end{array}$  & - \\
    \hline
        228 & $Fd\ovl{3}c$ & $\textcolor{red}{z_4}$ &  $\begin{array}{rrrr}
            & 2 & 4 & 2 \\
        \end{array}$ & $\begin{array}{rrrr}
            & 0 & 0 & 0 \\
        \end{array}$ & - \\
    \hline
        229 & $Im\ovl{3}m$ & $(z_{\rm 2w} , \textcolor{red}{z_8})$ & $\begin{array}{rrrrrrrr}
            & 660 & 692 & 660 & 746 & 660 & 692 & 660  \\
            640 & 660 & 696 & 660 & 644 & 660 & 696 & 660  \\
        \end{array}$ & $\begin{array}{rrrrrrrr}
              & 95 & 216 & 95 & 1124 & 95 & 216 & 95  \\
            8 & 95 & 194 & 95 & 32 & 95 & 194 & 95  \\
        \end{array}$ & $\begin{array}{rrrrrrrr}
            & 50 & 52 & 50 & 14 & 50 & 52 & 50  \\
            52 & 50 & 14 & 50 & 64 & 50 & 14 & 50  \\
        \end{array}$ \\
    \hline
        230 & $Ia\ovl{3}d$ & $\textcolor{red}{z_4}$ &  $\begin{array}{rrrr}
            & 0 & 2 & 0 \\
        \end{array}$  & $\begin{array}{rrrr}
            & 0 & 0 & 0 \\
        \end{array}$  & - \\
    \hline
    \end{tabular}
    \caption{Summary of CTFB constructions in 3D double-valued space groups with TRS. 
    Odd values of the red SIs correspond to the strong $\mathbb{Z}_2$ states.
    In the last three columns, we only list results corresponding to non-trivial SIs, and SIs are sorted in ascending order. 
    In searching for $\mclB + \Delta\mclB$ combinations, we restrict $d_{\mclB} \le 8$ and $D_{\Delta \mclB} \le 16$. 
    Explicit constructions for minimal models are provided in \cref{{sec:3D-double-T}}. 
    For space groups \#205, \#206, \#220, \#223, \#224, \#227, \#228, \#230, we have not found CTFB constructions (namely, combination of $\mclB + \Delta\mclB$) for $d_{\mclB} \le 8$ until increasing $D_{\Delta \mclB}$ to 32. }
    \label{tab:DoubleTR_summary}
\end{table}

\begin{table}[h!] \footnotesize
    \centering
    \begin{tabular}{l|l|l|r|r|r}
    \hline
        ID & SG & SI & Number of $\mclB$ within $d_{\mclB} \le 8$ & Number of $\mclB + \Delta\mclB$ within $D_{\Delta \mclB} \le 16$ & Minimal band dimension $d$ \\
    \hline
        200 & $Pm\bar{3}$ & $(z_{2w}, \textcolor{red}{z_4})$ & 
        $\begin{array}{rrrr}   
         & 4289 & 4328 & 4289 \\
         4232 & 4289 & 4232 & 4289 
        \end{array}$  & $\begin{array}{rrrr}
            & {303} & {576} & {303} \\
        {120} & {303} & {120} & {303}
        \end{array}$  & $\begin{array}{rrrr} 
        & {16} & {20} & {16} \\
        {14} & {16} & {14} & {16}
        \end{array}$   \\
     \hline
        201 & $Pn\bar{3}$ & $\textcolor{red}{\eta_{2I}'}$ & $\begin{array}{r} 
        600
        \end{array}$  & $\begin{array}{rr}
        & 72       
        \end{array}$  & $\begin{array}{r} 
        28
        \end{array}$   \\
    \hline
        202 & $Fm\bar{3}$ & $\textcolor{red}{z_4}$ & $\begin{array}{rrrr} 
        & 7941 & 8058 & 7941
        \end{array}$ & $\begin{array}{rrrr} 
        & 285 & 384 & 285 
        \end{array}$ & $\begin{array}{rrrr} 
        14 & 18 & 14
        \end{array}$  \\
    \hline
        203 & $Fd\bar{3}$ & $\textcolor{red}{\eta_{2I}'}$ & $\begin{array}{rr} 
        & 2730 
        \end{array}$ & $\begin{array}{rr}
        & 204 
        \end{array}$ & $\begin{array}{rr} 
        & 28
        \end{array}$  \\
    \hline
        204 & $Im\bar{3}$ & $\textcolor{red}{z_4}$ & $\begin{array}{rrrr} 
        & 1798 & 1816 & 1798
        \end{array}$ & $\begin{array}{rrrr}
        & 210 & 420 & 210 
        \end{array}$ & $\begin{array}{rrrr} 
        & 14 & 14 & 14
        \end{array}$  \\
    \hline
        205 & $Pa\bar{3}$ & $\textcolor{red}{\eta_{2I}'}$ & $\begin{array}{rr} 
        & 120
        \end{array}$ & $\begin{array}{rr}
        & 0 
        \end{array}$ & - \\
    \hline
        206 & $Ia\bar{3}$ & $\textcolor{red}{\eta_{2I}'}$ & $\begin{array}{rr} 
        & 360 
        \end{array}$ & $\begin{array}{rr}
        & 0
        \end{array}$ & - \\        
    \hline
        215 & $P\bar{4}3m$ & $\textcolor{red}{z_2}$ & $\begin{array}{rr}
        & 648 \end{array}$ & $\begin{array}{rr} 1720 
        \end{array}$ & 14 \\    
    \hline
        216 & $F\bar{4}3m$ & $\textcolor{red}{z_2}$ & $\begin{array}{rr}
        & 816 \end{array}$ & $\begin{array}{rr} 444 
        \end{array}$ & 6 \\
    \hline
        217 & $I\bar{4}3m$ & $\textcolor{red}{z_2}$ & $\begin{array}{rr}
        & 536 \end{array}$ & $\begin{array}{rr} & 14960 
        \end{array}$ & 14 \\
    \hline
        218 & $P\bar{4}3n$ & $\textcolor{red}{z_2}$ & $\begin{array}{rr}
        & 648 \end{array}$ & $\begin{array}{rr} & 1720 
        \end{array}$ & 22 \\
    \hline
        219 & $F\bar{4}3c$ & $\textcolor{red}{z_2}$ & $\begin{array}{rr}
        & 816 \end{array}$ & $\begin{array}{rr} & 444 
        \end{array}$ & 22 \\
    \hline
        220 & $I\bar{4}3d$ & $\textcolor{red}{z_2}$ & $\begin{array}{rr}
        & 16 \end{array}$ & $\begin{array}{rr} & 0 
        \end{array}$ & - \\
    \hline
        221 & $Pm\bar{3}m$ & ${(z_{4m,\pi}^+,\textcolor{red}{z_8})}$ & 
        {$\begin{array}{rrrrrrrr} 
             & 3452 & 3488 & 3452 & 3396 & 3452 & 3488 & 3452 \\
        3394 & 3449 & 3394 & 3452 & 3418 & 3449 & 3418 & 3452 \\
        3510 & 3449 & 3488 & 3449 & 3510 & 3449 & 3488 & 3449 \\
        3394 & 3452 & 3418 & 3449 & 3418 & 3452 & 3394 & 3449 \end{array}$} &
        {$\begin{array}{rrrrrrrr}  
           & 112 & 142 & 112 & 464 & 112 & 142 & 112\\
        16 & 16 & 16 & 112 & 20 & 16 & 20 & 112\\
        28 & 16 & 142 & 16 & 28 & 16 & 142 & 16\\
        16 & 112 & 20 & 16 & 20 & 112 & 16 & 16\\
        \end{array}$}
        & {$\begin{array}{rrrrrrrr}
           & 26&  28&  26&  56&  26&  28& 26\\
        26 & 44&  26&  26&  34&  44&  34& 26\\
        36 & 44&  28&  44&  36&  44&  28& 44\\
        26 & 26&  34&  44&  34&  26&  26& 44
        \end{array}$} \\
    \hline
        222 & $Pn\bar{3}n$ & $\textcolor{red}{\eta_{2I}'}$ & $\begin{array}{rr}
        & 88 
        \end{array}$ & $\begin{array}{rr} & 8 \end{array}$ & $\begin{array}{rr}
        & 52 
        \end{array}$\\
    \hline
        223 & $Pm\bar{3}n$ & $\textcolor{red}{z_4}$ & $\begin{array}{rrrr}
        & 82 & 86 & 82 
        \end{array}$ & $\begin{array}{rrrr} & 0 & 0 & {0}
        \end{array}$ & - \\
    \hline
        224 & $Pn\bar{3}m$ & $\textcolor{red}{\eta_{2I}'}$ & $\begin{array}{rr}
        & 80 \end{array}$ & $\begin{array}{rr} & 0
        \end{array}$ & - \\
    \hline
        225 & $Fm\bar{3}m$ & $\textcolor{red}{z_8}$ & 
        $\begin{array}{rrrrrrrr}
        & 2179 & 2195 & 2179 & 2276 & 2179 & 2195 & 2179
        \end{array}$ & $\begin{array}{rrrrrrrr} 
        46 & 48 & 46 & 156 & 46 & 48 & 46
        \end{array}$ & 
        $\begin{array}{rrrrrrrr}
        & 34 & 36 & 34 & 42 & 34 & 36 & 34
        \end{array}$\\
    \hline
        226 & $Fm\bar{3}c$ & $\textcolor{red}{z_4}$ & $\begin{array}{rrrr} & 3182 & 3234 & 3182 \end{array}$
        & $\begin{array}{rrrr} & 190 & 216 & 190 \end{array}$ & $\begin{array}{rrrr} & 26 & 42 & 26 \end{array}$\\
    \hline
        227 & $Fd\bar{3}m$ & $\textcolor{red}{\eta_{2I}'}$ &  
        $\begin{array}{rr}
        & 320 \end{array}$ 
        & $\begin{array}{rr} & 0  \end{array}$
        & - \\
    \hline
        228 & $Fd\bar{3}c$ & $\textcolor{red}{\eta_{2I}'}$ & $\begin{array}{rr}
        & 128 \end{array}$ & $\begin{array}{rr} & 0 \end{array}$
        & - \\
    \hline
        229 & $Im\bar{3}m$ & $\textcolor{red}{z_8}$ & $\begin{array}{rrrrrrrr} 1320 & 1388 & 1320 & 1390 & 1320 & 1388 & 1320 \end{array}$ & $\begin{array}{rrrrrrrr} & 418 & 842 & 418 & 2108 & 418 & 842 & 418 \end{array}$
        & 
        $\begin{array}{rrrrrrrr}
        & 26 & 14 & 26 & 14 & 26 & 14 & 26
        \end{array}$ \\
    \hline 
        230 & $Ia\bar{3}d$ & $\textcolor{red}{\eta_{2I}'}$ & $\begin{array}{rr} 16 \end{array}$ & $\begin{array}{rr} & 0 \end{array}$ 
        & - \\
    \hline
    \end{tabular}
    \caption{Summary of CTFB constructions in 3D double-valued space groups without TRS. 
    Odd values of the red SIs correspond to the axion states.
    In the last three columns, we only list results corresponding to non-trivial SIs, and SIs are sorted in ascending order. 
    In searching for $\mclB + \Delta\mclB$ combinations, we restrict $d_{\mclB} \le 8$ and $D_{\Delta \mclB} \le 16$. 
    The zero entries in $\mclB + \Delta\mclB$ in space groups \#205, \#206, \#220, \#223, \#224, \#227, \#228, \#230 may become finite as the cutoff $D_{\Delta \mclB}$ is further enlarged.  
    Explicit constructions for minimal models are provided in \cref{sec:3D-double-noT}.
    \label{tab:DoubleNoTR_summary}}
\end{table}

\clearpage

\section{The minimal construction of CTFBs}
\label{sec:CTFB-constructions}

The minimal CTFB constructions for 2D wallpaper groups without TRS are listed in \cref{sec:2D-wallpaper}, 
while those for 2D double-valued layer groups with TRS are given in \cref{sec:2D-layer}. 
For 3D double-valued space groups, the constructions with TRS are listed in \cref{sec:3D-double-T}, 
and those without TRS in \cref{sec:3D-double-noT}. 
For a given SI in the convention of Refs.~\cite{song2018quantitative,elcoro_magnetic_2021}, the explicit minimal realization $\mathcal{BR}_L \boxminus \mathcal{BR}_{\widetilde{L}}$ is tabulated.

We apply the algorithm described in \cref{app:algr-workflow} to minimize the band number of the bipartite construction. 
In particular, we search the integer-valued free parameters $t_p$ ($p>r$) in the range $[-4,4]$. 

\subsection{2D wallpaper groups without TRS} 
\label{sec:2D-wallpaper}
\renewcommand{\arraystretch}{1.1}
\begin{longtable}{cccc cccc c}
\caption{Wallpaper group $p3$.}\\
\label{tab:SG143_DoubleTR} \\
\hline\hline
\multicolumn{9}{l}{ SI: $(z_{{3R}}) = (1)$ }\\
\hline
\multicolumn{7}{l}{\footnotesize $\mathcal{BR}_L \boxminus \mathcal{BR}_{\widetilde{L}}: (2 {\Gamma}_{2} \boxminus {\Gamma}_{3} \; ; {\rK}_{1} \; ; {\rKA}_{1} \; ; {\rM}_{1})$} & \multicolumn{2}{r}{\footnotesize $(n_b,\mathrm{N})=(1,3)$} \\
\hline
{\footnotesize $\mathrm{[{^2{E}}]_{1a}}$} & {\footnotesize $\mathrm{[{^1{E}}]_{1a}}$} & {\footnotesize $\mathrm{[{{A}_{1}}]_{1a}}$} & {\footnotesize $\mathrm{[{^2{E}}]_{1b}}$} & {\footnotesize $\mathrm{[{^1{E}}]_{1b}}$} & {\footnotesize $\mathrm{[{{A}_{1}}]_{1b}}$} & {\footnotesize $\mathrm{[{^2{E}}]_{1c}}$} & {\footnotesize $\mathrm{[{^1{E}}]_{1c}}$} & {\footnotesize $\mathrm{[{{A}_{1}}]_{1c}}$}\\
\hline
 $0$ & $\ovl{1}$ & $0$ & $1$ & $0$ & $0$ & $1$ & $0$ & $0$\\
\hline
\multicolumn{7}{l}{\footnotesize $\mathcal{BR}_L \boxminus \mathcal{BR}_{\widetilde{L}}: (2 {\Gamma}_{1} \boxminus {\Gamma}_{2} \; ; {\rK}_{2} \; ; {\rKA}_{1} \; ; {\rM}_{1})$} & \multicolumn{2}{r}{\footnotesize $(n_b,\mathrm{N})=(1,3)$} \\
\hline
{\footnotesize $\mathrm{[{^2{E}}]_{1a}}$} & {\footnotesize $\mathrm{[{^1{E}}]_{1a}}$} & {\footnotesize $\mathrm{[{{A}_{1}}]_{1a}}$} & {\footnotesize $\mathrm{[{^2{E}}]_{1b}}$} & {\footnotesize $\mathrm{[{^1{E}}]_{1b}}$} & {\footnotesize $\mathrm{[{{A}_{1}}]_{1b}}$} & {\footnotesize $\mathrm{[{^2{E}}]_{1c}}$} & {\footnotesize $\mathrm{[{^1{E}}]_{1c}}$} & {\footnotesize $\mathrm{[{{A}_{1}}]_{1c}}$}\\
\hline
\end{longtable}

\vspace{0.5cm}

\renewcommand{\arraystretch}{1.1}
\begin{longtable}{cccccccccc}
\caption{Wallpaper group $p4$.}\\
\label{tab:SG75_DoubleTR}\\
\hline\hline
\multicolumn{10}{l}{{$\bm{\mathrm{SI}: (z_{_{4R}}) = (1)}$}}\\
\hline
\multicolumn{8}{l}{\footnotesize $\mathcal{BR}_L \boxminus \mathcal{BR}_{\widetilde{L}}: (2 {\Gamma}_{3} \boxminus {\Gamma}_{2} \; ; {\rM}_{1} \; ; {\rX}_{1})$} & \multicolumn{2}{r}{\footnotesize $(n_b,\mathrm{N})=(1,5)$} \\
\hline
{\footnotesize $\mathrm{[{^2{E}}]_{1a}}$} & {\footnotesize $\mathrm{[{^1{E}}]_{1a}}$} & {\footnotesize $\mathrm{[{{A}}]_{1a}}$} & {\footnotesize $\mathrm{[{{B}}]_{1a}}$} & {\footnotesize $\mathrm{[{^2{E}}]_{1b}}$} & {\footnotesize $\mathrm{[{^1{E}}]_{1b}}$} & {\footnotesize $\mathrm{[{{A}}]_{1b}}$} & {\footnotesize $\mathrm{[{{B}}]_{1b}}$} & {\footnotesize $\mathrm{[{{A}}]_{2c}}$} & {\footnotesize $\mathrm{[{{B}}]_{2c}}$}\\
\hline
 $1$ & $0$ & $1$ & $0$ & $1$ & $0$ & $0$ & $0$ & $\ovl{1}$ & $0$\\
 $0$ & $\ovl{1}$ & $0$ & $\ovl{1}$ & $1$ & $0$ & $0$ & $0$ & $0$ & $1$\\
\hline
\multicolumn{8}{l}{\footnotesize $\mathcal{BR}_L \boxminus \mathcal{BR}_{\widetilde{L}}: (2 {\Gamma}_{4} \boxminus {\Gamma}_{1} \; ; {\rM}_{2} \; ; {\rX}_{1})$} & \multicolumn{2}{r}{\footnotesize $(n_b,\mathrm{N})=(1,5)$} \\
\hline
\end{longtable}

\vspace{0.5cm}

\renewcommand{\arraystretch}{1.1}
\begin{longtable}{ccccccccccc}
\caption{Wallpaper group $p6$.}\\
\label{tab:SG168_DoubleTR}\\
\hline\hline
\multicolumn{11}{l}{{$\bm{\mathrm{SI}: (z_{_{6R}}) = (1)}$}}\\
\hline
\multicolumn{9}{l}{\footnotesize $\mathcal{BR}_L \boxminus \mathcal{BR}_{\widetilde{L}}: (2 {\Gamma}_{4} \boxminus {\Gamma}_{5} \; ; {\rK}_{1} \; ; {\rM}_{1})$} & \multicolumn{2}{r}{\footnotesize $(n_b,\mathrm{N})=(1,7)$} \\
\hline
{\footnotesize $\mathrm{[{^2{E}_{1}}]_{1a}}$} & {\footnotesize $\mathrm{[{^2{E}_{2}}]_{1a}}$} & {\footnotesize $\mathrm{[{^1{E}_{1}}]_{1a}}$} & {\footnotesize $\mathrm{[{^1{E}_{2}}]_{1a}}$} & {\footnotesize $\mathrm{[{{A}}]_{1a}}$} & {\footnotesize $\mathrm{[{{B}}]_{1a}}$} & {\footnotesize $\mathrm{[{^2{E}}]_{2b}}$} & {\footnotesize $\mathrm{[{^1{E}}]_{2b}}$} & {\footnotesize $\mathrm{[{{A}_{1}}]_{2b}}$} & {\footnotesize $\mathrm{[{{A}}]_{3c}}$} & {\footnotesize $\mathrm{[{{B}}]_{3c}}$}\\
\hline
 $0$ & $1$ & $0$ & $0$ & $1$ & $0$ & $1$ & $0$ & $0$ & $\ovl{1}$ & $0$\\
\hline
\multicolumn{9}{l}{\footnotesize $\mathcal{BR}_L \boxminus \mathcal{BR}_{\widetilde{L}}: (2 {\Gamma}_{6} \boxminus {\Gamma}_{1} \; ; {\rK}_{2} \; ; {\rM}_{1})$} & \multicolumn{2}{r}{\footnotesize $(n_b,\mathrm{N})=(1,7)$} \\
\hline
{\footnotesize $\mathrm{[{^2{E}_{1}}]_{1a}}$} & {\footnotesize $\mathrm{[{^2{E}_{2}}]_{1a}}$} & {\footnotesize $\mathrm{[{^1{E}_{1}}]_{1a}}$} & {\footnotesize $\mathrm{[{^1{E}_{2}}]_{1a}}$} & {\footnotesize $\mathrm{[{{A}}]_{1a}}$} & {\footnotesize $\mathrm{[{{B}}]_{1a}}$} & {\footnotesize $\mathrm{[{^2{E}}]_{2b}}$} & {\footnotesize $\mathrm{[{^1{E}}]_{2b}}$} & {\footnotesize $\mathrm{[{{A}_{1}}]_{2b}}$} & {\footnotesize $\mathrm{[{{A}}]_{3c}}$} & {\footnotesize $\mathrm{[{{B}}]_{3c}}$}\\
\hline
\end{longtable}

\vspace{0.5cm}

\clearpage

\subsection{2D layer groups with TRS} 
\label{sec:2D-layer}
\renewcommand{\arraystretch}{1.1}
\begin{longtable}{cccccccccc}
\caption{Layer group $\mathrm{{LG}51}$ ($p4/m$).}\\
\label{tab:{SG}83_DoubleTR}\\
\hline\hline
\multicolumn{10}{l}{{$\bm{\mathrm{SI}: (C_{_m}) = (1)}$}}\\
\hline
\multicolumn{8}{l}{\footnotesize $\mathcal{BR}_L \boxminus \mathcal{BR}_{\widetilde{L}}: (2 \ovl{\Gamma}_{10}\ovl{\Gamma}_{12} \boxminus \ovl{\Gamma}_{6}\ovl{\Gamma}_{8} \; ; \ovl{\rM}_{5}\ovl{\rM}_{7} \; ; \ovl{\rX}_{3}\ovl{\rX}_{4})$} & \multicolumn{2}{r}{\footnotesize $(n_b,\mathrm{N})=(2,10)$} \\
\hline
{\scriptsize $\mathrm{[{}^{1}\ovl{\mathrm{E}}_{\mathrm{1g}}{}^{2}\ovl{\mathrm{E}}_{\mathrm{1g}}]_{1a}}$} & {\scriptsize $\mathrm{[{}^{1}\ovl{\mathrm{E}}_{\mathrm{1u}}{}^{2}\ovl{\mathrm{E}}_{\mathrm{1u}}]_{1a}}$} & {\scriptsize $\mathrm{[{}^{1}\ovl{\mathrm{E}}_{\mathrm{2g}}{}^{2}\ovl{\mathrm{E}}_{\mathrm{2g}}]_{1a}}$} & {\scriptsize $\mathrm{[{}^{1}\ovl{\mathrm{E}}_{\mathrm{2u}}{}^{2}\ovl{\mathrm{E}}_{\mathrm{2u}}]_{1a}}$} & {\scriptsize $\mathrm{[{}^{1}\ovl{\mathrm{E}}_{\mathrm{1g}}{}^{2}\ovl{\mathrm{E}}_{\mathrm{1g}}]_{1c}}$} & {\scriptsize $\mathrm{[{}^{1}\ovl{\mathrm{E}}_{\mathrm{1u}}{}^{2}\ovl{\mathrm{E}}_{\mathrm{1u}}]_{1c}}$} & {\scriptsize $\mathrm{[{}^{1}\ovl{\mathrm{E}}_{\mathrm{2g}}{}^{2}\ovl{\mathrm{E}}_{\mathrm{2g}}]_{1c}}$} & {\scriptsize $\mathrm{[{}^{1}\ovl{\mathrm{E}}_{\mathrm{2u}}{}^{2}\ovl{\mathrm{E}}_{\mathrm{2u}}]_{1c}}$} & {\scriptsize $\mathrm{[{}^{1}\ovl{\mathrm{E}}_{\mathrm{g}}{}^{2}\ovl{\mathrm{E}}_{\mathrm{g}}]_{2e}}$} & {\scriptsize $\mathrm{[{}^{1}\ovl{\mathrm{E}}_{\mathrm{u}}{}^{2}\ovl{\mathrm{E}}_{\mathrm{u}}]_{2e}}$}\\
\hline
 $0$ & $1$ & $1$ & $0$ & $0$ & $1$ & $0$ & $0$ & $\ovl{1}$ & $0$\\
 $\ovl{1}$ & $0$ & $0$ & $\ovl{1}$ & $0$ & $1$ & $0$ & $0$ & $0$ & $1$\\
\hline
\end{longtable}

\vspace{0.5cm}

\renewcommand{\arraystretch}{1.1}
\begin{longtable}{cccccc}
\caption{Layer group $\mathrm{{LG}52}$ ($p4/n$).}\\
\label{tab:{SG}85_DoubleTR}\\
\hline\hline
\multicolumn{6}{l}{{$\bm{\mathrm{SI}: (\delta_t) = (1)}$}}\\
\hline
\multicolumn{4}{l}{\footnotesize $\mathcal{BR}_L \boxminus \mathcal{BR}_{\widetilde{L}}: (3 \ovl{\Gamma}_{5}\ovl{\Gamma}_{7} \boxminus \ovl{\Gamma}_{10}\ovl{\Gamma}_{12} \; ; \ovl{\rM}_{3}\ovl{\rM}_{4} \; ; \ovl{\rX}_{2}\ovl{\rX}_{2})$} & \multicolumn{2}{r}{\footnotesize $(n_b,\mathrm{N})=(4,20)$} \\
\hline
{\scriptsize $\mathrm{[{}^{1}\ovl{\mathrm{E}}_{\mathrm{1}}{}^{2}\ovl{\mathrm{E}}_{\mathrm{1}}]_{2a}}$} & {\scriptsize $\mathrm{[{}^{1}\ovl{\mathrm{E}}_{\mathrm{2}}{}^{2}\ovl{\mathrm{E}}_{\mathrm{2}}]_{2a}}$} & {\scriptsize $\mathrm{[{}^{1}\ovl{\mathrm{E}}_{\mathrm{1}}{}^{2}\ovl{\mathrm{E}}_{\mathrm{1}}]_{2c}}$} & {\scriptsize $\mathrm{[{}^{1}\ovl{\mathrm{E}}_{\mathrm{2}}{}^{2}\ovl{\mathrm{E}}_{\mathrm{2}}]_{2c}}$} & {\scriptsize $\mathrm{[\ovl{\mathrm{A}}_{\mathrm{g}}\ovl{\mathrm{A}}_{\mathrm{g}}]_{4d}}$} & {\scriptsize $\mathrm{[\ovl{\mathrm{A}}_{\mathrm{u}}\ovl{\mathrm{A}}_{\mathrm{u}}]_{4d}}$}\\
\hline
 $0$ & $1$ & $0$ & $2$ & $0$ & $\ovl{1}$\\
 $0$ & $1$ & $\ovl{2}$ & $0$ & $1$ & $0$\\
 $\ovl{1}$ & $0$ & $\ovl{1}$ & $1$ & $1$ & $0$\\
\hline
\end{longtable}

\vspace{0.5cm}

\renewcommand{\arraystretch}{1.1}
\begin{longtable}{cccccccccc}
\caption{Layer group $\mathrm{{LG}61}$ ($p4/mmm$).}\\
\label{tab:{SG}123_DoubleTR}\\
\hline\hline
\multicolumn{10}{l}{{$\bm{\mathrm{SI}: (C_{_m}) = (1)}$}}\\
\hline
\multicolumn{8}{l}{\footnotesize $\mathcal{BR}_L \boxminus \mathcal{BR}_{\widetilde{L}}: (2 \ovl{\Gamma}_{8} \boxminus \ovl{\Gamma}_{7} \; ; \ovl{\rM}_{6} \; ; \ovl{\rX}_{5})$} & \multicolumn{2}{r}{\footnotesize $(n_b,\mathrm{N})=(2,10)$} \\
\hline
{\scriptsize $\mathrm{[\ovl{\mathrm{E}}_{\mathrm{1g}}]_{1a}}$} & {\scriptsize $\mathrm{[\ovl{\mathrm{E}}_{\mathrm{1u}}]_{1a}}$} & {\scriptsize $\mathrm{[\ovl{\mathrm{E}}_{\mathrm{2g}}]_{1a}}$} & {\scriptsize $\mathrm{[\ovl{\mathrm{E}}_{\mathrm{2u}}]_{1a}}$} & {\scriptsize $\mathrm{[\ovl{\mathrm{E}}_{\mathrm{1g}}]_{1c}}$} & {\scriptsize $\mathrm{[\ovl{\mathrm{E}}_{\mathrm{1u}}]_{1c}}$} & {\scriptsize $\mathrm{[\ovl{\mathrm{E}}_{\mathrm{2g}}]_{1c}}$} & {\scriptsize $\mathrm{[\ovl{\mathrm{E}}_{\mathrm{2u}}]_{1c}}$} & {\scriptsize $\mathrm{[\ovl{\mathrm{E}}_{\mathrm{g}}]_{2e}}$} & {\scriptsize $\mathrm{[\ovl{\mathrm{E}}_{\mathrm{u}}]_{2e}}$}\\
\hline
 $0$ & $0$ & $1$ & $1$ & $0$ & $0$ & $0$ & $1$ & $\ovl{1}$ & $0$\\
 $\ovl{1}$ & $\ovl{1}$ & $0$ & $0$ & $0$ & $0$ & $0$ & $1$ & $0$ & $1$\\
\hline
\multicolumn{8}{l}{\footnotesize $\mathcal{BR}_L \boxminus \mathcal{BR}_{\widetilde{L}}: (2 \ovl{\Gamma}_{9} \boxminus \ovl{\Gamma}_{6} \; ; \ovl{\rM}_{7} \; ; \ovl{\rX}_{5})$} & \multicolumn{2}{r}{\footnotesize $(n_b,\mathrm{N})=(2,10)$} \\
\hline
\end{longtable}

\vspace{0.5cm}

\renewcommand{\arraystretch}{1.1}
\begin{longtable}{cccccc}
\caption{Layer group $\mathrm{{LG}62}$ ($p4/nbm$).}\\
\label{tab:{SG}125_DoubleTR}\\
\hline\hline
\multicolumn{6}{l}{{$\bm{\mathrm{SI}: (\delta_t) = (1)}$}}\\
\hline
\multicolumn{4}{l}{\footnotesize $\mathcal{BR}_L \boxminus \mathcal{BR}_{\widetilde{L}}: (3 \ovl{\Gamma}_{6} \boxminus \ovl{\Gamma}_{8} \; ; \ovl{\rM}_{5} \; ; \ovl{\rX}_{3}\ovl{\rX}_{4})$} & \multicolumn{2}{r}{\footnotesize $(n_b,\mathrm{N})=(4,20)$} \\
\hline
{\scriptsize $\mathrm{[\ovl{\mathrm{E}}_{\mathrm{1}}]_{2a}}$} & {\scriptsize $\mathrm{[\ovl{\mathrm{E}}_{\mathrm{2}}]_{2a}}$} & {\scriptsize $\mathrm{[\ovl{\mathrm{E}}_{\mathrm{1}}]_{2c}}$} & {\scriptsize $\mathrm{[\ovl{\mathrm{E}}_{\mathrm{2}}]_{2c}}$} & {\scriptsize $\mathrm{[{}^{1}\ovl{\mathrm{E}}_{\mathrm{g}}{}^{2}\ovl{\mathrm{E}}_{\mathrm{g}}]_{4e}}$} & {\scriptsize $\mathrm{[{}^{1}\ovl{\mathrm{E}}_{\mathrm{u}}{}^{2}\ovl{\mathrm{E}}_{\mathrm{u}}]_{4e}}$}\\
\hline
 $0$ & $1$ & $0$ & $2$ & $0$ & $\ovl{1}$\\
 $0$ & $1$ & $\ovl{2}$ & $0$ & $1$ & $0$\\
 $\ovl{1}$ & $0$ & $\ovl{1}$ & $1$ & $1$ & $0$\\
\hline
\end{longtable}

\vspace{0.5cm}

\renewcommand{\arraystretch}{1.1}
\begin{longtable}{cccccc}
\caption{Layer group $\mathrm{{LG}63}$ ($p4/mbm$).}\\
\label{tab:{SG}127_DoubleTR}\\
\hline\hline
\multicolumn{6}{l}{{$\bm{\mathrm{SI}: (C_{_m}) = (2)}$}}\\
\hline
\multicolumn{4}{l}{\footnotesize $\mathcal{BR}_L \boxminus \mathcal{BR}_{\widetilde{L}}: (4 \ovl{\Gamma}_{8} \boxminus 2 \ovl{\Gamma}_{6} \; ; \ovl{\rM}_{6}\ovl{\rM}_{7} \; ; \ovl{\rX}_{3}\ovl{\rX}_{4})$} & \multicolumn{2}{r}{\footnotesize $(n_b,\mathrm{N})=(4,20)$} \\
\hline
{\scriptsize $\mathrm{[{}^{1}\ovl{\mathrm{E}}_{\mathrm{1g}}{}^{2}\ovl{\mathrm{E}}_{\mathrm{1g}}]_{2a}}$} & {\scriptsize $\mathrm{[{}^{1}\ovl{\mathrm{E}}_{\mathrm{1u}}{}^{2}\ovl{\mathrm{E}}_{\mathrm{1u}}]_{2a}}$} & {\scriptsize $\mathrm{[{}^{1}\ovl{\mathrm{E}}_{\mathrm{2g}}{}^{2}\ovl{\mathrm{E}}_{\mathrm{2g}}]_{2a}}$} & {\scriptsize $\mathrm{[{}^{1}\ovl{\mathrm{E}}_{\mathrm{2u}}{}^{2}\ovl{\mathrm{E}}_{\mathrm{2u}}]_{2a}}$} & {\scriptsize $\mathrm{[\ovl{\mathrm{E}}_{\mathrm{g}}]_{2c}}$} & {\scriptsize $\mathrm{[\ovl{\mathrm{E}}_{\mathrm{u}}]_{2c}}$}\\
\hline
 $0$ & $\ovl{1}$ & $\ovl{1}$ & $1$ & $0$ & $2$\\
 $1$ & $0$ & $0$ & $2$ & $\ovl{2}$ & $0$\\
\hline
\multicolumn{4}{l}{\footnotesize $\mathcal{BR}_L \boxminus \mathcal{BR}_{\widetilde{L}}: (4 \ovl{\Gamma}_{8} \boxminus 2 \ovl{\Gamma}_{7} \; ; \ovl{\rM}_{6}\ovl{\rM}_{7} \; ; \ovl{\rX}_{3}\ovl{\rX}_{4})$} & \multicolumn{2}{r}{\footnotesize $(n_b,\mathrm{N})=(4,20)$} \\
\hline
\end{longtable}

\vspace{0.5cm}

\renewcommand{\arraystretch}{1.1}
\begin{longtable}{cccccc}
\caption{Layer group $\mathrm{{LG}64}$ ($p4/nmm$).}\\
\label{tab:{SG}129_DoubleTR}\\
\hline\hline
\multicolumn{6}{l}{{$\bm{\mathrm{SI}: (\delta_t) = (1)}$}}\\
\hline
\multicolumn{4}{l}{\footnotesize $\mathcal{BR}_L \boxminus \mathcal{BR}_{\widetilde{L}}: (3 \ovl{\Gamma}_{6} \boxminus \ovl{\Gamma}_{8} \; ; \ovl{\rM}_{5} \; ; \ovl{\rX}_{3}\ovl{\rX}_{4})$} & \multicolumn{2}{r}{\footnotesize $(n_b,\mathrm{N})=(4,20)$} \\
\hline
{\scriptsize $\mathrm{[\ovl{\mathrm{E}}_{\mathrm{1}}]_{2a}}$} & {\scriptsize $\mathrm{[\ovl{\mathrm{E}}_{\mathrm{2}}]_{2a}}$} & {\scriptsize $\mathrm{[\ovl{\mathrm{E}}_{\mathrm{1}}]_{2c}}$} & {\scriptsize $\mathrm{[\ovl{\mathrm{E}}_{\mathrm{2}}]_{2c}}$} & {\scriptsize $\mathrm{[{}^{1}\ovl{\mathrm{E}}_{\mathrm{g}}{}^{2}\ovl{\mathrm{E}}_{\mathrm{g}}]_{4d}}$} & {\scriptsize $\mathrm{[{}^{1}\ovl{\mathrm{E}}_{\mathrm{u}}{}^{2}\ovl{\mathrm{E}}_{\mathrm{u}}]_{4d}}$}\\
\hline
 $0$ & $2$ & $0$ & $1$ & $0$ & $\ovl{1}$\\
 $\ovl{1}$ & $1$ & $\ovl{1}$ & $0$ & $1$ & $0$\\
 $\ovl{2}$ & $0$ & $0$ & $1$ & $1$ & $0$\\
\hline
\end{longtable}

\vspace{0.5cm}

\renewcommand{\arraystretch}{1.1}
\begin{longtable}{cccccccc}
\caption{Layer group $\mathrm{{LG}66}$ ($p\bar{3}$).}\\
\label{tab:{SG}147_DoubleTR}\\
\hline\hline
\multicolumn{8}{l}{{$\bm{\mathrm{SI}: (\delta_t) = (1)}$}}\\
\hline
\multicolumn{6}{l}{\footnotesize $\mathcal{BR}_L \boxminus \mathcal{BR}_{\widetilde{L}}: (2 \ovl{\Gamma}_{5}\ovl{\Gamma}_{6} \boxminus \ovl{\Gamma}_{8}\ovl{\Gamma}_{9} \; ; \ovl{\rK}_{5}\ovl{\rK}_{6} \; ; \ovl{\rM}_{2}\ovl{\rM}_{2})$} & \multicolumn{2}{r}{\footnotesize $(n_b,\mathrm{N})=(2,10)$} \\
\hline
{\scriptsize $\mathrm{[{}^{1}\ovl{\mathrm{E}}_{\mathrm{g}}{}^{2}\ovl{\mathrm{E}}_{\mathrm{g}}]_{1a}}$} & {\scriptsize $\mathrm{[{}^{1}\ovl{\mathrm{E}}_{\mathrm{u}}{}^{2}\ovl{\mathrm{E}}_{\mathrm{u}}]_{1a}}$} & {\scriptsize $\mathrm{[\ovl{\mathrm{E}}_{\mathrm{g}}\ovl{\mathrm{E}}_{\mathrm{g}}]_{1a}}$} & {\scriptsize $\mathrm{[\ovl{\mathrm{E}}_{\mathrm{u}}\ovl{\mathrm{E}}_{\mathrm{u}}]_{1a}}$} & {\scriptsize $\mathrm{[{}^{1}\ovl{\mathrm{E}}{}^{2}\ovl{\mathrm{E}}]_{d}}$} & {\scriptsize $\mathrm{[\ovl{\mathrm{E}}\ovl{\mathrm{E}}]_{d}}$} & {\scriptsize $\mathrm{[\ovl{\mathrm{A}}_{\mathrm{g}}\ovl{\mathrm{A}}_{\mathrm{g}}]_{3e}}$} & {\scriptsize $\mathrm{[\ovl{\mathrm{A}}_{\mathrm{u}}\ovl{\mathrm{A}}_{\mathrm{u}}]_{3e}}$}\\
\hline
 $0$ & $\ovl{1}$ & $\ovl{1}$ & $0$ & $0$ & $0$ & $1$ & $0$\\
\hline
\multicolumn{6}{l}{\footnotesize $\mathcal{BR}_L \boxminus \mathcal{BR}_{\widetilde{L}}: (2 \ovl{\Gamma}_{8}\ovl{\Gamma}_{9} \boxminus \ovl{\Gamma}_{5}\ovl{\Gamma}_{6} \; ; \ovl{\rK}_{5}\ovl{\rK}_{6} \; ; \ovl{\rM}_{3}\ovl{\rM}_{3})$} & \multicolumn{2}{r}{\footnotesize $(n_b,\mathrm{N})=(2,10)$} \\
\hline
\end{longtable}

\vspace{0.5cm}

\renewcommand{\arraystretch}{1.1}
\begin{longtable}{cccccccc}
\caption{Layer group $\mathrm{{LG}71}$ ($p\bar{3}1m$).}\\
\label{tab:{SG}162_DoubleTR}\\
\hline\hline
\multicolumn{8}{l}{{$\bm{\mathrm{SI}: (\delta_t) = (1)}$}}\\
\hline
\multicolumn{6}{l}{\footnotesize $\mathcal{BR}_L \boxminus \mathcal{BR}_{\widetilde{L}}: (2 \ovl{\Gamma}_{9} \boxminus \ovl{\Gamma}_{8} \; ; \ovl{\rK}_{6} \; ; \ovl{\rM}_{3}\ovl{\rM}_{4})$} & \multicolumn{2}{r}{\footnotesize $(n_b,\mathrm{N})=(2,10)$} \\
\hline
{\scriptsize $\mathrm{[{}^{1}\ovl{\mathrm{E}}_{\mathrm{g}}{}^{2}\ovl{\mathrm{E}}_{\mathrm{g}}]_{1a}}$} & {\scriptsize $\mathrm{[{}^{1}\ovl{\mathrm{E}}_{\mathrm{u}}{}^{2}\ovl{\mathrm{E}}_{\mathrm{u}}]_{1a}}$} & {\scriptsize $\mathrm{[\ovl{\mathrm{E}}_{\mathrm{1g}}]_{1a}}$} & {\scriptsize $\mathrm{[\ovl{\mathrm{E}}_{\mathrm{1u}}]_{1a}}$} & {\scriptsize $\mathrm{[{}^{1}\ovl{\mathrm{E}}{}^{2}\ovl{\mathrm{E}}]_{2c}}$} & {\scriptsize $\mathrm{[\ovl{\mathrm{E}}_{\mathrm{1}}]_{2c}}$} & {\scriptsize $\mathrm{[{}^{1}\ovl{\mathrm{E}}_{\mathrm{g}}{}^{2}\ovl{\mathrm{E}}_{\mathrm{g}}]_{3f}}$} & {\scriptsize $\mathrm{[{}^{1}\ovl{\mathrm{E}}_{\mathrm{u}}{}^{2}\ovl{\mathrm{E}}_{\mathrm{u}}]_{3f}}$}\\
\hline
 $0$ & $\ovl{1}$ & $\ovl{1}$ & $0$ & $0$ & $0$ & $0$ & $1$\\
\hline
\multicolumn{6}{l}{\footnotesize $\mathcal{BR}_L \boxminus \mathcal{BR}_{\widetilde{L}}: (2 \ovl{\Gamma}_{8} \boxminus \ovl{\Gamma}_{9} \; ; \ovl{\rK}_{6} \; ; \ovl{\rM}_{5}\ovl{\rM}_{6})$} & \multicolumn{2}{r}{\footnotesize $(n_b,\mathrm{N})=(2,10)$} \\
\hline
\end{longtable}

\vspace{0.5cm}

\renewcommand{\arraystretch}{1.1}
\begin{longtable}{cccccccc}
\caption{Layer group $\mathrm{{LG}72}$ ($p\bar{3}m1$).}\\
\label{tab:{SG}164_DoubleTR}\\
\hline\hline
\multicolumn{8}{l}{{$\bm{\mathrm{SI}: (\delta_t) = (1)}$}}\\
\hline
\multicolumn{6}{l}{\footnotesize $\mathcal{BR}_L \boxminus \mathcal{BR}_{\widetilde{L}}: (2 \ovl{\Gamma}_{9} \boxminus \ovl{\Gamma}_{8} \; ; \ovl{\rK}_{6} \; ; \ovl{\rM}_{3}\ovl{\rM}_{4})$} & \multicolumn{2}{r}{\footnotesize $(n_b,\mathrm{N})=(2,10)$} \\
\hline
{\scriptsize $\mathrm{[{}^{1}\ovl{\mathrm{E}}_{\mathrm{g}}{}^{2}\ovl{\mathrm{E}}_{\mathrm{g}}]_{1a}}$} & {\scriptsize $\mathrm{[{}^{1}\ovl{\mathrm{E}}_{\mathrm{u}}{}^{2}\ovl{\mathrm{E}}_{\mathrm{u}}]_{1a}}$} & {\scriptsize $\mathrm{[\ovl{\mathrm{E}}_{\mathrm{1g}}]_{1a}}$} & {\scriptsize $\mathrm{[\ovl{\mathrm{E}}_{\mathrm{1u}}]_{1a}}$} & {\scriptsize $\mathrm{[{}^{1}\ovl{\mathrm{E}}{}^{2}\ovl{\mathrm{E}}]_{2d}}$} & {\scriptsize $\mathrm{[\ovl{\mathrm{E}}_{\mathrm{1}}]_{2d}}$} & {\scriptsize $\mathrm{[{}^{1}\ovl{\mathrm{E}}_{\mathrm{g}}{}^{2}\ovl{\mathrm{E}}_{\mathrm{g}}]_{3e}}$} & {\scriptsize $\mathrm{[{}^{1}\ovl{\mathrm{E}}_{\mathrm{u}}{}^{2}\ovl{\mathrm{E}}_{\mathrm{u}}]_{3e}}$}\\
\hline
 $0$ & $\ovl{1}$ & $\ovl{1}$ & $0$ & $0$ & $0$ & $0$ & $1$\\
\hline
\multicolumn{6}{l}{\footnotesize $\mathcal{BR}_L \boxminus \mathcal{BR}_{\widetilde{L}}: (2 \ovl{\Gamma}_{8} \boxminus \ovl{\Gamma}_{9} \; ; \ovl{\rK}_{6} \; ; \ovl{\rM}_{5}\ovl{\rM}_{6})$} & \multicolumn{2}{r}{\footnotesize $(n_b,\mathrm{N})=(2,10)$} \\
\hline
\end{longtable}

\vspace{0.5cm}

\renewcommand{\arraystretch}{1.1}
\begin{longtable}{ccccccccc}
\caption{Layer group $\mathrm{{LG}74}$ ($p\bar{6}$).}\\
\label{tab:{SG}174_DoubleTR}\\
\hline\hline
\multicolumn{9}{l}{{$\bm{\mathrm{SI}: (C_{_m}) = (1)}$}}\\
\hline
\multicolumn{7}{l}{\footnotesize $\mathcal{BR}_L \boxminus \mathcal{BR}_{\widetilde{L}}: (2 \ovl{\Gamma}_{12}\ovl{\Gamma}_{9} \boxminus \ovl{\Gamma}_{10}\ovl{\Gamma}_{11} \; ; \ovl{\rK}_{7} \oplus \ovl{\rK}_{8} \; ; \ovl{\rKA}_{7} \oplus \ovl{\rKA}_{8} \; ; \ovl{\rM}_{3}\ovl{\rM}_{4})$} & \multicolumn{2}{r}{\footnotesize $(n_b,\mathrm{N})=(2,6)$} \\
\hline
{\scriptsize $\mathrm{[{}^{1}\ovl{\mathrm{E}}_{\mathrm{1}}{}^{2}\ovl{\mathrm{E}}_{\mathrm{1}}]_{1a}}$} & {\scriptsize $\mathrm{[{}^{1}\ovl{\mathrm{E}}_{\mathrm{2}}{}^{2}\ovl{\mathrm{E}}_{\mathrm{2}}]_{1a}}$} & {\scriptsize $\mathrm{[{}^{1}\ovl{\mathrm{E}}_{\mathrm{3}}{}^{2}\ovl{\mathrm{E}}_{\mathrm{3}}]_{1a}}$} & {\scriptsize $\mathrm{[{}^{1}\ovl{\mathrm{E}}_{\mathrm{1}}{}^{2}\ovl{\mathrm{E}}_{\mathrm{1}}]_{1c}}$} & {\scriptsize $\mathrm{[{}^{1}\ovl{\mathrm{E}}_{\mathrm{2}}{}^{2}\ovl{\mathrm{E}}_{\mathrm{2}}]_{1c}}$} & {\scriptsize $\mathrm{[{}^{1}\ovl{\mathrm{E}}_{\mathrm{3}}{}^{2}\ovl{\mathrm{E}}_{\mathrm{3}}]_{1c}}$} & {\scriptsize $\mathrm{[{}^{1}\ovl{\mathrm{E}}_{\mathrm{1}}{}^{2}\ovl{\mathrm{E}}_{\mathrm{1}}]_{1e}}$} & {\scriptsize $\mathrm{[{}^{1}\ovl{\mathrm{E}}_{\mathrm{2}}{}^{2}\ovl{\mathrm{E}}_{\mathrm{2}}]_{1e}}$} & {\scriptsize $\mathrm{[{}^{1}\ovl{\mathrm{E}}_{\mathrm{3}}{}^{2}\ovl{\mathrm{E}}_{\mathrm{3}}]_{1e}}$}\\
\hline
 $0$ & $0$ & $\ovl{1}$ & $0$ & $1$ & $0$ & $0$ & $1$ & $0$\\
\hline
\multicolumn{7}{l}{\footnotesize $\mathcal{BR}_L \boxminus \mathcal{BR}_{\widetilde{L}}: (2 \ovl{\Gamma}_{7}\ovl{\Gamma}_{8} \boxminus \ovl{\Gamma}_{12}\ovl{\Gamma}_{9} \; ; \ovl{\rK}_{8} \oplus \ovl{\rK}_{9} \; ; \ovl{\rKA}_{8} \oplus \ovl{\rKA}_{9} \; ; \ovl{\rM}_{3}\ovl{\rM}_{4})$} & \multicolumn{2}{r}{\footnotesize $(n_b,\mathrm{N})=(2,6)$} \\
\hline
\end{longtable}

\vspace{0.5cm}

\renewcommand{\arraystretch}{1.1}
\begin{longtable}{ccccccccccc}
\caption{Layer group $\mathrm{{LG}75}$ ($p6/m$).}\\
\label{tab:{SG}175_DoubleTR}\\
\hline\hline
\multicolumn{11}{l}{{$\bm{\mathrm{SI}: (C_{_m}) = (1)}$}}\\
\hline
\multicolumn{9}{l}{\footnotesize $\mathcal{BR}_L \boxminus \mathcal{BR}_{\widetilde{L}}: (2 \ovl{\Gamma}_{16}\ovl{\Gamma}_{17} \boxminus \ovl{\Gamma}_{10}\ovl{\Gamma}_{11} \; ; \ovl{\rK}_{7}\ovl{\rK}_{8} \; ; \ovl{\rM}_{3}\ovl{\rM}_{4})$} & \multicolumn{2}{r}{\footnotesize $(n_b,\mathrm{N})=(2,14)$} \\
\hline
{\scriptsize $\mathrm{[{}^{1}\ovl{\mathrm{E}}_{\mathrm{1g}}{}^{2}\ovl{\mathrm{E}}_{\mathrm{1g}}]_{1a}}$} & {\scriptsize $\mathrm{[{}^{1}\ovl{\mathrm{E}}_{\mathrm{1u}}{}^{2}\ovl{\mathrm{E}}_{\mathrm{1u}}]_{1a}}$} & {\scriptsize $\mathrm{[{}^{1}\ovl{\mathrm{E}}_{\mathrm{2g}}{}^{2}\ovl{\mathrm{E}}_{\mathrm{2g}}]_{1a}}$} & {\scriptsize $\mathrm{[{}^{1}\ovl{\mathrm{E}}_{\mathrm{2u}}{}^{2}\ovl{\mathrm{E}}_{\mathrm{2u}}]_{1a}}$} & {\scriptsize $\mathrm{[{}^{1}\ovl{\mathrm{E}}_{\mathrm{3g}}{}^{2}\ovl{\mathrm{E}}_{\mathrm{3g}}]_{1a}}$} & {\scriptsize $\mathrm{[{}^{1}\ovl{\mathrm{E}}_{\mathrm{3u}}{}^{2}\ovl{\mathrm{E}}_{\mathrm{3u}}]_{1a}}$} & {\scriptsize $\mathrm{[{}^{1}\ovl{\mathrm{E}}_{\mathrm{1}}{}^{2}\ovl{\mathrm{E}}_{\mathrm{1}}]_{2c}}$} & {\scriptsize $\mathrm{[{}^{1}\ovl{\mathrm{E}}_{\mathrm{2}}{}^{2}\ovl{\mathrm{E}}_{\mathrm{2}}]_{2c}}$} & {\scriptsize $\mathrm{[{}^{1}\ovl{\mathrm{E}}_{\mathrm{3}}{}^{2}\ovl{\mathrm{E}}_{\mathrm{3}}]_{2c}}$} & {\scriptsize $\mathrm{[{}^{1}\ovl{\mathrm{E}}_{\mathrm{g}}{}^{2}\ovl{\mathrm{E}}_{\mathrm{g}}]_{3f}}$} & {\scriptsize $\mathrm{[{}^{1}\ovl{\mathrm{E}}_{\mathrm{u}}{}^{2}\ovl{\mathrm{E}}_{\mathrm{u}}]_{3f}}$}\\
\hline
 $1$ & $0$ & $0$ & $0$ & $0$ & $1$ & $0$ & $1$ & $0$ & $\ovl{1}$ & $0$\\
\hline
\end{longtable}

\vspace{0.5cm}

\renewcommand{\arraystretch}{1.1}
\begin{longtable}{ccccccccc}
\caption{Layer group $\mathrm{{LG}78}$ ($p\bar{6}m2$).}\\
\label{tab:{SG}187_DoubleTR}\\
\hline\hline
\multicolumn{9}{l}{{$\bm{\mathrm{SI}: (C_{_m}) = (1)}$}}\\
\hline
\multicolumn{7}{l}{\footnotesize $\mathcal{BR}_L \boxminus \mathcal{BR}_{\widetilde{L}}: (2 \ovl{\Gamma}_{9} \boxminus \ovl{\Gamma}_{8} \; ; \ovl{\rK}_{7} \oplus \ovl{\rK}_{8} \; ; \ovl{\rM}_{5})$} & \multicolumn{2}{r}{\footnotesize $(n_b,\mathrm{N})=(2,6)$} \\
\hline
{\scriptsize $\mathrm{[\ovl{\mathrm{E}}_{\mathrm{1}}]_{1a}}$} & {\scriptsize $\mathrm{[\ovl{\mathrm{E}}_{\mathrm{2}}]_{1a}}$} & {\scriptsize $\mathrm{[\ovl{\mathrm{E}}_{\mathrm{3}}]_{1a}}$} & {\scriptsize $\mathrm{[\ovl{\mathrm{E}}_{\mathrm{1}}]_{1c}}$} & {\scriptsize $\mathrm{[\ovl{\mathrm{E}}_{\mathrm{2}}]_{1c}}$} & {\scriptsize $\mathrm{[\ovl{\mathrm{E}}_{\mathrm{3}}]_{1c}}$} & {\scriptsize $\mathrm{[\ovl{\mathrm{E}}_{\mathrm{1}}]_{1e}}$} & {\scriptsize $\mathrm{[\ovl{\mathrm{E}}_{\mathrm{2}}]_{1e}}$} & {\scriptsize $\mathrm{[\ovl{\mathrm{E}}_{\mathrm{3}}]_{1e}}$}\\
\hline
 $0$ & $\ovl{1}$ & $0$ & $1$ & $0$ & $0$ & $1$ & $0$ & $0$\\
\hline
\multicolumn{7}{l}{\footnotesize $\mathcal{BR}_L \boxminus \mathcal{BR}_{\widetilde{L}}: (2 \ovl{\Gamma}_{8} \boxminus \ovl{\Gamma}_{7} \; ; \ovl{\rK}_{8} \oplus \ovl{\rK}_{9} \; ; \ovl{\rM}_{5})$} & \multicolumn{2}{r}{\footnotesize $(n_b,\mathrm{N})=(2,6)$} \\
\hline
{\scriptsize $\mathrm{[\ovl{\mathrm{E}}_{\mathrm{1}}]_{1a}}$} & {\scriptsize $\mathrm{[\ovl{\mathrm{E}}_{\mathrm{2}}]_{1a}}$} & {\scriptsize $\mathrm{[\ovl{\mathrm{E}}_{\mathrm{3}}]_{1a}}$} & {\scriptsize $\mathrm{[\ovl{\mathrm{E}}_{\mathrm{1}}]_{1c}}$} & {\scriptsize $\mathrm{[\ovl{\mathrm{E}}_{\mathrm{2}}]_{1c}}$} & {\scriptsize $\mathrm{[\ovl{\mathrm{E}}_{\mathrm{3}}]_{1c}}$} & {\scriptsize $\mathrm{[\ovl{\mathrm{E}}_{\mathrm{1}}]_{1e}}$} & {\scriptsize $\mathrm{[\ovl{\mathrm{E}}_{\mathrm{2}}]_{1e}}$} & {\scriptsize $\mathrm{[\ovl{\mathrm{E}}_{\mathrm{3}}]_{1e}}$}\\
\hline
\end{longtable}

\vspace{0.5cm}

\renewcommand{\arraystretch}{1.1}
\begin{longtable}{cccccc}
\caption{Layer group $\mathrm{{LG}79}$ ($p\bar{6}2m$).}\\
\label{tab:{SG}189_DoubleTR}\\
\hline\hline
\multicolumn{6}{l}{{$\bm{\mathrm{SI}: (C_{_m}) = (1)}$}}\\
\hline
\multicolumn{4}{l}{\footnotesize $\mathcal{BR}_L \boxminus \mathcal{BR}_{\widetilde{L}}: (2 \ovl{\Gamma}_{8} \boxminus \ovl{\Gamma}_{9} \; ; \ovl{\rK}_{7} \; ; \ovl{\rKA}_{7} \; ; \ovl{\rM}_{5})$} & \multicolumn{2}{r}{\footnotesize $(n_b,\mathrm{N})=(2,6)$} \\
\hline
{\scriptsize $\mathrm{[\ovl{\mathrm{E}}_{\mathrm{1}}]_{1a}}$} & {\scriptsize $\mathrm{[\ovl{\mathrm{E}}_{\mathrm{2}}]_{1a}}$} & {\scriptsize $\mathrm{[\ovl{\mathrm{E}}_{\mathrm{3}}]_{1a}}$} & {\scriptsize $\mathrm{[{}^{1}\ovl{\mathrm{E}}_{\mathrm{1}}{}^{2}\ovl{\mathrm{E}}_{\mathrm{1}}]_{2c}}$} & {\scriptsize $\mathrm{[{}^{1}\ovl{\mathrm{E}}_{\mathrm{2}}{}^{2}\ovl{\mathrm{E}}_{\mathrm{2}}]_{2c}}$} & {\scriptsize $\mathrm{[{}^{1}\ovl{\mathrm{E}}_{\mathrm{3}}{}^{2}\ovl{\mathrm{E}}_{\mathrm{3}}]_{2c}}$}\\
\hline
 $\ovl{1}$ & $0$ & $0$ & $0$ & $1$ & $0$\\
\hline
\multicolumn{4}{l}{\footnotesize $\mathcal{BR}_L \boxminus \mathcal{BR}_{\widetilde{L}}: (2 \ovl{\Gamma}_{9} \boxminus \ovl{\Gamma}_{7} \; ; \ovl{\rK}_{8} \; ; \ovl{\rKA}_{8} \; ; \ovl{\rM}_{5})$} & \multicolumn{2}{r}{\footnotesize $(n_b,\mathrm{N})=(2,6)$} \\
\hline
{\scriptsize $\mathrm{[\ovl{\mathrm{E}}_{\mathrm{1}}]_{1a}}$} & {\scriptsize $\mathrm{[\ovl{\mathrm{E}}_{\mathrm{2}}]_{1a}}$} & {\scriptsize $\mathrm{[\ovl{\mathrm{E}}_{\mathrm{3}}]_{1a}}$} & {\scriptsize $\mathrm{[{}^{1}\ovl{\mathrm{E}}_{\mathrm{1}}{}^{2}\ovl{\mathrm{E}}_{\mathrm{1}}]_{2c}}$} & {\scriptsize $\mathrm{[{}^{1}\ovl{\mathrm{E}}_{\mathrm{2}}{}^{2}\ovl{\mathrm{E}}_{\mathrm{2}}]_{2c}}$} & {\scriptsize $\mathrm{[{}^{1}\ovl{\mathrm{E}}_{\mathrm{3}}{}^{2}\ovl{\mathrm{E}}_{\mathrm{3}}]_{2c}}$}\\
\hline
\end{longtable}

\vspace{0.5cm}

\renewcommand{\arraystretch}{1.1}
\begin{longtable}{ccccccccccc}
\caption{Layer group $\mathrm{{LG}80}$ ($p6/mmm$).}\\
\label{tab:{SG}191_DoubleTR}\\
\hline\hline
\multicolumn{11}{l}{{$\bm{\mathrm{SI}: (C_{_m}) = (1)}$}}\\
\hline
\multicolumn{9}{l}{\footnotesize $\mathcal{BR}_L \boxminus \mathcal{BR}_{\widetilde{L}}: (2 \ovl{\Gamma}_{12} \boxminus \ovl{\Gamma}_{9} \; ; \ovl{\rK}_{7} \; ; \ovl{\rM}_{5})$} & \multicolumn{2}{r}{\footnotesize $(n_b,\mathrm{N})=(2,14)$} \\
\hline
{\scriptsize $\mathrm{[\ovl{\mathrm{E}}_{\mathrm{1g}}]_{1a}}$} & {\scriptsize $\mathrm{[\ovl{\mathrm{E}}_{\mathrm{1u}}]_{1a}}$} & {\scriptsize $\mathrm{[\ovl{\mathrm{E}}_{\mathrm{2g}}]_{1a}}$} & {\scriptsize $\mathrm{[\ovl{\mathrm{E}}_{\mathrm{2u}}]_{1a}}$} & {\scriptsize $\mathrm{[\ovl{\mathrm{E}}_{\mathrm{3g}}]_{1a}}$} & {\scriptsize $\mathrm{[\ovl{\mathrm{E}}_{\mathrm{3u}}]_{1a}}$} & {\scriptsize $\mathrm{[\ovl{\mathrm{E}}_{\mathrm{1}}]_{2c}}$} & {\scriptsize $\mathrm{[\ovl{\mathrm{E}}_{\mathrm{2}}]_{2c}}$} & {\scriptsize $\mathrm{[\ovl{\mathrm{E}}_{\mathrm{3}}]_{2c}}$} & {\scriptsize $\mathrm{[\ovl{\mathrm{E}}_{\mathrm{g}}]_{3f}}$} & {\scriptsize $\mathrm{[\ovl{\mathrm{E}}_{\mathrm{u}}]_{3f}}$}\\
\hline
 $0$ & $1$ & $0$ & $0$ & $1$ & $0$ & $0$ & $1$ & $0$ & $\ovl{1}$ & $0$\\
\hline
\multicolumn{9}{l}{\footnotesize $\mathcal{BR}_L \boxminus \mathcal{BR}_{\widetilde{L}}: (2 \ovl{\Gamma}_{11} \boxminus \ovl{\Gamma}_{7} \; ; \ovl{\rK}_{8} \; ; \ovl{\rM}_{5})$} & \multicolumn{2}{r}{\footnotesize $(n_b,\mathrm{N})=(2,14)$} \\
\hline
\end{longtable}

\vspace{0.5cm}
\clearpage

\subsection{3D Double group with TRS}
\label{sec:3D-double-T}


\renewcommand{\arraystretch}{1.1}
\begin{longtable}{cccccccccccc}
\caption{Space group $\mathrm{{SG}200}$ ($Pm\bar{3}$).}\\
\label{tab:{SG}200_DoubleTR}\\
\hline\hline
\multicolumn{12}{l}{{$\bm{\mathrm{SI}: (z_{2w}, z_{4}) = (1, 0)}$}}\\
\hline
\multicolumn{10}{l}{\footnotesize $\mathcal{BR}_L \boxminus \mathcal{BR}_{\widetilde{L}}: (2 \ovl{\Gamma}_{5} \boxminus \ovl{\Gamma}_{8} \; ; 2 \ovl{\rR}_{8} \boxminus \ovl{\rR}_{5} \; ; \ovl{\rM}_{6} \; ; \ovl{\rX}_{5})$} & \multicolumn{2}{r}{\footnotesize $(n_b,\mathrm{N})=(2,14)$} \\
\hline
{\footnotesize $[{}^{1}\ovl{\mathrm{F}}_{\mathrm{g}}{}^{2}\ovl{\mathrm{F}}_{\mathrm{g}}]_{1a}$} & {\footnotesize $[{}^{1}\ovl{\mathrm{F}}_{\mathrm{u}}{}^{2}\ovl{\mathrm{F}}_{\mathrm{u}}]_{1a}$} & {\footnotesize $[\ovl{\mathrm{E}}_{\mathrm{g}}]_{1a}$} & {\footnotesize $[\ovl{\mathrm{E}}_{\mathrm{u}}]_{1a}$} & {\footnotesize $[{}^{1}\ovl{\mathrm{F}}_{\mathrm{g}}{}^{2}\ovl{\mathrm{F}}_{\mathrm{g}}]_{1b}$} & {\footnotesize $[{}^{1}\ovl{\mathrm{F}}_{\mathrm{u}}{}^{2}\ovl{\mathrm{F}}_{\mathrm{u}}]_{1b}$} & {\footnotesize $[\ovl{\mathrm{E}}_{\mathrm{g}}]_{1b}$} & {\footnotesize $[\ovl{\mathrm{E}}_{\mathrm{u}}]_{1b}$} & {\footnotesize $[\ovl{\mathrm{E}}_{\mathrm{g}}]_{3c}$} & {\footnotesize $[\ovl{\mathrm{E}}_{\mathrm{u}}]_{3c}$} & {\footnotesize $[\ovl{\mathrm{E}}_{\mathrm{g}}]_{3d}$} & {\footnotesize $[\ovl{\mathrm{E}}_{\mathrm{u}}]_{3d}$}\\
\hline
 $0$ & $0$ & $0$ & $0$ & $0$ & $1$ & $2$ & $0$ & $0$ & $0$ & $0$ & $\ovl{1}$\\
 $0$ & $0$ & $0$ & $0$ & $\ovl{1}$ & $0$ & $1$ & $\ovl{1}$ & $0$ & $0$ & $1$ & $0$\\
\hline
\multicolumn{10}{l}{\footnotesize $\mathcal{BR}_L \boxminus \mathcal{BR}_{\widetilde{L}}: (2 \ovl{\Gamma}_{8} \boxminus \ovl{\Gamma}_{5} \; ; 2 \ovl{\rR}_{5} \boxminus \ovl{\rR}_{8} \; ; \ovl{\rM}_{5} \; ; \ovl{\rX}_{6})$} & \multicolumn{2}{r}{\footnotesize $(n_b,\mathrm{N})=(2,14)$} \\
\hline
\end{longtable}

\vspace{0.5cm}


\renewcommand{\arraystretch}{1.1}
\begin{longtable}{ccccccccccc}
\caption{Space group $\mathrm{{SG}201}$ ($Pn\bar{3}$).}\\
\label{tab:{SG}201_DoubleTR}\\
\hline\hline
\multicolumn{11}{l}{{$\bm{\mathrm{SI}: (z_{2w}, z_{4}) = (1, 0)}$}}\\
\hline
\multicolumn{9}{l}{\footnotesize $\mathcal{BR}_L \boxminus \mathcal{BR}_{\widetilde{L}}: (3 \ovl{\Gamma}_{8} \boxminus \ovl{\Gamma}_{5} \; ; 3 \ovl{\rR}_{5} \boxminus \ovl{\rR}_{8} \; ; \ovl{\rM}_{3}\ovl{\rM}_{4} \; ; \ovl{\rX}_{3}\ovl{\rX}_{4})$} & \multicolumn{2}{r}{\footnotesize $(n_b,\mathrm{N})=(4,28)$} \\
\hline
{\footnotesize $[{}^{1}\ovl{\mathrm{F}}{}^{2}\ovl{\mathrm{F}}]_{2a}$} & {\footnotesize $[\ovl{\mathrm{E}}]_{2a}$} & {\footnotesize $[{}^{1}\ovl{\mathrm{E}}_{\mathrm{g}}{}^{2}\ovl{\mathrm{E}}_{\mathrm{g}}]_{4b}$} & {\footnotesize $[{}^{1}\ovl{\mathrm{E}}_{\mathrm{u}}{}^{2}\ovl{\mathrm{E}}_{\mathrm{u}}]_{4b}$} & {\footnotesize $[\ovl{\mathrm{E}}_{\mathrm{g}}\ovl{\mathrm{E}}_{\mathrm{g}}]_{4b}$} & {\footnotesize $[\ovl{\mathrm{E}}_{\mathrm{u}}\ovl{\mathrm{E}}_{\mathrm{u}}]_{4b}$} & {\footnotesize $[{}^{1}\ovl{\mathrm{E}}_{\mathrm{g}}{}^{2}\ovl{\mathrm{E}}_{\mathrm{g}}]_{4c}$} & {\footnotesize $[{}^{1}\ovl{\mathrm{E}}_{\mathrm{u}}{}^{2}\ovl{\mathrm{E}}_{\mathrm{u}}]_{4c}$} & {\footnotesize $[\ovl{\mathrm{E}}_{\mathrm{g}}\ovl{\mathrm{E}}_{\mathrm{g}}]_{4c}$} & {\footnotesize $[\ovl{\mathrm{E}}_{\mathrm{u}}\ovl{\mathrm{E}}_{\mathrm{u}}]_{4c}$} & {\footnotesize $[\ovl{\mathrm{E}}]_{6d}$}\\
\hline
 $0$ & $\ovl{1}$ & $0$ & $0$ & $0$ & $0$ & $0$ & $2$ & $0$ & $\ovl{1}$ & $0$\\
\hline
\multicolumn{9}{l}{\footnotesize $\mathcal{BR}_L \boxminus \mathcal{BR}_{\widetilde{L}}: (3 \ovl{\Gamma}_{5} \boxminus \ovl{\Gamma}_{8} \; ; 3 \ovl{\rR}_{8} \boxminus \ovl{\rR}_{5} \; ; \ovl{\rM}_{3}\ovl{\rM}_{4} \; ; \ovl{\rX}_{3}\ovl{\rX}_{4})$} & \multicolumn{2}{r}{\footnotesize $(n_b,\mathrm{N})=(4,28)$} \\
\hline
\end{longtable}

\vspace{0.5cm}

\renewcommand{\arraystretch}{1.1}
\begin{longtable}{cccccccccccc}
\caption{Space group $\mathrm{{SG}202}$ ($Fm\bar{3}$).}\\
\label{tab:{SG}202_DoubleTR}\\
\hline\hline
\multicolumn{12}{l}{{$\bm{\mathrm{SI}: (z_{4}) = (1)}$}}\\
\hline
\multicolumn{10}{l}{\footnotesize $\mathcal{BR}_L \boxminus \mathcal{BR}_{\widetilde{L}}: (4 \ovl{\Gamma}_{5} \boxminus \ovl{\Gamma}_{8} \; ; 3 \ovl{\rX}_{6} \; ; 3 \ovl{\rL}_{5}\ovl{\rL}_{6} \; ; 3 \ovl{\rW}_{5})$} & \multicolumn{2}{r}{\footnotesize $(n_b,\mathrm{N})=(6,26)$} \\
\hline
{\footnotesize $[{}^{1}\ovl{\mathrm{F}}_{\mathrm{g}}{}^{2}\ovl{\mathrm{F}}_{\mathrm{g}}]_{4a}$} & {\footnotesize $[{}^{1}\ovl{\mathrm{F}}_{\mathrm{u}}{}^{2}\ovl{\mathrm{F}}_{\mathrm{u}}]_{4a}$} & {\footnotesize $[\ovl{\mathrm{E}}_{\mathrm{g}}]_{4a}$} & {\footnotesize $[\ovl{\mathrm{E}}_{\mathrm{u}}]_{4a}$} & {\footnotesize $[{}^{1}\ovl{\mathrm{F}}_{\mathrm{g}}{}^{2}\ovl{\mathrm{F}}_{\mathrm{g}}]_{4b}$} & {\footnotesize $[{}^{1}\ovl{\mathrm{F}}_{\mathrm{u}}{}^{2}\ovl{\mathrm{F}}_{\mathrm{u}}]_{4b}$} & {\footnotesize $[\ovl{\mathrm{E}}_{\mathrm{g}}]_{4b}$} & {\footnotesize $[\ovl{\mathrm{E}}_{\mathrm{u}}]_{4b}$} & {\footnotesize $[{}^{1}\ovl{\mathrm{F}}{}^{2}\ovl{\mathrm{F}}]_{8c}$} & {\footnotesize $[\ovl{\mathrm{E}}]_{8c}$} & {\footnotesize $[{}^{1}\ovl{\mathrm{E}}_{\mathrm{g}}{}^{2}\ovl{\mathrm{E}}_{\mathrm{g}}]_{24d}$} & {\footnotesize $[{}^{1}\ovl{\mathrm{E}}_{\mathrm{u}}{}^{2}\ovl{\mathrm{E}}_{\mathrm{u}}]_{24d}$}\\
\hline
 $\ovl{1}$ & $0$ & $2$ & $\ovl{1}$ & $\ovl{1}$ & $0$ & $0$ & $0$ & $0$ & $0$ & $1$ & $0$\\
\hline
\multicolumn{10}{l}{\footnotesize $\mathcal{BR}_L \boxminus \mathcal{BR}_{\widetilde{L}}: (4 \ovl{\Gamma}_{5} \boxminus \ovl{\Gamma}_{8} \; ; 3 \ovl{\rX}_{6} \; ; 2 \ovl{\rL}_{5}\ovl{\rL}_{6} \oplus \ovl{\rL}_{8}\ovl{\rL}_{9} \; ; 3 \ovl{\rW}_{5})$} & \multicolumn{2}{r}{\footnotesize $(n_b,\mathrm{N})=(6,26)$} \\
\hline
\end{longtable}

\vspace{0.5cm}

\renewcommand{\arraystretch}{1.1}
\begin{longtable}{cccccccccccc}
\caption{Space group $\mathrm{{SG}203}$ ($Fd\bar{3}$).}\\
\label{tab:{SG}203_DoubleTR}\\
\hline\hline
\multicolumn{12}{l}{{$\bm{\mathrm{SI}: (z_{4}) = (1)}$}}\\
\hline
\multicolumn{10}{l}{\footnotesize $\mathcal{BR}_L \boxminus \mathcal{BR}_{\widetilde{L}}: (3 \ovl{\Gamma}_{5} \boxminus \ovl{\Gamma}_{8} \; ; \ovl{\rX}_{3}\ovl{\rX}_{4} \; ; 2 \ovl{\rL}_{5}\ovl{\rL}_{6} \; ; \ovl{\rW}_{2}\ovl{\rW}_{4} \oplus \ovl{\rW}_{3}\ovl{\rW}_{5})$} & \multicolumn{2}{r}{\footnotesize $(n_b,\mathrm{N})=(4,28)$} \\
\hline
{\footnotesize $[{}^{1}\ovl{\mathrm{F}}{}^{2}\ovl{\mathrm{F}}]_{8a}$} & {\footnotesize $[\ovl{\mathrm{E}}]_{8a}$} & {\footnotesize $[{}^{1}\ovl{\mathrm{F}}{}^{2}\ovl{\mathrm{F}}]_{8b}$} & {\footnotesize $[\ovl{\mathrm{E}}]_{8b}$} & {\footnotesize $[{}^{1}\ovl{\mathrm{E}}_{\mathrm{g}}{}^{2}\ovl{\mathrm{E}}_{\mathrm{g}}]_{16c}$} & {\footnotesize $[{}^{1}\ovl{\mathrm{E}}_{\mathrm{u}}{}^{2}\ovl{\mathrm{E}}_{\mathrm{u}}]_{16c}$} & {\footnotesize $[\ovl{\mathrm{E}}_{\mathrm{g}}\ovl{\mathrm{E}}_{\mathrm{g}}]_{16c}$} & {\footnotesize $[\ovl{\mathrm{E}}_{\mathrm{u}}\ovl{\mathrm{E}}_{\mathrm{u}}]_{16c}$} & {\footnotesize $[{}^{1}\ovl{\mathrm{E}}_{\mathrm{g}}{}^{2}\ovl{\mathrm{E}}_{\mathrm{g}}]_{16d}$} & {\footnotesize $[{}^{1}\ovl{\mathrm{E}}_{\mathrm{u}}{}^{2}\ovl{\mathrm{E}}_{\mathrm{u}}]_{16d}$} & {\footnotesize $[\ovl{\mathrm{E}}_{\mathrm{g}}\ovl{\mathrm{E}}_{\mathrm{g}}]_{16d}$} & {\footnotesize $[\ovl{\mathrm{E}}_{\mathrm{u}}\ovl{\mathrm{E}}_{\mathrm{u}}]_{16d}$}\\
\hline
 $0$ & $0$ & $0$ & $\ovl{1}$ & $1$ & $0$ & $\ovl{1}$ & $0$ & $1$ & $0$ & $0$ & $0$\\
 $0$ & $\ovl{1}$ & $0$ & $0$ & $1$ & $0$ & $\ovl{1}$ & $0$ & $1$ & $0$ & $0$ & $0$\\
\hline
\end{longtable}

\vspace{0.5cm}

\renewcommand{\arraystretch}{1.1}
\begin{longtable}{cccccccccc}
\caption{Space group $\mathrm{{SG}204}$ ($Im\bar{3}$).}\\
\label{tab:{SG}204_DoubleTR}\\
\hline\hline
\multicolumn{10}{l}{{$\bm{\mathrm{SI}: (z_{2w}, z_{4}) = (1, 0)}$}}\\
\hline
\multicolumn{8}{l}{\footnotesize $\mathcal{BR}_L \boxminus \mathcal{BR}_{\widetilde{L}}: (3 \ovl{\Gamma}_{8} \boxminus \ovl{\Gamma}_{5} \; ; 3 \ovl{\rH}_{5} \boxminus \ovl{\rH}_{8} \; ; 2 \ovl{\rP}_{5} \; ; \ovl{\rN}_{3}\ovl{\rN}_{4} \oplus \ovl{\rN}_{5}\ovl{\rN}_{6})$} & \multicolumn{2}{r}{\footnotesize $(n_b,\mathrm{N})=(4,28)$} \\
\hline
{\footnotesize $[{}^{1}\ovl{\mathrm{F}}_{\mathrm{g}}{}^{2}\ovl{\mathrm{F}}_{\mathrm{g}}]_{2a}$} & {\footnotesize $[{}^{1}\ovl{\mathrm{F}}_{\mathrm{u}}{}^{2}\ovl{\mathrm{F}}_{\mathrm{u}}]_{2a}$} & {\footnotesize $[\ovl{\mathrm{E}}_{\mathrm{g}}]_{2a}$} & {\footnotesize $[\ovl{\mathrm{E}}_{\mathrm{u}}]_{2a}$} & {\footnotesize $[\ovl{\mathrm{E}}_{\mathrm{g}}]_{6b}$} & {\footnotesize $[\ovl{\mathrm{E}}_{\mathrm{u}}]_{6b}$} & {\footnotesize $[{}^{1}\ovl{\mathrm{E}}_{\mathrm{g}}{}^{2}\ovl{\mathrm{E}}_{\mathrm{g}}]_{8c}$} & {\footnotesize $[{}^{1}\ovl{\mathrm{E}}_{\mathrm{u}}{}^{2}\ovl{\mathrm{E}}_{\mathrm{u}}]_{8c}$} & {\footnotesize $[\ovl{\mathrm{E}}_{\mathrm{g}}\ovl{\mathrm{E}}_{\mathrm{g}}]_{8c}$} & {\footnotesize $[\ovl{\mathrm{E}}_{\mathrm{u}}\ovl{\mathrm{E}}_{\mathrm{u}}]_{8c}$}\\
\hline
 $0$ & $0$ & $\ovl{1}$ & $\ovl{1}$ & $0$ & $0$ & $0$ & $2$ & $0$ & $\ovl{1}$\\
\hline
\multicolumn{8}{l}{\footnotesize $\mathcal{BR}_L \boxminus \mathcal{BR}_{\widetilde{L}}: (3 \ovl{\Gamma}_{5} \boxminus \ovl{\Gamma}_{8} \; ; 3 \ovl{\rH}_{8} \boxminus \ovl{\rH}_{5} \; ; 2 \ovl{\rP}_{5} \; ; \ovl{\rN}_{3}\ovl{\rN}_{4} \oplus \ovl{\rN}_{5}\ovl{\rN}_{6})$} & \multicolumn{2}{r}{\footnotesize $(n_b,\mathrm{N})=(4,28)$} \\
\hline
\end{longtable}

\vspace{0.5cm}

\renewcommand{\arraystretch}{1.1}
\begin{longtable}{cccccccccc}
\caption{Space group $\mathrm{{SG}215}$ ($P\bar{4}3m$).}\\
\label{tab:{SG}215_DoubleTR}\\
\hline\hline
\multicolumn{10}{l}{{$\bm{\mathrm{SI}: (z_{2}) = (1)}$}}\\
\hline
\multicolumn{8}{l}{\footnotesize $\mathcal{BR}_L \boxminus \mathcal{BR}_{\widetilde{L}}: (2 \ovl{\Gamma}_{7} \boxminus \ovl{\Gamma}_{6} \; ; \ovl{\rR}_{6} \; ; \ovl{\rM}_{6} \; ; \ovl{\rX}_{6})$} & \multicolumn{2}{r}{\footnotesize $(n_b,\mathrm{N})=(2,14)$} \\
\hline
{\footnotesize $[\ovl{\mathrm{E}}_{\mathrm{1}}]_{1a}$} & {\footnotesize $[\ovl{\mathrm{E}}_{\mathrm{2}}]_{1a}$} & {\footnotesize $[\ovl{\mathrm{F}}]_{1a}$} & {\footnotesize $[\ovl{\mathrm{E}}_{\mathrm{1}}]_{1b}$} & {\footnotesize $[\ovl{\mathrm{E}}_{\mathrm{2}}]_{1b}$} & {\footnotesize $[\ovl{\mathrm{F}}]_{1b}$} & {\footnotesize $[\ovl{\mathrm{E}}_{\mathrm{1}}]_{3c}$} & {\footnotesize $[\ovl{\mathrm{E}}_{\mathrm{2}}]_{3c}$} & {\footnotesize $[\ovl{\mathrm{E}}_{\mathrm{1}}]_{3d}$} & {\footnotesize $[\ovl{\mathrm{E}}_{\mathrm{2}}]_{3d}$}\\
\hline
 $0$ & $1$ & $0$ & $0$ & $0$ & $0$ & $0$ & $1$ & $\ovl{1}$ & $0$\\
 $0$ & $1$ & $0$ & $0$ & $0$ & $0$ & $\ovl{1}$ & $0$ & $0$ & $1$\\
\hline
\multicolumn{8}{l}{\footnotesize $\mathcal{BR}_L \boxminus \mathcal{BR}_{\widetilde{L}}: (2 \ovl{\Gamma}_{6} \boxminus \ovl{\Gamma}_{7} \; ; \ovl{\rR}_{7} \; ; \ovl{\rM}_{6} \; ; \ovl{\rX}_{6})$} & \multicolumn{2}{r}{\footnotesize $(n_b,\mathrm{N})=(2,14)$} \\
\hline
\end{longtable}

\vspace{0.5cm}

\renewcommand{\arraystretch}{1.1}
\begin{longtable}{cccccccccccc}
\caption{Space group $\mathrm{{SG}216}$ ($F\bar{4}3m$).}\\
\label{tab:{SG}216_DoubleTR}\\
\hline\hline
\multicolumn{12}{l}{{$\bm{\mathrm{SI}: (z_{2}) = (1)}$}}\\
\hline
\multicolumn{10}{l}{\footnotesize $\mathcal{BR}_L \boxminus \mathcal{BR}_{\widetilde{L}}: (2 \ovl{\Gamma}_{6} \boxminus \ovl{\Gamma}_{7} \; ; \ovl{\rX}_{6} \; ; \ovl{\rL}_{6} \; ; \ovl{\rW}_{5} \oplus \ovl{\rW}_{7})$} & \multicolumn{2}{r}{\footnotesize $(n_b,\mathrm{N})=(2,6)$} \\
\hline
{\footnotesize $[\ovl{\mathrm{E}}_{\mathrm{1}}]_{4a}$} & {\footnotesize $[\ovl{\mathrm{E}}_{\mathrm{2}}]_{4a}$} & {\footnotesize $[\ovl{\mathrm{F}}]_{4a}$} & {\footnotesize $[\ovl{\mathrm{E}}_{\mathrm{1}}]_{4b}$} & {\footnotesize $[\ovl{\mathrm{E}}_{\mathrm{2}}]_{4b}$} & {\footnotesize $[\ovl{\mathrm{F}}]_{4b}$} & {\footnotesize $[\ovl{\mathrm{E}}_{\mathrm{1}}]_{4c}$} & {\footnotesize $[\ovl{\mathrm{E}}_{\mathrm{2}}]_{4c}$} & {\footnotesize $[\ovl{\mathrm{F}}]_{4c}$} & {\footnotesize $[\ovl{\mathrm{E}}_{\mathrm{1}}]_{4d}$} & {\footnotesize $[\ovl{\mathrm{E}}_{\mathrm{2}}]_{4d}$} & {\footnotesize $[\ovl{\mathrm{F}}]_{4d}$}\\
\hline
 $0$ & $0$ & $0$ & $1$ & $0$ & $0$ & $1$ & $0$ & $0$ & $0$ & $\ovl{1}$ & $0$\\
 $0$ & $0$ & $0$ & $1$ & $0$ & $0$ & $0$ & $\ovl{1}$ & $0$ & $1$ & $0$ & $0$\\
 $0$ & $0$ & $0$ & $0$ & $\ovl{1}$ & $0$ & $1$ & $0$ & $0$ & $1$ & $0$ & $0$\\
\hline
\end{longtable}

\vspace{0.5cm}

\renewcommand{\arraystretch}{1.1}
\begin{longtable}{ccccccc}
\caption{Space group $\mathrm{{SG}217}$ ($I\bar{4}3m$).}\\
\label{tab:{SG}217_DoubleTR}\\
\hline\hline
\multicolumn{7}{l}{{$\bm{\mathrm{SI}: (z_{2}) = (1)}$}}\\
\hline
\multicolumn{5}{l}{\footnotesize $\mathcal{BR}_L \boxminus \mathcal{BR}_{\widetilde{L}}: (2 \ovl{\Gamma}_{7} \boxminus \ovl{\Gamma}_{6} \; ; \ovl{\rH}_{6} \; ; \ovl{\rP}_{6} \; ; \ovl{\rPA}_{6} \; ; \ovl{\rN}_{5})$} & \multicolumn{2}{r}{\footnotesize $(n_b,\mathrm{N})=(2,22)$} \\
\hline
{\footnotesize $[\ovl{\mathrm{E}}_{\mathrm{1}}]_{2a}$} & {\footnotesize $[\ovl{\mathrm{E}}_{\mathrm{2}}]_{2a}$} & {\footnotesize $[\ovl{\mathrm{F}}]_{2a}$} & {\footnotesize $[\ovl{\mathrm{E}}_{\mathrm{1}}]_{6b}$} & {\footnotesize $[\ovl{\mathrm{E}}_{\mathrm{2}}]_{6b}$} & {\footnotesize $[{}^{1}\ovl{\mathrm{E}}_{\mathrm{1}}{}^{2}\ovl{\mathrm{E}}_{\mathrm{1}}]_{12d}$} & {\footnotesize $[{}^{1}\ovl{\mathrm{E}}_{\mathrm{2}}{}^{2}\ovl{\mathrm{E}}_{\mathrm{2}}]_{12d}$}\\
\hline
 $0$ & $0$ & $\ovl{1}$ & $\ovl{1}$ & $0$ & $0$ & $1$\\
\hline
\multicolumn{5}{l}{\footnotesize $\mathcal{BR}_L \boxminus \mathcal{BR}_{\widetilde{L}}: (2 \ovl{\Gamma}_{6} \boxminus \ovl{\Gamma}_{7} \; ; \ovl{\rH}_{7} \; ; \ovl{\rP}_{6} \; ; \ovl{\rPA}_{6} \; ; \ovl{\rN}_{5})$} & \multicolumn{2}{r}{\footnotesize $(n_b,\mathrm{N})=(2,22)$} \\
\hline
{\footnotesize $[\ovl{\mathrm{E}}_{\mathrm{1}}]_{2a}$} & {\footnotesize $[\ovl{\mathrm{E}}_{\mathrm{2}}]_{2a}$} & {\footnotesize $[\ovl{\mathrm{F}}]_{2a}$} & {\footnotesize $[\ovl{\mathrm{E}}_{\mathrm{1}}]_{6b}$} & {\footnotesize $[\ovl{\mathrm{E}}_{\mathrm{2}}]_{6b}$} & {\footnotesize $[{}^{1}\ovl{\mathrm{E}}_{\mathrm{1}}{}^{2}\ovl{\mathrm{E}}_{\mathrm{1}}]_{12d}$} & {\footnotesize $[{}^{1}\ovl{\mathrm{E}}_{\mathrm{2}}{}^{2}\ovl{\mathrm{E}}_{\mathrm{2}}]_{12d}$}\\
\hline
\end{longtable}

\vspace{0.5cm}

\renewcommand{\arraystretch}{1.1}
\begin{longtable}{ccccccc}
\caption{Space group $\mathrm{{SG}218}$ ($P\bar{4}3n$).}\\
\label{tab:{SG}218_DoubleTR}\\
\hline\hline
\multicolumn{7}{l}{{$\bm{\mathrm{SI}: (z_{2}) = (1)}$}}\\
\hline
\multicolumn{5}{l}{\footnotesize $\mathcal{BR}_L \boxminus \mathcal{BR}_{\widetilde{L}}: (3 \ovl{\Gamma}_{6} \boxminus \ovl{\Gamma}_{7} \; ; \ovl{\rR}_{6}\ovl{\rR}_{7} \; ; \ovl{\rM}_{6} \oplus \ovl{\rM}_{7} \; ; \ovl{\rX}_{6}\ovl{\rX}_{7})$} & \multicolumn{2}{r}{\footnotesize $(n_b,\mathrm{N})=(4,28)$} \\
\hline
{\footnotesize $[{}^{1}\ovl{\mathrm{F}}{}^{2}\ovl{\mathrm{F}}]_{2a}$} & {\footnotesize $[\ovl{\mathrm{E}}]_{2a}$} & {\footnotesize $[\ovl{\mathrm{E}}]_{6b}$} & {\footnotesize $[{}^{1}\ovl{\mathrm{E}}_{\mathrm{1}}{}^{2}\ovl{\mathrm{E}}_{\mathrm{1}}]_{6c}$} & {\footnotesize $[{}^{1}\ovl{\mathrm{E}}_{\mathrm{2}}{}^{2}\ovl{\mathrm{E}}_{\mathrm{2}}]_{6c}$} & {\footnotesize $[{}^{1}\ovl{\mathrm{E}}_{\mathrm{1}}{}^{2}\ovl{\mathrm{E}}_{\mathrm{1}}]_{6d}$} & {\footnotesize $[{}^{1}\ovl{\mathrm{E}}_{\mathrm{2}}{}^{2}\ovl{\mathrm{E}}_{\mathrm{2}}]_{6d}$}\\
\hline
 $0$ & $1$ & $0$ & $1$ & $0$ & $0$ & $\ovl{1}$\\
 $0$ & $1$ & $0$ & $0$ & $\ovl{1}$ & $1$ & $0$\\
\hline
\multicolumn{5}{l}{\footnotesize $\mathcal{BR}_L \boxminus \mathcal{BR}_{\widetilde{L}}: (3 \ovl{\Gamma}_{7} \boxminus \ovl{\Gamma}_{6} \; ; \ovl{\rR}_{6}\ovl{\rR}_{7} \; ; \ovl{\rM}_{6} \oplus \ovl{\rM}_{7} \; ; \ovl{\rX}_{6}\ovl{\rX}_{7})$} & \multicolumn{2}{r}{\footnotesize $(n_b,\mathrm{N})=(4,28)$} \\
\hline
\end{longtable}

\vspace{0.5cm}

\renewcommand{\arraystretch}{1.1}
\begin{longtable}{cccccccc}
\caption{Space group $\mathrm{{SG}219}$ ($F\bar{4}3c$).}\\
\label{tab:{SG}219_DoubleTR}\\
\hline\hline
\multicolumn{8}{l}{{$\bm{\mathrm{SI}: (z_{2}) = (1)}$}}\\
\hline
\multicolumn{6}{l}{\footnotesize $\mathcal{BR}_L \boxminus \mathcal{BR}_{\widetilde{L}}: (3 \ovl{\Gamma}_{6} \boxminus \ovl{\Gamma}_{7} \; ; \ovl{\rX}_{6} \oplus \ovl{\rX}_{7} \; ; \ovl{\rL}_{6}\ovl{\rL}_{6} \; ; \ovl{\rW}_{5}\ovl{\rW}_{6} \oplus \ovl{\rW}_{7}\ovl{\rW}_{8})$} & \multicolumn{2}{r}{\footnotesize $(n_b,\mathrm{N})=(4,28)$} \\
\hline
{\footnotesize $[{}^{1}\ovl{\mathrm{F}}{}^{2}\ovl{\mathrm{F}}]_{8a}$} & {\footnotesize $[\ovl{\mathrm{E}}]_{8a}$} & {\footnotesize $[{}^{1}\ovl{\mathrm{F}}{}^{2}\ovl{\mathrm{F}}]_{8b}$} & {\footnotesize $[\ovl{\mathrm{E}}]_{8b}$} & {\footnotesize $[{}^{1}\ovl{\mathrm{E}}_{\mathrm{1}}{}^{2}\ovl{\mathrm{E}}_{\mathrm{1}}]_{24c}$} & {\footnotesize $[{}^{1}\ovl{\mathrm{E}}_{\mathrm{2}}{}^{2}\ovl{\mathrm{E}}_{\mathrm{2}}]_{24c}$} & {\footnotesize $[{}^{1}\ovl{\mathrm{E}}_{\mathrm{1}}{}^{2}\ovl{\mathrm{E}}_{\mathrm{1}}]_{24d}$} & {\footnotesize $[{}^{1}\ovl{\mathrm{E}}_{\mathrm{2}}{}^{2}\ovl{\mathrm{E}}_{\mathrm{2}}]_{24d}$}\\
\hline
 $0$ & $1$ & $0$ & $0$ & $1$ & $0$ & $0$ & $\ovl{1}$\\
 $0$ & $0$ & $0$ & $1$ & $1$ & $0$ & $0$ & $\ovl{1}$\\
 $0$ & $1$ & $0$ & $0$ & $0$ & $\ovl{1}$ & $1$ & $0$\\
\hline
\end{longtable}

\vspace{0.5cm}

\renewcommand{\arraystretch}{1.1}
\begin{longtable}{cccccccccccccccccccc}
\caption{Space group $\mathrm{{SG}221}$ ($Pm\bar{3}m$).}\\
\label{tab:{SG}221_DoubleTR}\\
\hline\hline
\multicolumn{20}{l}{{$\bm{\mathrm{SI}: (z_{4m,\pi}, z_{8}) = (1, 0)}$}}\\
\hline
\multicolumn{18}{l}{\scriptsize $\mathcal{BR}_L \boxminus \mathcal{BR}_{\widetilde{L}}: (2 \ovl{\Gamma}_{8} \boxminus \ovl{\Gamma}_{6} \; ; 2 \ovl{\rR}_{6} \boxminus \ovl{\rR}_{8} \; ; \ovl{\rM}_{7} \; ; \ovl{\rX}_{8})$} & \multicolumn{2}{r}{\scriptsize $(n_b,\mathrm{N})=(2,26)$} \\
\hline
{\tiny $[\ovl{\mathrm{E}}_{\mathrm{1g}}]_{1a}$} & {\tiny $[\ovl{\mathrm{E}}_{\mathrm{1u}}]_{1a}$} & {\tiny $[\ovl{\mathrm{E}}_{\mathrm{2g}}]_{1a}$} & {\tiny $[\ovl{\mathrm{E}}_{\mathrm{2u}}]_{1a}$} & {\tiny $[\ovl{\mathrm{F}}_{\mathrm{g}}]_{1a}$} & {\tiny $[\ovl{\mathrm{F}}_{\mathrm{u}}]_{1a}$} & {\tiny $[\ovl{\mathrm{E}}_{\mathrm{1g}}]_{1b}$} & {\tiny $[\ovl{\mathrm{E}}_{\mathrm{1u}}]_{1b}$} & {\tiny $[\ovl{\mathrm{E}}_{\mathrm{2g}}]_{1b}$} & {\tiny $[\ovl{\mathrm{E}}_{\mathrm{2u}}]_{1b}$} & {\tiny $[\ovl{\mathrm{F}}_{\mathrm{g}}]_{1b}$} & {\tiny $[\ovl{\mathrm{F}}_{\mathrm{u}}]_{1b}$} & {\tiny $[\ovl{\mathrm{E}}_{\mathrm{1g}}]_{3c}$} & {\tiny $[\ovl{\mathrm{E}}_{\mathrm{1u}}]_{3c}$} & {\tiny $[\ovl{\mathrm{E}}_{\mathrm{2g}}]_{3c}$} & {\tiny $[\ovl{\mathrm{E}}_{\mathrm{2u}}]_{3c}$} & {\tiny $[\ovl{\mathrm{E}}_{\mathrm{1g}}]_{3d}$} & {\tiny $[\ovl{\mathrm{E}}_{\mathrm{1u}}]_{3d}$} & {\tiny $[\ovl{\mathrm{E}}_{\mathrm{2g}}]_{3d}$} & {\tiny $[\ovl{\mathrm{E}}_{\mathrm{2u}}]_{3d}$}\\
\hline
 $0$ & $0$ & $0$ & $0$ & $0$ & $0$ & $0$ & $1$ & $\ovl{1}$ & $0$ & $0$ & $\ovl{1}$ & $\ovl{1}$ & $0$ & $1$ & $0$ & $0$ & $1$ & $0$ & $0$\\
 $0$ & $0$ & $0$ & $0$ & $0$ & $0$ & $0$ & $1$ & $0$ & $1$ & $1$ & $0$ & $\ovl{1}$ & $0$ & $0$ & $\ovl{1}$ & $0$ & $1$ & $0$ & $0$\\
 $0$ & $0$ & $0$ & $0$ & $0$ & $0$ & $\ovl{1}$ & $0$ & $0$ & $1$ & $0$ & $\ovl{1}$ & $0$ & $1$ & $0$ & $\ovl{1}$ & $0$ & $1$ & $0$ & $0$\\
\hline
\end{longtable}

\vspace{0.5cm}

\renewcommand{\arraystretch}{1.1}
\begin{longtable}{ccccccccccc}
\caption{Space group $\mathrm{{SG}222}$ ($Pn\bar{3}n$).}\\
\label{tab:{SG}222_DoubleTR}\\
\hline\hline
\multicolumn{11}{l}{{$\bm{\mathrm{SI}: (z_{4}) = (1)}$}}\\
\hline
\multicolumn{9}{l}{\footnotesize $\mathcal{BR}_L \boxminus \mathcal{BR}_{\widetilde{L}}: (5 \ovl{\Gamma}_{8} \boxminus 3 \ovl{\Gamma}_{6} \; ; \ovl{\rR}_{5} \; ; \ovl{\rM}_{5} \; ; \ovl{\rX}_{5}\ovl{\rX}_{7})$} & \multicolumn{2}{r}{\footnotesize $(n_b,\mathrm{N})=(4,100)$} \\
\hline
{\scriptsize $[\ovl{\mathrm{E}}_{\mathrm{1}}]_{2a}$} & {\scriptsize $[\ovl{\mathrm{E}}_{\mathrm{2}}]_{2a}$} & {\scriptsize $[\ovl{\mathrm{F}}]_{2a}$} & {\scriptsize $[\ovl{\mathrm{E}}_{\mathrm{1}}]_{6b}$} & {\scriptsize $[\ovl{\mathrm{E}}_{\mathrm{2}}]_{6b}$} & {\scriptsize $[{}^{1}\ovl{\mathrm{E}}_{\mathrm{g}}{}^{2}\ovl{\mathrm{E}}_{\mathrm{g}}]_{8c}$} & {\scriptsize $[{}^{1}\ovl{\mathrm{E}}_{\mathrm{u}}{}^{2}\ovl{\mathrm{E}}_{\mathrm{u}}]_{8c}$} & {\scriptsize $[\ovl{\mathrm{E}}_{\mathrm{g}}\ovl{\mathrm{E}}_{\mathrm{g}}]_{8c}$} & {\scriptsize $[\ovl{\mathrm{E}}_{\mathrm{u}}\ovl{\mathrm{E}}_{\mathrm{u}}]_{8c}$} & {\scriptsize $[{}^{1}\ovl{\mathrm{E}}_{\mathrm{1}}{}^{2}\ovl{\mathrm{E}}_{\mathrm{1}}]_{12d}$} & {\scriptsize $[{}^{1}\ovl{\mathrm{E}}_{\mathrm{2}}{}^{2}\ovl{\mathrm{E}}_{\mathrm{2}}]_{12d}$}\\
\hline
 $1$ & $0$ & $0$ & $0$ & $0$ & $0$ & $2$ & $1$ & $0$ & $\ovl{2}$ & $0$\\
 $1$ & $0$ & $0$ & $0$ & $0$ & $\ovl{2}$ & $0$ & $0$ & $\ovl{1}$ & $0$ & $2$\\
\hline
\multicolumn{9}{l}{\footnotesize $\mathcal{BR}_L \boxminus \mathcal{BR}_{\widetilde{L}}: (5 \ovl{\Gamma}_{9} \boxminus 3 \ovl{\Gamma}_{7} \; ; \ovl{\rR}_{5} \; ; \ovl{\rM}_{5} \; ; \ovl{\rX}_{6}\ovl{\rX}_{8})$} & \multicolumn{2}{r}{\footnotesize $(n_b,\mathrm{N})=(4,100)$} \\
\hline
\end{longtable}

\vspace{0.5cm}

\renewcommand{\arraystretch}{1.1}
\begin{longtable}{ccccccccccccccccc}
\caption{Space group $\mathrm{{SG}225}$ ($Fm\bar{3}m$).}\\
\label{tab:{SG}225_DoubleTR}\\
\hline\hline
\multicolumn{17}{l}{{$\bm{\mathrm{SI}: (z_{8}) = (1)}$}}\\
\hline
\multicolumn{15}{l}{\footnotesize $\mathcal{BR}_L \boxminus \mathcal{BR}_{\widetilde{L}}: (4 \ovl{\Gamma}_{6} \boxminus \ovl{\Gamma}_{8} \; ; 3 \ovl{\rX}_{8} \; ; \ovl{\rL}_{8} \oplus 2 \ovl{\rL}_{9} \; ; 3 \ovl{\rW}_{6})$} & \multicolumn{2}{r}{\footnotesize $(n_b,\mathrm{N})=(6,34)$} \\
\hline
{\scriptsize $[\ovl{\mathrm{E}}_{\mathrm{1g}}]_{4a}$} & {\scriptsize $[\ovl{\mathrm{E}}_{\mathrm{1u}}]_{4a}$} & {\scriptsize $[\ovl{\mathrm{E}}_{\mathrm{2g}}]_{4a}$} & {\scriptsize $[\ovl{\mathrm{E}}_{\mathrm{2u}}]_{4a}$} & {\scriptsize $[\ovl{\mathrm{F}}_{\mathrm{g}}]_{4a}$} & {\scriptsize $[\ovl{\mathrm{F}}_{\mathrm{u}}]_{4a}$} & {\scriptsize $[\ovl{\mathrm{E}}_{\mathrm{1g}}]_{4b}$} & {\scriptsize $[\ovl{\mathrm{E}}_{\mathrm{1u}}]_{4b}$} & {\scriptsize $[\ovl{\mathrm{E}}_{\mathrm{2g}}]_{4b}$} & {\scriptsize $[\ovl{\mathrm{E}}_{\mathrm{2u}}]_{4b}$} & {\scriptsize $[\ovl{\mathrm{F}}_{\mathrm{g}}]_{4b}$} & {\scriptsize $[\ovl{\mathrm{F}}_{\mathrm{u}}]_{4b}$} & {\scriptsize $[\ovl{\mathrm{E}}_{\mathrm{1}}]_{8c}$} & {\scriptsize $[\ovl{\mathrm{E}}_{\mathrm{2}}]_{8c}$} & {\scriptsize $[\ovl{\mathrm{F}}]_{8c}$} & {\scriptsize $[\ovl{\mathrm{E}}_{\mathrm{g}}]_{24d}$} & {\scriptsize $[\ovl{\mathrm{E}}_{\mathrm{u}}]_{24d}$}\\
\hline
 $0$ & $0$ & $0$ & $\ovl{1}$ & $0$ & $1$ & $2$ & $0$ & $0$ & $0$ & $0$ & $1$ & $2$ & $0$ & $0$ & $0$ & $\ovl{1}$\\
 $0$ & $0$ & $0$ & $\ovl{1}$ & $\ovl{1}$ & $0$ & $2$ & $0$ & $0$ & $0$ & $\ovl{1}$ & $0$ & $1$ & $\ovl{1}$ & $0$ & $1$ & $0$\\
\hline
\multicolumn{15}{l}{\footnotesize $\mathcal{BR}_L \boxminus \mathcal{BR}_{\widetilde{L}}: (4 \ovl{\Gamma}_{6} \boxminus \ovl{\Gamma}_{8} \; ; 3 \ovl{\rX}_{8} \; ; 2 \ovl{\rL}_{8} \oplus \ovl{\rL}_{9} \; ; 2 \ovl{\rW}_{6} \oplus \ovl{\rW}_{7})$} & \multicolumn{2}{r}{\footnotesize $(n_b,\mathrm{N})=(6,34)$} \\
\hline
\end{longtable}

\vspace{0.5cm}

\renewcommand{\arraystretch}{1.1}
\begin{longtable}{ccccccccccccc}
\caption{Space group $\mathrm{{SG}226}$ ($Fm\bar{3}c$).}\\
\label{tab:{SG}226_DoubleTR}\\
\hline\hline
\multicolumn{13}{l}{{$\bm{\mathrm{SI}: (z_{8}) = (1)}$}}\\
\hline
\multicolumn{11}{l}{\footnotesize $\mathcal{BR}_L \boxminus \mathcal{BR}_{\widetilde{L}}: (5 \ovl{\Gamma}_{6} \boxminus \ovl{\Gamma}_{8} \; ; \ovl{\rX}_{7} \oplus 3 \ovl{\rX}_{9} \; ; 2 \ovl{\rL}_{5}\ovl{\rL}_{6} \; ; 2 \ovl{\rW}_{6}\ovl{\rW}_{7})$} & \multicolumn{2}{r}{\footnotesize $(n_b,\mathrm{N})=(8,56)$} \\
\hline
{\scriptsize $[\ovl{\mathrm{E}}_{\mathrm{1}}]_{8a}$} & {\scriptsize $[\ovl{\mathrm{E}}_{\mathrm{2}}]_{8a}$} & {\scriptsize $[\ovl{\mathrm{F}}]_{8a}$} & {\scriptsize $[{}^{1}\ovl{\mathrm{F}}_{\mathrm{g}}{}^{2}\ovl{\mathrm{F}}_{\mathrm{g}}]_{8b}$} & {\scriptsize $[{}^{1}\ovl{\mathrm{F}}_{\mathrm{u}}{}^{2}\ovl{\mathrm{F}}_{\mathrm{u}}]_{8b}$} & {\scriptsize $[\ovl{\mathrm{E}}_{\mathrm{g}}]_{8b}$} & {\scriptsize $[\ovl{\mathrm{E}}_{\mathrm{u}}]_{8b}$} & {\scriptsize $[\ovl{\mathrm{E}}_{\mathrm{1}}]_{24c}$} & {\scriptsize $[\ovl{\mathrm{E}}_{\mathrm{2}}]_{24c}$} & {\scriptsize $[{}^{1}\ovl{\mathrm{E}}_{\mathrm{1g}}{}^{2}\ovl{\mathrm{E}}_{\mathrm{1g}}]_{24d}$} & {\scriptsize $[{}^{1}\ovl{\mathrm{E}}_{\mathrm{1u}}{}^{2}\ovl{\mathrm{E}}_{\mathrm{1u}}]_{24d}$} & {\scriptsize $[{}^{1}\ovl{\mathrm{E}}_{\mathrm{2g}}{}^{2}\ovl{\mathrm{E}}_{\mathrm{2g}}]_{24d}$} & {\scriptsize $[{}^{1}\ovl{\mathrm{E}}_{\mathrm{2u}}{}^{2}\ovl{\mathrm{E}}_{\mathrm{2u}}]_{24d}$}\\
\hline
 $1$ & $0$ & $0$ & $\ovl{1}$ & $0$ & $1$ & $\ovl{1}$ & $1$ & $\ovl{1}$ & $1$ & $0$ & $0$ & $0$\\
 $0$ & $\ovl{1}$ & $0$ & $\ovl{1}$ & $0$ & $2$ & $0$ & $1$ & $\ovl{1}$ & $1$ & $0$ & $0$ & $0$\\
 $1$ & $0$ & $0$ & $0$ & $1$ & $2$ & $0$ & $0$ & $\ovl{2}$ & $1$ & $0$ & $0$ & $0$\\
\hline
\end{longtable}

\vspace{0.5cm}

\renewcommand{\arraystretch}{1.2}
\begin{longtable}{cccccccccccccccc}
\caption{Space group $\mathrm{{SG}229}$ ($Im\bar{3}m$).}\\
\label{tab:{SG}229_DoubleTR}\\
\hline\hline
\multicolumn{16}{l}{{$\bm{\mathrm{SI}: (z_{2w}, z_{8}) = (1, 0)}$}}\\
\hline
\multicolumn{14}{l}{\footnotesize $\mathcal{BR}_L \boxminus \mathcal{BR}_{\widetilde{L}}: (3 \ovl{\Gamma}_{8} \boxminus \ovl{\Gamma}_{6} \; ; 3 \ovl{\rH}_{6} \boxminus \ovl{\rH}_{8} \; ; \ovl{\rP}_{6} \oplus \ovl{\rP}_{7} \; ; \ovl{\rN}_{5} \oplus \ovl{\rN}_{6})$} & \multicolumn{2}{r}{\footnotesize $(n_b,\mathrm{N})=(4,52)$} \\
\hline
{\scriptsize $[\ovl{\mathrm{E}}_{\mathrm{1g}}]_{2a}$} & {\scriptsize $[\ovl{\mathrm{E}}_{\mathrm{1u}}]_{2a}$} & {\scriptsize $[\ovl{\mathrm{E}}_{\mathrm{2g}}]_{2a}$} & {\scriptsize $[\ovl{\mathrm{E}}_{\mathrm{2u}}]_{2a}$} & {\scriptsize $[\ovl{\mathrm{F}}_{\mathrm{g}}]_{2a}$} & {\scriptsize $[\ovl{\mathrm{F}}_{\mathrm{u}}]_{2a}$} & {\scriptsize $[\ovl{\mathrm{E}}_{\mathrm{1g}}]_{6b}$} & {\scriptsize $[\ovl{\mathrm{E}}_{\mathrm{1u}}]_{6b}$} & {\scriptsize $[\ovl{\mathrm{E}}_{\mathrm{2g}}]_{6b}$} & {\scriptsize $[\ovl{\mathrm{E}}_{\mathrm{2u}}]_{6b}$} & {\scriptsize $[{}^{1}\ovl{\mathrm{E}}_{\mathrm{g}}{}^{2}\ovl{\mathrm{E}}_{\mathrm{g}}]_{8c}$} & {\scriptsize $[{}^{1}\ovl{\mathrm{E}}_{\mathrm{u}}{}^{2}\ovl{\mathrm{E}}_{\mathrm{u}}]_{8c}$} & {\scriptsize $[\ovl{\mathrm{E}}_{\mathrm{1g}}]_{8c}$} & {\scriptsize $[\ovl{\mathrm{E}}_{\mathrm{1u}}]_{8c}$} & {\scriptsize $[\ovl{\mathrm{E}}_{\mathrm{1}}]_{12d}$} & {\scriptsize $[\ovl{\mathrm{E}}_{\mathrm{2}}]_{12d}$}\\
\hline
 $1$ & $1$ & $0$ & $0$ & $0$ & $0$ & $0$ & $0$ & $0$ & $0$ & $1$ & $0$ & $0$ & $2$ & $\ovl{2}$ & $0$\\
 $0$ & $0$ & $\ovl{1}$ & $\ovl{1}$ & $0$ & $0$ & $0$ & $0$ & $0$ & $0$ & $0$ & $\ovl{1}$ & $0$ & $2$ & $\ovl{1}$ & $1$\\
 $1$ & $1$ & $0$ & $0$ & $0$ & $0$ & $0$ & $0$ & $0$ & $0$ & $0$ & $\ovl{1}$ & $\ovl{2}$ & $0$ & $0$ & $2$\\
\hline
\end{longtable}

\vspace{0.5cm}
\clearpage

\subsection{3D Double group without TRS}
\label{sec:3D-double-noT}
\input{3D_double_NoTR}


\end{document}